\documentclass[aps,prd,nofootinbib,tightenlines,notitlepage, floatfix,superscriptaddress,twocolumn,showkeys]{revtex4-1}
\usepackage[utf8]{inputenc}
\usepackage{slashed}
\usepackage{float}
\usepackage{amsmath,bm}
\allowdisplaybreaks 
\def\bea#1\eea{\begin{align}#1\end{align}} 

\newcommand{\bef}{\begin{figure}\centering}
\newcommand{\eef}{\end{figure}}
\usepackage{graphicx}
\usepackage{natbib}
\usepackage{float}
\floatstyle{plaintop}
\restylefloat{table}
\usepackage{color}
\usepackage[dvipsnames]{xcolor}
\usepackage[normalem]{ulem}

\newcommand{\beq}{\begin{equation}}
\newcommand{\eeq}{\end{equation}}
\def\bea#1\eea{\begin{align}#1\end{align}}
\newcommand{\be}{\begin{eqnarray}}
\newcommand{\ee}{\end{eqnarray}}

\newcommand{\txt}[1]{\textrm{#1}}
\usepackage{hyperref}
\newcommand{\snn}{\sqrt{s_{\rm{NN}}}}
\usepackage{comment}

\newcommand{\infrac}[2]{{#1}/{#2}}

\newcommand{\inparr}[2]{\partial{#1}/\partial{#2}}

\graphicspath{{./Figures/}}

\begin{document}
\title{Structure in the speed of sound: From neutron stars to heavy-ion collisions }

\author{Nanxi Yao}
\affiliation{Illinois Center for Advanced Studies of the Universe, Department of Physics, University of Illinois at Urbana-Champaign, Urbana, Illinois 61801, USA}

\author{Agnieszka Sorensen}
\affiliation{Institute for Nuclear Theory, University of Washington, Box 351550, Seattle, Washington 98195, USA}

\author{Veronica Dexheimer}
\affiliation{Department of Physics, Kent State University, Kent, Ohio 44243, USA}

\author{Jacquelyn Noronha-Hostler}
\affiliation{Illinois Center for Advanced Studies of the Universe, Department of Physics, University of Illinois at Urbana-Champaign, Urbana, Illinois 61801, USA}

\date{\today}

\begin{abstract}
From the observation of both heavy neutron stars and light ones with small radii, one anticipates a steep rise in the speed of sound of nuclear matter  as a function of baryon density up to values close to the causal limit. A question follows whether such behavior of the speed of sound in neutron-rich matter is compatible with the equation of state extracted from low-energy heavy-ion collisions. In this work, we consider a family of neutron-star equations of state characterized by a steep rise in the speed of sound, and use the symmetry energy expansion to obtain equations of state applicable to the almost-symmetric nuclear matter created in heavy-ion collisions. We then compare collective flow data from low-energy heavy-ion experiments with results of simulations obtained using the hadronic transport code \texttt{SMASH} with the mean-field potential reproducing the density-dependence of the speed of sound. We show that equations of state featuring a peak in the speed of sound squared occurring at densities between $2$--$3$ times the saturation density of normal nuclear matter, producing neutron stars of nearly $M_{\txt{max}}\approx 2.5 M_{\odot}$, are consistent with heavy-ion collision data.
\end{abstract}

\maketitle

\section{Introduction}

The theoretical description of the strong interactions, quantum chromodynamics (QCD), presents a rich phase structure when one varies temperature $T$ and baryon density $n_B$ \cite{Baym:2017whm,Ratti:2018ksb,Bzdak:2019pkr,Dexheimer:2020zzs,An:2021wof}. Lattice QCD calculations show that at large $T$ and vanishing $n_B$ there is a crossover transition from deconfined, color-carrying quarks and gluons into colorless hadrons~\cite{Aoki:2006we}. Using relativistic heavy-ion collisions ~(HIC) one can explore this phase transition experimentally, and it has been confirmed that the equation of state~(EOS) extracted from a Bayesian analysis at high center-of-mass collision energies, $\sqrt{s_{\txt{NN}}}=200 ~ \txt{GeV}$ and $\sqrt{s_{\txt{NN}}}=2.76~ \txt{TeV}$, reproduces the lattice QCD results~\cite{Pratt:2015zsa}. At lower collision energies, $\sqrt{s_{\txt{NN}}} < 200~ \txt{GeV}$, baryon stopping becomes more relevant, leading to finite values of $n_B$ (and consequently finite values of the baryon chemical potential $\mu_B$) characterizing the matter produced in the collisions. First-principle lattice QCD calculations cannot directly calculate the EOS in this regime; however, there are various expansion schemes, using, e.g., the Taylor series, that can extend the lattice QCD results into the finite-$\mu_B$ region~\cite{Parotto:2018pwx,Monnai:2019hkn,Noronha-Hostler:2019ayj,Karthein:2021nxe}. Thus, up to approximately $\mu_B/T\approx 3$, the EOS can be calculated directly from QCD. 
At moderate $T$ and low $n_B$, the hadron resonance gas (HRG) model can be used to reasonably describe the QCD phase diagram~\cite{Alba:2017mqu,Vovchenko:2017xad,Bellwied:2019pxh,Borsanyi:2021sxv,Borsanyi:2022qlh}. When additional interaction terms are incorporated in the HRG, it can also reasonably describe larger $n_B$~\cite{Vovchenko:2016rkn}. 
However, as one approaches larger and larger $n_B$, theoretical models become significantly less constrained and more uncertainties arise. Here, the exception is the region of the phase diagram relevant to the ordinary nuclear matter, both around the saturation density and around the nuclear liquid-gas phase transition, where nuclear matter properties are relatively well known (see Fig.~1 of Ref.~\cite{MUSES:2023hyz}).

In this regime, one approach to extracting the EOS is based on comparisons of measurements from low-energy heavy-ion collisions to simulations using hadronic transport models (see, e.g.,~\cite{Danielewicz:2002pu}); here, the approximate collision energy range considered is $0.15~\txt{GeV} \leq E_{\txt{kin}} \leq 10~\txt{GeV/nucleon}$, or equivalently $1.95~\rm{GeV} \leq \snn \leq 4.7~\rm{GeV}$, where $E_{\txt{kin}}$ is the incident kinetic energy per nucleon (excluding the rest mass) in the fixed-target frame and $\snn$ is the center-of-mass collision energy\footnote{Explicitly, $E_{\rm{kin}}$ and $\sqrt{s_{\rm{NN}}}$ are related to each other through $E_{\rm{kin}} = E_{\rm{lab}} - m_N = \infrac{(\sqrt{s_{\rm{NN}}})^2}{(2m_N)} - 2m_N$, where $E_{\rm{lab}}$ is the (total, i.e., including mass) energy of an incident nucleon.}.
This approach, of course, implicitly assumes that hadrons are the correct degrees of freedom, which may be questioned given the high values of $T$ recently measured by the HADES experiment at $\sqrt{s_{\txt{NN}}} = 2.4 ~ \txt{GeV}$~\cite{HADES:2019auv}. 
Additionally, the heavy-ion EOS extracted in~\cite{Danielewicz:2002pu} (in the $T=0$ limit) appears to be in tension with the significantly stiffer EOS that may be required to support heavy neutron stars (NS)~\cite{Fattoyev:2020cws}. 
In a different approach, ideal hydrodynamic simulations using an EOS derived from a model with a transition between hadronic and quark degrees of freedom also appears to reasonably describe low-energy ($1 ~\txt{AGeV} \leq E_{\txt{kin}} \leq 10 ~ \txt{AGeV}$, or equivalently $2.3~\rm{GeV}\leq \snn \leq 4.7~\rm{GeV}$) flow measurements~\cite{Spieles:2020zaa}.

Recent X-ray observations of isolated neutron stars and posteriors of the tidal deformability obtained from neutron-star inspirals have led to significantly tighter constraints on the EOS at $T=0$ (see~\cite{MUSES:2023hyz} for an in-depth discussion).  
Specifically, the requirement that neutron stars reach a maximum mass of at least $M_{\txt{max}}\geq 2 M_{\odot} $~\cite{science.1233232,Fonseca:2021wxt,Romani:2021xmb} leads to a speed of sound squared $c_s^2$ that surpasses the conformal limit of $1/3$~\cite{Bedaque:2014sqa,Tews:2018kmu,Fujimoto:2019hxv,Altiparmak:2022bke,Han:2022rug}. 
This picture may be further strengthened by the GW190814 binary merger observation featuring a mysterious secondary compact object of mass $M\approx 2.6 M_{\odot}$ ~\cite{LIGOScientific:2020zkf}, which falls into the mass gap between the heaviest known neutron stars and the lightest known black holes~\cite{Bailyn:1997xt,Ozel:2010su,deSa:2022qny,Ye:2022qoe,Farah:2021qom}. 
Unfortunately, it was not possible to measure the tidal deformability~$\Lambda$ for GW190814, which would have been a clear signature of a neutron star ($\Lambda>0$) or a black hole ($\Lambda=0$). In fact, for extremely massive neutron stars, the tidal deformability is anticipated to be likewise extremely small, \mbox{$\Lambda \approx 2$--$20$}~\cite{Tan:2020ics,Tan:2021ahl}, which will only be possible to determine with next generation gravitational wave detectors~\cite{Carson:2019rjx}. 

There are additional constraints relevant for the $T=0$ EOS of neutron stars. 
Using measurements of X-ray pulses and comparing them to models of X-ray emission, NASA's NICER has extracted the neutron-star radius posteriors from two separate pulsars, J0030 and J0740, yielding the values of the equatorial radius $R_{\txt{eq}} = 11.52$--$14.26~ \txt{km}$ and $R_{\txt{eq}} = 11.41$--$16.30~ \txt{km}$, respectively~\cite{Riley:2019yda,Miller:2019cac, Riley:2021pdl,Miller:2021qha}, with more observations on the way. 
The LIGO-Virgo Collaboration (LVC) has measured gravitational waves from a variety of compact objects ~\cite{LIGOScientific:2021psn}, however, the only one with a tidal deformability posterior is GW170817~\cite{TheLIGOScientific:2017qsa} which limits the stiffness of the low density EOS~\cite{LIGOScientific:2018cki}.  
Together, the constraints from NICER and LIGO-Virgo do not allow for an EOS that is consistently too stiff (it would not pass the low-mass--radius or tidal deformability constraints) or too soft (it would not reproduce $2 M_\odot$ stars); instead, EOSs which are soft at low densities and stiff at high densities appear to be preferred. 
A good example of such an EOS is a quarkyonic matter EOS~\cite{McLerran:2018hbz,Zhao:2020dvu,Sen:2020qcd,Duarte:2020xsp,Sen:2020peq} which is soft at low densities and features a large peak in $c_s^2$, nearly approaching the causal limit, at intermediate densities. 
If neutron stars with masses up to $M\approx 2.6 M_{\odot}$ are observed, then one can anticipate that the speed of sound approaches the speed of light~\cite{Tan:2020ics,Tan:2021ahl}, i.e., $c_s^2=dp/d\varepsilon\rightarrow 1$, where $p$ is the pressure and $\varepsilon$ is the energy density. 
Moreover, it may be possible to measure such a steep rise in $c_s^2$ directly from the binary Love relations~\cite{Tan:2021nat}. 
Interestingly, there are also nuclear physics motivations for rises and falls in $c_s^2$, including certain hyperon interactions, quarkyonic matter, quark-hadron crossovers, and approach to asymptotic freedom at very large densities~\cite{Dexheimer:2014pea,Dutra:2015hxa,McLerran:2018hbz,Jakobus:2020nxw,Alford:2017qgh,Zacchi:2015oma,Alvarez-Castillo:2018pve,Li:2019fqe,Wang:2019npj,Fadafa:2019euu,Xia:2019xax,Yazdizadeh:2019ivy,Shahrbaf:2019vtf,Zacchi:2019ayh,Zhao:2020dvu,Lopes:2020rqn,Blaschke:2020qrs,Duarte:2020xsp,Rho:2020eqo,Marczenko:2020wlc,Minamikawa:2020jfj,Hippert:2021gfs,Pisarski:2021aoz,Sen:2020qcd,Stone:2021ngh,Ferreira:2020kvu,Kapusta:2021ney,Kojo:2021ugu,Somasundaram:2021ljr,Zuo:2022rks,Ivanytskyi:2022oxv,Fraga:2022yls,Kovacs:2021ger,Rho:2022wco,Most:2022wgo,Shao:2023pqu,Kumar:2023qyu,Issifu:2023ovi,Yamamoto:2023osc,Kouno:2023ygw}.

Although neutron-star mergers and heavy-ion collisions have been simulated with the same EOS~\cite{Most:2022wgo}, a large peak in $c_s^2$ has not been explored much in heavy-ion collision studies, where most of the focus so far~\cite{Ratti:2018ksb,Bzdak:2019pkr,Dexheimer:2020zzs} has been on EOSs with the QCD critical point followed by a first-order phase transition, producing a valley in $c_s^2$ (or, if one considers the spinodal region, even negative $c_s^2$). 
Thus, the vast majority of efforts have considered a small $c_s^2$ that approaches zero, not one that approaches the causal limit, $c_s^2\approx 1$. 
The one recent exception is Ref.~\cite{Oliinychenko:2022uvy}, where a large peak in $c_s^2$ was considered within the hadronic transport code \texttt{SMASH}~\cite{Weil:2016zrk}  and, in fact, a preference was seen for the location of the peak to be around $2$--$3 n_{\rm{sat}}$, where $n_{\rm{sat}}$ is the nuclear saturation density. 
In that work, the connection to the neutron-star EOS was not thoroughly explored, which we address here.

In order to better understand the interplay between heavy-ion collision and neutron-star studies, it is important to understand how direct comparisons can be made between these fields (see~\cite{Lovato:2022vgq,Sorensen:2023zkk,Aryal:2020ocm} for further discussion). 
Heavy-ion collisions involve matter probing different regions of the QCD phase diagram than neutron stars. 
While neutron stars are electrically neutral to remain stable, $n_Q=0$, heavy-ion collisions have a finite net charge density $n_Q$, which in terms of the fraction of protons to nucleons in the initial state can be expressed as $Y_{Q} = Z/N_B =n_Q/n_B$; here, the charge fraction $Y_Q$ is constant throughout the collision. 
More generally, $Y_{Q,\rm{QCD}}^{\rm{const}}=n_{Q,\rm{QCD}}/n_B$, where $n_{Q,\rm{QCD}}$ is the net charge ratio \emph{of hadrons}. 
Note that while $Y_{Q,\rm{QCD}}=Y_{Q}$ and 
\begin{equation}
n_Q = n_{Q,\rm{QCD}}\ ,
\end{equation}
holds in heavy-ion collisions, it is not true for neutron stars due to a nonzero net lepton number,
\begin{equation}
\label{eqn:betaEQ}
n_Q = n_{Q,\rm{QCD}}+n_{Q,\rm{lep}}\ .
\end{equation}
Because heavy-ion collisions are performed with ordinary heavy nuclei, the initial conditions involve only protons and neutrons and one can write $Y_{Q,\rm{QCD}}=n_p/(n_p+n_n)$, where $n_p$ and $n_n$ are proton and neutron density, respectively. 
Similarly, the same approximate formula holds for neutron stars as long as one does not include hyperons and/or deconfined quarks within the star.

Lighter nuclei tend to be closer to symmetric nuclear matter (SNM) for which $Y_{Q,\rm{QCD}}^{\txt{const}}\approx 0.5$, while heavier nuclei must be more neutron-rich for stability such that, e.g., uranium ($A=238$) is characterized by $Y_{Q,\rm{QCD}}^{\txt{const}}\approx 0.38$. 
On the other hand, most theoretical calculations find $Y_{Q,\rm{QCD}}^{\txt{\txt{NS}}}\lesssim 0.1$ for neutron-star cores~\cite{Gao:2017mnl}, and the neutron-star matter is referred to as asymmetric nuclear matter (ANM). 
We stress here that while the number of positively charged particles (e.g., protons) in neutron stars is not zero, it is exactly balanced by negatively charged particles; for example, assuming neutron, proton, and electron degrees of freedom (also known as {\textit{npe}} matter), one has $n_Q = n_p + n_e = 0$, where $n_e<0$ is the electron charge density. 
Further differences between symmetric and asymmetric matter arise from strange degrees of freedom, but we leave the study of strangeness-related effects for a future work. 

The energy per baryon for ANM can be written as~\cite{Bombaci:1991zz}
\begin{equation}\label{eqn:symExpan}
\frac{E_{\rm{ANM}}}{N_B} = \frac{E_{\rm{SNM}}}{N_B}+ E_{\txt{\txt{sym}}} \delta ^2+\mathcal{O}(\delta^4) \ ,
\end{equation}
where $E$ is the energy, $N_B$ is the baryon number, $E_{\txt{\txt{sym}}}$ is known as the symmetry energy, and
\begin{equation}\label{eqn:delta}
\delta \equiv (n_n - n_p)/(n_n + n_p) = 1 - 2Y_{Q,\rm{QCD}}
\end{equation}
is known as the asymmetry parameter
such that for SNM $\delta=0$ and for pure neutron matter $\delta=1$.  
Note that the second equality in Eq.~(\ref{eqn:delta}) only holds for nonstrange, e.g., \textit{npe} matter (see Appendix~\ref{app:delta_definition}). 
It is common to use a Taylor expansion of the symmetry energy around SNM, characterized by $\delta = 0$ or equivalently $Y_{Q,\rm{QCD}}^{\txt{const}}=0.5$, to obtain a description of the ANM present in neutron stars, characterized by $Y_{Q,\rm{QCD}}^{\txt{NS}}\approx 0.1$. Moreover, nuclear experiments only provide information about the symmetry energy around $n_{\rm{sat}}$~\cite{Sorensen:2023zkk,MUSES:2023hyz}. 
Thus, the symmetry energy $E_{\rm{sym}}$ at a given baryon number density $n_B \neq n_{\rm{sat}}$ is obtained based on another expansion around its value at $n_{\rm{sat}}$~\cite{Baldo:2016jhp} using its first few derivatives with respect to $n_B$.
Many of such approaches have studied EOSs that do not include exotic degrees of freedom such as hyperons and quarks (although there are exceptions~\cite{Lee:2014nya,Bedaque:2014ada}) nor do they describe phase transitions. 
Consequently, these studies deal with rather smooth EOSs that are likely to be easily expanded using a Taylor series. In the case of an EOS including a large peak in $c_s^2$ or other sharp features, we want to test both the applicability of the symmetry energy expansion as well as its influence on the magnitude and location of a peak in  $c_s^2$. 
For instance, one can certainly wonder whether a peak in $c_s^2$ occurring, e.g., for $Y_{Q,\rm{QCD}} \approx 0.1$ may disappear entirely for $Y_{Q,\rm{QCD}} \approx 0.5$ (see, for instance, Fig.~3 of~\cite{Zhang:2022sep}).

We can also reconsider Eq.~(\ref{eqn:symExpan}) in the limit where ANM becomes pure neutron matter (PNM).  
In this limit, $\delta=1$ and the equation simplifies to 
\begin{equation}
   \frac{ E_{\rm{PNM}}}{N_B} = \frac{E_{\rm{SNM}}}{N_B} + E_{\txt{\txt{sym}}} \ ,
\end{equation}
such that {$E_{\txt{\txt{sym}}}$} is the difference between energy per baryon in PNM and SNM.  

In this paper, we consider EOSs from~\cite{Tan:2020ics,Tan:2021ahl} that have been shown to support extremely heavy neutron stars with maximum masses in the range $M_{\txt{max}}\approx 2$--$3 M_\odot$ while satisfying all known astrophysical constraints. To convert these EOSs to EOSs for SNM or nearly SNM, we subtract the lepton contributions and then, using the symmetry energy expansion (and applying known bounds for the coefficients of the expansion), we construct new EOSs characterized by $Y_{Q,\rm{QCD}}^{\txt{const}}\approx 0.5$ which can be used in heavy-ion collision simulations. 
We find that large peaks in the $c_s^2$ of a NS EOS can significantly limit the parameter space of the coefficients in the symmetry energy expansion due to the fact that many parameter sets lead to acausal speeds of sound when $Y_{Q,\rm{QCD}}$ is increased. 
We find bounds on the symmetry energy expansion coefficients by ensuring causality, $c_s^2 \leq 1$, as well  stability, $c_s^2 (n_B \geq 0.9 n_{\rm{sat}}) \geq 0$, of the SNM EOS for densities larger than those corresponding to the spinodal region of the nuclear-liquid gas phase transition. 
We also explore a more accurate symmetry energy expansion that does not assume exact SNM ($Y_{Q,\rm{QCD}}^{\txt{const}} = 0.5$), but rather allows one to freely choose $Y_{Q,\rm{QCD}}^{\txt{const}}$  and thus to adjust the EOS for different species of the colliding ions. 

Finally, using the hadronic transport code \texttt{SMASH}~\cite{Weil:2016zrk} (version 2.1~\cite{smash_version_2.1}) with relativistic vector density functional mean-field potentials~\cite{Sorensen:2020ygf} parametrized to reproduce an arbitrary behavior of $c_s^2$ as a function of $n_B$ (as described in~\cite{Oliinychenko:2022uvy}), we use the obtained set of SNM EOSs to compute flow observables from simulations of heavy-ion collisions and compare them with experimental data. 
Through that, we provide a proof-of-principle demonstration of a way to confront the ever-better constrained neutron-star EOSs with heavy-ion collision measurements, which in the future may lead to setting meaningful bounds on the EOS of dense nuclear matter based on combined information from neutron star and heavy-ion collision studies.

In this work, we find a number of key takeaways which lead us to conclude that future studies on the connection between neutron stars and heavy-ion collisions are not only well-motivated, but possible within the currently existing or soon-to-be-developed approaches:
\begin{itemize}
    \item {\bf Methodology for converting from a neutron star~(NS) EOS into a heavy-ion collision~(HIC) EOS} We delineate the multi-step process needed when translating a neutron star (NS) EOS into one that can be used as input for hadronic transport codes. The steps (in order) are: determine the charge fraction, remove the lepton contribution from the EOS, apply the symmetry energy expansion, and finally calculate all remaining thermodynamic observables. 
    \item {\bf Consistency of ultramassive neutron stars and heavy-ion data} In a recent work~\cite{Fattoyev:2020cws} it was argued that massive neutron stars are excluded because their EOSs are inconsistent with the ``heavy-ion data''\footnote{To be precise, the result obtained in~\cite{Danielewicz:2002pu} is not data, but rather it is an EOS extracted from a comparison to data using a specific model under certain assumptions.
    }. 
    However, here we find a counterexample: we show that a NS EOS with a large peak in the speed of sound squared occurring between $n_B\approx 2$--$3 n_{\rm{sat}}$, which leads to a very heavy neutron star (nearly $M\approx 2.5 M_{\odot}$), is in fact consistent with an SNM EOS describing heavy-ion observables from modern measurements~\cite{HADES:2022osk,STAR:2020dav,STAR:2021yiu}. 
    \item {\bf Probability distributions for the coefficients of the symmetry energy expansion} Using only causality and stability constraints for the speed of sound, we find preferences for small values of $L_{\rm{sym}}$, $K_{\rm{sym}}<0$, and $J_{\rm{sym}}>0$.  
    While some variations exist between the different NS EOSs considered, these generic features remain the same.
    \item {\bf Applicability of the symmetry energy expansion} Using a test case of the CMF EOS~\cite{Dexheimer:2008ax}, we find that if $Y_{Q,\rm{QCD}}$ and the symmetry energy coefficients are known, the symmetry energy expansion can reasonably convert a NS EOS into a SNM EOS.  
    However, without the knowledge of $Y_{Q,\rm{QCD}}$ more uncertainty appears when converting between SNM and $\beta$-equilibrium. 
    While it is possible to apply an expansion scheme for $Y_{Q,\rm{QCD}}$, which we do here, uncertainties in the symmetry energy coefficients lead to large uncertainties in the converted EOS at high baryon densities.
    Our finding emphasizes the need for developing new methods to extract $Y_{Q,\rm{QCD}}$ from astrophysical observations as well as determining the precise symmetry energy expansion coefficients.
\end{itemize}

\section{Equation of state}\label{sec:EOS}

The primary probes of the QCD phase diagram are heavy-ion collisions, reaching large temperatures $T\approx 50$--$600$ MeV and a range of baryon densities (depending on the beam kinetic energy in the fixed-target frame $E_{\txt{kin}}$), and neutron stars (either isolated, accreting matter from a companion, or those in merging binaries), reaching lower temperatures $T\approx 0$--$70$~MeV, high densities, and very high isospin asymmetries. In heavy-ion collisions at high beam energies, quarks and gluons become deconfined producing a strongly-interacting quark gluon plasma (QGP), described by relativistic hydrodynamics, followed by a phase transition into an interacting hadron gas, described by hadronic transport models. As the beam energy is lowered, the lifetime of the hydrodynamics phase shrinks relative to the hadron gas phase~\cite{Auvinen:2013sba}. Eventually, at low enough beam energies, one expects a regime in which the dynamics can be described exclusively by hadronic degrees of freedom. The majority of hadron transport codes that model this low-beam-energy regime use mean-field potentials dependent on vector baryon density; such potentials do not have a complex temperature dependence (that is, a dependence beyond the basic, even if nontrivial, effects due to increasing particle momenta at high temperatures) and, therefore, one can essentially use a $T=0$ EOS as an input in hadronic transport. However, one cannot directly apply the neutron star (NS) EOS to heavy-ion collisions because $Y_{Q,\rm{QCD}}$ is quite different in the two systems. Thus,  we describe below the procedure used to switch from the $T=0$ NS EOS to the $T=0$ heavy-ion collision (HIC) EOS. Note that this is a standard procedure when done in the opposite order~\cite{Bombaci:1991zz,Zhang:2018vrx}, that is, when converting from SNM to ANM.

\subsection{Neutron-star EOS}
\label{sec:neutron_star_EOS}

We begin our discussion with three EOSs that have been previously shown in Refs.~\cite{Tan:2020ics,Tan:2021ahl,Tan:2021nat} to produce heavy neutron stars compatible with X-ray observations from NICER J0740~\cite{Riley:2021pdl,Miller:2021qha} and J0030~\cite{Riley:2019yda,Miller:2019cac}, which provide posteriors for the mass-radius relation for two different neutron stars, and with gravitational wave extraction of the tidal deformability from a binary neutron-star coalescence observed in GW170817~\cite{TheLIGOScientific:2017qsa,LIGOScientific:2018cki}.  
We are especially interested in an EOS that can support even heavier neutron stars in light of recent observations of mystery objects that fall within the mass gap between the heaviest known neutron star and the lightest known black hole. For instance, GW190814 observed the coalescence of a heavy black hole with a mystery object of mass $M = 2.5$--$2.67$~M$_\odot$~\cite{LIGOScientific:2020zkf}, V723 Mon is a dark object with mass of $M\geq 2.91\pm0.08$~M$_\odot $~\cite{Jayasinghe:2021uqb}, and the recently measured pulsars PSR J1810+1744 and PSR J0952-0607 have inferred masses $M > 2.19$~M$_\odot$~\cite{Romani:2021xmb} and $2.35\pm 0.17$~M$_\odot$~\cite{Romani:2022jhd}, respectively.

\begin{figure*}
\centering
\begin{tabular}{ccc}
\hspace{-.2cm}\includegraphics[width=0.3\linewidth]{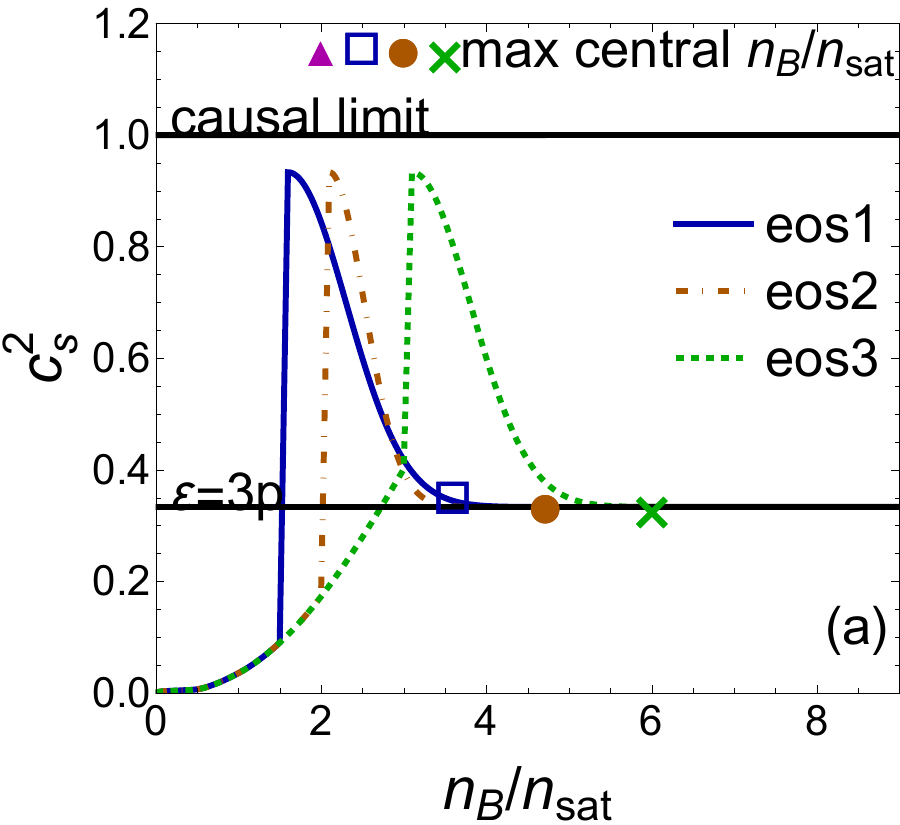} & 
\hspace{.5cm}\includegraphics[width=0.3\linewidth]{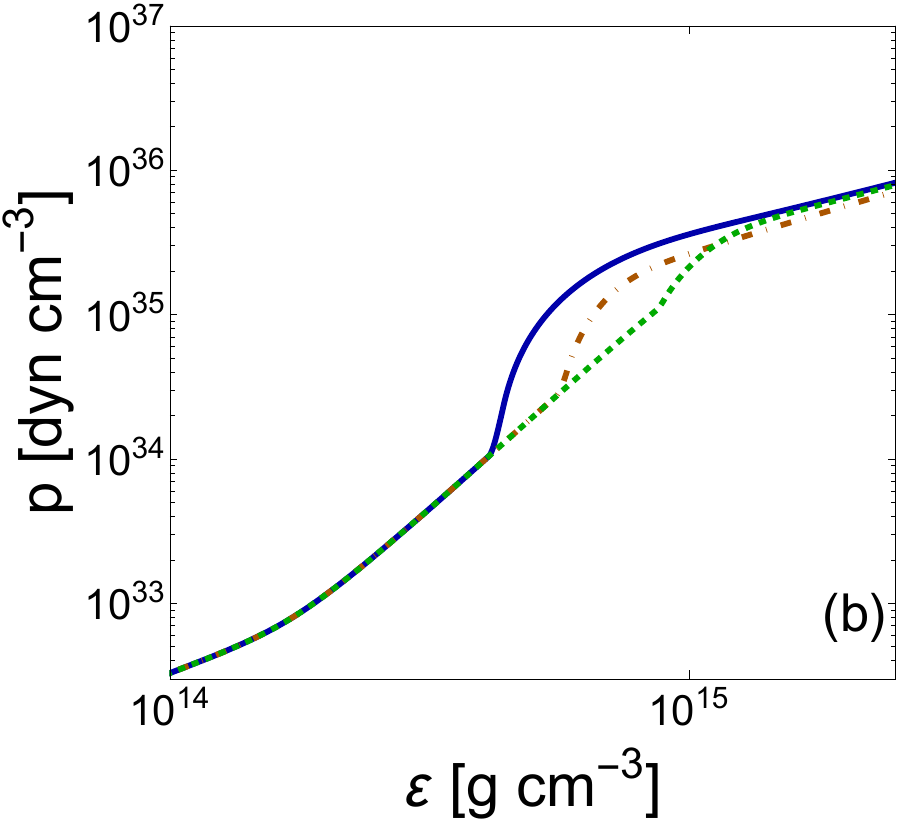} &
\hspace{.7cm}\includegraphics[width=0.3\linewidth]{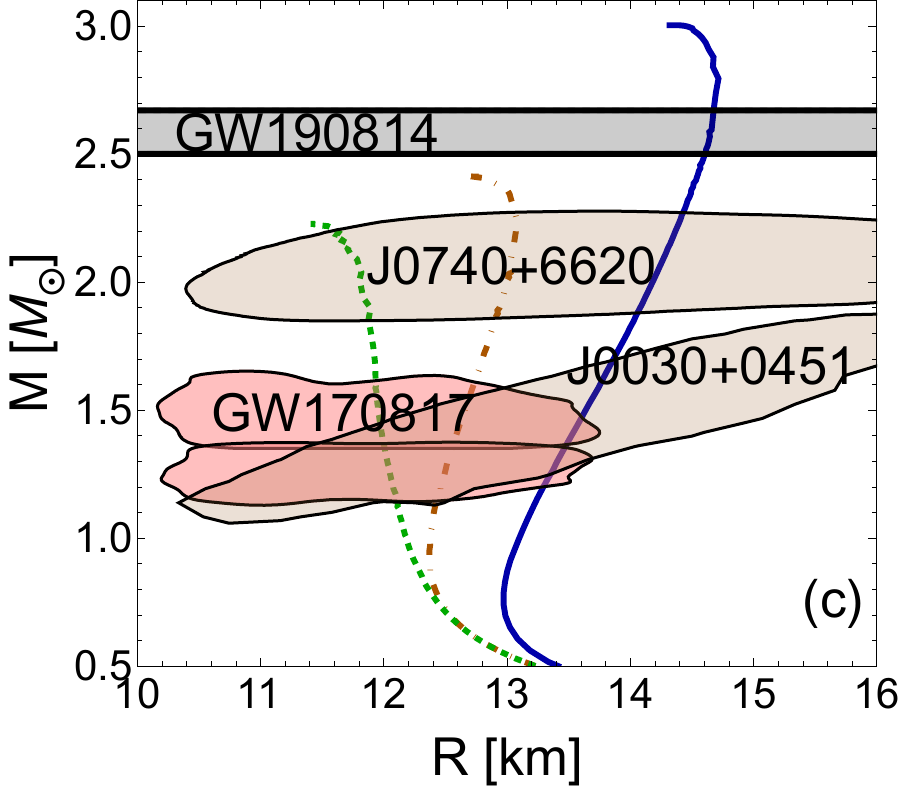} 
\end{tabular}
\caption{\textit{Left}: The speed of sound squared vs.\ normalized baryon density for three NS EOSs exhibiting a very steep rise (peak) in the speed sound at different densities. The symbols mark the central density of the most massive stable star for each of the EOSs. \textit{Middle}: Pressure vs.\ energy density for the same EOSs. The upper limit of the energy density range shown is set at the maximum central density for eos3. \textit{Right}: Mass-radius curves corresponding to the three NS EOSs compared to 90\% confidence regions of the LIGO/Virgo~\cite{LIGOScientific:2018cki} and NICER~\cite{Riley:2019yda,Miller:2019cac,Riley:2021pdl,Miller:2021qha} posteriors.}
\label{fig:NS_EOS}
\end{figure*}

Extensive details about the construction of these EOSs, as well as the link to the open-source code that generates them can be found in~\cite{Tan:2021ahl}. Here, we provide a brief overview of the technique. At low densities, the 
Togashi EOS \cite{Togashi:2013bfg,Togashi:2017mjp,Baym:2019iky}
is adopted up to some chosen switching point $n_{\txt{sw}}$, after which a functional form of the EOS reproducing a given shape of the speed of sound $c_s^2$ is used. Motivated by quarkyonic matter~\cite{McLerran:2018hbz,Zhao:2020dvu,Sen:2020qcd,Duarte:2020xsp,Sen:2020peq} and hyperon EOSs~\cite{Stone:2021ngh}, a large peak in $c_s^2$ is introduced, implemented by using a linear function with a large slope, describing the steep rise in $c_s^2$ as a function of $n_B$ up to the maximal value of $c_s^2|_{\txt{max}}$, followed by an exponential decay to the conformal limit $c_s^2\rightarrow 1/3$. While a large family of EOSs that can produce heavy neutron stars was developed in~\cite{Tan:2021ahl}, here we specifically only select three EOSs, including two extreme EOSs, to give a proof-of-principle of our method. We have checked our results against pQCD constraints extended to lower densities~\cite{Komoltsev:2021jzg} at their maximum central baryon density (similar to what was done in~\cite{Somasundaram:2022ztm,Mroczek:2023zxo}), and note that all EOSs easily fall within the allowed band.

In the left panel of Fig.~\ref{fig:NS_EOS}, one can see the three NS EOSs considered in this work. 
The distinguishing feature of eos1, eos2, and eos3 is a peak created by a very steep rise in $c_s^2$ starting at $n_B=\{1.5,~2.0,~3.0\}~n_{\txt{sat}}$, respectively, followed by an exponential relaxation to the conformal limit. These particular EOSs are chosen to explore the question whether extremely massive neutron stars (or, equivalently, extreme peaks in the speed of sound) are excluded based on the inconsistency of the corresponding EOSs with inferences from heavy-ion collision measurements.
In all cases, the maximum mass of the stellar sequence occurs at relatively low densities. 
The early peak in $c_s^2$, characterizing eos1, leads to a maximum central density of $n_B^{\txt{max}}\approx 3.5 n_{\txt{sat}}$, the peak at intermediate $n_B$ corresponding to eos2 allows for a central density of $n_B^{\txt{max}}\approx\ 4.5 n_{\txt{sat}}$, and the late peak in $c_s^2$ used in eos3 results in a maximum central density of almost $n_B^{\txt{max}}\approx 6 n_{\txt{sat}}$. 
The middle panel of Fig.~\ref{fig:NS_EOS} shows the corresponding dependence of pressure on energy density. 
There, the peaks in $c_s^2$ as a function of $n_B$ lead to corresponding changes in the slope of the pressure as a function of the energy density.
In the right panel of Fig.~\ref{fig:NS_EOS}, where we show the mass-radius sequence calculated using the Tolman-Oppenheimer-Volkoff equation~\cite{Oppenheimer:1939ne,Tolman:1939jz} for all three EOSs together with relevant constraints, one can see how the three considered locations of a peak in $c_s^2$ influence the mass-radius curves. We note that one should use some caution in considering the allowed mass-radius region extracted from GW170817, as there is some dependence on the choice of the EOS used to obtain the constraint, in particular on whether one uses a spectral EOS vs.\ universal relations~\cite{LIGOScientific:2018cki} (also see Ref.~\cite{Landry:2020vaw} for a discussion of Gaussian processes and Ref.~\cite{Tan:2021nat} for a discussion of binary Love relations); we also note here that spectral and polytropic EOSs cannot describe nontrivial structures in $c_s^2$. This dependence of the extracted mass-radius posterior from gravitational wave data on the EOS and its effect on the extracted maximum mass is discussed in ~\cite{Tan:2021ahl}, see in particular Fig.~8.

The results from the right panel of Fig.~\ref{fig:NS_EOS} make it clear why these three specific EOSs were chosen for the current study. One can see that eos1 is a very extreme EOS that is near the large-radius edge of the posterior distribution for extracted mass-radius region of GW170817. In consequence, it produces a very large maximum mass of $3 M_{\odot}$.  In contrast, eos3 is positioned towards the other end of the posterior distributions of observed neutron stars and has a maximum mass of $\approx 2.2 M_{\odot}$. This value still corresponds to a large maximum mass, but also happens to fall within recent bounds based on merging systems of binary neutron stars and quasi-universal relations~\cite{Nathanail:2021tay}. In between the extremes of eos1 and eos3, we have eos2 that fits right through the center of all astrophysical constraints and reaches to nearly $2.5 M_\odot$ for its maximum mass. With these three EOSs, we can explore the extreme variants of heavy neutron stars with eos1 and eos3, while eos2 leads to more moderate results.

\subsection{Subtracting the lepton contribution}\label{sec:leptons}

Neutron stars must be electrically neutral to be stable.  Thus, the net hadronic (or quark) contribution to electric charge, $n_{Q,\rm{QCD}}$, must be counterbalanced by the net lepton contribution to electric charge $n_{Q,\rm{lep}}$, as shown in Eq.~\ref{eqn:betaEQ},
where in most cases the hadrons considered in neutron stars provide only a positive contribution to $n_Q^{\rm{QCD}}$ (i.e., protons) and the leptons always contribute negatively (i.e., electrons).  However, if strange particles are present, the situation becomes complicated by the fact that, e.g., sigmas can be positive, negative, or neutral. Additionally, deconfined quarks present challenges related to fractional charges.  

In the case of an EOS for neutron stars (specifically, fully evolved neutron stars that are not undergoing a merger), one considers the EOS in $\beta$ equilibrium. 
Consequently, for a simple EOS with contributions from only neutrons, protons, and electrons, the following interaction is in weak equilibrium
\begin{equation}
p+e^- \leftrightarrow n+\nu  ~,
\end{equation}
such that the chemical potentials can be related as
\begin{equation}
\mu_p+\mu_e=\mu_n \ .
\end{equation}
Note that since the neutrinos can easily escape in this case, they do not contribute to the chemical potentials. From this reaction, it is clear that for an EOS in $\beta$ equilibrium the number of protons must equal the number of electrons such that Eq.~(\ref{eqn:betaEQ}) holds. Thus, the EOS has two separate contributions from baryons and leptons such that
\begin{eqnarray}
p&=& p_{\rm{QCD}}+p_{\rm{lep}}~,\\
\varepsilon&=&\varepsilon_{\rm{QCD}}+\varepsilon_{lep} \ .
\end{eqnarray}
The lepton contribution can be described by a noninteracting quantum gas of leptons (i.e., an free Fermi gas), which leads to analytic expressions for thermodynamic variables at $T=0$,
\begin{align}
& n_X  = \frac{k_F^3}{3 \pi^2} ~, \nonumber\\
& \epsilon_X=\frac{1}{\pi^2}\Bigg[\left(\frac{1}{8}m_X^2k_{F_X}+\frac{1}{4}k_{F_X}^3\right)\sqrt{m_X^2+k_{F_X}^2}
\nonumber\\
&\hspace{15mm}-~ \frac{1}{8}m_X^4\ln{\frac{k_{F_X}+\sqrt{m_X^2+k_{F_X}^2}}{m_X}}\Bigg]\ ,
\nonumber\\
&p_X=\frac{1}{3}\frac{1}{\pi^2}\Bigg[\left(\frac{1}{4}k_{F_X}^3-\frac{3}{8}m_X^2k_{F_X}\right)\sqrt{m_X^2+k_{F_X}^2}
\nonumber\\
&\hspace{15mm}+~\frac{3}{8}m_i^4\ln{\frac{k_F+\sqrt{m_X^2+k_{F_X}^2}}{m_X}}\Bigg]\ ,
\label{eqn:ideal}
\end{align}
where $X=e^-,\mu^-,\tau^-$ (although $\tau^-$'s are not considered within neutron stars due to their large mass), $k_{F_X}$ is the Fermi momentum of the lepton (that is, the point where the energy is equal to the chemical potential $E_{F_X}=\mu_X$), $m_X$ is the mass of the lepton, and we have taken \mbox{$\hbar=c=1$}. 

In this paper, we assume that only electrons are present (although it would not be difficult to include muon contributions as well, see~\cite{deTovar:2021sjo} for a discussion of muons in the context of the symmetry energy). Here, we are considering functional forms of the EOS without specific microscopic information. Therefore, we must rely on a functional form of $Y_{Q,\rm{QCD}}$ to determine the appropriate $n_Q^{\rm{lep}}$. With charge neutrality $n_Q = 0$, we can rearrange Eq.~(\ref{eqn:betaEQ}),
\begin{equation}
n_{Q,\rm{lep}}=-n_B Y_{Q,\rm{QCD}} \ ,
\end{equation}
where the negative sign cancels out once one recalls that electrons have a negative charge.
Then, given a $Y_{Q,\rm{QCD}}$, for each density $n_B$ we can obtain the corresponding $n_{Q,\rm{lep}}$, and from Eqs.~(\ref{eqn:ideal}) we can solve for the electron energy density and pressure at that $n_B$.

\begin{figure}[t!]
\centering
\includegraphics[width=\linewidth]{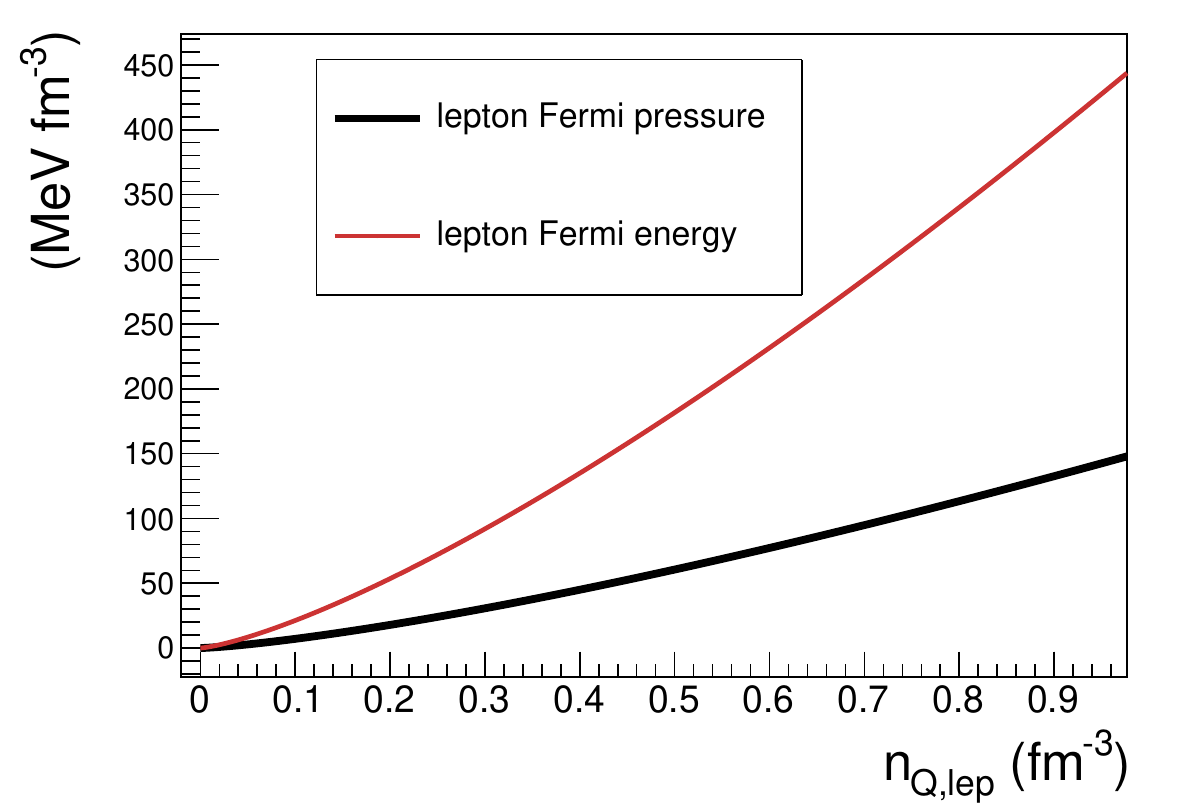} \caption{
Comparison of $p$ and $\varepsilon$ for a free Fermi gas of electrons. 
}
\label{fig:CMFleptonpvse}
\end{figure}

Finally, we obtain the QCD contribution to the EOS
\begin{eqnarray}
p_{\rm{QCD}}&=&p- p_{\rm{lep}}~,\\
\varepsilon_{\rm{QCD}}&=&\varepsilon-\varepsilon_{\rm{lep}}~,
\end{eqnarray}
which is what we use in the following section when considering the symmetry energy expansion. In Fig.~\ref{fig:CMFleptonpvse}, we plot the contributions of the leptons for both the pressure and energy density against the corresponding electric charge density, $n_{Q,{\rm{lep}}}$. We see that for a fixed value of $n_{Q,{\rm{lep}}}$, the pressure contribution from leptons is smaller than the energy density contribution. Thus, when $p_{\rm{lep}}$ and $\varepsilon_{\rm{lep}}$ are subtracted from the EOS, it has the effect of a slight stiffening of the EOS (because more energy density is subtracted than pressure).

\begin{figure}[t!]
\centering
\includegraphics[width=\linewidth]{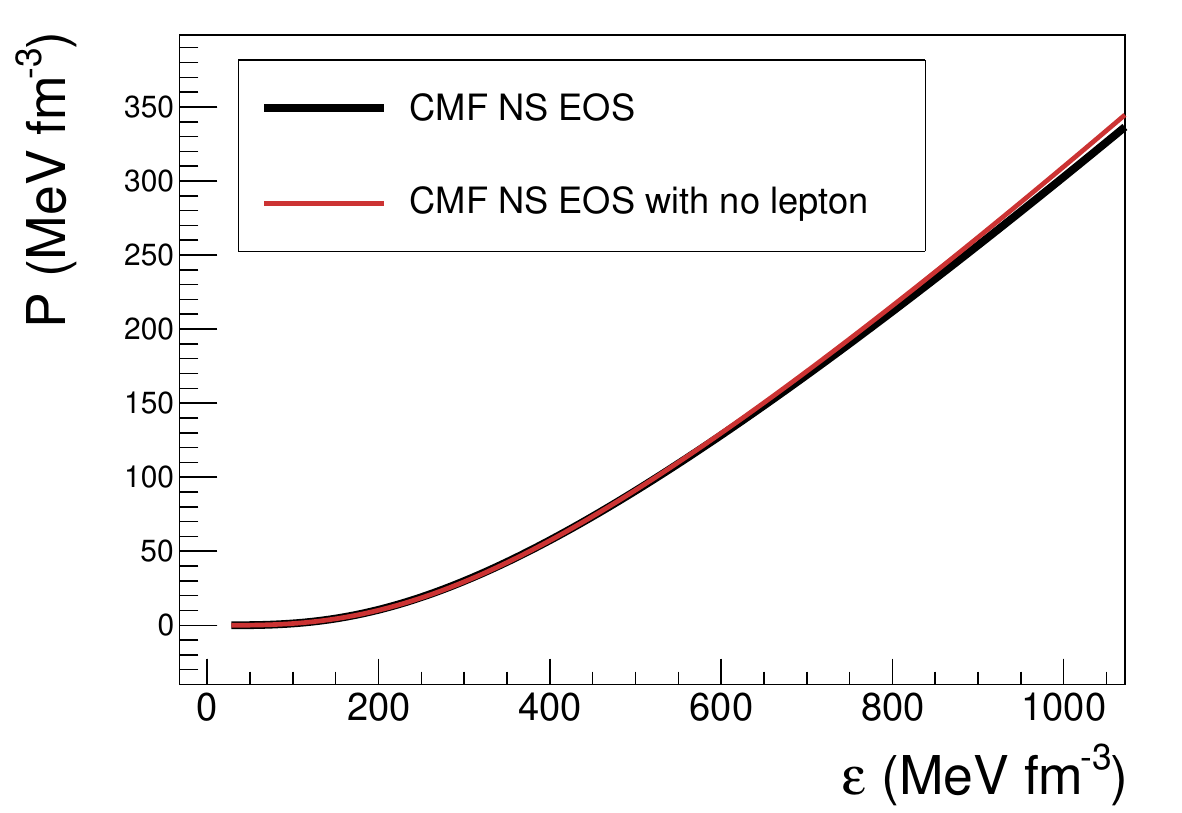} \caption{Comparison between the original CMF NS EOS and the 
CMF NS EOS without the lepton contribution. The corresponding $n_{B}$ range for this plot is 0--5$n_{\rm{sat}}$.
}
\label{fig:CMFlepton}
\end{figure}

As a proof of principle, we take the hadronic CMF NS EOS \#2 (with only neutrons, protons, and electrons) at $T=0$, assuming $\beta$ equilibrium, from~\cite{Dexheimer:2008ax}, where we know $Y_{Q,\rm{QCD}}$ exactly, so it is straightforward to subtract the lepton contribution. In Fig.~\ref{fig:CMFlepton}, we compare the original CMF NS EOS to the CMF NS EOS without leptons in the range of pressures and energy densities relevant for neutron stars. 
The lepton contribution is comparatively very small. Thus, it is natural to ask whether it is relevant to subtract it when performing the symmetry energy expansion?  
As it turns out, while the contribution to the EOS is quite small, it does play a significant role when it comes to reproducing saturation properties, which we discuss in more detail in the next section.

\subsection{Symmetry energy expansion}

\begin{table*}
\centering
\begin{tabular}{c|c|c|c}
Coefficient & Definition & Range & References\\
\toprule
$E_{\txt{sym,sat}}$ & $\big(\frac{E_{\rm{PNM}}-E_{\rm{SNM}}}{N_B}\big)_{n_{\rm{sat}}}$ &  $31.7 \pm 3.2$ [MeV] & Multiple data analyses from nuclear physics and astrophysics \cite{Li:2019xxz} \\
\hline
$L_{\txt{\txt{sym,sat}}}$  & $3n_{\txt{sat}}\big(\frac{dE_{\txt{sym},2}}{dn_B}\big)_{n_{\rm{sat}}}$ & $58.7 \pm 28.1$ [MeV] & Multiple data analyses from nuclear physics and astrophysics \cite{Li:2019xxz}  \\
\hline
$K_{\txt{\txt{sym,sat}}}$ & $9n_{\txt{sat}}^2\big(\frac{d^2E_{\txt{sym},2}}{dn_B^2}\big)_{n_{\txt{sat}}}$& $106\pm 37$ [MeV] & PREXII \cite{PREX:2021umo,Reed:2021nqk}\\
 & &  $-120_{-100}^{+80}$ [MeV]  & Bayesian analyses inferred from GW170817 and PSR J0030+0451\cite{Xie:2020tdo} \\
\hline
$J_{\txt{sym,sat}}$  &  $27n_{\txt{sat}}^3\big(\frac{d^3 E_{\txt{sym},2}}{dn_B^3}\big)_{n_{\txt{sat}}}$ &$300\pm 500$ [MeV] & Many-body nuclear theory \cite{Tews:2016jhi}  \\
\end{tabular}
\caption{Definitions of the symmetry energy expansion coefficients, as used in Eqs.\ \eqref{eq3} and \eqref{eqn:eHIC}, together with constraints on their values based on experimental data \cite{Li:2019xxz}. }
\label{tab:coef}
\end{table*}

\subsubsection{Connecting neutron-star EOSs with the heavy-ion collision domain}

To make the connection between the EOS in HIC, where the isospin asymmetry is small or even taken to be zero, and the EOS in NS, in which there is a large isospin asymmetry, we make use of the definition of the symmetry energy expansion from Eq.~(\ref{eqn:symExpan}).
There, the asymmetry parameter $\delta$, originally defined for neutron-star matter consisting only of nucleons in terms of neutron and proton number densities as $\delta = (n_{n}-n_{p})/(n_{n}+n_{p})$, can be defined in terms of the charge fraction as $\delta=1-2 Y_{Q,\rm{QCD}}$ (see Eq.~\eqref{eqn:delta} and the corresponding comments in the text), which is the form that we use in this work.
The advantage of the latter expression is that it can be applied to both nonstrange and strange matter, as well as to matter consisting of hadrons and/or quarks. $Y_{Q,\rm{QCD}}$ is also provided as an independent variable by several EOS repositories, such as CompOSE~\cite{CompOSECoreTeam:2022ddl}; see Appendix~\ref{app:delta_definition} for more details. 

We note here that the nucleon energy $E$, following the notation used in studies devoted to the symmetry energy, does not include the rest mass energy; in contrast, in heavy-ion collision studies, it is customary to use total energy density $\varepsilon$ which does include contributions from the rest mass. The two quantities are then connected by
\begin{equation}\label{eqn:E2vare}
\varepsilon=n_B\left[\frac{E}{N_B}+m_N \right]\ ,
\end{equation} 
where $m_N$ is the average mass of the nucleons. 
 
The symmetry energy $E_{\rm{sym}}$ at a given baryon number density $n_B$ can be expanded around its value at $n_{\rm{sat}}$~\cite{Baldo:2016jhp}. Inserting this expansion in Eq.~\ref{eqn:symExpan}, using Eq.~\ref{eqn:delta}, and renaming ANM$\to$NS,QCD and SNM$\to$HIC,sym, we obtain
\begin{widetext}
\begin{equation}
\frac{E_{\txt{HIC,sym}}}{N_B}=\frac{E_{\txt{NS,QCD}}}{N_B}-\left[E_{\txt{sym,sat}}+\frac{L_{\txt{sym,sat}}}{3}\left(\frac{n_B}{n_{0}}-1\right)+\frac{K_{\txt{sym,sat}}}{18}\left(\frac{n_B}{n_{0}}-1\right)^{2}+\frac{J_{\txt{sym,sat}}}{162}\left(\frac{n_B}{n_{0}}-1\right)^{3}\right]\big(1-2Y_{Q,QCD}\big)^2\ ,
\label{eq3}
\end{equation}
\end{widetext}
where $E_{\txt{sym,sat}}$ is the symmetry energy at $n_{\rm{sat}}$, $L_{\txt{sym,sat}}$ is the slope of the symmetry energy at $n_{\rm{sat}}$, and $K_{\txt{sym,sat}}$ and $J_{\txt{sym,sat}}$ are higher-order coefficients, also taken at $n_{\rm{sat}}$; the definitions of these coefficients and their ranges are given in Tab.~\ref{tab:coef}. Astrophysical constraints on these coefficients have also been found~\cite{Zhang:2018vrx}, although we do not take these specific constraints into account in this work. Additionally, a recent large-scale study on the interplay of these coefficients can be found in Ref.~\cite{Sun:2023xkg} which finds similar ranges of these coefficients as the ones that we use in this work. As explained above, the energy that enters into the symmetry energy expansion is \emph{only} the QCD energy such that one must first subtract the lepton contribution. 
See also~\cite{Imam:2021dbe} for an alternative expansion scheme. 

Applying Eq.~(\ref{eqn:E2vare}), we can rewrite Eq.~\eqref{eq3} in a form more commonly used in high-energy physics,
\begin{widetext}
\begin{equation}\label{symexpand}    \varepsilon_{\txt{HIC,sym}}=\varepsilon_{\txt{NS,QCD}} - n_B\left[E_{\txt{sym,sat}}+\frac{L_{\txt{sym,sat}}}{3}\left(\frac{n_B}{n_{\rm{sat}}}-1\right)+\frac{K_{\txt{sym,sat}}}{18}\left(\frac{n_B}{n_{\rm{sat}}}-1\right)^{2}+\frac{J_{\txt{sym,sat}}}{162}\left(\frac{n_B}{n_{\rm{sat}}}-1\right)^{3}\right]  \big(1-2Y_{Q,QCD}\big)^2 \ ,
\end{equation}
\end{widetext}
where the mass contribution cancels on each side.

As previously mentioned, collisions of heavy ions do not typically involve exactly symmetric nuclei so that $\infrac{n_p}{n_B} \neq 0.5$. Additionally, unlike in neutron stars, $Y_{Q,\rm{QCD}}$ is not a function of baryon density and instead it only depends on the choice of nuclei, $Y_{Q,\rm{QCD}}^{\txt{const}}=Z/N_B$. This definition only works for collisions of identical nuclei, e.g., Au--Au collisions, and assumes that the collisions are such that the particular value of $Z/N_B$ describes well the fraction of nucleons participating in the collision. If we instead considered collisions of unlike nuclei, e.g., $^{16}$O--Au, then the picture would be more complex. However, in this paper we only consider collisions of identical nuclei and assume that $Y_{Q,\rm{QCD}}$ describes the nucleons participating in the collision well; as we discuss later, tiny differences in $Y_{Q,\rm{QCD}}$ due to fluctuations in the number of participating protons and neutrons are too small to lead to significant effects on the EOS. 

To investigate asymmetric HIC matter, introducing $Y_{Q,\rm{QCD}}^{\txt{const}}$, we can write a symmetry energy expansion for heavy-ion collisions (again following Eq.~\eqref{eqn:symExpan}, where we now rename ANM$\to$HIC,asym),
\begin{equation}
\varepsilon_{\substack{\txt{HIC,asym}}}=\varepsilon_{\substack{\txt{HIC,sym}}}+n_{B} E_{sym}\big(1-2Y_{Q,\rm{QCD}}^{\txt{const}}\big)^{2}\ ,
\end{equation}
where $\epsilon_{\txt{HIC}, \txt{sym}}$ can be identified with the term on the left-hand side of Eq.~\eqref{symexpand}.
After some rearrangements, we eventually obtain
\begin{widetext}
\begin{eqnarray}\label{eqn:eHIC}
\varepsilon_{\txt{HIC}, \txt{asym}}&=&\varepsilon_{\txt{NS,\rm{QCD}}}-4n_{B}\left[E_{\txt{sym,sat}}+\frac{L_{\txt{sym,sat}}}{3}\left(\frac{n_B}{n_{\rm{sat}}}-1\right)+\frac{K_{\txt{sym,sat}}}{18}\left(\frac{n_B}{n_{\rm{sat}}}-1\right)^{2}+\frac{J_{\txt{sym,sat}}}{162}\left(\frac{n_B}{n_{\rm{sat}}}-1\right)^{3}\right]\nonumber\\
&\times&\left[\Big(Y_{Q,\rm{QCD}}^{\txt{const}}-Y_{Q,\rm{QCD}}\Big)+ \Big(Y_{Q,\rm{QCD}}^{2}-\left(Y_{Q,\rm{QCD}}^{\txt{const}}\right)^{2}\Big)\right]\ ,
\end{eqnarray}
\end{widetext}
where we once again stress that $Y_{Q,\rm{QCD}}$ has a dependence on $n_B$ which comes from the fact that the NS EOS is an EOS for $\beta$-equilibrated matter, and $Y_{Q,\rm{QCD}}^{\txt{const}}$ is simply equal to $Z/N_B$ of the colliding nuclei. Note that by varying the value of $Y_{Q,\rm{QCD}}^{\txt{const}}$, one can effectively study the EOS at $T=0$ in a 2D space of possible values of $n_B$ and $Y_{Q,\rm{QCD}}^{\txt{const}}$, with particular values of $Y_{Q,\rm{QCD}}^{\txt{const}}$ corresponding to slices of the $\left\{n_B,Y_{Q,\rm{QCD}}\right\}$ space along lines of constant $Y_{Q,\rm{QCD}}$.

We constrain the density dependence of $Y_{Q,\rm{QCD}}$ by requiring that it describes the fraction of charge along the line of $\beta$ equilibrium in the $\left\{n_B,Y_{Q,\rm{QCD}}\right\}$ 
space. Within a microscopic model, one would, of course, be able to extract this dependence directly. However, among the microscopic models that reproduce large peaks in the speed of sound, very few results are provided for neutron-star matter, and even these do not include data for $Y_{Q,\rm{QCD}}$ or $Y_{\rm{lep}}$. 
Thus, we do not have the desired guidance from microscopic models on the functional form of $Y_{Q,\rm{QCD}}(n_B)$ and must use other approaches.
In this work, we want to be able to determine $Y_{Q,\rm{QCD}}$ from the functional form of the NS EOS. In principle, $Y_{Q,\rm{QCD}}$ may include contributions from charged hadrons other than protons (e.g. $\Delta$'s and $\Sigma$'s) or from quarks. However, we anticipate that protons are still likely the dominant contribution to electric charge in the relevant regime (assuming no quark phase), so that here we set $Y_{Q,\rm{QCD}}\equiv  Y_{\mathrm{p}}$; in future work, we may explore alternative schemes. Thus, we can use the fact that in the symmetry energy expansion, the proton fraction $Y_p$ can be expanded around its value at the saturation density such that
\begin{eqnarray}\label{long}
\hspace{-5mm} Y_{Q,\rm{QCD}}(n_B)&\equiv &Y_p(n_B)\nonumber\\
&=&\frac{1}{16}\left[8-\frac{\pi^{4/3}n_{B}}{2^{1/3}X}+\left(\frac{\pi}{2}\right)^{2/3}\frac{X}{E_{sym}^3}\right] ~,
\end{eqnarray} 
where X is given by
\begin{equation}\label{eqn:X_esym}
    X=
(-24 E_{\rm{sym}}^6 n_B+\sqrt{2} \sqrt{288 E_{\rm{sym}}^{12} n_B^2+\pi^2 E_{\rm{sym}}^9 n_B^3})^{1 / 3}.
\end{equation} 
The derivation of the above result as well as the range of possible values of $Y_{p}$ obtained given the allowed ranges in the expansion coefficients $E_{\txt{sym,sat}}$, $L_{\txt{sym,sat}}$, and $K_{\txt{sym,sat}}$ can be found in Appendix~\ref{app:YQ}. Note that while we based this derivation on ideas from Ref.~\cite{Mendes:2021tos}, we relax the approximation that $Y_p\ll 1$ to find the exact solutions (among which we only then consider the real solution). We also include all 4 symmetry energy expansion coefficients (Ref.~\cite{Mendes:2021tos} included only 3) to ensure validity up to higher~$n_B$. From this point on, we will refer to our expansion as $ Y_{Q,\rm{QCD}}(n_B)$ instead of $Y_p$. 

\begin{figure}[t!]
\centering
\includegraphics[width=\linewidth]{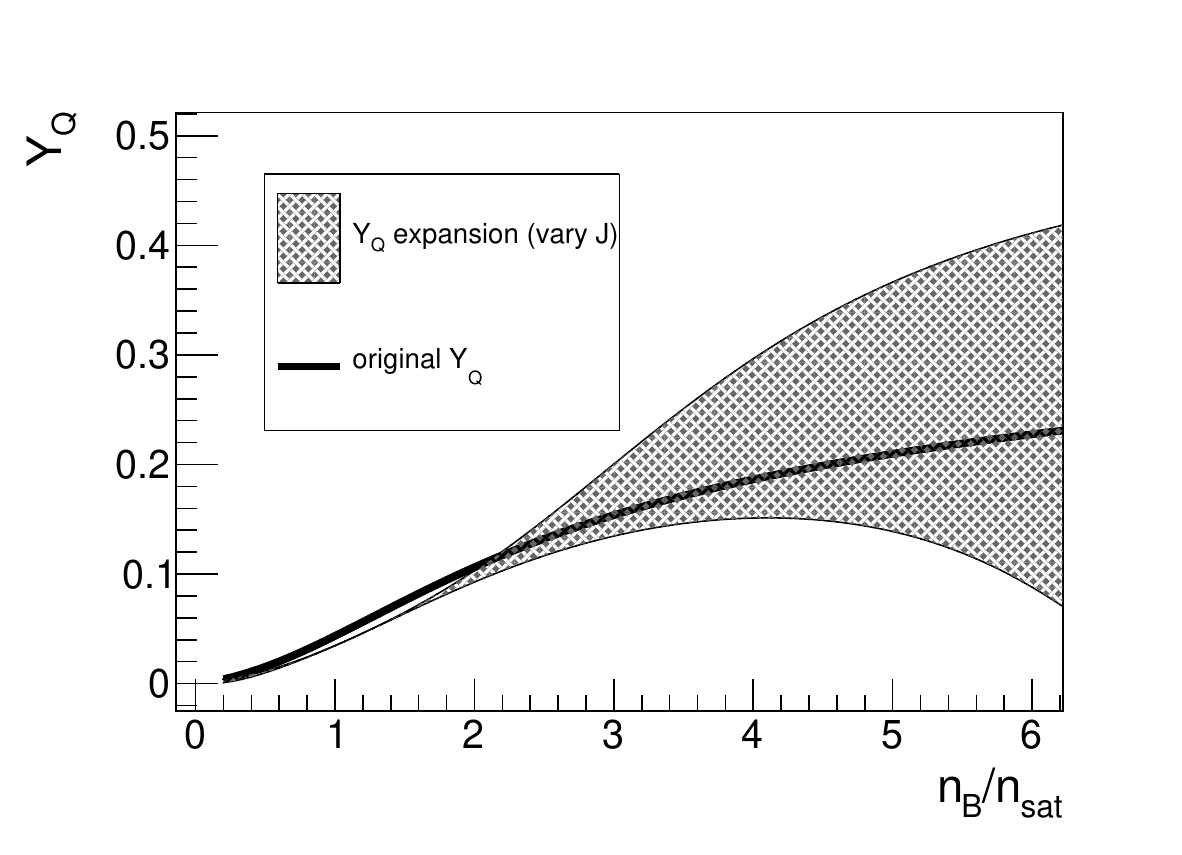} \caption{
Comparison of the $Y_Q$ directly from the CMF model \cite{Dexheimer:2008ax} with the $Y_Q$ expansion from Eq.\ (\ref{long}) with the coefficients $E_{\rm{sym}}$, $L_{\rm{sym}}$, $K_{\rm{sym}}$ given directly from CMF.  Since $J_{\rm{sym}}$ is unknown from CMF the band demonstrates the variation of $J_{\rm{sym}}$ and the corresponding uncertainty.
}
\label{fig:CMFYQvsExpan}
\end{figure}

To test the validity of $ Y_{Q,\rm{QCD}}(n_B)$ expansion in Eq.~(\ref{long}), we compare values of $ Y_{Q,\rm{QCD}}(n_B)$ extracted directly from CMF (assuming just protons, neutrons, and electrons) to values obtained from the expansion using the symmetry energy coefficients from CMF: $E_{\rm{sym,sat}}=30.45$ MeV, $L_{\rm{sym,sat}}=90.43$ MeV, $K_{\rm{sym,sat}}=25.72$ MeV. Because $J_{\rm{sym,sat}}$ has not yet been calculated within CMF, we vary its value between -200 and 700 MeV to obtain a range of solutions and understand the uncertainty in $ Y_{Q,\rm{QCD}}(n_B)$ when $J_{\rm{sym,sat}}$ is unknown.
The results are shown in Fig.~\ref{fig:CMFYQvsExpan}. We find that the original $ Y_{Q,\rm{QCD}}(n_B)$ from CMF and the one we obtain from the expansion using the exact symmetry energy coefficients from CMF are similar.  
Up to $n_B\lesssim 2 n_{\rm{sat}}$, the influence of $J_{\rm{sym,sat}}$ is negligible, and even up to $n_B\lesssim 3 n_{sat}$ it still plays a relatively small role. However, at higher baryons densities there is a significant amount of uncertainty due to the lack of knowledge about $J_{\rm{sym,sat}}$.

At this point we should note that for a small subset of symmetry energy coefficients (and at large values of $n_B$), we occasionally obtain values of $ Y_{Q,\rm{QCD}}(n_B)$ that are imaginary. One can see from Eq.~(\ref{eqn:X_esym}) that this can occur when the total symmetry energy is negative, i.e., when $E_{\rm{sym}}<0$. Since $K_{\rm{sym,sat}}$ and $J_{\rm{sym,sat}}$ are allowed to be both positive and negative, then it is indeed possible for this to occur at large enough $n_B$ (since $K_{\rm{sym,sat}}$ and $J_{\rm{sym,sat}}$ are higher-order terms). In that case, we simply discard all those energy expansion coefficients that lead to imaginary $ Y_{Q,\rm{QCD}}(n_B)$ for $n_B<6n_{\rm{sat}}$. 

In summary, our approach for obtaining the HIC EOS from a given NS EOS is based on the symmetry energy expansion of the energy density, Eq.~\eqref{eqn:eHIC}, as well as on the symmetry energy expansion of the charge fraction $Y_{Q,\rm{QCD}}$, Eq.~\eqref{long}. Within the validity of the Taylor expansion, this approach would be very well-defined if the exact values of the symmetry energy expansion coefficients $\left\{E_{\txt{sym,sat}},L_{\txt{sym,sat}},K_{\txt{sym,sat}}, J_{\txt{sym,sat}}\right\}$  were known. Alas, putting meaningful constraints on the values of the symmetry energy coefficients is the subject of ongoing vigorous research. Therefore, in this work we are going to concentrate on exploring the space of possible HIC EOSs, as obtained from NS EOSs, given the possible ranges of values of $\left\{E_{\txt{sym,sat}},L_{\txt{sym,sat}},K_{\txt{sym,sat}}, J_{\txt{sym,sat}}\right\}$.

In Ref.~\cite{Li:2019xxz}, values of the symmetry energy expansion coefficients are extracted from the 2016 survey of 53 analyses from both nuclear experiments and astrophysical observations. The central values for the first two expansion coefficients are $E_{\txt{sym,sat}}=31.7 \pm 3.2 ~ \mathrm{MeV}$ and $L_{\txt{sym,sat}}=58.7 \pm 28.1 ~\mathrm{MeV}$ \cite{Li:2019xxz}. However, the recent PREXII results~\cite{PREX:2021umo,Reed:2021nqk} are in tension with many of the previous measurements of $L_{\txt{sym,sat}}$. Thus, in this paper we consider ranges of priors on the first two symmetry coefficients to be: $E_{\txt{sym,sat}}=25 $ to $40$ MeV and $L_{\txt{sym,sat}}=30$ to $136$ MeV. The higher order coefficients are poorly known, with values for the curvature $K_{\txt{sym,sat}}$ and skewness $J_{\txt{sym,sat}}$ restricted to be $-400 \leq K_{\txt{sym,sat}} \leq 100 ~\mathrm{MeV}$ and $-200 \leq J_{\txt{sym,sat}} \leq 800 ~\mathrm{MeV}$, respectively, based on a Bayesian analysis of astrophysics data and on nuclear many-body theory~\cite{Li:2019xxz}. We summarize our priors for all four symmetry energy expansion coefficients in Tab.~\ref{tab:coef}.

Note that we use a broad prior for $E_{\rm{sym,sat}}$ both because it has been considered in previous studies~\cite{Neill:2022psd}, as well as because we want to entertain the possibility of large values for $L_{\txt{sym,sat}}$ that would be compatible with PREXII. Generally, $E_{\rm{sym,sat}}$ and $L_{\txt{sym,sat}}$ are positively correlated such that large values of $E_{\rm{sym,sat}}$ are needed to obtain large $L_{\txt{sym,sat}}$; see~\cite{Lattimer:2014sga} for detailed discussion involving theory and experiments.

\subsubsection{Other thermodynamic quantities}

\begin{table*}
\centering
\begin{tabular}{l|l|c|c}
NS   &\ \  $\varepsilon +p=n_B\mu_B$ &\ \  $\mu_B=\frac{d\varepsilon}{dn_B}\big|_{n_Q=\rm{const}}$ &\hspace{1mm}  $\mu_Q=\frac{d\varepsilon}{d(n_BY_{Q,\rm{QCD}})}\big|_{n_B=\rm{const}}$\\
\hline
HIC,sym  &\ \  $\varepsilon +p=n_B\mu_B$ &\ \  $\mu_B=\frac{d\varepsilon}{dn_B}\big|_{n_Q=\rm{const}}$&\\
\hline
 HIC,asym    &\ \  $\varepsilon +p=n_B\mu_B+n_Q\mu_Q$ &\ \  $\mu_B=\frac{d\varepsilon}{dn_B}\big|_{n_Q=\rm{const}}$ &\hspace{1mm}  $\mu_Q=\frac{d\varepsilon}{d(n_BY_{Q,\rm{QCD}}^{\txt{const}})}\big|_{n_B=\rm{const}}$
\end{tabular}
\caption{Gibbs free energy density equations at $T=0$ for different axes of the QCD phase diagram.}
\label{tab:gibbs}
\end{table*}

Besides the energy density $\varepsilon$, we need a number of other thermodynamic quantities to obtain the full EOS to be used in heavy-ion collision simulations. These can be derived directly from thermodynamic relations using the Gibbs free energy density,
\begin{equation}
\varepsilon +p-sT=n_B\mu_B+n_Q\mu_Q+n_S\mu_S\ ,
\end{equation}
where $s$ is the entropy, $\mu_Q$ is the charge chemical potential, $n_S$ is the strangeness density, and $\mu_S$ is the strangeness chemical potential.
Generally, to study the QCD phase diagram, we assume that there are no significant magnetic fields and that the energy scale is not large enough to produce charm, bottom, or top quarks in any relevant amounts. Additionally, for $T=0$ the entropy term vanishes. Moreover, in heavy-ion collisions strangeness is conserved such that $n_S=0$. Neutron stars do not conserve strangeness, but it is unlikely that strange quarks and hadrons are in thermodynamic equilibrium so that the associated strangeness chemical potential $\mu_S$ is likely equal zero. We then obtain a simplified equation,
\begin{equation}
    \varepsilon +p=n_B\mu_B+n_Q\mu_Q\ .
    \label{pressures}
\end{equation}
As already discussed, neutron stars are electrically neutral and in $\beta$-equilibrium, so the additional lepton contribution satisfies $n_{Q,\rm{QCD}}\mu_{Q, \rm{QCD}} + n_{Q,\rm{lep}}\mu_{Q,\rm{lep}} = 0$, while isospin symmetry arguments allow one to put $\mu_Q = 0$ for exactly SNM~\cite{Aryal:2020ocm}. However, in heavy-ion collisions matter is not exactly symmetric, such that the $n_Q\mu_Q$ term is nonzero. Thus we are dealing with three scenarios summarized in Tab.~\ref{tab:gibbs}.

In our work, we first obtain the energy density $\varepsilon_{\txt{HIC}}$ from the symmetry energy expansion (note that below, we drop the ``NS'' and ``HIC'' subscripts since the used equations are generic). When making the connection between the NS and the HIC EOS, one has the choice to connect them at the same values of $n_B$ and then obtain the associated $\mu_B$, or to connect them at the same values of $\mu_B$ and then obtain the associated $n_B$. Since the symmetry energy expansion explicitly depends on $n_B$, we choose $n_B$ as the primary variable in our procedure. If $Y_{Q,\rm{QCD}}^{\txt{const}}\neq 0.5$, then the system is also described by a nonzero charge chemical potential,
\begin{equation}\label{eqn:muQ}
\mu_Q=\frac{d\varepsilon}{dn_Q}\Bigg|_{n_B=const}\ ,
\end{equation}
which, given $Y_{Q,\rm{QCD}}=n_Q/n_B$, can be rewritten as
\begin{equation}
   \mu_Q= \frac{d\varepsilon}{d(n_BY_{Q,\rm{QCD}})}\Bigg|_{n_B=const}\ .
\end{equation}
Finally, one then uses the appropriate Gibbs free energy equation (see Tab.~\ref{tab:gibbs}) to solve for the pressure. We note that for SNM, an equivalent expression for pressure is
\begin{equation}\label{pressureeq}
    p=n_B^2 \frac{d(\varepsilon/n_B)}{dn_B} ~.
\end{equation}
With these thermodynamic variables, we can then calculate the speed of sound
\begin{equation}\label{cs2eq}
c_s^2 = \left(\frac{dp}{d\varepsilon}\right)_{T=0}\ .
\end{equation}
Note that at finite $T$, if the speed of sound is calculated to be used in, e.g., hydrodynamic evolution, one must consider the appropriate trajectory through the QCD phase diagram. This is typically done by following isentropes such that the entropy per baryon (or entropy density per baryon density) $s/n_B=\txt{const}$.

\subsubsection{Effect of $Y_{Q,\rm{QCD}}^{\rm{const}}$}

\begin{figure}[b]
\centering
\includegraphics[width=\linewidth]{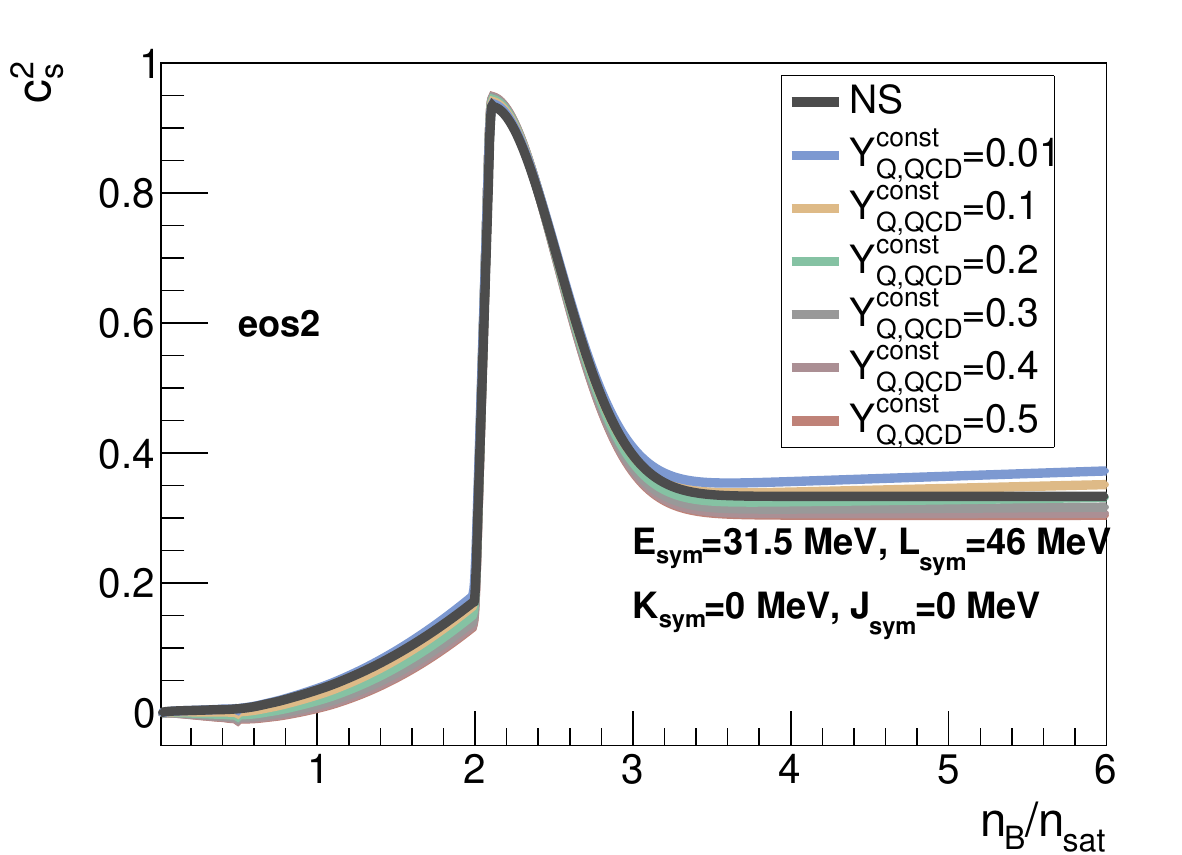} 
\caption{Speed of sound squared as a function of normalized baryon density, obtained from eos2 (peak at medium density) using different values of $Y_{Q,\rm{QCD}}^{\txt{const}}$ and the central values of $E_{\txt{sym,sat}}$ and $L_{\txt{sym,sat}}$ from Ref.~\cite{Li:2019xxz}. 
}
\label{fig:YQexpan}
\end{figure}

Now that we have described our methodology, we can apply the symmetry energy expansion to the three NS EOSs we are considering in this work (see Fig.~\ref{fig:NS_EOS} and the corresponding discussion in the text). To demonstrate the effect of the symmetry energy expansion on the EOS, we detail the necessary steps with eos2 as an example, however, the same results can be found for eos1 and eos3 in Appendix~\ref{sec:SymEn_13}.
Using Eq.~(\ref{eqn:eHIC}), Tab.~\ref{tab:gibbs}, and Eq.~\ref{cs2eq}, we can construct the EOS across different slices of constant $Y_{Q,\rm{QCD}}^{\txt{const}}$.  In Fig.~\ref{fig:YQexpan}, we demonstrate this procedure by showing the calculated $c_s^2$ for different values of $Y_{Q,\rm{QCD}}^{\txt{const}}$ and comparing to our original NS EOS; here, we used the following expansion coefficients: $E_{\txt{sym,sat}}=31.5$ MeV, $L_{\txt{sym,sat}}=46$ MeV, $K_{\txt{sym,sat}}=0$ MeV, and $J_{\txt{sym,sat}}=0$ MeV. We find that $Y_{Q,\rm{QCD}}$ most strongly affects $c_s^2$ at $n_B\gtrsim 3\ n_{\txt{sat}}$. It is also interesting to note that slices of constant $Y_{Q,\rm{QCD}}^{\txt{const}}$ indicate a fairly similar functional form despite differences in their magnitudes, which we believe is a consequence of using the Taylor expansion. What is quite surprising is that the large peak in $c_s^2$ is clearly preserved both in terms of magnitude and location.

Based on Fig.~\ref{fig:YQexpan}, we find that a realistic heavy-ion $Y_{Q,\rm{QCD}}^{\txt{const}} \approx 0.4$ is nearly identical to that of exactly SNM, $Y_{Q,\rm{QCD}}^{\txt{const}} = 0.5$. 
Thus, for the rest of this work, we use the simpler assumption that $Y_{Q,\rm{QCD}}^{\txt{const}}=0.5$. 
The reason for this is twofold: First, we can then avoid calculating  $\mu_Q$, which significantly reduces the computational cost when sampling across all 4 coefficients of the symmetry energy expansion. 
Second, $Y_{Q,\rm{QCD}}^{\txt{const}} = 0.5$ is the maximally different value as compared with values of $Y_Q$ in the NS EOS. Since we want to test causality and stability of the EOSs obtained through the expansion, all $Y_{Q,\rm{QCD}}^{\txt{const}}$ slices along the expansion must be stable and causal for an EOS to be valid. In particular, by testing $Y_{Q,\rm{QCD}}^{\txt{const}}=0.5$ we are able to put the most stringent constraint on the EOS.

We note that results in Fig.~\ref{fig:YQexpan} are obtained using central values of both $E_{\txt{sym,sat}}$ and $L_{\txt{sym,sat}}$, and values of $K_{\txt{sym,sat}}$ and $J_{\txt{sym,sat}}$ that lead to reasonable results. It is natural to ask how much of an effect other combinations of these coefficients have? Could combinations of $\left\{E_{\txt{sym,sat}},L_{\txt{sym,sat}},K_{\txt{sym,sat}}, J_{\txt{sym,sat}}\right\}$ completely change the location of the peak in $c_s^2$ or its overall magnitude? Could they make the EOS unstable or acausal? To explore these questions, we conduct a systematic study of the symmetry energy expansion coefficients and their effect on the obtained HIC EOS. 

\subsection{Proof-of-principle comparison and saturation properties}

Here we demonstrate (again using the CMF NS EOS at $T=0$, as in Sec.~\ref{sec:leptons}) the effectiveness of our approach.  Taking only the QCD contributions to the CMF NS EOS, we apply the symmetry energy expansion to obtain the converted EOS. In the case of CMF, we know precisely the correct values of $Y_{Q,\rm{QCD}}$ as well as the coefficients $E_{\txt{sym,sat}}$, $L_{\txt{sym,sat}}$, $K_{\txt{sym,sat}}$. Furthermore, we also know directly from CMF the correct EOS for SNM, so that we can also compare our expanded EOS to the one calculated directly in the CMF model. As it was done for results shown in Fig.~\ref{fig:CMFYQvsExpan}, we vary $J_{\txt{sym,sat}}$ in our calculations. 

\begin{figure}[t!]
\centering
\includegraphics[width=\linewidth]{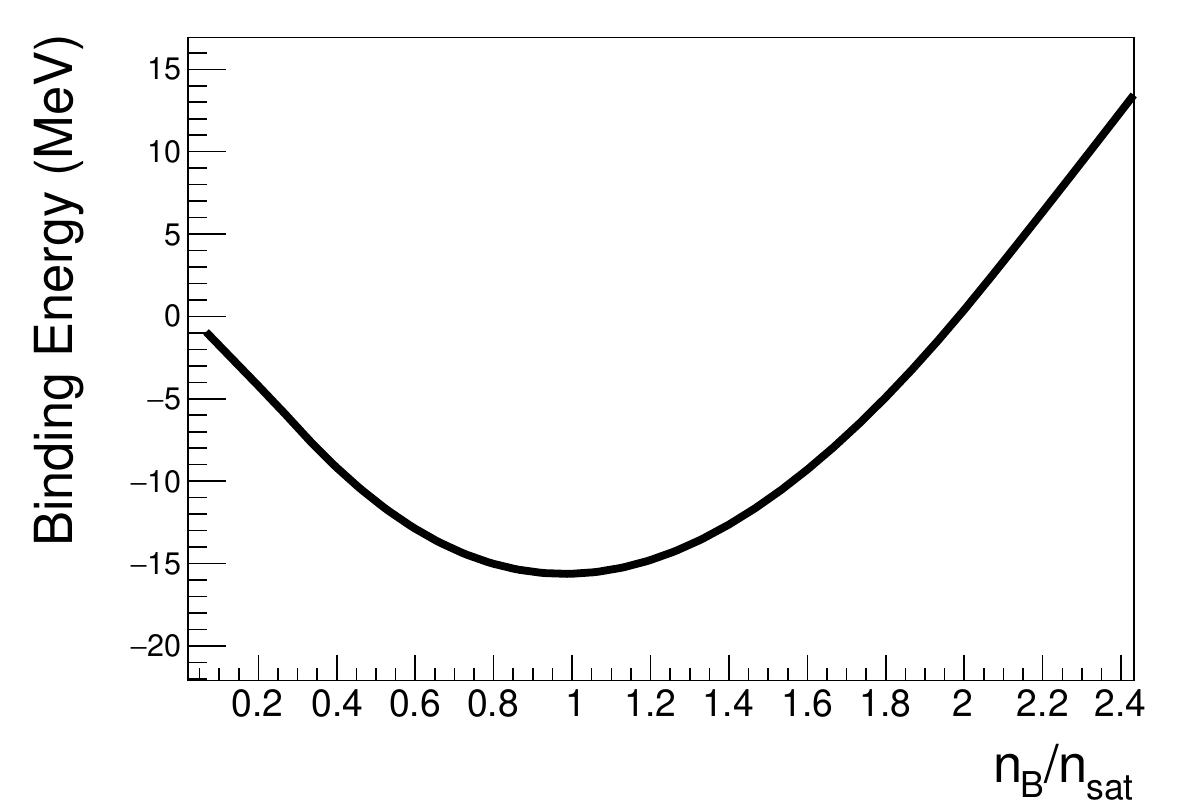} \caption{Result of the conversion of the 
CMF NS EOS to $Y^{\rm{const}}_{Q,\rm{QCD}}=0.5$ using the exact $Y_Q(n_B)$ from CMF. We find that the converted EOS shows the correct binding energy per nucleon at saturation.}
\label{fig:CMFlepton2}
\end{figure}

To begin our proof-of-principle comparison, we calculate the binding energy using the EOS expanded with the exact $Y_{Q,\rm{QCD}}(n_B)$ function from CMF. In Fig.~\ref{fig:CMFlepton2}, we show how the binding energy per nucleon for the converted CMF EOS reproduces the correct binding energy at saturation, 
\begin{equation}
    B=\frac{\varepsilon_{\rm{HIC,sym}}}{n_B}-m_N=-16~\rm{MeV}~,
\end{equation}
using $m_N=938.919$ MeV for the mass of the nucleon.
\begin{figure}[t]
\centering
\includegraphics[width=\linewidth]{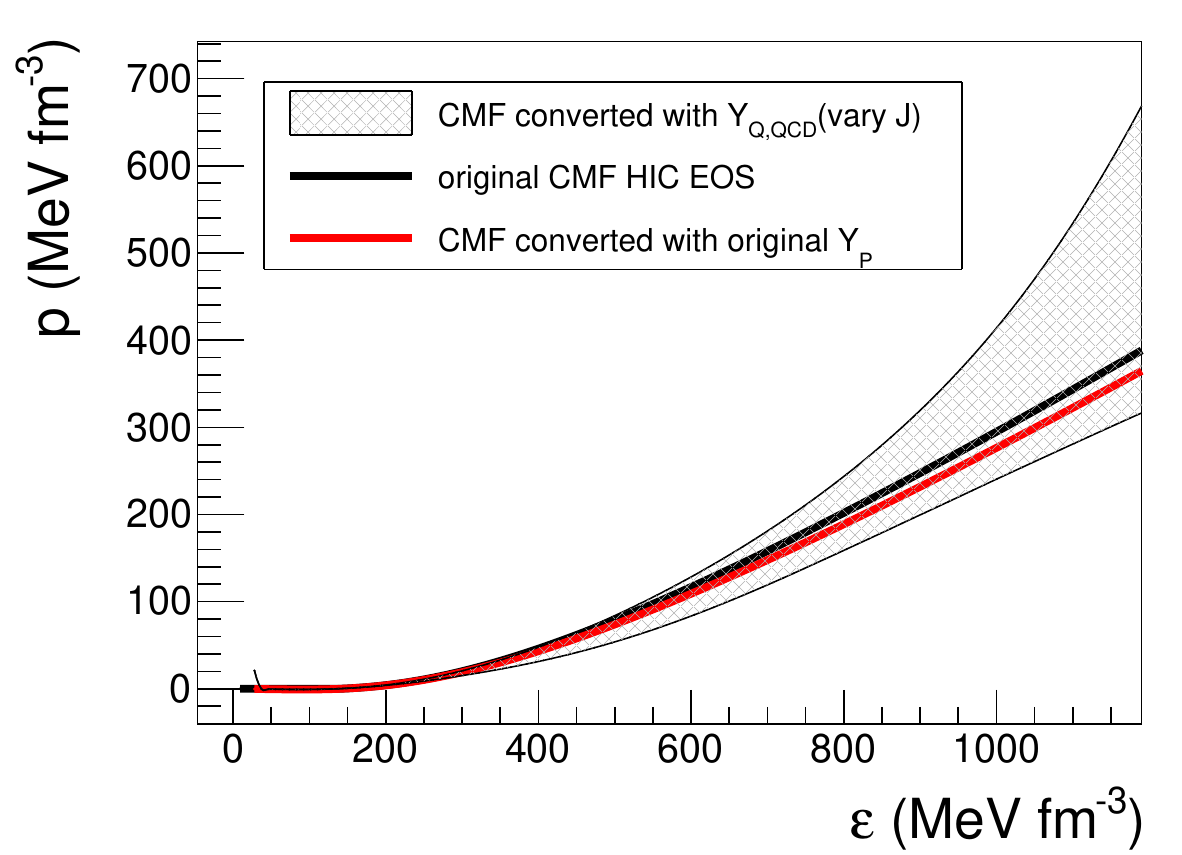} 
\caption{Comparison between the original CMF EOS for $Y^{\rm{const}}_{Q,\rm{QCD}}=0.5$, the converted CMF EOS (from CMF NS EOS into HIC EOS at $Y^{\rm{const}}_{Q,\rm{QCD}}=0.5$)  using the symmetry energy expansion, and $Y_{Q,\rm{QCD}}$ obtained directly from CMF, and the converted CMF EOS using the symmetry energy expansion and $Y_{Q,\rm{QCD}}$ from Eq.~(\ref{long}). The plot covers a density range of $n_B=[0$--$5] n_{\rm{sat}}$ .}
\label{fig:CMF2}
\end{figure}
We note here that it is only possible to obtain the correct binding energy after one subtracts the lepton contribution from the CMF NS EOS, even though at $n_{\rm{sat}}$, the corresponding $Y_Q$ is quite small~\cite{Drischler:2020fvz}: on the order of $Y_Q\approx 0.05$.

Next, we convert the CMF NS EOS using the symmetry energy expansion, both for the exact $Y_{Q,\rm{QCD}}$ obtained from CMF and for $Y_{Q,\rm{QCD}}$ from the expansion in Eq.~(\ref{long}). As before, we take the exact coefficients for $E_{\txt{sym,sat}}$, $L_{\txt{sym,sat}}$, and $K_{\txt{sym,sat}}$, and vary $J_{sat,sym}$. In Fig.~\ref{fig:CMF2}, we show the result for the converted CMF HIC EOS vs.\ the original CMF HIC EOS for SNM. We find that using the exact $Y_{Q,\rm{QCD}}$, one obtains a very good reproduction of the CMF EOS for SNM. When using the expansion of $Y_{Q,\rm{QCD}}$ from Eq.~(\ref{long}), more uncertainty occurs given that $J_{\rm{sat,sym}}$ is unknown, however, this only becomes a significant effect at high densities. Thus, we find that our approach is reasonable (especially for low densities), even though there is still significant uncertainty in the large density regime.

\section{Constraints on the symmetry energy expansion}
\label{sec:constraints_on_the_symmetry_energy_expansion}

The general methodology used in this paper is to take EOSs shown to be consistent with neutron-star observations and convert them into EOSs corresponding to a different value of $Y_{Q,\rm{QCD}}$ while ensuring that the EOSs remain causal and stable. The EOSs we use here are not (by far) the only possible EOSs that fit within the constraints set by neutron star observations, however, they do both obey NICER and LIGO/Virgo constraints and support heavy neutron stars with masses $M\geq 2 M_{\odot}$ (even up to $M\geq 3 M_{\odot}$). For each NS EOS, the conversion to a chosen $Y_{Q,\rm{QCD}}$, given many allowed choices of the coefficients $\left\{E_{\txt{sym,sat}},L_{\txt{sym,sat}},K_{\txt{sym,sat}}, J_{\txt{sym,sat}}\right\}$, yields different possible HIC EOSs, some of which lead to acausal solutions $c_s^2>1$ or unstable solutions \mbox{$c_s^2<0$}\footnote[2]{While an EOS including an unstable region (where $c_s^2 <0$) is perfectly well-defined, and in fact necessary for the description of the spinodal region of a first-order phase transition, the symmetry energy expansion is a simple Taylor series that cannot be expected to reliably describe the nonmonotonic behavior of $c_s^2$, occurring in this case, at large distances from $n_{\rm{sat}}$.}. Here, we only apply the stability constraint at $n_B>0.9n_{\rm{sat}}$ to preserve the description of the first-order nuclear liquid-gas phase transition at lower densities; likewise, we only apply the stability constraint up to a chosen density value, $n_{\txt{cut}}\approx 6n_{\rm{sat}}$, which generally covers the range of densities relevant to both heavy-ion collisions and neutron stars. Additionally, we also perform a check of the properties of the converted EOS at saturation. The saturation density, where the pressure turns from negative to positive, is required to be in the range of 0.14 to 0.18 $\rm{fm}^{-3}$. The binding energy at saturation, $B = \varepsilon_{\rm{HIC,sym}}/n_B-m_N$, is required to be in the range of $-14$ to $-18$ MeV. The incompressibility, $K_0=9\frac{\partial p}{\partial n_B}|_{n_B=n_0}$, is required to be larger than $200$ MeV; since the values of $K_0$ tends to be small in our converted EOSs, there is no need to set a constraint on the upper value of $K_0$. 

\begin{figure}[t]
\centering
\includegraphics[width=\linewidth]{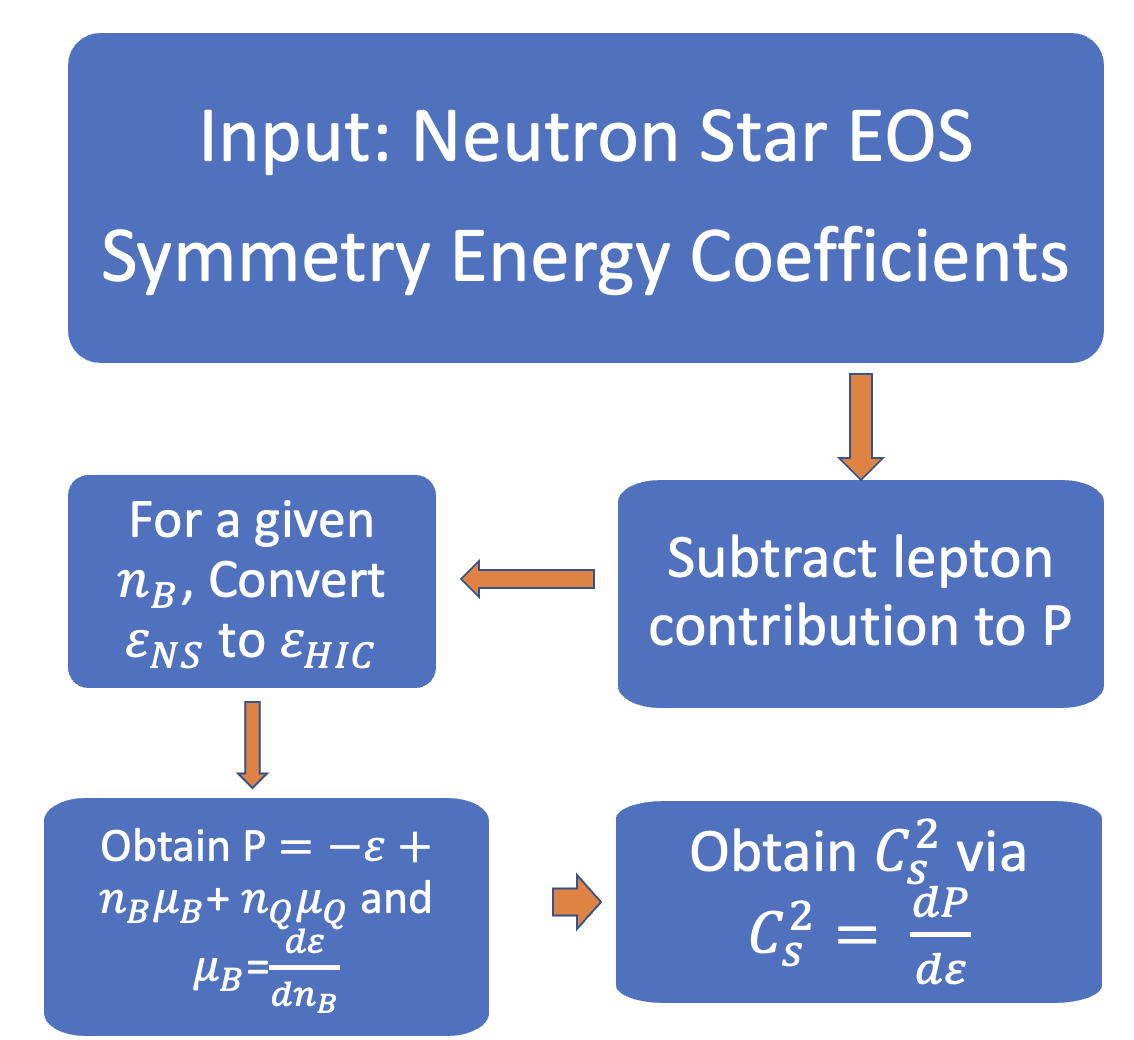}
\caption{Algorithm used to convert a NS EOS at $T=0$ into a HIC,sym EOS at $T=0$.} 
\label{fig:algo}
\end{figure}

Once the acausal and unstable EOSs, as well as EOSs which do not exhibit reasonable saturation properties, are removed, the remaining EOSs can be used to define a band of viable EOSs. Moreover, we can study whether there is a preference for certain values of $\left\{E_{\txt{sym,sat}},L_{\txt{sym,sat}},K_{\txt{sym,sat}}, J_{\txt{sym,sat}}\right\}$ that produce physical, causal, and stable EOSs.  
A summary of our algorithm is shown in Fig.~\ref{fig:algo}, where the crucial inputs are the NS EOS and the symmetry energy expansion coefficients. The precise thermodynamic conversion from neutron stars to heavy-ion collisions was discussed already in Sec.~\ref{sec:EOS}, however, we briefly summarize it here. The neutron-star energy density (the NS EOS) and the symmetry energy expansion coefficients are inputs used in Eq.~(\ref{eqn:eHIC}) (or, equivalently, Eq.~(\ref{symexpand})) to obtain the corresponding HIC energy density (the HIC EOS). Because we use Eq.~(\ref{long}) for $Y_{Q,\rm{QCD}}$ in Eq.\ (\ref{eqn:eHIC}) (or Eq.~(\ref{symexpand})), which tends to lead to a divergent behavior when $n_B\rightarrow 0$, we enforce a maximum value $Y_{Q,\rm{QCD}} \leq 1$. With the obtained $\varepsilon_{\txt{HIC}}$, we calculate $\mu_B$ using $\mu_B = d\varepsilon_{\txt{HIC}}/dn_B|_{n_Q=const}$,
which then allows us to obtain the pressure from Eq.~(\ref{pressures}). Finally, we obtain $c^{2}_{s}$ from Eq.~(\ref{cs2eq}).

\subsection{$c_s^2$ from NS to HIC}
\label{cs2_NS_to_HIC}

We start with a discussion about the current knowledge of the symmetry energy expansion coefficients, summarized in Tab.~\ref{tab:coef}. In~\cite{Li:2019xxz}, the magnitudes of $E_{\txt{sym,sat}} = 31.6 \pm 2.7 $ MeV and $L_{\txt{sym,sat}}=58.9 \pm 16$ MeV are obtained from 28 analyses of both nuclear experiments and astrophysics observations. As mentioned before, PREXII results demonstrate a much larger $L_{\txt{sym,sat}}=106\pm 37$ MeV~\cite{PREX:2021umo,Reed:2021nqk}. We are not aware of any experimental constraints on~$K_{\txt{sym,sat}}$ and~$J_{\txt{sym,sat}}$ and, therefore, must rely on theoretical estimates. Thus, to provide a wide prior and ensure that we are not strongly biasing our work, we explore wide ranges of values of the symmetry energy expansion coefficients, summarized in Tab.~\ref{tab:symmetry_energy_expansion_coefficients_ranges}.

\begin{table}
\centering
\begin{tabular}{c|r|r}
Coefficient & ~~~ Range&  ~~~Step size \\
\toprule
$E_{\txt{sym,sat}}$ \ \ \   & 27 -- 40 &  1 MeV\\
\hline
$L_{\txt{sym,sat}}$ \ \ \   & 20 -- 130 &  10 MeV \\
\hline
$K_{\txt{sym,sat}}$ \ \ \    & -220 -- 180 &  50 MeV \\
\hline
$J_{\txt{sym,sat}}$ \ \ \    & -200 -- 800 &  100 MeV\\
\end{tabular}
\caption{Ranges of values of the symmetry energy expansion coefficients used in the following study. 
}
\label{tab:symmetry_energy_expansion_coefficients_ranges}
\end{table}

\begin{figure}[b]
\centering
\includegraphics[width=\linewidth]{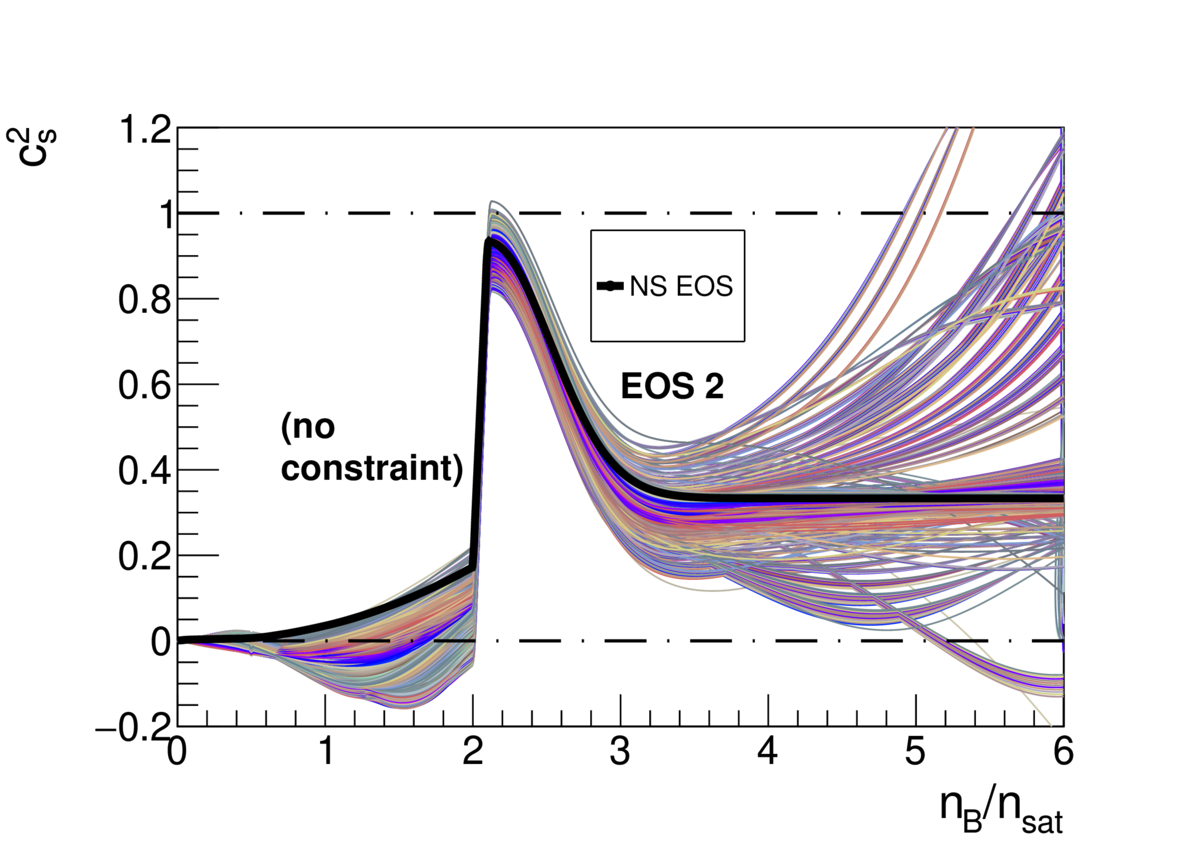}
\caption{Speed of sound squared as a function of normalized baryon density, calculated for HIC EOSs obtained from a NS EOS with a peak in $c_s^2$ at $n_B = 2n_{\rm{sat}}$ (eos2) by using the symmetry energy expansion and varying all 4 symmetry energy expansion coefficients over their allowed ranges. The original NS EOS is shown in black. Here, we show EOSs for all symmetry energy coefficient combinations without applying causality and stability constraints. The horizontal dot-dashed lines show constrains on causality and stability that we enforce in the following figures.
} 
\label{fig:Precausality}
\end{figure}

\begin{figure}[t]
\centering
\includegraphics[width=1\linewidth]{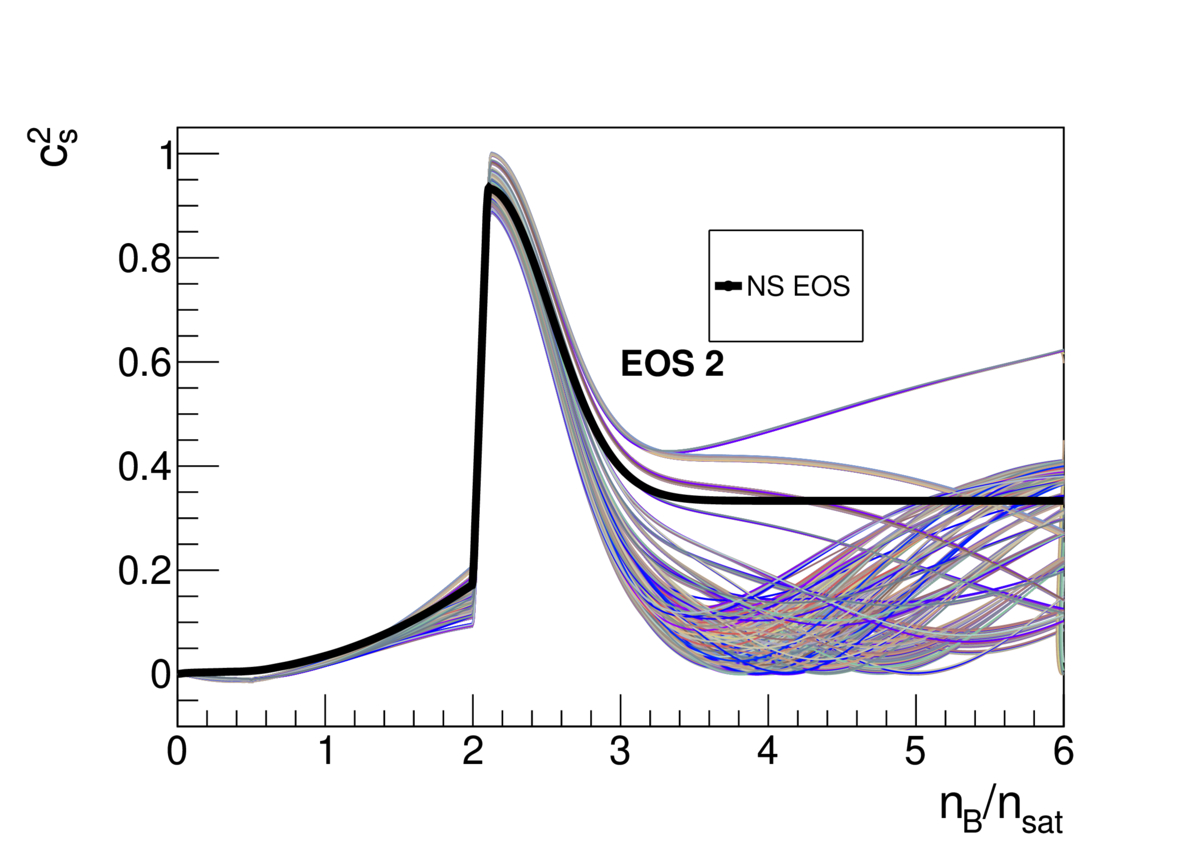}\\
\includegraphics[width=1\linewidth]{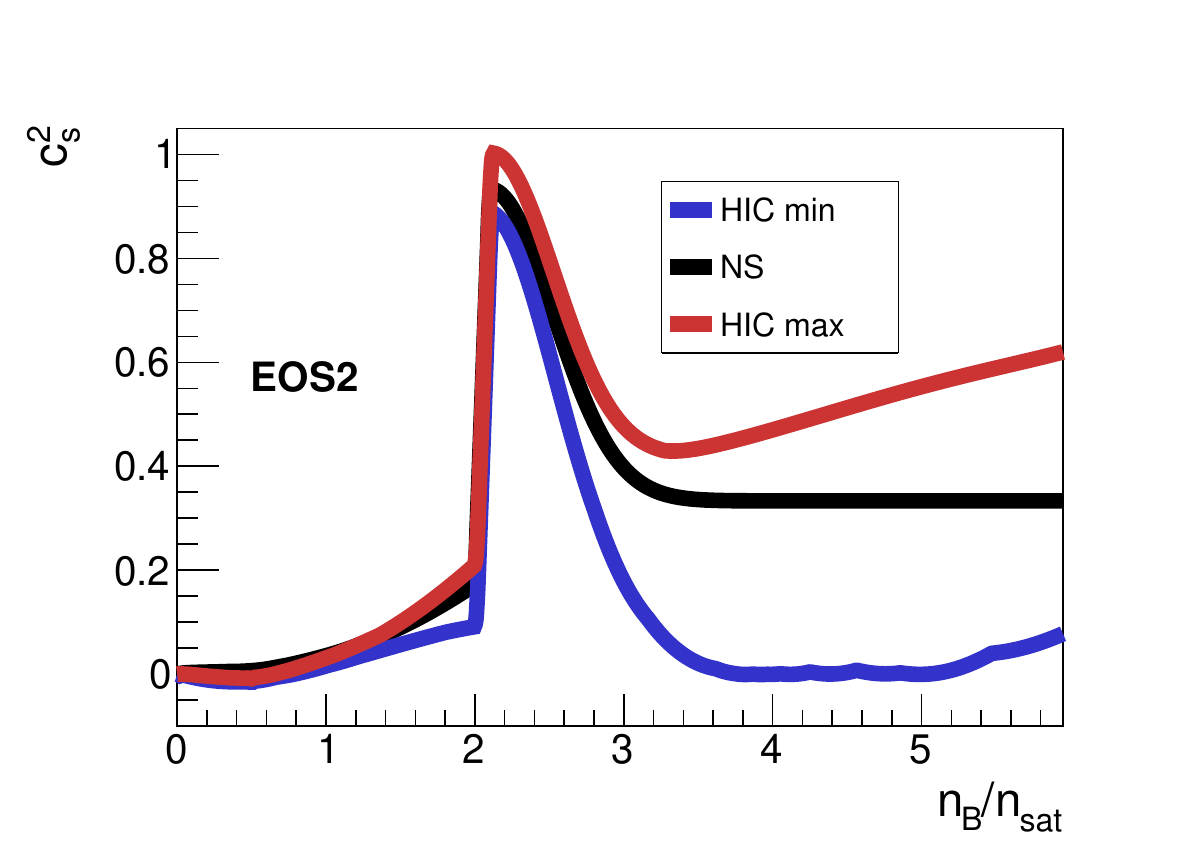}     
\caption{\textit{Top}: Same as in Fig.~\ref{fig:Precausality}, but only showing possible combinations of symmetry energy expansion coefficients which lead to results that are causal, stable, and satisfy nuclear matter properties at saturation density. \textit{Bottom}: The minimum and maximum bands encompassing all obtained HIC EOSs shown in the top panel.
}
\label{fig:meddens}
\end{figure}

Next, we take all possible combinations of $\left\{E_{\txt{sym,sat}},L_{\txt{sym,sat}},K_{\txt{sym,sat}}, J_{\txt{sym,sat}}\right\}$ from Tab.~\ref{tab:symmetry_energy_expansion_coefficients_ranges}, use them in Eq.~\eqref{eqn:eHIC} (where we take $n_{\rm{sat}} = 0.160~\rm{fm}^{-3}$) to convert the chosen NS (QCD only) EOS (eos1, eos2, or eos3) into the HIC,sym EOS, and evaluate the resulting EOSs for causality and stability up to $n_{\txt{cut}}$, as well as check the saturation properties. Using our priors on the symmetry energy coefficients, we obtain $c_s^2$ for SNM from eos2 as shown in Fig.~\ref{fig:Precausality}. In the figure, no causality, stability, or saturation properties constraints have been applied and one can see that at high densities, $n_B\gtrsim 3.5 n_{\rm{sat}}$, some of the symmetry energy expansion coefficients lead to $c_s^2$ that are either negative (unstable) or larger than 1 (acausal). The constraints of causality and stability (that we will enforce in the following) are shown by horizontal dot-dashed lines.

By constraining the possible combinations of the symmetry expansion coefficients to only those that preserve the causality, stability, and saturation properties, we can then determine the underlying distribution of the viable parameter space for $\left\{E_{\txt{sym}},L_{\txt{sym}},K_{\txt{sym}}, J_{\txt{sym}}\right\}$; we stress here that each of the considered NS EOSs features a large peak in the speed of sound, and, therefore, our constraint on the symmetry energy coefficients is based on the assumption that they should allow one to convert such an EOS from $Y_{Q} = Y_{Q,\rm{QCD}}^{\txt{NS}}$ to $Y_{Q,\rm{QCD}}^{\txt{const}} \approx 0.5$ while preserving causality, stability, and properties at $n_{\rm{sat}}$.

The dependence of $c^{2}_{s}$ on $n_B$ for SNM ($Y_{Q,\rm{QCD}}^{\txt{const}}=0.5$) EOSs, converted from eos2, that fulfill causality, stability, and nuclear matter properties at saturation density ($-18~\rm{MeV} < B <-14~\rm{MeV}$ and $K_0 > 200~\rm{MeV}$), is shown in Fig.~\ref{fig:meddens}. The top panel demonstrates the structure of the individual EOS from various sets of the symmetry energy expansion coefficients. The bottom panel compares the original NS EOS (eos2), shown in black, to the extracted maximum (red) and minimum (blue) bands which form an envelope over all EOSs shown in the top panel. To extract the bands, negative values of $c_s^2<0$ and acausal values of $c_s^2>1$, which may still occur for $n_B > n_{\txt{cut}}$, are set to 0 and 1, respectively.

Interestingly, regardless of the choice of symmetry energy expansion coefficients, the peak in $c_s^2$ for HIC EOSs is located in the exact same density region as for the NS EOS (here, eos2). Depending on the particular set of the expansion coefficients used, there is a (rather slight) shift in the overall magnitude of the peak. Following the peak, there is a huge variation in the HIC EOSs, rendering them nearly unconstrained past the peak. This wide spread in the EOSs for densities above the peak is not surprising, since it reflects the spread in the allowed values of the symmetry energy expansion coefficients, in particular the higher-order ones. In addition, if the peak in the NS EOS occurs due to the appearance of strange baryons or quarks, then one would require an altered symmetry energy expansion to account for these new degrees of freedom.
We note that we purposely incorporated a large range of $E_{\txt{sym},\rm{sat}}$ and $L_{\rm{sym,sat}}$, see Tab.~\ref{tab:symmetry_energy_expansion_coefficients_ranges}, which is likely beyond what one would expect from experimental constraints, see Tab.~\ref{tab:coef}. However, we verified that even if we include tighter constraints on $E_{\txt{sym},\rm{sat}}$, we find similar results.

\begin{figure*}[t!]
\centering
\includegraphics[width=0.8\linewidth]{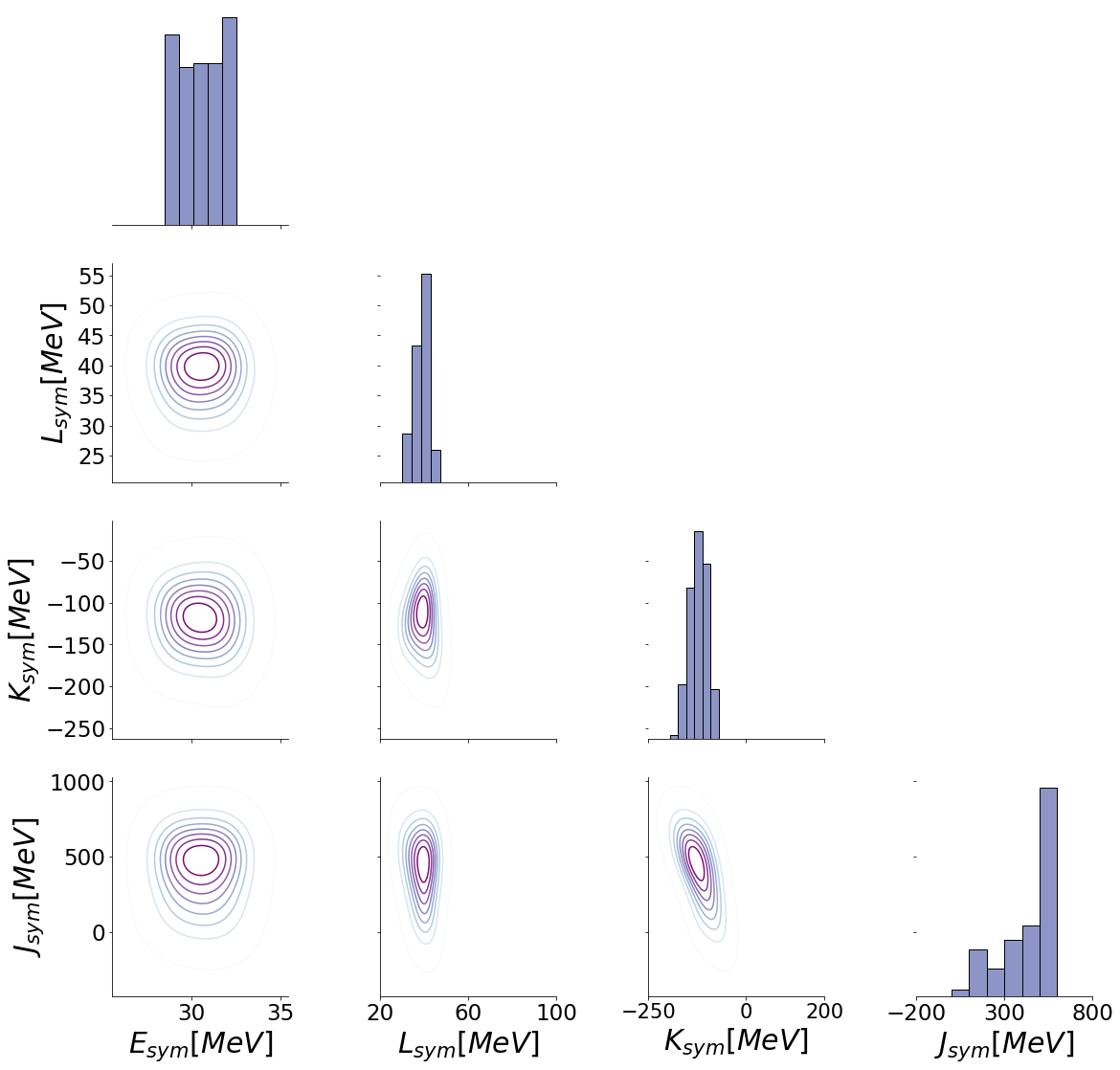}
\caption{Corner plot for the symmetry energy expansion coefficients $\left\{E_{\txt{sym,sat}},L_{\txt{sym,sat}},K_{\txt{sym,sat}}, J_{\txt{sym,sat}}\right\}$ as constrained by demanding causality, stability, and reasonable nuclear matter saturation properties for HIC EOSs obtained from eos2.
}
\label{fig:corner_med}
\end{figure*}

In Fig.~\ref{fig:corner_med}, we show a corner plot for the distributions of combinations of $\left\{E_{\txt{sym,sat}},L_{\txt{sym,sat}},K_{\txt{sym,sat}}, J_{\txt{sym,sat}}\right\}$ leading to causal, stable, and physical (satisfying saturation properties) HIC EOSs from eos2. We emphasize that this corner plot is only based upon ensuring causality and stability from $0.9\ n_{\txt{sat}}$ up to $n_{\rm{cut}} = 6\ n_{\txt{sat}}$ as well as satisfying saturation properties. We find a number of interesting features. For instance, we find a preference for a small $L_{\txt{sym,sat}}$. We also find a preference for a negative value of $K_{\txt{sym,sat}}$, while $J_{\txt{sym,sat}}$ appears to have a preference for a (large) positive value.

\begin{figure}[t]
\centering
\includegraphics[width=0.99\linewidth]{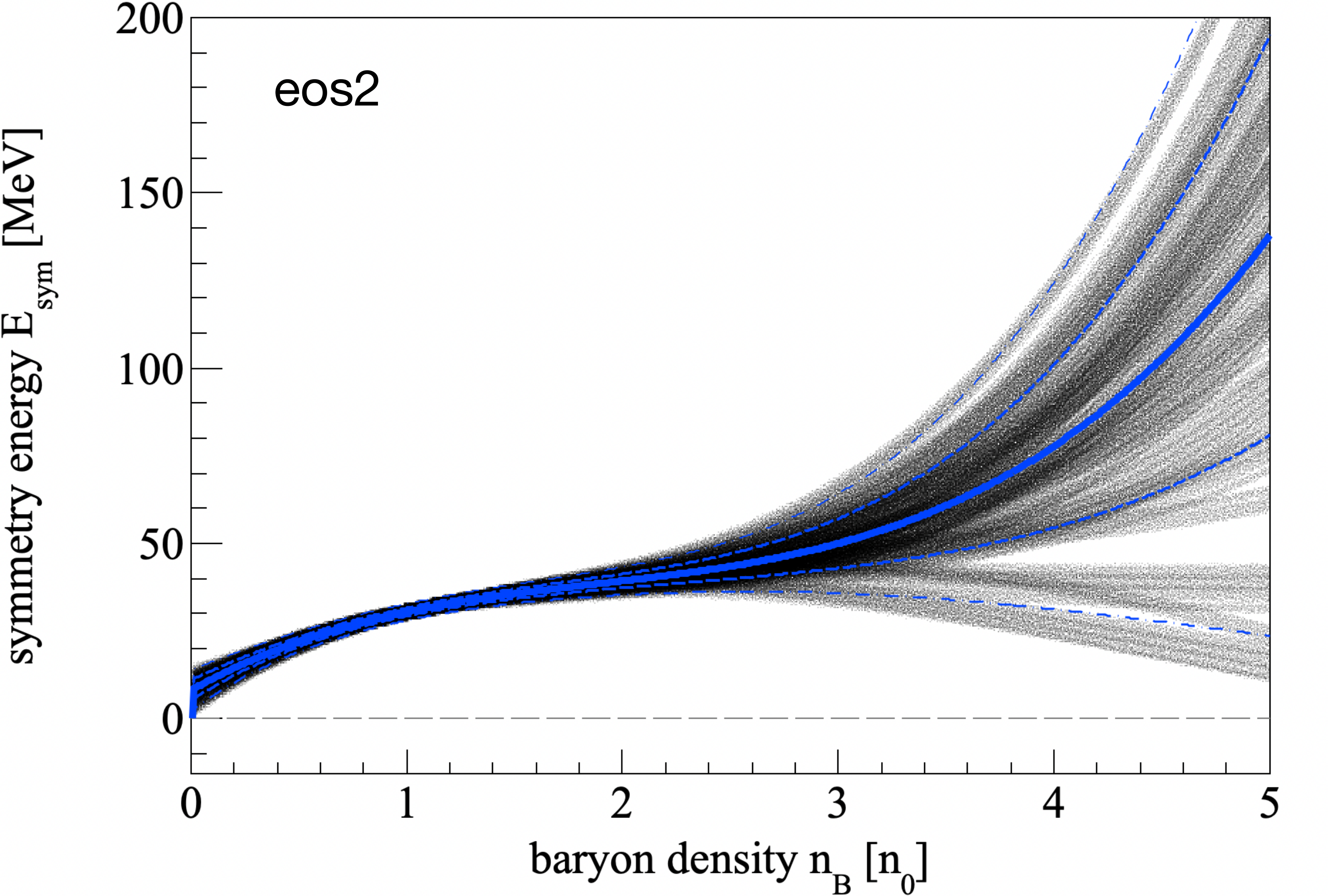} 
\caption{Scatter plot of the symmetry energy $E_{\rm{sym}}$ as a function of baryon density $n_B$, obtained using all sets of coefficients for which the converted eos2 HIC EOS satisfies the physical, causality, and stability constraints (see Fig.~\ref{fig:corner_med}). The thick solid line corresponds to the mean symmetry energy, while the thick (thin) dashed line corresponds to the $\pm1\sigma$ ($\pm2\sigma$) contour around the mean. 
}
\label{fig:symmetry_energy_scatter_eos2}
\end{figure}

\begin{figure}[t]
\centering
\includegraphics[width=0.99\linewidth]{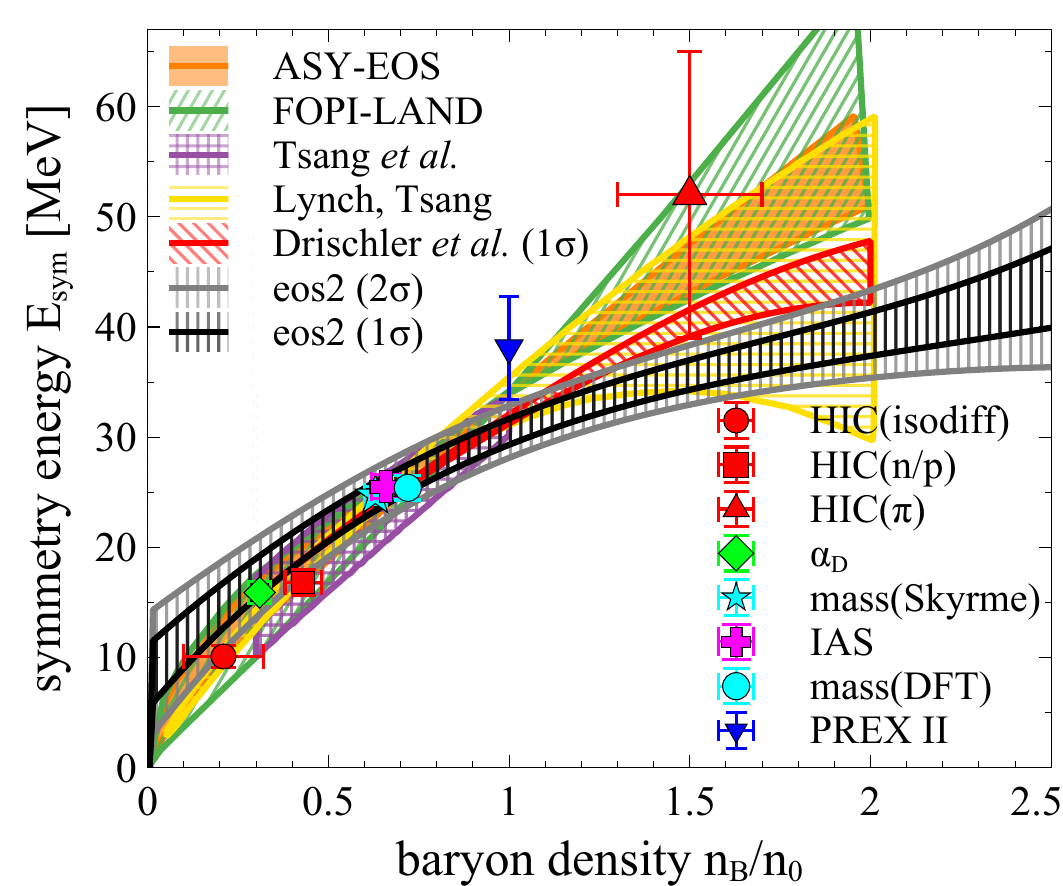} 
\caption{The $\pm1\sigma$ (region with black vertical stripes) and $\pm2\sigma$ (region with gray vertical stripes) contours of the symmetry energy $E_{\rm{sym}}$ as a function of baryon density $n_B$, obtained in this work as in Fig.~\ref{fig:symmetry_energy_scatter_eos2}, compared with selected constraints: N3LO chiral effective field theory calculations (band labeled as ``Drischler et al.'')~\cite{Drischler:2020yad}, constraints from comparisons of experimental data to results of transport simulations of heavy-ion collisions (bands labeled as ``ASY-EOS''~\cite{Russotto:2016ucm}, ``FOPI-LAND''~\cite{Russotto:2011hq}, and ``Tsang \textit{et al.}''~\cite{Tsang:2008fd}  and symbols labeled as ``HIC(isodiff)'', ``HIC(n/p)'', and ``HIC($\pi$)''~\cite{Lynch:2021xkq}), constraints obtained in Ref.~\cite{Lynch:2021xkq} based on a novel interpretation of analyses of dipole polarizability $\alpha_D$~\cite{Zhang:2015ava} (green diamond), of nuclear masses in DFTs~\cite{Kortelainen:2010hv,Kortelainen:2011ft} (cyan dot symbol) and in Skyrme models~\cite{Brown:2013mga} (cyan star symbol), of Isobaric Analog States (IAS) energies~\cite{Danielewicz:2013upa} (magenta plus symbol), and of PREX-II experiment~\cite{PREX:2021umo} (blue inverted triangle symbol), as well as the $68\%$ confidence region consistent with the best fit of experimental data points (band labeled ``Lynch, Tsang'')~\cite{Lynch:2021xkq}. Figure adapted from~\cite{Sorensen:2023zkk}.
}
\label{fig:symmetry_energy_constraints}
\end{figure}

Using all of the sets of coefficients producing physical, causal, and stable HIC EOSs from eos2, we investigate the corresponding values of the symmetry energy $E_{\rm{sym}}$ as a function of baryon density. A scatter plot of all such obtained symmetry energy values is shown in Fig.~\ref{fig:symmetry_energy_scatter_eos2} along with the $\pm1\sigma$ and $\pm2\sigma$ contours around the mean symmetry energy for each value of $n_B$.
As could be seen already in Fig.~\ref{fig:corner_med}, the requirement of a physical HIC EOS yields a rather tight constraint on the symmetry energy in the vicinity of $n_{\rm{sat}}$ or, equivalently, on $E_{\rm{sym,sat}}$ and $L_{\rm{sym, sat}}$ as the lowest expansion coefficients are most important for the expansion in this region. This is, in fact, expected given that for densities below $2n_{\rm{sat}}$, eos2 simply reproduces the Togashi EOS which is characterized by particular values of $E_{\rm{sym,sat}}$ and $L_{\rm{sym, sat}}$: because we require that the saturation properties are reasonably reproduced, the accepted sets of symmetry energy expansion coefficients cannot contain $E_{\rm{sym,sat}}$ and $L_{\rm{sym, sat}}$ which differ wildly from those of Togashi. On the other hand, the constraints of causality and stability apply at higher densities and result in a strong preference for negative values of $K_{\rm{sym, sat}}$ and positive values of $J_{\rm{sym, sat}}$. As a result, the rise of the symmetry energy with density at $n_B > n_{\rm{sat}}$ is first suppressed, showing the influence of $K_{\rm{sym, sat}}$, and then accelerates again when $J_{\rm{sym, sat}}$ starts to dominate, producing a characteristic inflection as a function of~$n_B$. 
One might consider also this feature to be a trivial consequence of the applied constraints and the fact that the symmetry energy is modeled with a polynomial: it is indeed intuitive that if the higher-order coefficients, $K_{\rm{sym,sat}}$ and $J_{\rm{sym,sat}}$, have opposite signs, then altogether their contributions lead to more moderate, and therefore causal and stable, behavior than in the case if they were either both positive (and therefore more likely to lead to acausal behavior at moderate densities) or negative (and therefore more likely to lead to unstable behavior at moderate densities).
However, it is interesting to note that a similar structure with a stalled rise above $n_{\rm{sat}}$, suggesting $K_{\rm{sym, sat}} < 0$, has been obtained both in chiral effective field theory calculations and in a recent fit to experimental data points, both of which are shown as bands labeled as ``Drischler et al.''~\cite{Drischler:2020yad} and ``Lynch, Tsang''~\cite{Lynch:2021xkq}, respectively, in Fig.~\ref{fig:symmetry_energy_constraints}. Therefore, our results for $K_{\rm{sym, sat}}$ and $J_{\rm{sym, sat}}$ might point toward robust features of the symmetry energy at higher densities.
Finally, we note that results corresponding to those shown in Fig.~\ref{fig:symmetry_energy_scatter_eos2}, but obtained for eos1 and eos3, are very similar, as can be seen in Appendix~\ref{sec:SymEn_13}.

\subsection{The limit of vanishing $n_B$}
\label{sec:the_limit_of_vanishing_nB}

The very low density limit of NS EOS represents the crust and it contains quite different degrees of freedom than the matter created in heavy-ion collisions (beyond containing a degenerate Fermi gas of electrons). As the density increases within the crust, the influence of the nuclei becomes larger, firstly contributing primarily to the energy density but eventually, depending on the lattice structure that they form, also providing a substantial contribution to the pressure. For each $n_B$, a specific nucleus  minimizes the free energy. As one increases $n_B$, heavier and heavier nuclei that are also more neutron-rich dominate the EOS. Eventually, the neutron drip line is reached around $n_B \approx 10^{-4}$ $n_{\rm{sat}}$, at which point the nuclear matter EOS has inputs from a combination of nuclei, electrons, and free neutrons. Further details can be found, e.g., in \cite{Baym:1971pw} and in various textbooks \cite{1986bhwd.book.....S,1997csnp.book.....G}.

Systems produced in heavy-ion collisions contain matter originating from the collision of two heavy, fully-ionized nuclei. These events are extremely short-lived, such that the only relevant interactions are those on the time scales of the strong force. While leptons (and antileptons) are produced in the collisions, they generally immediately leave the collision region without further interactions and thus do not contribute to the EOS. On the other hand, light nuclei are not only produced in the collision region~\cite{Dorso:1994js,Puri:1996qv}, but also contribute to its dynamics; this remains true even for the highest energy collisions~\cite{ALICE:2015wav,STAR:2019sjh}. 
Moreover, in recent years production of various light hypernuclei \cite{ALICE:2015oer} and antihypernuclei \cite{STAR:2010gyg,STAR:2023fbc2} has been measured in heavy-ion collisions.
Nevertheless, the produced matter at low temperatures and moderate densities is predominantly composed of protons and neutrons.

The single-particle potentials used in simulations of heavy-ion collisions depend on the baryon (and, in cases where isospin dependence is included, also isospin) number of the affected particles, and one can parametrize the underlying HIC EOS starting from the simple concept of uniform nuclear matter, that is matter composed of only protons and neutrons. In this situation, the energy per baryon $\varepsilon/n_B$ converges to the nucleon mass in the limit of vanishing $n_B$. This, however, is not the case for an EOS obtained through the symmetry energy expansion of a NS EOS. The reasons for this are twofold: First, as already explained above, the two EOSs employ different degrees of freedom. Second, the expansion is only strictly well-defined around $n_{\rm{sat}}$ and, consequently, does not necessarily lead to reasonable results as $n_B\rightarrow 0$; in particular, in our approach the approximation for $Y_Q$ diverges in this limit (see Fig.~\ref{fig:YQ} in Appendix~\ref{app:YQ}), leading to an EOS incompatible with the assumptions used in modeling the EOS for heavy-ion collisions. For this reason, we choose to only use the HIC EOS obtained from a symmetry energy expansion of a NS EOS above some chosen low density $n_B^{\rm{low}}$ where its behavior is reasonable in the context of heavy-ion collisions. For $n_B<n_B^{\rm{low}}$, we smoothly match the converted EOS to an EOS obtained from the density functional (VDF) model \cite{Sorensen:2020ygf} EOS whose behavior as $n_B \to 0$ is compatible with the basic assumptions about uniform nuclear matter. In particular, such an EOS then satisfies two necessary conditions at $n_B\rightarrow 0$: 
\begin{enumerate}
    \item the energy density per baryon density, $\varepsilon/n_B$, is equal to the mass of the nucleon,
    \begin{equation}
        \frac{\varepsilon_{\rm{HIC,sym}}}{n_B}\Big|_{n_B\rightarrow 0} = m_N ~,
    \end{equation}
    \item the speed of sound squared is zero,
    \begin{equation}
        c^2_s\Big|_{n_B\rightarrow 0} = 0~.
    \end{equation}
\end{enumerate}

The two above conditions can be achieved by smoothly matching the VDF EOS and our converted EOS at $n_B^{\rm{low}}=0.5n_{\rm{sat}}$. The smooth matching is done using a hyperbolic tangent function; additionally, we also ensure that certain properties of SNM are the same between the VDF and our converted EOS. Note that only SNM is relevant since this matching is done at $Y_Q^{\rm{QCD}}=0.5$. The properties that we ensure are identical between the two matched EOSs are:
 \begin{itemize}
 \item the saturation density, $n_{\rm{sat}}$ 
     \item the binding energy of SNM at $n_{\rm{sat}}$, $B = \epsilon_{\rm{SNM}}(n_{\rm{sat}})/n_{\rm{sat}} - m_N$ 
     \item the nuclear incompressibility, $K_{\rm{SNM}}(n_{\rm{sat}}) = 9\frac{dp_{\rm{SNM}}}{dn_B}|_{n_{\rm{sat}}}$ 
 \end{itemize}
We begin with a NS EOS that has been converted into a SNM EOS, $\varepsilon_{\rm{HIC,sym}}$. We then calculate $\left\{n_{\rm{sat}},E_{0, \rm{SNM}},K_{\rm{SNM}}(n_{\rm{sat}})\right\}$ for the given converted EOS. These are then used as input parameters for parametrizing the VDF EOS for SNM, denoted here as $\varepsilon_{\rm{VDF}}$.

Having in this way obtained the converted EOS, $\varepsilon_{\rm{HIC,sym}}$, and the VDF EOS with the same saturation properties, $\varepsilon_{\rm{VDF}}$, we can combine the two using a smooth matching function as follows:
\begin{equation}
\varepsilon(n_B)_{\rm{fin}}=s(n_B)\varepsilon_{\rm{HIC,sym}}(n_B)+\left[1-s(n_B)\right]\varepsilon_{\rm{VDF}}(n_B) ~,
\end{equation}
where $\varepsilon(n_B)_{\rm{fin}}$ is our final energy density that is used as the basis for the \texttt{SMASH} input. While the smoothing function can be defined in a number of ways, here we use a hyperbolic tangent, i.e.,
\begin{equation}
    s(n_B)=0.5+0.5 \tanh\big[(n_B-n_{B}^{\rm{low}})/a\big] ~,
\end{equation}
where $n_{B}^{\rm{low}}$ was already defined above as the matching density and $a$ is a free parameter that determines the region in $n_B$ where the transition from one EOS to another is smoothed over; a larger (smaller) value of $a$ leads to a smoother (sharper) transition. Here we used $n_{B}^{\rm{low}}=0.5n_{\rm{sat}}$ and $a=0.03\; fm^{-3}$, which we have tested to provide a reasonably smooth transition between the EOSs. 

\begin{figure}
\centering
\includegraphics[width=0.95\linewidth]{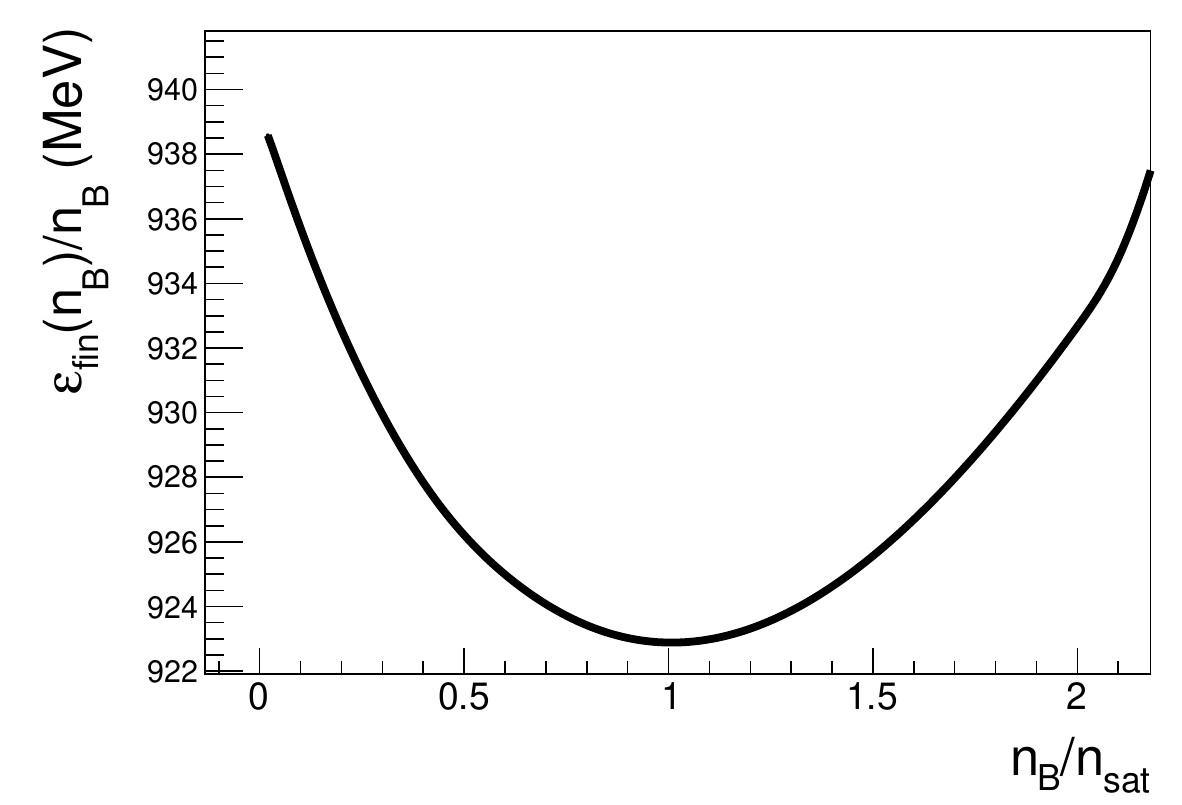}\\
\includegraphics[width=\linewidth]{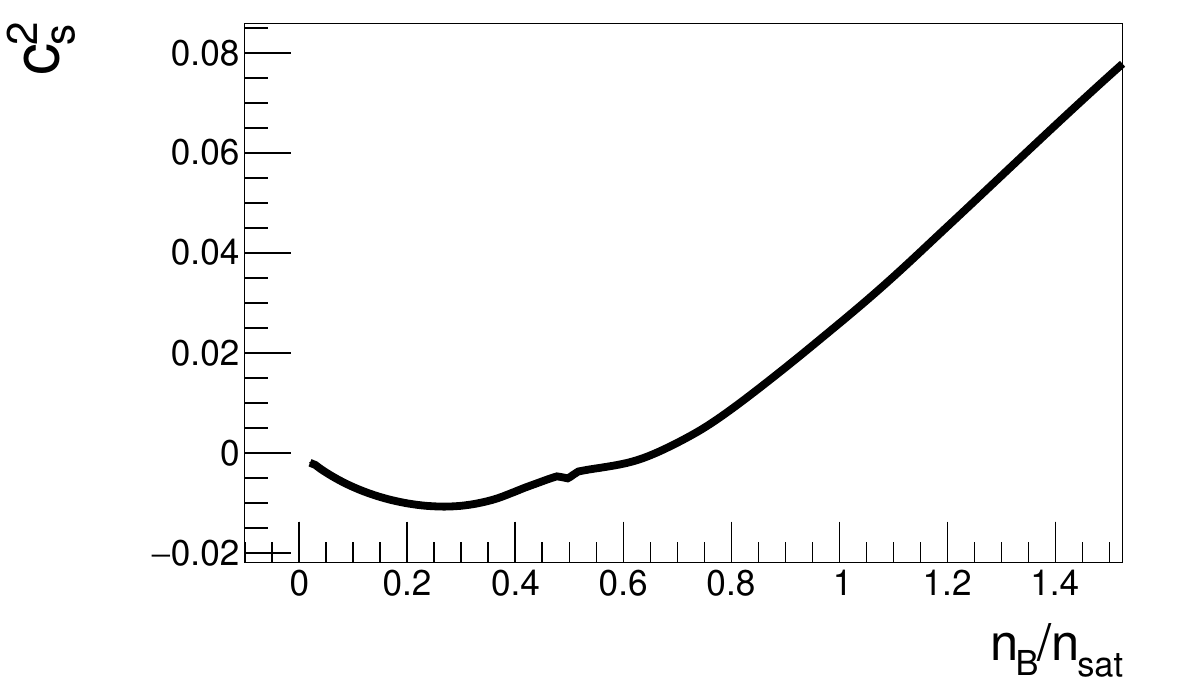} 
\caption{\textit{Top}: An example of an energy per particle, $\varepsilon(n_B)_{\rm{fin}}/n_B$, of the final EOS obtained from matching our converted EOS, $\varepsilon_{\rm{HIC,sym}}(n_B)$, with a VDF EOS, $\varepsilon_{\rm{VDF}}(n_B)$, to ensure reasonable properties at vanishing baryon densities. \textit{Bottom}: $c_s^2$ obtained from the final EOS shown in the top panel.
}
\label{fig:vdfmatch}
\end{figure}
An example of a single realization of the final EOS, $\varepsilon(n_B)_{\rm{fin}}$, is shown in the top panel of Fig.~\ref{fig:vdfmatch}. In this example, the matched EOSs have the following common saturation properties: $n_{\rm{sat}}=0.1616\;\rm{fm}^{-3}$, $E_{0, \rm{SNM}}(n_{\rm{sat}})=-16.0352 $ MeV, and $K_{\rm{SNM}}(n_{\rm{sat}})=218.367 $ MeV. The pressure and the speed of sound squared are then calculated from $\varepsilon(n_B)_{\rm{fin}}$ using the thermodynamic relations from Eqs.~(\ref{pressureeq}) and (\ref{cs2eq}). In Fig.~\ref{fig:vdfmatch}, we can see that as $n_B \rightarrow 0$, we have $\varepsilon/n_{B} \rightarrow m_N $ (top panel) and $c_s^2\rightarrow 0$ (bottom panel).

\section{Comparisons to heavy-ion collective flow data}

The most energetic heavy-ion collisions, obtained with beams at the center-of-mass energy of $\sqrt{s_{\txt{NN}}}=5.02$ TeV (available at the Large Hadron Collider), can reach temperatures up to approximately $T_{\txt{max}}\approx 650 $ MeV\footnote[3]{This number comes from simulations using v-USPhydro~\cite{Noronha-Hostler:2013gga,Noronha-Hostler:2014dqa}, e.g., by looking at the temperature profiles from the simulations performed in Ref.~\cite{Alba:2017hhe}.}. At moderate values of rapidity, almost all of the particles measured in these events are created from the energy deposited in the collision region rather than coming from the original nuclei. Because quarks and antiquarks are always produced in pairs, the created systems have very low net baryon or net strangeness density. Such matter is similar to matter that filled the Universe in the first microseconds after the Big Bang, but different from the dense nuclear matter in neutron stars and supernovae explosions. Only once the beam energy is lowered, a relevant number of baryon charges originating from the colliding nuclei is trapped in or is otherwise affected by the expanding collision region, such that the evolving systems are characterized by a finite net baryon density $n_B$ at moderate values of rapidity. Note, however, that while matter produced in these experiments can be characterized by a comparable range of probed baryon densities, it is still very different from the neutron star matter both in terms of $Y_Q$ and strangeness content.

Lowering the collision energy not only increases $n_B$ in the collision region, but also simultaneously decreases $T_{\txt{max}}$. Eventually, at relatively low beam energies, a significant fraction of the probed matter is a compressed material of the original nuclei, and one reaches a maximum accessible net baryon density $n_{B,\txt{max}}$. If one continues to lower the beam energy past this point, both $T_{\txt{max}}$ and $n_{B,\txt{max}}$ decrease. Note that in this discussion, we purposefully do not include exact values of $\sqrt{s_{\txt{NN}}}$ where these effects occur, nor can we easily discuss the $T_{\txt{max}}$ and $n_{B,\txt{max}}$ probed across the full range of $\sqrt{s_{\txt{NN}}}$ at this time. The reason behind this choice is the fact that both $T_{\txt{max}}$ and $n_{B,\txt{max}}$ depend on the dense nuclear matter EOS, which at this moment still carries a significant amount of uncertainties at densities $n_B \gtrsim 2n_0$, reached at mid to lower values of $\sqrt{s_{\txt{NN}}}$~\cite{Danielewicz:2002pu, Oliinychenko:2022uvy,Sorensen:2023zkk}. Moreover, identifying the optimal simulation framework for these beam energies is currently still the subject of vigorous research. 

At high $\sqrt{s_{\txt{NN}}}$, the reached $T_{\txt{max}}$ are larger than the pseudocritical temperature of the QCD phase transition calculated within lattice QCD, $T_{\txt{pc}}^{(\txt{LQCD})}\approx 155$ MeV~\cite{Borsanyi:2010bp}, and experimental analyses clearly suggest that in these collisions quarks and gluons form a Quark-Gluon Plasma (QGP)~\cite{BRAHMS:2004adc,PHOBOS:2004zne,PHENIX:2004vcz,STAR:2005gfr}, that is a strongly-interacting fluid of deconfined quarks and gluons. The dynamics of these systems is very well described by relativistic viscous hydrodynamics, which reproduces the experimental data and makes precise predictions across a range of $\sqrt{s_{\txt{NN}}}$~\cite{Noronha-Hostler:2015uye, Niemi:2015voa, Adam:2015ptt}. As the system cools and expands, the temperature eventually drops to a value at which a crossover transition to a gas of hadrons occurs; on the simulation level, this is where one should switch from a hydrodynamic description to one utilizing hadronic degrees of freedom, e.g., a hadronic transport code. In practice, for very high $\sqrt{s_{\txt{NN}}}$, the hadron gas phase is a subdominant effect compared with the hydrodynamic phase when looking at flow observables~\cite{Ryu:2012at}. Nevertheless, meaningful comparisons with the experimental data still require hydrodynamics to be coupled to a hadronic afterburner to reproduce, e.g., particle spectra~\cite{Teaney:2000cw,Teaney:2001av}, the transverse momentum and particle species dependence of the elliptic flow~\cite{PHENIX:2004vcz}, or the centrality and rapidity dependence of the elliptic flow~\cite{Hirano:2005xf}.

At low $\sqrt{s_{\txt{NN}}}$, significantly high values of $n_B$ are reached while values of $T_{\txt{max}}$ are lower. 
In this case, there is an important interplay between the QGP phase and the hadronic phase. Because $T_{\txt{max}}$ decreases, the fraction of the total duration of the collision described by hydrodynamics becomes smaller, and at the same time the hadronic phase starts to play a significant role~\cite{Auvinen:2013sba}. Moreover, as the collision energy is lowered, collision systems take increasingly longer times to arrive at near-equilibrium, necessary for a hydrodynamic description to be applicable. As a result, in this energy range heavy-ion collisions are best described by hybrid models, combining relativistic hydrodynamics with a hadronic transport. Current relativistic viscous hydrodynamic calculations coupled with the hadronic transport code \texttt{SMASH}~\cite{Weil:2016zrk} provide a reasonable description of particle production down to $\sqrt{s_{\txt{NN}}}=4.3$~GeV ($E_{\txt{kin}} = 8$ AGeV)~\cite{Schafer:2021csj}, while another recent study found that relativistic viscous hydrodynamics calculations using MC Glauber with baryon stopping for the initial state and \texttt{UrQMD} for the hadronic afterburner can also reproduce experimental data down to $\sqrt{s_{\txt{NN}}}=7.7 $ GeV~\cite{Shen:2022oyg}. At the same time, at these and lower beam energies $\sqrt{s_{\txt{NN}}} \approx 2.0$--$10$ GeV, heavy-ion collision systems can be well modeled using only hadronic transport simulations, which in particular can describe the out-of-equilibrium stages of the evolution that have to be included if the system evolves through unstable regions of the phase diagram (e.g., the coexistence region of a first-order phase transition). Consequently, these codes are commonly employed to reproduce experimental data in that region (see Ref.~\cite{TMEP:2022xjg} for an overview of models used, and Ref.~\cite{Sorensen:2023zkk} for a discussion of challenges and opportunities in hadronic transport theory). 

Presently, uncertainties remain regarding the appropriate $\sqrt{s_{\txt{NN}}}$ at which one should switch from one simulation approach to the other~\cite{Inghirami:2022afu}. Additional questions have been recently raised in view of the HADES experiment measurement of the average temperature $T_{\txt{avg}}\approx 71~\rm{MeV}$ of systems created in Au+Au collisions at $\sqrt{s_{\txt{NN}}}=2.4$~GeV~\cite{HADES:2019auv}. Even though this average temperature is much lower than the pseudocritical temperature at $\mu_B = 0$ as obtained in lattice QCD, $T_{\txt{avg}}<T^{\rm{LQCD}}_{\rm{pc}}$, it is high enough to ask whether a QGP is produced at these beam energies~\cite{Spieles:2020zaa}, given that the QCD transition temperature $T_{\txt{QCD}}$ decreases with increasing $n_B$~\cite{Bellwied:2015rza,Borsanyi:2020fev}. Consequently, experiments at low values of $\sqrt{s_{\txt{NN}}}$ have recently garnered a significant amount of attention within the field~\cite{Dexheimer:2020zzs}. This interest is very timely, as data from the Fixed Target campaign of the Beam Energy Scan (BES) program at the Relativistic Heavy Ion Collider (RHIC), exploring $\sqrt{s_{\txt{NN}}} \in (3.0, 7.7) ~ \txt{GeV}$, is being analyzed. At the same time, future experiments, such as the Compressed Baryonic Matter (CBM) experiment~\cite{Friese:2006dj,Tahir:2005zz,Lutz:2009ff,Durante:2019hzd} at the Facility for Antiproton and Ion Research (FAIR), are scheduled to go online within several years. Both of these efforts will further explore the above questions, among others.  

Overall, constraints on the nuclear matter EOS can be obtained by comparisons of simulation results to experimental observables. These include measurements of the collective flow, that is the Fourier coefficients of the angular momentum distribution in the transverse plane $dN/d\phi$, where the $n$-th coefficient is given by
\begin{eqnarray}
v_n \equiv \frac{\int d\phi ~ \cos(n\phi) ~ \frac{dN}{d\phi}}{\int d\phi ~  \frac{dN}{d\phi}} \ .
\end{eqnarray}
In particular, the slope of the directed flow $v_1$, $dv_1/dy\big|_{y=0}$, and the elliptic flow $v_2$ at midrapidity, $v_2(y=0)$, are shown to be very sensitive to the EOS by numerous hydrodynamic~\cite{Stoecker:1980vf,Ollitrault:1992bk,Rischke:1995pe,Stoecker:2004qu,Brachmann:1999xt,Csernai:1999nf,Ivanov:2014ioa} and hadronic transport~\cite{Hartnack:1994ce,Li:1998ze,Danielewicz:2002pu,LeFevre:2015paj,Wang:2018hsw,Nara:2021fuu,Oliinychenko:2022uvy,Steinheimer:2022gqb} models. 
Notably, as demonstrated in Ref.~\cite{Zhang:2018wlk}, for $\sqrt{s_{\txt{NN}}} \lesssim 6.4~\rm{GeV}$ spectators play a crucial role in the development of flow observables. Because most of the state-of-the-art hydrodynamic codes are intended for the description of heavy-ion collisions at very high energies, where the spectators do not influence the evolution of the system at midrapidity, these codes neglect the spectators. As a result, the majority of modern hydrodynamic codes are not currently applicable for modeling flow observables at $\sqrt{s_{\txt{NN}}} \lesssim 6.4~\rm{GeV}$ without substantial modifications. 

As noted above, in this collision energy range hadronic transport simulations are well-positioned to capture most of the relevant physics driving the influence of the EOS on experimental measurements. In addition to describing the evolution of the entire heavy-ion collision system including the spectators, hadronic transport also naturally includes transport of conserved charges: baryon number $B$, strangeness $S$, and isospin $I_3$ (or, equivalently, electric charge $Q$). Moreover, because hadronic transport simulations do not require the assumption of near-equilibrium to be fulfilled, they can also describe an evolution through unstable regions of the phase diagram, which may be necessary for systems evolving at temperatures lower than the critical temperature and densities higher than the critical density of the conjectured QCD first-order phase transition. 

Below, we use extremely soft and extremely stiff EOSs, chosen from the families of EOSs obtained for eos1, eos2, and eos3 as described in Section \ref{sec:constraints_on_the_symmetry_energy_expansion}, in hadronic transport simulations at center-of-mass collision energes $\sqrt{s_{\rm{NN}}} = 2.4,~3.0, ~4.5~\txt{GeV}$, and we compare the results to data from modern heavy-ion collision experiments.

\subsection{Simulation framework}

To simulate collisions at low $\sqrt{s_{\txt{NN}}}$, we use a hadronic transport model \texttt{SMASH}~\cite{Weil:2016zrk} (version 2.1~\cite{smash_version_2.1}). In \texttt{SMASH}, the dynamics of the system is simulated by numerically solving the relativistic Boltzmann-Uehling-Uhlenbeck (BUU) equation for the evolution of the one-body distribution function of the $n$-th particle species $f_n(t, \bm{x}, \bm{p})$,
\begin{align}
\bigg[ \frac{\partial}{\partial t} + \frac{d\bm{x}}{dt} \frac{\partial}{\partial \bm{x}} +  \frac{d\bm{p}}{dt} \frac{\partial}{\partial \bm{p}}   \bigg] f_n (t, \bm{x}, \bm{p}) = I_{\rm{coll}}~.
\label{eq:BUU_eq}
\end{align}
In the above equation, the left-hand side is equal zero in the absence of particle-particle collisions and decays. This follows directly from the Liouville theorem, and can be understood intuitively by noting that the total time derivative $df_n(t, \bm{x}, \bm{p})/dt$, i.e., the left-hand side of Eq.~\eqref{eq:BUU_eq}, describes how the value of $f_n$ changes in time as seen by an observer who travels along with the system on its trajectory $\big(\bm{x}(t), \bm{p}(t)\big)$. Without collisions and decays, which can either remove or add particles to a given subvolume of the phase space, the distribution function remains constant in the subvolume defined by trajectories taken by the particles. If collisions and decays are allowed, then the corresponding changes in the distribution function are accounted for through the right-hand side of Eq.~\eqref{eq:BUU_eq}, known as the collision term. This term includes, at least in principle, all possible cross sections and decay channels between different particle species. Thus in a system with $N$ possible particle species, one considers a set of $N$ coupled differential equations of the form given in Eq.~\eqref{eq:BUU_eq}. We note here that although Eq.~\eqref{eq:BUU_eq} is not written in a manifestly covariant way, it can be shown that it is relativistically covariant.

The numerical approach to solving Eq.\ \eqref{eq:BUU_eq} is based on the method of test particles, within which the continuous distribution $f_n(t, \bm{x}, \bm{p})$ for a system of $A$ particles of species $n$ is approximated by a discrete distribution of a large number $N = N_T A$ of test particles with phase space coordinates $(t, \bm{x}_i, \bm{p}_i)$, 
\begin{align}
f_n(t, \bm{x}, \bm{p}) \approx \frac{1}{N_T} \sum_{i=1}^{N} \delta\Big( \bm{x} - \bm{x}_i(t)\Big) \delta\Big( \bm{p} - \bm{p}_i(t)\Big) ~;
\label{eq:test_particle_method}
\end{align}
here, $N_T$ is the number of test particles per particle and the factor of $1/N_T$ in the above equation is introduced to preserve the normalization of $f$ reflecting the number of real particles present in the system. The trajectories of the test particles in the phase space are determined by their velocities and forces acting upon them,
\begin{align}
& \frac{d\bm{x}_i}{dt} = \frac{\bm{p}_i}{E_i}~, 
\label{eq:test_particles_EOMs_1} \\
& \frac{d\bm{p}_i}{dt} = \bm{F}_i = - \bm{\nabla} U_i~.
\label{eq:test_particles_EOMs_2}
\end{align}
The test particles are also allowed to undergo scatterings and decays. Since the system of test particles is ``oversampled'' with respect to the real system by a factor of $N_T$, preserving the average physical number of scatterings per test particle requires scaling the cross sections according to $\sigma/N_T$, where $\sigma$ is the physical cross section. For a given phase space distribution at the initial time $t_0$, $f_n(t_0, \bm{x}, \bm{p})$, approximating $f_n(t_0, \bm{x}, \bm{p})$ by a distribution of a (large) number of test particles as prescribed in Eq.~\eqref{eq:test_particle_method} and evolving them according to Eqs.~(\ref{eq:test_particles_EOMs_1}-\ref{eq:test_particles_EOMs_2}) as well as performing scatterings and decays as described above effectively solves the BUU equation, Eq.~\eqref{eq:BUU_eq}. This can be intuitively understood by realizing that such evolved test particles can at any time $t$ be used to form the discrete distribution approximating the true distribution $f_n(t, \bm{x}, \bm{p})$, as shown in Eq.~\eqref{eq:test_particle_method}, thus yielding the time evolution of the continuous phase space distribution $f_n(t, \bm{x}, \bm{p})$ as long as the evolution of each test particle correctly mimics the evolution of a ``real'' particle under given conditions.

It is important to note that the accuracy of the approximation used in Eq.~\eqref{eq:test_particles_EOMs_2} is directly related to the number of used test particles or, equivalently, to the used number of test particles per particle $N_T$. Larger values of $N_T$ lead to better accuracy, but in practice the choice of $N_T$ is constrained by numerical costs; while the optimal value of $N_T$ may depend on the particular situation at hand, it is fairly well-established that this number should be in the hundreds or even thousands (see, e.g., Ref.~\cite{Oliinychenko:2022uvy}). 

The information about the EOS of the modeled system enters transport simulations through Eq.~\eqref{eq:test_particles_EOMs_2} (and, in case of potentials leading to a development of an effective mass, also indirectly through Eq.~\eqref{eq:test_particles_EOMs_1}), in which the single-particle potential $U_i$
is usually modeled as a mean-field potential. Commonly used mean-field potentials range from simple parametrizations reproducing the known properties of nuclear matter around the saturation density, such as the well-known nonrelativistic Skyrme parametrizations \cite{Dutra:2012mb}, to more complex potentials such as the relativistic potentials obtained from the Vector Density Functional (VDF) model~\cite{Sorensen:2020ygf}, allowing one to describe nontrivial features of the EOS at high densities (e.g., a phase transition). 

\begin{table}[]
    \centering
    \begin{tabular}{l|c|c|c|c}
    EOS & $n_{\rm{sat}}$ [fm$^{-3}$] & $B$ [MeV] & $K_0$ [MeV] & $c_s^2(n_B = n_{\rm{sat}})$ \vspace{0.0mm} \\
    \toprule
    eos1 min & 0.175 & -14.6 & 200.5 & 0.024 \\
    \hline
    eos1 max & 0.171 & -17.8 & 325.9 & 0.039 \\
    \hline
    eos2 min & 0.167 & -14.6 & 206.7 & 0.025 \\
    \hline
    eos2 max & 0.161 & -16.9 & 214.8 & 0.026 \\
    \hline
    eos3 min & 0.153 & -14.8 & 220.2 & 0.027 \\
    \hline
    eos3 max & 0.162 & -16.5 & 201.7 & 0.024
    \end{tabular}
    \caption{Properties of symmetric nuclear matter at saturation density for each of the considered EOSs: saturation density $n_{\rm{sat}}$, binding energy $B$, incompressibility $K_0$, and the speed of sound squared at saturation $c_s^2(n_B=n_{\rm{sat}})$.
    }
    \label{tab:converted_EOS_properties}
\end{table}

In this work, we choose to use the approach proposed in Ref.~\cite{Oliinychenko:2022uvy}, where the single-particle potentials used in transport simulations are parametrized to reproduce a given dependence of the speed sound squared at zero temperature on baryon density, $c_s^2(n_B)$. Within each of the families of EOSs obtained by converting eos1, eos2, and eos3 (all of which satisfy neutron-star constraints, see Section~\ref{sec:neutron_star_EOS}) from neutron star to heavy-ion collision conditions (see Section~\ref{sec:constraints_on_the_symmetry_energy_expansion}), we consider two limiting cases: an EOS that most accurately traces the minimum band encompassing all converted EOSs and an EOS that most accurately traces the corresponding maximum band (see Fig.~\ref{fig:meddens}). The basic properties of symmetric nuclear matter at saturation density for each of the considered EOSs are given in Table~\ref{tab:converted_EOS_properties}. Note that, with the exception of ``eos1 max'', these EOSs satisfy reasonable constraints on the incompressibility $K_0 = 230 \pm 30~\rm{MeV}$~\cite{Dutra:2012mb}. Using the fact that $K_0 = 9 \mu_B c_s^2$ and assuming $\mu_B(n_B = n_{\rm{sat}}) = 922~\rm{MeV}$, the above constraints on $K_0$ further lead to $c_s^2(n_B = n_{\rm{sat}}) = 0.028 \pm 0.004$, which is again satisfied by all considered EOSs with the exception of ``eos1 max''. To ensure compatibility of the considered EOSs with the parametrization of the single-particle potentials used in \texttt{SMASH}, for low values of the density we smoothly match each of the chosen six converted EOSs with a VDF EOS according to the procedure described in Sec.~\ref{sec:the_limit_of_vanishing_nB}. 
We stress here that this procedure preserves the two most important features of the converted EOSs: the location of the peak in $c_s^2$ and the behavior at high densities driven by the used values of the symmetry energy expansion coefficients.
Finally, for each of such obtained EOSs we extract the profile $c_s^2(n_B)$ which is used as an input to parametrize the single-particle potentials employed in \texttt{SMASH}.

We note here that although the input EOSs are constructed at $T=0$, they are still applicable in heavy-ion collision simulations at finite $T$ given that the mean-field potentials used are of the vector density type: such potentials do not depend on the temperature and, therefore, can be fixed at $T=0$, while at the same time they support nontrivial changes in the EOS with increasing $T$ due to contributions arising from the kinetic part of the EOS. We elaborate on this point in Appendix~\ref{sec:T-dependence}.

\subsection{Results}

For each of the extracted $c_s^2(n_B)$ profiles, we run \texttt{SMASH} simulations at a series of beam energies corresponding to most recent experimental results. In order to assess the effect of different EOSs on the collision dynamics, we then analyze the simulation data to obtain predictions for the slope of the directed flow $dv_1/dy'\big|_{y'=0}$ and the elliptic flow at midrapity $v_2(y'=0)$. These predictions, obtained using appropriate impact parameter ranges and transverse momentum cuts (see Table~\ref{tab:parameters_of_experimental_measurements}), can be compared directly to experimental data obtained by the HADES~\cite{HADES:2022osk} and STAR~\cite{STAR:2020dav,STAR:2021yiu} experiments.

\begin{table}[t]
    \centering
    \begin{tabular}{c|c|c|c|c}
    Data & $\snn$ [GeV] &  $b$ [fm] & $p_T$ cut [GeV] \\
    \toprule
    HADES    & 2.4   &  4.7 -- 8.1 & [0.2, 2.0] \\
    \hline
    STAR     & 3.0   &  4.7 -- 9.3 & [0.4, 2.0] \\
    \hline
    STAR     &  4.5   & $v_1$: 4.7 -- 7.4 & [0.4, 2.0] \\
     &    & $v_2$: 0.0 -- 8.10 &
    \end{tabular}
    \caption{Comparison of different experimental conditions for flow measurements \cite{HADES:2022osk,STAR:2020dav,STAR:2021yiu}: the center-of-mass collision energy $\snn$, considered range of the impact parameter $b$, and cuts on the proton transverse momentum $p_T$.
    }
    \label{tab:parameters_of_experimental_measurements}
\end{table}

\begin{figure*}[t]
    \centering
        \begin{tabular}{cc}
          \includegraphics[width=\linewidth]{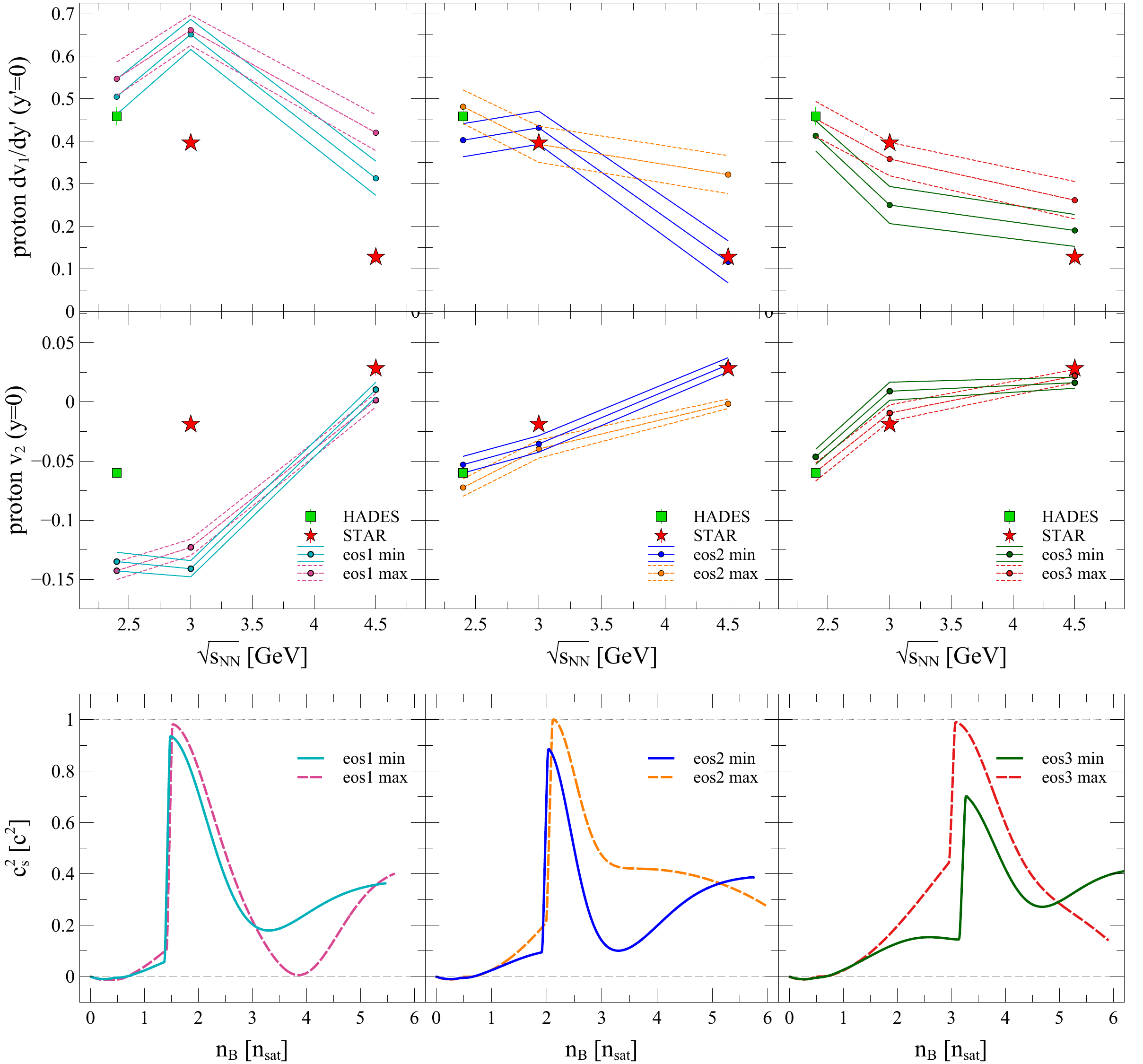}
        \end{tabular}
    \caption{Results of our study for EOSs representing two limiting cases, ``min'' (solid lines) and ``max'' (dashed lines), chosen from the family of EOSs obtained by converting eos1 (left column), eos2 (middle column), and eos3 (right column). \textit{Top:} Simulation results for the slope of the directed flow of protons at midrapidity $dv1/dy'\big|_{y'=0}$ as a function of the center-of-mass collision energy $\snn$. Also shown are measurements from the HADES~\cite{HADES:2022osk} and STAR~\cite{STAR:2020dav,STAR:2021yiu} experiments. Note that the results are expressed using scaled rapidity $y' = y / y_0$, where $y_0$ is the beam rapidity.
    \textit{Middle:} Simulation results for the elliptic flow at midrapity $v_2(y=0)$ as a function of the center-of-mass collision energy $\snn$.
    \textit{Bottom:} The speed of sound squared $c_s^2$ as a function of density in units of $n_{\rm{sat}}$. Note that while the peak of the speed of sound is put at a specific position for eos1, eos2, and eos3 (at $n_B = 1.5,~ 2.0,$ and $3.0~n_{\rm{sat}}^{(0)}$, respectively, where $n_{\rm{sat}}^{(0)}$ is the saturation density of the underlying NS EOS), the value of $n_{\rm{sat}}$ after conversion depends on the symmetry energy expansion coefficients, leading to slight discrepancies between the position of the $c_s^2$ peak for the ``min'' and ``max'' curves.
    }
    \label{fig:HIC_flow_and_EOSs}
\end{figure*}

The results of our simulations are shown in Fig.~\ref{fig:HIC_flow_and_EOSs}. Overall, we find that the experimental data are described best by the EOS labeled as ``eos2 min'' (i.e., corresponding to an EOS converted from eos2 that was most accurately tracing the minimum band encompassing all converted EOSs based on eos2, see Fig.~\ref{fig:meddens} and the bottom middle panel in Fig.~\ref{fig:HIC_flow_and_EOSs}). In the case of the EOSs labeled as ``eos2 max'', the EOS is seen to be too stiff at baryon densities around $3$--$4n_{\rm{sat}}$, probed by collisions at $\snn=4.5~\rm{GeV}$ (see, e.g., Refs.~\cite{Oliinychenko:2022uvy,Sorensen:2023zkk}), thus leading to an overprediction of the values of $dv_1/dy'\big|_{y'=0}$ and underprediction of $v_2(y'=0)$.
Results for the minimum and maximum EOSs chosen from the family of EOSs converted from eos1 (``eos1 min'' and ``eos1 max'') strongly indicate that a large peak in $c_s^2$ starting at $n_B = 1.5n_{\rm{sat}}$ is excluded by experimental data. Additionally, we note here that the large difference between the incompressibilities of ``eos1 min'' and ``eos1 max'' does not lead to large differences in results, in particular those corresponding to the HADES experiment which performs collisions at the lowest considered beam energy and thus probes the smallest range of densities, underscoring the fact that the behavior at saturation density is not probed well even at these relatively low energies. Similarly, results for ``eos3 min'' and ``eos3 max'' suggest that a large peak in $c_s^2$ starting at $n_B = 3n_{\rm{sat}}$ is unlikely: even though ``eos3 max'' describes the $v_2(y'=0)$ data well, it overpredicts values of $dv_1/dy'\big|_{y'=0}$ at higher energies, suggesting it is too stiff at high densities; conversely, ``eos3 min'' underpredicts $dv_1/dy'\big|_{y'=0}$ at $\snn = 3~\rm{GeV}$, suggesting it is too soft at the corresponding densities. This contrasts with the results obtained using ``eos2 min'', which consistently describe both the $dv_1/dy'\big|_{y'=0}$  and $v_2(y'=0)$ data. As can be seen in the bottom middle panel in Fig.~\ref{fig:HIC_flow_and_EOSs}, the ``eos2 min'' EOS corresponds to a large peak in $c_s^2$ starting at $n_B = 2.0n_{\rm{sat}}$ and symmetry energy expansion coefficients equal $E_{\rm{sym}}=28.5~\rm{MeV}$, $L_{\rm{sym}}=37.5 ~\rm{MeV}$, 
$K_{\rm{sym}}=-95 ~\rm{MeV}$, and
$J_{\rm{sym}}=600~\rm{MeV}$, which are well within the most preferred ranges of symmetry energy coefficients as identified by constraints based on causality, see Fig.~\ref{fig:corner_med}. We also note that the EOS labeled as ``eos2 min'' is qualitatively similar to the constraint on the EOS obtained in Ref.~\cite{Oliinychenko:2022uvy} through a Bayesian analysis of the STAR experiment data~\cite{STAR:2020dav,STAR:2021yiu}, which likewise exhibits a sharp peak between $2.0$--$3.0n_{\rm{sat}}$ and a significant softening between $3.0$--$4.0n_{\rm{sat}}$. At the same time, another recent Bayesian analysis~\cite{OmanaKuttan:2022aml} of heavy-ion collision data supports high values of $c_s^2$ up to $n_B = 3.0$--$4.0n_{\rm{sat}}$.

\begin{figure}[t]
    \centering
    \includegraphics[width=\linewidth]{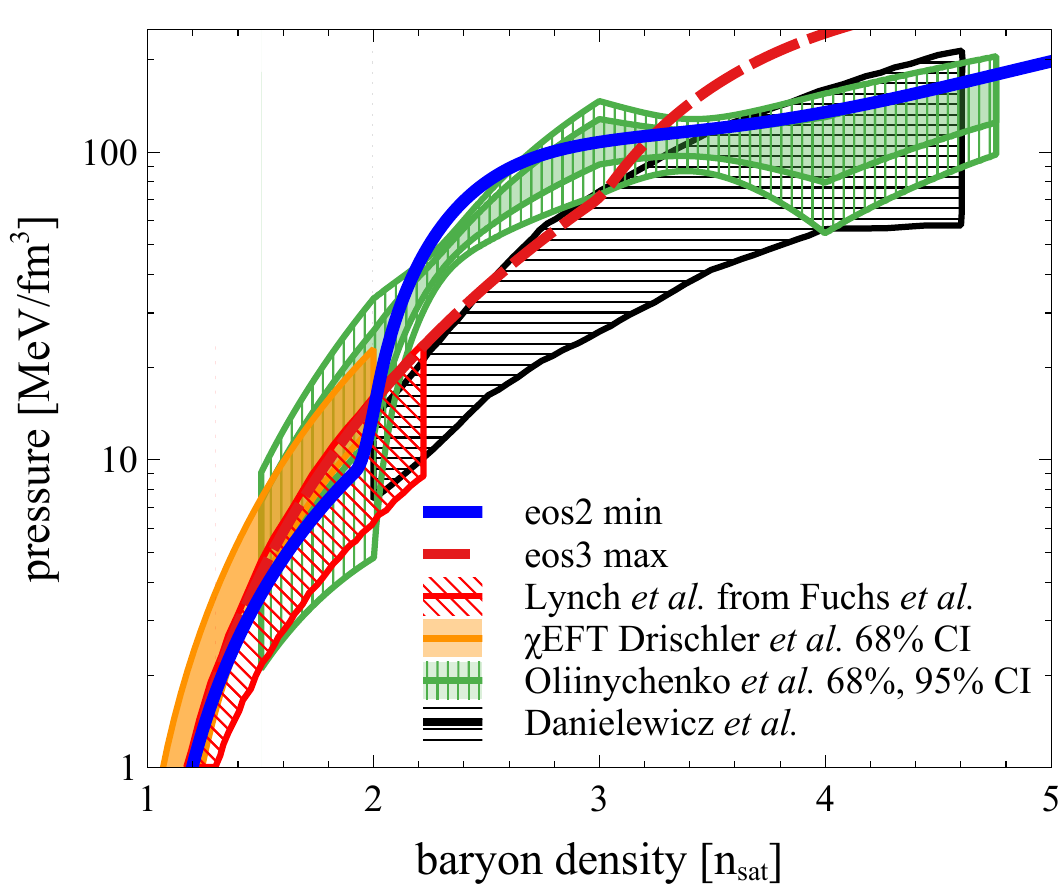}
    \caption{Pressure at zero temperature as a function of baryon density for two EOSs considered in this work, ``eos2 min'' (solid blue line) and ``eos3 max'' (dashed red line), as well as for constraints obtained from next-to-next-to-next-to-leading order chiral effective field theory calculations~\cite{Drischler:2020yad} (solid orange region), comparisons of transport model simulations to heavy-ion collision measurements~\cite{Danielewicz:2002pu} (region with black horizontal stripes), and a Bayesian analysis of modern heavy-ion collision measurements~\cite{Oliinychenko:2022uvy} (region with vertical green stripes, where the shaded (unshaded) contour corresponds to the 68\% (95\%) confidence interval).
    }
    \label{fig:combined_constraints_EOSs}
\end{figure}

In Fig.~\ref{fig:combined_constraints_EOSs}, we compare pressure as a function of baryon density for ``eos2 min'' and ``eos3 max'' against constraints from chiral effective field theory~\cite{Drischler:2020yad} and constraints extracted from comparisons to heavy-ion observables~\cite{Danielewicz:2002pu,Oliinychenko:2022uvy}. We note that the fact that ``eos3 max'' largely aligns with results from Ref.~\cite{Danielewicz:2002pu} (region with black horizontal stripes) is not that surprising given that in both cases, the EOS describes some, but not all of the considered experimental data. Additionally, the excellent agreement of ``eos2 min'' with results of Bayesian analysis of heavy-ion flow observables from Ref.~\cite{Oliinychenko:2022uvy} (region with vertical green stripes) is also to be expected due to the fact that both that study and this work used the same simulation framework. Here, however, the ``eos2 min'' EOS has been shown to not only describe heavy-ion measurements, but also satisfy all constraints from neutron-star observations as well as support extremely heavy neutron stars. Thus our results show that NS EOS with large peaks in the speed of sound can be compatible with EOS extracted from heavy-ion data as long as the simulation framework used for that extraction allows for similarly nontrivial behavior of the EOS with density.

Still, the simulation results must be evaluated with caution. The EOS parametrization used in our study does not include a momentum dependent potential, which is expected to be repulsive for nucleon kinetic energies of most relevance for the explored collision energies~\cite{Botermans:1990qi,Sammarruca:2021bpn}. Consequently, our simulations miss an additional source of repulsion between baryons. Including this additional repulsion would generally lead to larger values of $dv_1/dy'\big|_{y'=0}$ and smaller (more negative) values of $v_2(y=0)$, however, it is difficult to estimate the magnitude of this effect. Ongoing work on including momentum dependent potentials in the simulation framework will enable addressing this problem in the future. Additionally, we also did not include effects due to isospin (i.e., we considered EOSs of exactly isospin-symmetric matter even though systems created in heavy-ion collisions are characterized by $Y_{Q,\rm{QCD}}^{\rm{const}} \neq 0.5$), which, however, are expected to be small at the considered energies~\cite{Sorensen:2023zkk}; this can be also seen from results shown in Fig.~\ref{fig:YQexpan}. Still, it is worth noting that the isospin-related effects in heavy-ion collisions, although small, are the subject of active research~\cite{Sorensen:2023zkk, FRIB400, FRIB-TA_white_paper_motivations}, including its effect on the momentum dependence of the potential~\cite{Das:2002fr,Li:2003zg,Li:2018lpy}.

\section{Conclusions}

In this work, we considered three neutron-star EOSs containing a prominent structure (a large peak) in the speed of sound $c_s^2$ as a function of baryon density $n_B$. 
These EOSs can reproduce heavy neutron stars and fulfill all astrophysical observation constraints. For each of these EOSs, we then subtracted the contribution of leptons and explored allowed values of the expansion coefficients of the symmetry energy around saturation density to obtain families of EOSs corresponding to larger charge fractions, $Y_{Q,\rm{QCD}}\rightarrow 0.5$, and in particular to obtain EOSs corresponding to matter created in heavy-ion collisions at $T=0$. 
We found that the speed of sound squared for heavy-ion collision matter with $Y_Q=0.4$ is approximately the same as for $Y_Q=0.5$, and that the effect of using different expansion coefficients is much larger than the variation from using different charge fractions $Y_{Q,\rm{QCD}}$.

We also found that by just restricting the converted EOSs to satisfy $0\leq c_s^2 \leq 1$, we can constrain the allowed ranges of symmetry energy expansion coefficients. While we found almost no preference for $E_{\txt{sym,sat}}$ within the probed range, a small $L_{\txt{sym,sat}}$ was preferred in our study. Furthermore, we found a preference for the higher-order symmetry energy expansion coefficients to take values $K_{\txt{sym,sat}}\approx -100 $ MeV and $J_{\txt{sym,sat}} \gg 0$. Thus, our analysis provides a new proof-of-principle method for using astrophysics to constrain the nuclear EOS.

Next, we ran \texttt{SMASH} simulations of heavy-ion collisions at $\sqrt{s_{\txt{NN}}}=2.4,~3.0,~4.5$ GeV with example EOSs from the restricted families of converted EOSs for SNM. These EOSs were chosen to most accurately trace either the minimum or the maximum band encompassing all converted EOSs for a given NS EOS. We found that among the considered EOSs, the EOS most accurately reproducing the minimal profile of $c_s^2$ as a function of density (labeled as ``eos2 min'') describes the experimental data very well (see Fig.~\ref{fig:HIC_flow_and_EOSs}). This is consistent with the results of a Bayesian analysis of STAR flow data performed in Ref.~\cite{Oliinychenko:2022uvy} and supports an EOS with a sharp peak in the speed of sound for baryon densities $n_B \in (2,3)n_{\rm{sat}}$ followed by a substantial softening at larger densities. We also note that the ``eos2 min'' EOS corresponds to symmetry energy expansion coefficients equal $E_{\rm{sym}}=28.5~\rm{MeV}$, $L_{\rm{sym}}=37.5 ~\rm{MeV}$, 
$K_{\rm{sym}}=-95 ~\rm{MeV}$, and
$J_{\rm{sym}}=600~\rm{MeV}$, which are well within the most preferred ranges of symmetry energy coefficients as identified by constraints based on causality, see the discussion above and Fig.~\ref{fig:corner_med}. On the other hand, EOSs with a peak in $c_s^2$ starting at $n_B = 1.5n_{\rm{sat}}$ are strongly disfavored, while EOSs with a peak in $c_s^2$ starting at $n_B = 3.0 n_{\rm{sat}}$ are found to be unlikely due to the fact that, in contrast with ``eos2 min'', they do not lead to a consistent description of heavy-ion observables. Overall, given that ``eos2 min'' has been obtained from a NS EOS which satisfies all constraints from neutron-star observations and supports neutron stars with masses up to $2.5M_{\odot}$, our results show that NS EOS with large peaks in the speed of sound can be compatible with EOS extracted from heavy-ion data. However, let us note that our framework did not include momentum dependent terms in the mean-field potentials. Because of that, the simulations were missing an additional source of repulsion which would likely lead to a larger slope of the directed flow and a smaller (more negative) elliptic flow. Therefore, the above conclusions should be revisited within a framework including the influence of momentum dependent potentials. Nevertheless, from Fig.~\ref{fig:HIC_flow_and_EOSs} one can see that the expected corrections in flow would not favor the other EOSs over ``eos2 min''.

Our results demonstrate an algorithm that can be used in future studies: taking a single high-likelihood EOS constrained by neutron-star observations, converting it into a family of EOSs obtained by varying the coefficients in the symmetry energy expansion, and using limiting cases of 
these EOSs in hadronic transport simulations to further constrain the EOS with heavy-ion data. Ideally, one would go a step a further and combine constraining the EOS using neutron star and heavy-ion collision data within one comprehensive analysis.
However, recent studies that generate posteriors using Markov Chain Monte Carlo (MCMC) and neutron-star observations often take on the order of $10^5$ EOSs. One would then run a second MCMC for the symmetry energy expansion that could lead to another $10^4$ EOSs for each single neutron-star EOS, altogether leading to $10^9$ EOSs to test within a hadronic transport code which would have to be compared to experimental data at multiple beam energies. Currently, this is not remotely feasible: within a reasonable time, transport simulations using the framework employed in this study can test on the order of $10^1$ to $10^2$ EOSs (the latter case corresponding to about $10^6$ CPU hours of computation time). The numerical cost of these simulations will further increase by an order of magnitude with the inclusion of momentum dependent potentials. This presents an interesting numerical challenge that is important to overcome in the next decade, as new results begin to emerge from neutron star mergers with future upgraded gravitational wave detectors~\cite{Carson:2019rjx}, new and better constrained NICER results~\cite{Riley:2019yda,Miller:2019cac}, the awaited Beam Energy Scan II and fixed-target data from STAR~\cite{STARnote,Cebra:2014sxa}, the upcoming experimental campaign from HADES~\cite{HADES:2009aat}, the future FAIR facility at GSI~\cite{Friese:2006dj,Tahir:2005zz,Lutz:2009ff,Durante:2019hzd}, and the upcoming experiments colliding neutron-rich heavy nuclei at FRIB~\cite{FRIB400,FRIB-TA_white_paper_motivations}.

Additional challenges remain on the theoretical side.
For the symmetry energy expansion, one should think carefully about the inclusion of strangeness and quark degrees of freedom to improve the expansion at large $n_B$. Similarly, the expansion at large $n_B$ may become more reliable by including more terms in the symmetry energy expansion. Moreover, for hadronic transport models there are a number of theoretical and simulation developments required to correctly describe relevant aspects of a heavy-ion collision evolution, as detailed in \cite{Sorensen:2023zkk}. This includes, among others, momentum dependence of the potentials, in-medium cross sections, description of cluster production, subthreshold particle production, and off-shell propagation.

\section{Acknowledgments }

The authors want to thank Dmytro Oliinychenko for discussions of early results, Toru Kojo and Hajime Togashi for providing tables of the Togashi EOS, and Sanjay Reddy for insightful comments on the paper.

This work was supported in part by the NSF within the
framework of the MUSES Collaboration, under grant number OAC-2103680. 
J.N.H.\ acknowledges the support from
the US-DOE Nuclear Science Grant No. DE-SC0023861. 
V.D.\ acknowledges support from the National Science Foundation under grants PHY1748621 and
and NP3M PHY-2116686 and from the Fulbright Scholar Program. 
A.S.\ acknowledges support by the U.S.\ Department of Energy, Office of Science, Office of Nuclear Physics, under Grant No.\ DE-FG02-00ER41132. 
The authors also acknowledge support from the Illinois
Campus Cluster, a computing resource that is operated
by the Illinois Campus Cluster Program (ICCP) in conjunction with the National Center for Supercomputing
Applications (NCSA), and which is supported by funds
from the University of Illinois at Urbana-Champaign.

\bibliography{main,refNOTinspire}

\begin{thebibliography}{206}%
\makeatletter
\providecommand \@ifxundefined [1]{%
 \@ifx{#1\undefined}
}%
\providecommand \@ifnum [1]{%
 \ifnum #1\expandafter \@firstoftwo
 \else \expandafter \@secondoftwo
 \fi
}%
\providecommand \@ifx [1]{%
 \ifx #1\expandafter \@firstoftwo
 \else \expandafter \@secondoftwo
 \fi
}%
\providecommand \natexlab [1]{#1}%
\providecommand \enquote  [1]{``#1''}%
\providecommand \bibnamefont  [1]{#1}%
\providecommand \bibfnamefont [1]{#1}%
\providecommand \citenamefont [1]{#1}%
\providecommand \href@noop [0]{\@secondoftwo}%
\providecommand \href [0]{\begingroup \@sanitize@url \@href}%
\providecommand \@href[1]{\@@startlink{#1}\@@href}%
\providecommand \@@href[1]{\endgroup#1\@@endlink}%
\providecommand \@sanitize@url [0]{\catcode `\\12\catcode `\$12\catcode
  `\&12\catcode `\#12\catcode `\^12\catcode `\_12\catcode `\%12\relax}%
\providecommand \@@startlink[1]{}%
\providecommand \@@endlink[0]{}%
\providecommand \url  [0]{\begingroup\@sanitize@url \@url }%
\providecommand \@url [1]{\endgroup\@href {#1}{\urlprefix }}%
\providecommand \urlprefix  [0]{URL }%
\providecommand \Eprint [0]{\href }%
\providecommand \doibase [0]{http://dx.doi.org/}%
\providecommand \selectlanguage [0]{\@gobble}%
\providecommand \bibinfo  [0]{\@secondoftwo}%
\providecommand \bibfield  [0]{\@secondoftwo}%
\providecommand \translation [1]{[#1]}%
\providecommand \BibitemOpen [0]{}%
\providecommand \bibitemStop [0]{}%
\providecommand \bibitemNoStop [0]{.\EOS\space}%
\providecommand \EOS [0]{\spacefactor3000\relax}%
\providecommand \BibitemShut  [1]{\csname bibitem#1\endcsname}%
\let\auto@bib@innerbib\@empty
\bibitem [{\citenamefont {Baym}\ \emph {et~al.}(2018)\citenamefont {Baym},
  \citenamefont {Hatsuda}, \citenamefont {Kojo}, \citenamefont {Powell},
  \citenamefont {Song},\ and\ \citenamefont {Takatsuka}}]{Baym:2017whm}%
  \BibitemOpen
  \bibfield  {author} {\bibinfo {author} {\bibfnamefont {G.}~\bibnamefont
  {Baym}}, \bibinfo {author} {\bibfnamefont {T.}~\bibnamefont {Hatsuda}},
  \bibinfo {author} {\bibfnamefont {T.}~\bibnamefont {Kojo}}, \bibinfo {author}
  {\bibfnamefont {P.~D.}\ \bibnamefont {Powell}}, \bibinfo {author}
  {\bibfnamefont {Y.}~\bibnamefont {Song}}, \ and\ \bibinfo {author}
  {\bibfnamefont {T.}~\bibnamefont {Takatsuka}},\ }\href {\doibase
  10.1088/1361-6633/aaae14} {\bibfield  {journal} {\bibinfo  {journal} {Rept.
  Prog. Phys.}\ }\textbf {\bibinfo {volume} {81}},\ \bibinfo {pages} {056902}
  (\bibinfo {year} {2018})},\ \Eprint {http://arxiv.org/abs/1707.04966}
  {arXiv:1707.04966 [astro-ph.HE]} \BibitemShut {NoStop}%
\bibitem [{\citenamefont {Ratti}(2018)}]{Ratti:2018ksb}%
  \BibitemOpen
  \bibfield  {author} {\bibinfo {author} {\bibfnamefont {C.}~\bibnamefont
  {Ratti}},\ }\href {\doibase 10.1088/1361-6633/aabb97} {\bibfield  {journal}
  {\bibinfo  {journal} {Rept. Prog. Phys.}\ }\textbf {\bibinfo {volume} {81}},\
  \bibinfo {pages} {084301} (\bibinfo {year} {2018})},\ \Eprint
  {http://arxiv.org/abs/1804.07810} {arXiv:1804.07810 [hep-lat]} \BibitemShut
  {NoStop}%
\bibitem [{\citenamefont {Bzdak}\ \emph {et~al.}(2020)\citenamefont {Bzdak},
  \citenamefont {Esumi}, \citenamefont {Koch}, \citenamefont {Liao},
  \citenamefont {Stephanov},\ and\ \citenamefont {Xu}}]{Bzdak:2019pkr}%
  \BibitemOpen
  \bibfield  {author} {\bibinfo {author} {\bibfnamefont {A.}~\bibnamefont
  {Bzdak}}, \bibinfo {author} {\bibfnamefont {S.}~\bibnamefont {Esumi}},
  \bibinfo {author} {\bibfnamefont {V.}~\bibnamefont {Koch}}, \bibinfo {author}
  {\bibfnamefont {J.}~\bibnamefont {Liao}}, \bibinfo {author} {\bibfnamefont
  {M.}~\bibnamefont {Stephanov}}, \ and\ \bibinfo {author} {\bibfnamefont
  {N.}~\bibnamefont {Xu}},\ }\href {\doibase 10.1016/j.physrep.2020.01.005}
  {\bibfield  {journal} {\bibinfo  {journal} {Phys. Rept.}\ }\textbf {\bibinfo
  {volume} {853}},\ \bibinfo {pages} {1} (\bibinfo {year} {2020})},\ \Eprint
  {http://arxiv.org/abs/1906.00936} {arXiv:1906.00936 [nucl-th]} \BibitemShut
  {NoStop}%
\bibitem [{\citenamefont {Dexheimer}\ \emph {et~al.}(2021)\citenamefont
  {Dexheimer}, \citenamefont {Noronha}, \citenamefont {Noronha-Hostler},
  \citenamefont {Ratti},\ and\ \citenamefont {Yunes}}]{Dexheimer:2020zzs}%
  \BibitemOpen
  \bibfield  {author} {\bibinfo {author} {\bibfnamefont {V.}~\bibnamefont
  {Dexheimer}}, \bibinfo {author} {\bibfnamefont {J.}~\bibnamefont {Noronha}},
  \bibinfo {author} {\bibfnamefont {J.}~\bibnamefont {Noronha-Hostler}},
  \bibinfo {author} {\bibfnamefont {C.}~\bibnamefont {Ratti}}, \ and\ \bibinfo
  {author} {\bibfnamefont {N.}~\bibnamefont {Yunes}},\ }\href {\doibase
  10.1088/1361-6471/abe104} {\bibfield  {journal} {\bibinfo  {journal} {J.
  Phys. G}\ }\textbf {\bibinfo {volume} {48}},\ \bibinfo {pages} {073001}
  (\bibinfo {year} {2021})},\ \Eprint {http://arxiv.org/abs/2010.08834}
  {arXiv:2010.08834 [nucl-th]} \BibitemShut {NoStop}%
\bibitem [{\citenamefont {An}\ \emph {et~al.}(2022)\citenamefont {An} \emph
  {et~al.}}]{An:2021wof}%
  \BibitemOpen
  \bibfield  {author} {\bibinfo {author} {\bibfnamefont {X.}~\bibnamefont {An}}
  \emph {et~al.},\ }\href {\doibase 10.1016/j.nuclphysa.2021.122343} {\bibfield
   {journal} {\bibinfo  {journal} {Nucl. Phys. A}\ }\textbf {\bibinfo {volume}
  {1017}},\ \bibinfo {pages} {122343} (\bibinfo {year} {2022})},\ \Eprint
  {http://arxiv.org/abs/2108.13867} {arXiv:2108.13867 [nucl-th]} \BibitemShut
  {NoStop}%
\bibitem [{\citenamefont {Aoki}\ \emph {et~al.}(2006)\citenamefont {Aoki},
  \citenamefont {Endrodi}, \citenamefont {Fodor}, \citenamefont {Katz},\ and\
  \citenamefont {Szabo}}]{Aoki:2006we}%
  \BibitemOpen
  \bibfield  {author} {\bibinfo {author} {\bibfnamefont {Y.}~\bibnamefont
  {Aoki}}, \bibinfo {author} {\bibfnamefont {G.}~\bibnamefont {Endrodi}},
  \bibinfo {author} {\bibfnamefont {Z.}~\bibnamefont {Fodor}}, \bibinfo
  {author} {\bibfnamefont {S.~D.}\ \bibnamefont {Katz}}, \ and\ \bibinfo
  {author} {\bibfnamefont {K.~K.}\ \bibnamefont {Szabo}},\ }\href {\doibase
  10.1038/nature05120} {\bibfield  {journal} {\bibinfo  {journal} {Nature}\
  }\textbf {\bibinfo {volume} {443}},\ \bibinfo {pages} {675} (\bibinfo {year}
  {2006})},\ \Eprint {http://arxiv.org/abs/hep-lat/0611014}
  {arXiv:hep-lat/0611014} \BibitemShut {NoStop}%
\bibitem [{\citenamefont {Pratt}\ \emph {et~al.}(2015)\citenamefont {Pratt},
  \citenamefont {Sangaline}, \citenamefont {Sorensen},\ and\ \citenamefont
  {Wang}}]{Pratt:2015zsa}%
  \BibitemOpen
  \bibfield  {author} {\bibinfo {author} {\bibfnamefont {S.}~\bibnamefont
  {Pratt}}, \bibinfo {author} {\bibfnamefont {E.}~\bibnamefont {Sangaline}},
  \bibinfo {author} {\bibfnamefont {P.}~\bibnamefont {Sorensen}}, \ and\
  \bibinfo {author} {\bibfnamefont {H.}~\bibnamefont {Wang}},\ }\href {\doibase
  10.1103/PhysRevLett.114.202301} {\bibfield  {journal} {\bibinfo  {journal}
  {Phys. Rev. Lett.}\ }\textbf {\bibinfo {volume} {114}},\ \bibinfo {pages}
  {202301} (\bibinfo {year} {2015})},\ \Eprint
  {http://arxiv.org/abs/1501.04042} {arXiv:1501.04042 [nucl-th]} \BibitemShut
  {NoStop}%
\bibitem [{\citenamefont {Parotto}\ \emph {et~al.}(2020)\citenamefont
  {Parotto}, \citenamefont {Bluhm}, \citenamefont {Mroczek}, \citenamefont
  {Nahrgang}, \citenamefont {Noronha-Hostler}, \citenamefont {Rajagopal},
  \citenamefont {Ratti}, \citenamefont {Sch\"afer},\ and\ \citenamefont
  {Stephanov}}]{Parotto:2018pwx}%
  \BibitemOpen
  \bibfield  {author} {\bibinfo {author} {\bibfnamefont {P.}~\bibnamefont
  {Parotto}}, \bibinfo {author} {\bibfnamefont {M.}~\bibnamefont {Bluhm}},
  \bibinfo {author} {\bibfnamefont {D.}~\bibnamefont {Mroczek}}, \bibinfo
  {author} {\bibfnamefont {M.}~\bibnamefont {Nahrgang}}, \bibinfo {author}
  {\bibfnamefont {J.}~\bibnamefont {Noronha-Hostler}}, \bibinfo {author}
  {\bibfnamefont {K.}~\bibnamefont {Rajagopal}}, \bibinfo {author}
  {\bibfnamefont {C.}~\bibnamefont {Ratti}}, \bibinfo {author} {\bibfnamefont
  {T.}~\bibnamefont {Sch\"afer}}, \ and\ \bibinfo {author} {\bibfnamefont
  {M.}~\bibnamefont {Stephanov}},\ }\href {\doibase
  10.1103/PhysRevC.101.034901} {\bibfield  {journal} {\bibinfo  {journal}
  {Phys. Rev. C}\ }\textbf {\bibinfo {volume} {101}},\ \bibinfo {pages}
  {034901} (\bibinfo {year} {2020})},\ \Eprint
  {http://arxiv.org/abs/1805.05249} {arXiv:1805.05249 [hep-ph]} \BibitemShut
  {NoStop}%
\bibitem [{\citenamefont {Monnai}\ \emph {et~al.}(2019)\citenamefont {Monnai},
  \citenamefont {Schenke},\ and\ \citenamefont {Shen}}]{Monnai:2019hkn}%
  \BibitemOpen
  \bibfield  {author} {\bibinfo {author} {\bibfnamefont {A.}~\bibnamefont
  {Monnai}}, \bibinfo {author} {\bibfnamefont {B.}~\bibnamefont {Schenke}}, \
  and\ \bibinfo {author} {\bibfnamefont {C.}~\bibnamefont {Shen}},\ }\href
  {\doibase 10.1103/PhysRevC.100.024907} {\bibfield  {journal} {\bibinfo
  {journal} {Phys. Rev. C}\ }\textbf {\bibinfo {volume} {100}},\ \bibinfo
  {pages} {024907} (\bibinfo {year} {2019})},\ \Eprint
  {http://arxiv.org/abs/1902.05095} {arXiv:1902.05095 [nucl-th]} \BibitemShut
  {NoStop}%
\bibitem [{\citenamefont {Noronha-Hostler}\ \emph {et~al.}(2019)\citenamefont
  {Noronha-Hostler}, \citenamefont {Parotto}, \citenamefont {Ratti},\ and\
  \citenamefont {Stafford}}]{Noronha-Hostler:2019ayj}%
  \BibitemOpen
  \bibfield  {author} {\bibinfo {author} {\bibfnamefont {J.}~\bibnamefont
  {Noronha-Hostler}}, \bibinfo {author} {\bibfnamefont {P.}~\bibnamefont
  {Parotto}}, \bibinfo {author} {\bibfnamefont {C.}~\bibnamefont {Ratti}}, \
  and\ \bibinfo {author} {\bibfnamefont {J.~M.}\ \bibnamefont {Stafford}},\
  }\href {\doibase 10.1103/PhysRevC.100.064910} {\bibfield  {journal} {\bibinfo
   {journal} {Phys. Rev. C}\ }\textbf {\bibinfo {volume} {100}},\ \bibinfo
  {pages} {064910} (\bibinfo {year} {2019})},\ \Eprint
  {http://arxiv.org/abs/1902.06723} {arXiv:1902.06723 [hep-ph]} \BibitemShut
  {NoStop}%
\bibitem [{\citenamefont {Karthein}\ \emph {et~al.}(2021)\citenamefont
  {Karthein}, \citenamefont {Mroczek}, \citenamefont {Nava~Acuna},
  \citenamefont {Noronha-Hostler}, \citenamefont {Parotto}, \citenamefont
  {Price},\ and\ \citenamefont {Ratti}}]{Karthein:2021nxe}%
  \BibitemOpen
  \bibfield  {author} {\bibinfo {author} {\bibfnamefont {J.~M.}\ \bibnamefont
  {Karthein}}, \bibinfo {author} {\bibfnamefont {D.}~\bibnamefont {Mroczek}},
  \bibinfo {author} {\bibfnamefont {A.~R.}\ \bibnamefont {Nava~Acuna}},
  \bibinfo {author} {\bibfnamefont {J.}~\bibnamefont {Noronha-Hostler}},
  \bibinfo {author} {\bibfnamefont {P.}~\bibnamefont {Parotto}}, \bibinfo
  {author} {\bibfnamefont {D.~R.~P.}\ \bibnamefont {Price}}, \ and\ \bibinfo
  {author} {\bibfnamefont {C.}~\bibnamefont {Ratti}},\ }\href {\doibase
  10.1140/epjp/s13360-021-01615-5} {\bibfield  {journal} {\bibinfo  {journal}
  {Eur. Phys. J. Plus}\ }\textbf {\bibinfo {volume} {136}},\ \bibinfo {pages}
  {621} (\bibinfo {year} {2021})},\ \Eprint {http://arxiv.org/abs/2103.08146}
  {arXiv:2103.08146 [hep-ph]} \BibitemShut {NoStop}%
\bibitem [{\citenamefont {Alba}\ \emph {et~al.}(2017)\citenamefont {Alba} \emph
  {et~al.}}]{Alba:2017mqu}%
  \BibitemOpen
  \bibfield  {author} {\bibinfo {author} {\bibfnamefont {P.}~\bibnamefont
  {Alba}} \emph {et~al.},\ }\href {\doibase 10.1103/PhysRevD.96.034517}
  {\bibfield  {journal} {\bibinfo  {journal} {Phys. Rev. D}\ }\textbf {\bibinfo
  {volume} {96}},\ \bibinfo {pages} {034517} (\bibinfo {year} {2017})},\
  \Eprint {http://arxiv.org/abs/1702.01113} {arXiv:1702.01113 [hep-lat]}
  \BibitemShut {NoStop}%
\bibitem [{\citenamefont {Vovchenko}\ \emph
  {et~al.}(2017{\natexlab{a}})\citenamefont {Vovchenko}, \citenamefont
  {Pasztor}, \citenamefont {Fodor}, \citenamefont {Katz},\ and\ \citenamefont
  {Stoecker}}]{Vovchenko:2017xad}%
  \BibitemOpen
  \bibfield  {author} {\bibinfo {author} {\bibfnamefont {V.}~\bibnamefont
  {Vovchenko}}, \bibinfo {author} {\bibfnamefont {A.}~\bibnamefont {Pasztor}},
  \bibinfo {author} {\bibfnamefont {Z.}~\bibnamefont {Fodor}}, \bibinfo
  {author} {\bibfnamefont {S.~D.}\ \bibnamefont {Katz}}, \ and\ \bibinfo
  {author} {\bibfnamefont {H.}~\bibnamefont {Stoecker}},\ }\href {\doibase
  10.1016/j.physletb.2017.10.042} {\bibfield  {journal} {\bibinfo  {journal}
  {Phys. Lett. B}\ }\textbf {\bibinfo {volume} {775}},\ \bibinfo {pages} {71}
  (\bibinfo {year} {2017}{\natexlab{a}})},\ \Eprint
  {http://arxiv.org/abs/1708.02852} {arXiv:1708.02852 [hep-ph]} \BibitemShut
  {NoStop}%
\bibitem [{\citenamefont {Bellwied}\ \emph {et~al.}(2020)\citenamefont
  {Bellwied}, \citenamefont {Borsanyi}, \citenamefont {Fodor}, \citenamefont
  {Guenther}, \citenamefont {Noronha-Hostler}, \citenamefont {Parotto},
  \citenamefont {Pasztor}, \citenamefont {Ratti},\ and\ \citenamefont
  {Stafford}}]{Bellwied:2019pxh}%
  \BibitemOpen
  \bibfield  {author} {\bibinfo {author} {\bibfnamefont {R.}~\bibnamefont
  {Bellwied}}, \bibinfo {author} {\bibfnamefont {S.}~\bibnamefont {Borsanyi}},
  \bibinfo {author} {\bibfnamefont {Z.}~\bibnamefont {Fodor}}, \bibinfo
  {author} {\bibfnamefont {J.~N.}\ \bibnamefont {Guenther}}, \bibinfo {author}
  {\bibfnamefont {J.}~\bibnamefont {Noronha-Hostler}}, \bibinfo {author}
  {\bibfnamefont {P.}~\bibnamefont {Parotto}}, \bibinfo {author} {\bibfnamefont
  {A.}~\bibnamefont {Pasztor}}, \bibinfo {author} {\bibfnamefont
  {C.}~\bibnamefont {Ratti}}, \ and\ \bibinfo {author} {\bibfnamefont {J.~M.}\
  \bibnamefont {Stafford}},\ }\href {\doibase 10.1103/PhysRevD.101.034506}
  {\bibfield  {journal} {\bibinfo  {journal} {Phys. Rev. D}\ }\textbf {\bibinfo
  {volume} {101}},\ \bibinfo {pages} {034506} (\bibinfo {year} {2020})},\
  \Eprint {http://arxiv.org/abs/1910.14592} {arXiv:1910.14592 [hep-lat]}
  \BibitemShut {NoStop}%
\bibitem [{\citenamefont {Bors\'anyi}\ \emph {et~al.}(2021)\citenamefont
  {Bors\'anyi}, \citenamefont {Fodor}, \citenamefont {Guenther}, \citenamefont
  {Kara}, \citenamefont {Katz}, \citenamefont {Parotto}, \citenamefont
  {P\'asztor}, \citenamefont {Ratti},\ and\ \citenamefont
  {Szab\'o}}]{Borsanyi:2021sxv}%
  \BibitemOpen
  \bibfield  {author} {\bibinfo {author} {\bibfnamefont {S.}~\bibnamefont
  {Bors\'anyi}}, \bibinfo {author} {\bibfnamefont {Z.}~\bibnamefont {Fodor}},
  \bibinfo {author} {\bibfnamefont {J.~N.}\ \bibnamefont {Guenther}}, \bibinfo
  {author} {\bibfnamefont {R.}~\bibnamefont {Kara}}, \bibinfo {author}
  {\bibfnamefont {S.~D.}\ \bibnamefont {Katz}}, \bibinfo {author}
  {\bibfnamefont {P.}~\bibnamefont {Parotto}}, \bibinfo {author} {\bibfnamefont
  {A.}~\bibnamefont {P\'asztor}}, \bibinfo {author} {\bibfnamefont
  {C.}~\bibnamefont {Ratti}}, \ and\ \bibinfo {author} {\bibfnamefont {K.~K.}\
  \bibnamefont {Szab\'o}},\ }\href {\doibase 10.1103/PhysRevLett.126.232001}
  {\bibfield  {journal} {\bibinfo  {journal} {Phys. Rev. Lett.}\ }\textbf
  {\bibinfo {volume} {126}},\ \bibinfo {pages} {232001} (\bibinfo {year}
  {2021})},\ \Eprint {http://arxiv.org/abs/2102.06660} {arXiv:2102.06660
  [hep-lat]} \BibitemShut {NoStop}%
\bibitem [{\citenamefont {Borsanyi}\ \emph {et~al.}(2022)\citenamefont
  {Borsanyi}, \citenamefont {Guenther}, \citenamefont {Kara}, \citenamefont
  {Fodor}, \citenamefont {Parotto}, \citenamefont {Pasztor}, \citenamefont
  {Ratti},\ and\ \citenamefont {Szabo}}]{Borsanyi:2022qlh}%
  \BibitemOpen
  \bibfield  {author} {\bibinfo {author} {\bibfnamefont {S.}~\bibnamefont
  {Borsanyi}}, \bibinfo {author} {\bibfnamefont {J.~N.}\ \bibnamefont
  {Guenther}}, \bibinfo {author} {\bibfnamefont {R.}~\bibnamefont {Kara}},
  \bibinfo {author} {\bibfnamefont {Z.}~\bibnamefont {Fodor}}, \bibinfo
  {author} {\bibfnamefont {P.}~\bibnamefont {Parotto}}, \bibinfo {author}
  {\bibfnamefont {A.}~\bibnamefont {Pasztor}}, \bibinfo {author} {\bibfnamefont
  {C.}~\bibnamefont {Ratti}}, \ and\ \bibinfo {author} {\bibfnamefont {K.~K.}\
  \bibnamefont {Szabo}},\ }\href {\doibase 10.1103/PhysRevD.105.114504}
  {\bibfield  {journal} {\bibinfo  {journal} {Phys. Rev. D}\ }\textbf {\bibinfo
  {volume} {105}},\ \bibinfo {pages} {114504} (\bibinfo {year} {2022})},\
  \Eprint {http://arxiv.org/abs/2202.05574} {arXiv:2202.05574 [hep-lat]}
  \BibitemShut {NoStop}%
\bibitem [{\citenamefont {Vovchenko}\ \emph
  {et~al.}(2017{\natexlab{b}})\citenamefont {Vovchenko}, \citenamefont
  {Gorenstein},\ and\ \citenamefont {Stoecker}}]{Vovchenko:2016rkn}%
  \BibitemOpen
  \bibfield  {author} {\bibinfo {author} {\bibfnamefont {V.}~\bibnamefont
  {Vovchenko}}, \bibinfo {author} {\bibfnamefont {M.~I.}\ \bibnamefont
  {Gorenstein}}, \ and\ \bibinfo {author} {\bibfnamefont {H.}~\bibnamefont
  {Stoecker}},\ }\href {\doibase 10.1103/PhysRevLett.118.182301} {\bibfield
  {journal} {\bibinfo  {journal} {Phys. Rev. Lett.}\ }\textbf {\bibinfo
  {volume} {118}},\ \bibinfo {pages} {182301} (\bibinfo {year}
  {2017}{\natexlab{b}})},\ \Eprint {http://arxiv.org/abs/1609.03975}
  {arXiv:1609.03975 [hep-ph]} \BibitemShut {NoStop}%
\bibitem [{\citenamefont {Kumar}\ \emph
  {et~al.}(2023{\natexlab{a}})\citenamefont {Kumar} \emph
  {et~al.}}]{MUSES:2023hyz}%
  \BibitemOpen
  \bibfield  {author} {\bibinfo {author} {\bibfnamefont {R.}~\bibnamefont
  {Kumar}} \emph {et~al.} (\bibinfo {collaboration} {MUSES}),\ }\href@noop {}
  {\  (\bibinfo {year} {2023}{\natexlab{a}})},\ \Eprint
  {http://arxiv.org/abs/2303.17021} {arXiv:2303.17021 [nucl-th]} \BibitemShut
  {NoStop}%
\bibitem [{\citenamefont {Danielewicz}\ \emph {et~al.}(2002)\citenamefont
  {Danielewicz}, \citenamefont {Lacey},\ and\ \citenamefont
  {Lynch}}]{Danielewicz:2002pu}%
  \BibitemOpen
  \bibfield  {author} {\bibinfo {author} {\bibfnamefont {P.}~\bibnamefont
  {Danielewicz}}, \bibinfo {author} {\bibfnamefont {R.}~\bibnamefont {Lacey}},
  \ and\ \bibinfo {author} {\bibfnamefont {W.~G.}\ \bibnamefont {Lynch}},\
  }\href {\doibase 10.1126/science.1078070} {\bibfield  {journal} {\bibinfo
  {journal} {Science}\ }\textbf {\bibinfo {volume} {298}},\ \bibinfo {pages}
  {1592} (\bibinfo {year} {2002})},\ \Eprint
  {http://arxiv.org/abs/nucl-th/0208016} {arXiv:nucl-th/0208016} \BibitemShut
  {NoStop}%
\bibitem [{\citenamefont {Adamczewski-Musch}\ \emph {et~al.}(2019)\citenamefont
  {Adamczewski-Musch} \emph {et~al.}}]{HADES:2019auv}%
  \BibitemOpen
  \bibfield  {author} {\bibinfo {author} {\bibfnamefont {J.}~\bibnamefont
  {Adamczewski-Musch}} \emph {et~al.} (\bibinfo {collaboration} {HADES}),\
  }\href {\doibase 10.1038/s41567-019-0583-8} {\bibfield  {journal} {\bibinfo
  {journal} {Nature Phys.}\ }\textbf {\bibinfo {volume} {15}},\ \bibinfo
  {pages} {1040} (\bibinfo {year} {2019})}\BibitemShut {NoStop}%
\bibitem [{\citenamefont {Fattoyev}\ \emph {et~al.}(2020)\citenamefont
  {Fattoyev}, \citenamefont {Horowitz}, \citenamefont {Piekarewicz},\ and\
  \citenamefont {Reed}}]{Fattoyev:2020cws}%
  \BibitemOpen
  \bibfield  {author} {\bibinfo {author} {\bibfnamefont {F.~J.}\ \bibnamefont
  {Fattoyev}}, \bibinfo {author} {\bibfnamefont {C.~J.}\ \bibnamefont
  {Horowitz}}, \bibinfo {author} {\bibfnamefont {J.}~\bibnamefont
  {Piekarewicz}}, \ and\ \bibinfo {author} {\bibfnamefont {B.}~\bibnamefont
  {Reed}},\ }\href {\doibase 10.1103/PhysRevC.102.065805} {\bibfield  {journal}
  {\bibinfo  {journal} {Phys. Rev. C}\ }\textbf {\bibinfo {volume} {102}},\
  \bibinfo {pages} {065805} (\bibinfo {year} {2020})},\ \Eprint
  {http://arxiv.org/abs/2007.03799} {arXiv:2007.03799 [nucl-th]} \BibitemShut
  {NoStop}%
\bibitem [{\citenamefont {Spieles}\ and\ \citenamefont
  {Bleicher}(2020)}]{Spieles:2020zaa}%
  \BibitemOpen
  \bibfield  {author} {\bibinfo {author} {\bibfnamefont {C.}~\bibnamefont
  {Spieles}}\ and\ \bibinfo {author} {\bibfnamefont {M.}~\bibnamefont
  {Bleicher}},\ }\href {\doibase 10.1140/epjst/e2020-000102-4} {\bibfield
  {journal} {\bibinfo  {journal} {Eur. Phys. J. ST}\ }\textbf {\bibinfo
  {volume} {229}},\ \bibinfo {pages} {3537} (\bibinfo {year} {2020})},\ \Eprint
  {http://arxiv.org/abs/2006.01220} {arXiv:2006.01220 [nucl-th]} \BibitemShut
  {NoStop}%
\bibitem [{\citenamefont {Antoniadis}\ \emph {et~al.}(2013)\citenamefont
  {Antoniadis}, \citenamefont {Freire}, \citenamefont {Wex}, \citenamefont
  {Tauris}, \citenamefont {Lynch}, \citenamefont {van Kerkwijk}, \citenamefont
  {Kramer}, \citenamefont {Bassa}, \citenamefont {Dhillon}, \citenamefont
  {Driebe}, \citenamefont {Hessels}, \citenamefont {Kaspi}, \citenamefont
  {Kondratiev}, \citenamefont {Langer}, \citenamefont {Marsh}, \citenamefont
  {McLaughlin}, \citenamefont {Pennucci}, \citenamefont {Ransom}, \citenamefont
  {Stairs}, \citenamefont {van Leeuwen}, \citenamefont {Verbiest},\ and\
  \citenamefont {Whelan}}]{science.1233232}%
  \BibitemOpen
  \bibfield  {author} {\bibinfo {author} {\bibfnamefont {J.}~\bibnamefont
  {Antoniadis}}, \bibinfo {author} {\bibfnamefont {P.~C.~C.}\ \bibnamefont
  {Freire}}, \bibinfo {author} {\bibfnamefont {N.}~\bibnamefont {Wex}},
  \bibinfo {author} {\bibfnamefont {T.~M.}\ \bibnamefont {Tauris}}, \bibinfo
  {author} {\bibfnamefont {R.~S.}\ \bibnamefont {Lynch}}, \bibinfo {author}
  {\bibfnamefont {M.~H.}\ \bibnamefont {van Kerkwijk}}, \bibinfo {author}
  {\bibfnamefont {M.}~\bibnamefont {Kramer}}, \bibinfo {author} {\bibfnamefont
  {C.}~\bibnamefont {Bassa}}, \bibinfo {author} {\bibfnamefont {V.~S.}\
  \bibnamefont {Dhillon}}, \bibinfo {author} {\bibfnamefont {T.}~\bibnamefont
  {Driebe}}, \bibinfo {author} {\bibfnamefont {J.~W.~T.}\ \bibnamefont
  {Hessels}}, \bibinfo {author} {\bibfnamefont {V.~M.}\ \bibnamefont {Kaspi}},
  \bibinfo {author} {\bibfnamefont {V.~I.}\ \bibnamefont {Kondratiev}},
  \bibinfo {author} {\bibfnamefont {N.}~\bibnamefont {Langer}}, \bibinfo
  {author} {\bibfnamefont {T.~R.}\ \bibnamefont {Marsh}}, \bibinfo {author}
  {\bibfnamefont {M.~A.}\ \bibnamefont {McLaughlin}}, \bibinfo {author}
  {\bibfnamefont {T.~T.}\ \bibnamefont {Pennucci}}, \bibinfo {author}
  {\bibfnamefont {S.~M.}\ \bibnamefont {Ransom}}, \bibinfo {author}
  {\bibfnamefont {I.~H.}\ \bibnamefont {Stairs}}, \bibinfo {author}
  {\bibfnamefont {J.}~\bibnamefont {van Leeuwen}}, \bibinfo {author}
  {\bibfnamefont {J.~P.~W.}\ \bibnamefont {Verbiest}}, \ and\ \bibinfo {author}
  {\bibfnamefont {D.~G.}\ \bibnamefont {Whelan}},\ }\href {\doibase
  10.1126/science.1233232} {\bibfield  {journal} {\bibinfo  {journal}
  {Science}\ }\textbf {\bibinfo {volume} {340}},\ \bibinfo {pages} {1233232}
  (\bibinfo {year} {2013})},\ \Eprint
  {http://arxiv.org/abs/https://www.science.org/doi/pdf/10.1126/science.1233232}
  {https://www.science.org/doi/pdf/10.1126/science.1233232} \BibitemShut
  {NoStop}%
\bibitem [{\citenamefont {Fonseca}\ \emph {et~al.}(2021)\citenamefont {Fonseca}
  \emph {et~al.}}]{Fonseca:2021wxt}%
  \BibitemOpen
  \bibfield  {author} {\bibinfo {author} {\bibfnamefont {E.}~\bibnamefont
  {Fonseca}} \emph {et~al.},\ }\href {\doibase 10.3847/2041-8213/ac03b8}
  {\bibfield  {journal} {\bibinfo  {journal} {Astrophys. J. Lett.}\ }\textbf
  {\bibinfo {volume} {915}},\ \bibinfo {pages} {L12} (\bibinfo {year}
  {2021})},\ \Eprint {http://arxiv.org/abs/2104.00880} {arXiv:2104.00880
  [astro-ph.HE]} \BibitemShut {NoStop}%
\bibitem [{\citenamefont {Romani}\ \emph {et~al.}(2021)\citenamefont {Romani},
  \citenamefont {Kandel}, \citenamefont {Filippenko}, \citenamefont {Brink},\
  and\ \citenamefont {Zheng}}]{Romani:2021xmb}%
  \BibitemOpen
  \bibfield  {author} {\bibinfo {author} {\bibfnamefont {R.~W.}\ \bibnamefont
  {Romani}}, \bibinfo {author} {\bibfnamefont {D.}~\bibnamefont {Kandel}},
  \bibinfo {author} {\bibfnamefont {A.~V.}\ \bibnamefont {Filippenko}},
  \bibinfo {author} {\bibfnamefont {T.~G.}\ \bibnamefont {Brink}}, \ and\
  \bibinfo {author} {\bibfnamefont {W.}~\bibnamefont {Zheng}},\ }\href
  {\doibase 10.3847/2041-8213/abe2b4} {\bibfield  {journal} {\bibinfo
  {journal} {Astrophys. J. Lett.}\ }\textbf {\bibinfo {volume} {908}},\
  \bibinfo {pages} {L46} (\bibinfo {year} {2021})},\ \Eprint
  {http://arxiv.org/abs/2101.09822} {arXiv:2101.09822 [astro-ph.HE]}
  \BibitemShut {NoStop}%
\bibitem [{\citenamefont {Bedaque}\ and\ \citenamefont
  {Steiner}(2015{\natexlab{a}})}]{Bedaque:2014sqa}%
  \BibitemOpen
  \bibfield  {author} {\bibinfo {author} {\bibfnamefont {P.}~\bibnamefont
  {Bedaque}}\ and\ \bibinfo {author} {\bibfnamefont {A.~W.}\ \bibnamefont
  {Steiner}},\ }\href {\doibase 10.1103/PhysRevLett.114.031103} {\bibfield
  {journal} {\bibinfo  {journal} {Phys. Rev. Lett.}\ }\textbf {\bibinfo
  {volume} {114}},\ \bibinfo {pages} {031103} (\bibinfo {year}
  {2015}{\natexlab{a}})},\ \Eprint {http://arxiv.org/abs/1408.5116}
  {arXiv:1408.5116 [nucl-th]} \BibitemShut {NoStop}%
\bibitem [{\citenamefont {Tews}\ \emph {et~al.}(2018)\citenamefont {Tews},
  \citenamefont {Carlson}, \citenamefont {Gandolfi},\ and\ \citenamefont
  {Reddy}}]{Tews:2018kmu}%
  \BibitemOpen
  \bibfield  {author} {\bibinfo {author} {\bibfnamefont {I.}~\bibnamefont
  {Tews}}, \bibinfo {author} {\bibfnamefont {J.}~\bibnamefont {Carlson}},
  \bibinfo {author} {\bibfnamefont {S.}~\bibnamefont {Gandolfi}}, \ and\
  \bibinfo {author} {\bibfnamefont {S.}~\bibnamefont {Reddy}},\ }\href
  {\doibase 10.3847/1538-4357/aac267} {\bibfield  {journal} {\bibinfo
  {journal} {Astrophys. J.}\ }\textbf {\bibinfo {volume} {860}},\ \bibinfo
  {pages} {149} (\bibinfo {year} {2018})},\ \Eprint
  {http://arxiv.org/abs/1801.01923} {arXiv:1801.01923 [nucl-th]} \BibitemShut
  {NoStop}%
\bibitem [{\citenamefont {Fujimoto}\ \emph {et~al.}(2020)\citenamefont
  {Fujimoto}, \citenamefont {Fukushima},\ and\ \citenamefont
  {Murase}}]{Fujimoto:2019hxv}%
  \BibitemOpen
  \bibfield  {author} {\bibinfo {author} {\bibfnamefont {Y.}~\bibnamefont
  {Fujimoto}}, \bibinfo {author} {\bibfnamefont {K.}~\bibnamefont {Fukushima}},
  \ and\ \bibinfo {author} {\bibfnamefont {K.}~\bibnamefont {Murase}},\ }\href
  {\doibase 10.1103/PhysRevD.101.054016} {\bibfield  {journal} {\bibinfo
  {journal} {Phys. Rev. D}\ }\textbf {\bibinfo {volume} {101}},\ \bibinfo
  {pages} {054016} (\bibinfo {year} {2020})},\ \Eprint
  {http://arxiv.org/abs/1903.03400} {arXiv:1903.03400 [nucl-th]} \BibitemShut
  {NoStop}%
\bibitem [{\citenamefont {Altiparmak}\ \emph {et~al.}(2022)\citenamefont
  {Altiparmak}, \citenamefont {Ecker},\ and\ \citenamefont
  {Rezzolla}}]{Altiparmak:2022bke}%
  \BibitemOpen
  \bibfield  {author} {\bibinfo {author} {\bibfnamefont {S.}~\bibnamefont
  {Altiparmak}}, \bibinfo {author} {\bibfnamefont {C.}~\bibnamefont {Ecker}}, \
  and\ \bibinfo {author} {\bibfnamefont {L.}~\bibnamefont {Rezzolla}},\ }\href
  {\doibase 10.3847/2041-8213/ac9b2a} {\bibfield  {journal} {\bibinfo
  {journal} {Astrophys. J. Lett.}\ }\textbf {\bibinfo {volume} {939}},\
  \bibinfo {pages} {L34} (\bibinfo {year} {2022})},\ \Eprint
  {http://arxiv.org/abs/2203.14974} {arXiv:2203.14974 [astro-ph.HE]}
  \BibitemShut {NoStop}%
\bibitem [{\citenamefont {Han}\ \emph {et~al.}(2023)\citenamefont {Han},
  \citenamefont {Huang}, \citenamefont {Tang},\ and\ \citenamefont
  {Fan}}]{Han:2022rug}%
  \BibitemOpen
  \bibfield  {author} {\bibinfo {author} {\bibfnamefont {M.-Z.}\ \bibnamefont
  {Han}}, \bibinfo {author} {\bibfnamefont {Y.-J.}\ \bibnamefont {Huang}},
  \bibinfo {author} {\bibfnamefont {S.-P.}\ \bibnamefont {Tang}}, \ and\
  \bibinfo {author} {\bibfnamefont {Y.-Z.}\ \bibnamefont {Fan}},\ }\href
  {\doibase 10.1016/j.scib.2023.04.007} {\bibfield  {journal} {\bibinfo
  {journal} {Sci. Bull.}\ }\textbf {\bibinfo {volume} {68}},\ \bibinfo {pages}
  {913} (\bibinfo {year} {2023})},\ \Eprint {http://arxiv.org/abs/2207.13613}
  {arXiv:2207.13613 [astro-ph.HE]} \BibitemShut {NoStop}%
\bibitem [{\citenamefont {Abbott}\ \emph {et~al.}(2020)\citenamefont {Abbott}
  \emph {et~al.}}]{LIGOScientific:2020zkf}%
  \BibitemOpen
  \bibfield  {author} {\bibinfo {author} {\bibfnamefont {R.}~\bibnamefont
  {Abbott}} \emph {et~al.} (\bibinfo {collaboration} {LIGO Scientific,
  Virgo}),\ }\href {\doibase 10.3847/2041-8213/ab960f} {\bibfield  {journal}
  {\bibinfo  {journal} {Astrophys. J. Lett.}\ }\textbf {\bibinfo {volume}
  {896}},\ \bibinfo {pages} {L44} (\bibinfo {year} {2020})},\ \Eprint
  {http://arxiv.org/abs/2006.12611} {arXiv:2006.12611 [astro-ph.HE]}
  \BibitemShut {NoStop}%
\bibitem [{\citenamefont {Bailyn}\ \emph {et~al.}(1998)\citenamefont {Bailyn},
  \citenamefont {Jain}, \citenamefont {Coppi},\ and\ \citenamefont
  {Orosz}}]{Bailyn:1997xt}%
  \BibitemOpen
  \bibfield  {author} {\bibinfo {author} {\bibfnamefont {C.~D.}\ \bibnamefont
  {Bailyn}}, \bibinfo {author} {\bibfnamefont {R.~K.}\ \bibnamefont {Jain}},
  \bibinfo {author} {\bibfnamefont {P.}~\bibnamefont {Coppi}}, \ and\ \bibinfo
  {author} {\bibfnamefont {J.~A.}\ \bibnamefont {Orosz}},\ }\href {\doibase
  10.1086/305614} {\bibfield  {journal} {\bibinfo  {journal} {Astrophys. J.}\
  }\textbf {\bibinfo {volume} {499}},\ \bibinfo {pages} {367} (\bibinfo {year}
  {1998})},\ \Eprint {http://arxiv.org/abs/astro-ph/9708032}
  {arXiv:astro-ph/9708032} \BibitemShut {NoStop}%
\bibitem [{\citenamefont {Ozel}\ \emph {et~al.}(2010)\citenamefont {Ozel},
  \citenamefont {Psaltis}, \citenamefont {Narayan},\ and\ \citenamefont
  {McClintock}}]{Ozel:2010su}%
  \BibitemOpen
  \bibfield  {author} {\bibinfo {author} {\bibfnamefont {F.}~\bibnamefont
  {Ozel}}, \bibinfo {author} {\bibfnamefont {D.}~\bibnamefont {Psaltis}},
  \bibinfo {author} {\bibfnamefont {R.}~\bibnamefont {Narayan}}, \ and\
  \bibinfo {author} {\bibfnamefont {J.~E.}\ \bibnamefont {McClintock}},\ }\href
  {\doibase 10.1088/0004-637X/725/2/1918} {\bibfield  {journal} {\bibinfo
  {journal} {Astrophys. J.}\ }\textbf {\bibinfo {volume} {725}},\ \bibinfo
  {pages} {1918} (\bibinfo {year} {2010})},\ \Eprint
  {http://arxiv.org/abs/1006.2834} {arXiv:1006.2834 [astro-ph.GA]} \BibitemShut
  {NoStop}%
\bibitem [{\citenamefont {de~S\'a}\ \emph {et~al.}(2022)\citenamefont
  {de~S\'a}, \citenamefont {Bernardo}, \citenamefont {Bachega}, \citenamefont
  {Horvath}, \citenamefont {Rocha},\ and\ \citenamefont
  {Moraes}}]{deSa:2022qny}%
  \BibitemOpen
  \bibfield  {author} {\bibinfo {author} {\bibfnamefont {L.~M.}\ \bibnamefont
  {de~S\'a}}, \bibinfo {author} {\bibfnamefont {A.}~\bibnamefont {Bernardo}},
  \bibinfo {author} {\bibfnamefont {R.~R.~A.}\ \bibnamefont {Bachega}},
  \bibinfo {author} {\bibfnamefont {J.~E.}\ \bibnamefont {Horvath}}, \bibinfo
  {author} {\bibfnamefont {L.~S.}\ \bibnamefont {Rocha}}, \ and\ \bibinfo
  {author} {\bibfnamefont {P.~H. R.~S.}\ \bibnamefont {Moraes}},\ }\href
  {\doibase 10.3847/1538-4357/aca076} {\bibfield  {journal} {\bibinfo
  {journal} {Astrophys. J.}\ }\textbf {\bibinfo {volume} {941}},\ \bibinfo
  {pages} {130} (\bibinfo {year} {2022})},\ \Eprint
  {http://arxiv.org/abs/2211.01447} {arXiv:2211.01447 [astro-ph.HE]}
  \BibitemShut {NoStop}%
\bibitem [{\citenamefont {Ye}\ and\ \citenamefont
  {Fishbach}(2022)}]{Ye:2022qoe}%
  \BibitemOpen
  \bibfield  {author} {\bibinfo {author} {\bibfnamefont {C.}~\bibnamefont
  {Ye}}\ and\ \bibinfo {author} {\bibfnamefont {M.}~\bibnamefont {Fishbach}},\
  }\href {\doibase 10.3847/1538-4357/ac7f99} {\bibfield  {journal} {\bibinfo
  {journal} {Astrophys. J.}\ }\textbf {\bibinfo {volume} {937}},\ \bibinfo
  {pages} {73} (\bibinfo {year} {2022})},\ \Eprint
  {http://arxiv.org/abs/2202.05164} {arXiv:2202.05164 [astro-ph.HE]}
  \BibitemShut {NoStop}%
\bibitem [{\citenamefont {Farah}\ \emph {et~al.}(2022)\citenamefont {Farah},
  \citenamefont {Fishbach}, \citenamefont {Essick}, \citenamefont {Holz},\ and\
  \citenamefont {Galaudage}}]{Farah:2021qom}%
  \BibitemOpen
  \bibfield  {author} {\bibinfo {author} {\bibfnamefont {A.~M.}\ \bibnamefont
  {Farah}}, \bibinfo {author} {\bibfnamefont {M.}~\bibnamefont {Fishbach}},
  \bibinfo {author} {\bibfnamefont {R.}~\bibnamefont {Essick}}, \bibinfo
  {author} {\bibfnamefont {D.~E.}\ \bibnamefont {Holz}}, \ and\ \bibinfo
  {author} {\bibfnamefont {S.}~\bibnamefont {Galaudage}},\ }\href {\doibase
  10.3847/1538-4357/ac5f03} {\bibfield  {journal} {\bibinfo  {journal}
  {Astrophys. J.}\ }\textbf {\bibinfo {volume} {931}},\ \bibinfo {pages} {108}
  (\bibinfo {year} {2022})},\ \Eprint {http://arxiv.org/abs/2111.03498}
  {arXiv:2111.03498 [astro-ph.HE]} \BibitemShut {NoStop}%
\bibitem [{\citenamefont {Tan}\ \emph {et~al.}(2020)\citenamefont {Tan},
  \citenamefont {Noronha-Hostler},\ and\ \citenamefont {Yunes}}]{Tan:2020ics}%
  \BibitemOpen
  \bibfield  {author} {\bibinfo {author} {\bibfnamefont {H.}~\bibnamefont
  {Tan}}, \bibinfo {author} {\bibfnamefont {J.}~\bibnamefont
  {Noronha-Hostler}}, \ and\ \bibinfo {author} {\bibfnamefont {N.}~\bibnamefont
  {Yunes}},\ }\href {\doibase 10.1103/PhysRevLett.125.261104} {\bibfield
  {journal} {\bibinfo  {journal} {Phys. Rev. Lett.}\ }\textbf {\bibinfo
  {volume} {125}},\ \bibinfo {pages} {261104} (\bibinfo {year} {2020})},\
  \Eprint {http://arxiv.org/abs/2006.16296} {arXiv:2006.16296 [astro-ph.HE]}
  \BibitemShut {NoStop}%
\bibitem [{\citenamefont {Tan}\ \emph {et~al.}(2022{\natexlab{a}})\citenamefont
  {Tan}, \citenamefont {Dore}, \citenamefont {Dexheimer}, \citenamefont
  {Noronha-Hostler},\ and\ \citenamefont {Yunes}}]{Tan:2021ahl}%
  \BibitemOpen
  \bibfield  {author} {\bibinfo {author} {\bibfnamefont {H.}~\bibnamefont
  {Tan}}, \bibinfo {author} {\bibfnamefont {T.}~\bibnamefont {Dore}}, \bibinfo
  {author} {\bibfnamefont {V.}~\bibnamefont {Dexheimer}}, \bibinfo {author}
  {\bibfnamefont {J.}~\bibnamefont {Noronha-Hostler}}, \ and\ \bibinfo {author}
  {\bibfnamefont {N.}~\bibnamefont {Yunes}},\ }\href {\doibase
  10.1103/PhysRevD.105.023018} {\bibfield  {journal} {\bibinfo  {journal}
  {Phys. Rev. D}\ }\textbf {\bibinfo {volume} {105}},\ \bibinfo {pages}
  {023018} (\bibinfo {year} {2022}{\natexlab{a}})},\ \Eprint
  {http://arxiv.org/abs/2106.03890} {arXiv:2106.03890 [astro-ph.HE]}
  \BibitemShut {NoStop}%
\bibitem [{\citenamefont {Carson}\ \emph {et~al.}(2019)\citenamefont {Carson},
  \citenamefont {Chatziioannou}, \citenamefont {Haster}, \citenamefont {Yagi},\
  and\ \citenamefont {Yunes}}]{Carson:2019rjx}%
  \BibitemOpen
  \bibfield  {author} {\bibinfo {author} {\bibfnamefont {Z.}~\bibnamefont
  {Carson}}, \bibinfo {author} {\bibfnamefont {K.}~\bibnamefont
  {Chatziioannou}}, \bibinfo {author} {\bibfnamefont {C.-J.}\ \bibnamefont
  {Haster}}, \bibinfo {author} {\bibfnamefont {K.}~\bibnamefont {Yagi}}, \ and\
  \bibinfo {author} {\bibfnamefont {N.}~\bibnamefont {Yunes}},\ }\href
  {\doibase 10.1103/PhysRevD.99.083016} {\bibfield  {journal} {\bibinfo
  {journal} {Phys. Rev. D}\ }\textbf {\bibinfo {volume} {99}},\ \bibinfo
  {pages} {083016} (\bibinfo {year} {2019})},\ \Eprint
  {http://arxiv.org/abs/1903.03909} {arXiv:1903.03909 [gr-qc]} \BibitemShut
  {NoStop}%
\bibitem [{\citenamefont {Riley}\ \emph {et~al.}(2019)\citenamefont {Riley}
  \emph {et~al.}}]{Riley:2019yda}%
  \BibitemOpen
  \bibfield  {author} {\bibinfo {author} {\bibfnamefont {T.~E.}\ \bibnamefont
  {Riley}} \emph {et~al.},\ }\href {\doibase 10.3847/2041-8213/ab481c}
  {\bibfield  {journal} {\bibinfo  {journal} {Astrophys. J. Lett.}\ }\textbf
  {\bibinfo {volume} {887}},\ \bibinfo {pages} {L21} (\bibinfo {year}
  {2019})},\ \Eprint {http://arxiv.org/abs/1912.05702} {arXiv:1912.05702
  [astro-ph.HE]} \BibitemShut {NoStop}%
\bibitem [{\citenamefont {Miller}\ \emph {et~al.}(2019)\citenamefont {Miller}
  \emph {et~al.}}]{Miller:2019cac}%
  \BibitemOpen
  \bibfield  {author} {\bibinfo {author} {\bibfnamefont {M.~C.}\ \bibnamefont
  {Miller}} \emph {et~al.},\ }\href {\doibase 10.3847/2041-8213/ab50c5}
  {\bibfield  {journal} {\bibinfo  {journal} {Astrophys. J. Lett.}\ }\textbf
  {\bibinfo {volume} {887}},\ \bibinfo {pages} {L24} (\bibinfo {year}
  {2019})},\ \Eprint {http://arxiv.org/abs/1912.05705} {arXiv:1912.05705
  [astro-ph.HE]} \BibitemShut {NoStop}%
\bibitem [{\citenamefont {Riley}\ \emph {et~al.}(2021)\citenamefont {Riley}
  \emph {et~al.}}]{Riley:2021pdl}%
  \BibitemOpen
  \bibfield  {author} {\bibinfo {author} {\bibfnamefont {T.~E.}\ \bibnamefont
  {Riley}} \emph {et~al.},\ }\href {\doibase 10.3847/2041-8213/ac0a81}
  {\bibfield  {journal} {\bibinfo  {journal} {Astrophys. J. Lett.}\ }\textbf
  {\bibinfo {volume} {918}},\ \bibinfo {pages} {L27} (\bibinfo {year}
  {2021})},\ \Eprint {http://arxiv.org/abs/2105.06980} {arXiv:2105.06980
  [astro-ph.HE]} \BibitemShut {NoStop}%
\bibitem [{\citenamefont {Miller}\ \emph {et~al.}(2021)\citenamefont {Miller}
  \emph {et~al.}}]{Miller:2021qha}%
  \BibitemOpen
  \bibfield  {author} {\bibinfo {author} {\bibfnamefont {M.~C.}\ \bibnamefont
  {Miller}} \emph {et~al.},\ }\href {\doibase 10.3847/2041-8213/ac089b}
  {\bibfield  {journal} {\bibinfo  {journal} {Astrophys. J. Lett.}\ }\textbf
  {\bibinfo {volume} {918}},\ \bibinfo {pages} {L28} (\bibinfo {year}
  {2021})},\ \Eprint {http://arxiv.org/abs/2105.06979} {arXiv:2105.06979
  [astro-ph.HE]} \BibitemShut {NoStop}%
\bibitem [{\citenamefont {Abbott}\ \emph {et~al.}(2023)\citenamefont {Abbott}
  \emph {et~al.}}]{LIGOScientific:2021psn}%
  \BibitemOpen
  \bibfield  {author} {\bibinfo {author} {\bibfnamefont {R.}~\bibnamefont
  {Abbott}} \emph {et~al.} (\bibinfo {collaboration} {KAGRA, VIRGO, LIGO
  Scientific}),\ }\href {\doibase 10.1103/PhysRevX.13.011048} {\bibfield
  {journal} {\bibinfo  {journal} {Phys. Rev. X}\ }\textbf {\bibinfo {volume}
  {13}},\ \bibinfo {pages} {011048} (\bibinfo {year} {2023})},\ \Eprint
  {http://arxiv.org/abs/2111.03634} {arXiv:2111.03634 [astro-ph.HE]}
  \BibitemShut {NoStop}%
\bibitem [{\citenamefont {Abbott}\ \emph {et~al.}(2017)\citenamefont {Abbott}
  \emph {et~al.}}]{TheLIGOScientific:2017qsa}%
  \BibitemOpen
  \bibfield  {author} {\bibinfo {author} {\bibfnamefont {B.~P.}\ \bibnamefont
  {Abbott}} \emph {et~al.} (\bibinfo {collaboration} {LIGO Scientific,
  Virgo}),\ }\href {\doibase 10.1103/PhysRevLett.119.161101} {\bibfield
  {journal} {\bibinfo  {journal} {Phys. Rev. Lett.}\ }\textbf {\bibinfo
  {volume} {119}},\ \bibinfo {pages} {161101} (\bibinfo {year} {2017})},\
  \Eprint {http://arxiv.org/abs/1710.05832} {arXiv:1710.05832 [gr-qc]}
  \BibitemShut {NoStop}%
\bibitem [{\citenamefont {Abbott}\ \emph {et~al.}(2018)\citenamefont {Abbott}
  \emph {et~al.}}]{LIGOScientific:2018cki}%
  \BibitemOpen
  \bibfield  {author} {\bibinfo {author} {\bibfnamefont {B.~P.}\ \bibnamefont
  {Abbott}} \emph {et~al.} (\bibinfo {collaboration} {LIGO Scientific,
  Virgo}),\ }\href {\doibase 10.1103/PhysRevLett.121.161101} {\bibfield
  {journal} {\bibinfo  {journal} {Phys. Rev. Lett.}\ }\textbf {\bibinfo
  {volume} {121}},\ \bibinfo {pages} {161101} (\bibinfo {year} {2018})},\
  \Eprint {http://arxiv.org/abs/1805.11581} {arXiv:1805.11581 [gr-qc]}
  \BibitemShut {NoStop}%
\bibitem [{\citenamefont {McLerran}\ and\ \citenamefont
  {Reddy}(2019)}]{McLerran:2018hbz}%
  \BibitemOpen
  \bibfield  {author} {\bibinfo {author} {\bibfnamefont {L.}~\bibnamefont
  {McLerran}}\ and\ \bibinfo {author} {\bibfnamefont {S.}~\bibnamefont
  {Reddy}},\ }\href {\doibase 10.1103/PhysRevLett.122.122701} {\bibfield
  {journal} {\bibinfo  {journal} {Phys. Rev. Lett.}\ }\textbf {\bibinfo
  {volume} {122}},\ \bibinfo {pages} {122701} (\bibinfo {year} {2019})},\
  \Eprint {http://arxiv.org/abs/1811.12503} {arXiv:1811.12503 [nucl-th]}
  \BibitemShut {NoStop}%
\bibitem [{\citenamefont {Zhao}\ and\ \citenamefont
  {Lattimer}(2020)}]{Zhao:2020dvu}%
  \BibitemOpen
  \bibfield  {author} {\bibinfo {author} {\bibfnamefont {T.}~\bibnamefont
  {Zhao}}\ and\ \bibinfo {author} {\bibfnamefont {J.~M.}\ \bibnamefont
  {Lattimer}},\ }\href {\doibase 10.1103/PhysRevD.102.023021} {\bibfield
  {journal} {\bibinfo  {journal} {Phys. Rev. D}\ }\textbf {\bibinfo {volume}
  {102}},\ \bibinfo {pages} {023021} (\bibinfo {year} {2020})},\ \Eprint
  {http://arxiv.org/abs/2004.08293} {arXiv:2004.08293 [astro-ph.HE]}
  \BibitemShut {NoStop}%
\bibitem [{\citenamefont {Sen}\ and\ \citenamefont
  {Sivertsen}(2021)}]{Sen:2020qcd}%
  \BibitemOpen
  \bibfield  {author} {\bibinfo {author} {\bibfnamefont {S.}~\bibnamefont
  {Sen}}\ and\ \bibinfo {author} {\bibfnamefont {L.}~\bibnamefont
  {Sivertsen}},\ }\href {\doibase 10.3847/1538-4357/abff4c} {\bibfield
  {journal} {\bibinfo  {journal} {Astrophys. J.}\ }\textbf {\bibinfo {volume}
  {915}},\ \bibinfo {pages} {109} (\bibinfo {year} {2021})},\ \Eprint
  {http://arxiv.org/abs/2011.04681} {arXiv:2011.04681 [astro-ph.HE]}
  \BibitemShut {NoStop}%
\bibitem [{\citenamefont {Duarte}\ \emph {et~al.}(2020)\citenamefont {Duarte},
  \citenamefont {Hernandez-Ortiz},\ and\ \citenamefont
  {Jeong}}]{Duarte:2020xsp}%
  \BibitemOpen
  \bibfield  {author} {\bibinfo {author} {\bibfnamefont {D.~C.}\ \bibnamefont
  {Duarte}}, \bibinfo {author} {\bibfnamefont {S.}~\bibnamefont
  {Hernandez-Ortiz}}, \ and\ \bibinfo {author} {\bibfnamefont {K.~S.}\
  \bibnamefont {Jeong}},\ }\href {\doibase 10.1103/PhysRevC.102.025203}
  {\bibfield  {journal} {\bibinfo  {journal} {Phys. Rev. C}\ }\textbf {\bibinfo
  {volume} {102}},\ \bibinfo {pages} {025203} (\bibinfo {year} {2020})},\
  \Eprint {http://arxiv.org/abs/2003.02362} {arXiv:2003.02362 [nucl-th]}
  \BibitemShut {NoStop}%
\bibitem [{\citenamefont {Sen}\ and\ \citenamefont
  {Warrington}(2021)}]{Sen:2020peq}%
  \BibitemOpen
  \bibfield  {author} {\bibinfo {author} {\bibfnamefont {S.}~\bibnamefont
  {Sen}}\ and\ \bibinfo {author} {\bibfnamefont {N.~C.}\ \bibnamefont
  {Warrington}},\ }\href {\doibase 10.1016/j.nuclphysa.2020.122059} {\bibfield
  {journal} {\bibinfo  {journal} {Nucl. Phys. A}\ }\textbf {\bibinfo {volume}
  {1006}},\ \bibinfo {pages} {122059} (\bibinfo {year} {2021})},\ \Eprint
  {http://arxiv.org/abs/2002.11133} {arXiv:2002.11133 [nucl-th]} \BibitemShut
  {NoStop}%
\bibitem [{\citenamefont {Tan}\ \emph {et~al.}(2022{\natexlab{b}})\citenamefont
  {Tan}, \citenamefont {Dexheimer}, \citenamefont {Noronha-Hostler},\ and\
  \citenamefont {Yunes}}]{Tan:2021nat}%
  \BibitemOpen
  \bibfield  {author} {\bibinfo {author} {\bibfnamefont {H.}~\bibnamefont
  {Tan}}, \bibinfo {author} {\bibfnamefont {V.}~\bibnamefont {Dexheimer}},
  \bibinfo {author} {\bibfnamefont {J.}~\bibnamefont {Noronha-Hostler}}, \ and\
  \bibinfo {author} {\bibfnamefont {N.}~\bibnamefont {Yunes}},\ }\href
  {\doibase 10.1103/PhysRevLett.128.161101} {\bibfield  {journal} {\bibinfo
  {journal} {Phys. Rev. Lett.}\ }\textbf {\bibinfo {volume} {128}},\ \bibinfo
  {pages} {161101} (\bibinfo {year} {2022}{\natexlab{b}})},\ \Eprint
  {http://arxiv.org/abs/2111.10260} {arXiv:2111.10260 [astro-ph.HE]}
  \BibitemShut {NoStop}%
\bibitem [{\citenamefont {Dexheimer}\ \emph {et~al.}(2015)\citenamefont
  {Dexheimer}, \citenamefont {Negreiros},\ and\ \citenamefont
  {Schramm}}]{Dexheimer:2014pea}%
  \BibitemOpen
  \bibfield  {author} {\bibinfo {author} {\bibfnamefont {V.}~\bibnamefont
  {Dexheimer}}, \bibinfo {author} {\bibfnamefont {R.}~\bibnamefont
  {Negreiros}}, \ and\ \bibinfo {author} {\bibfnamefont {S.}~\bibnamefont
  {Schramm}},\ }\href {\doibase 10.1103/PhysRevC.91.055808} {\bibfield
  {journal} {\bibinfo  {journal} {Phys. Rev. C}\ }\textbf {\bibinfo {volume}
  {91}},\ \bibinfo {pages} {055808} (\bibinfo {year} {2015})},\ \Eprint
  {http://arxiv.org/abs/1411.4623} {arXiv:1411.4623 [astro-ph.HE]} \BibitemShut
  {NoStop}%
\bibitem [{\citenamefont {Dutra}\ \emph {et~al.}(2016)\citenamefont {Dutra},
  \citenamefont {Louren\c{c}o},\ and\ \citenamefont {Menezes}}]{Dutra:2015hxa}%
  \BibitemOpen
  \bibfield  {author} {\bibinfo {author} {\bibfnamefont {M.}~\bibnamefont
  {Dutra}}, \bibinfo {author} {\bibfnamefont {O.}~\bibnamefont {Louren\c{c}o}},
  \ and\ \bibinfo {author} {\bibfnamefont {D.~P.}\ \bibnamefont {Menezes}},\
  }\href {\doibase 10.1103/PhysRevC.93.025806} {\bibfield  {journal} {\bibinfo
  {journal} {Phys. Rev. C}\ }\textbf {\bibinfo {volume} {93}},\ \bibinfo
  {pages} {025806} (\bibinfo {year} {2016})},\ \bibinfo {note} {[Erratum:
  Phys.Rev.C 94, 049901 (2016)]},\ \Eprint {http://arxiv.org/abs/1510.02060}
  {arXiv:1510.02060 [astro-ph.HE]} \BibitemShut {NoStop}%
\bibitem [{\citenamefont {Jakobus}\ \emph {et~al.}(2021)\citenamefont
  {Jakobus}, \citenamefont {Motornenko}, \citenamefont {Gomes}, \citenamefont
  {Steinheimer},\ and\ \citenamefont {Stoecker}}]{Jakobus:2020nxw}%
  \BibitemOpen
  \bibfield  {author} {\bibinfo {author} {\bibfnamefont {P.}~\bibnamefont
  {Jakobus}}, \bibinfo {author} {\bibfnamefont {A.}~\bibnamefont {Motornenko}},
  \bibinfo {author} {\bibfnamefont {R.~O.}\ \bibnamefont {Gomes}}, \bibinfo
  {author} {\bibfnamefont {J.}~\bibnamefont {Steinheimer}}, \ and\ \bibinfo
  {author} {\bibfnamefont {H.}~\bibnamefont {Stoecker}},\ }\href {\doibase
  10.1140/epjc/s10052-020-08779-x} {\bibfield  {journal} {\bibinfo  {journal}
  {Eur. Phys. J. C}\ }\textbf {\bibinfo {volume} {81}},\ \bibinfo {pages} {41}
  (\bibinfo {year} {2021})},\ \Eprint {http://arxiv.org/abs/2004.07026}
  {arXiv:2004.07026 [nucl-th]} \BibitemShut {NoStop}%
\bibitem [{\citenamefont {Alford}\ and\ \citenamefont
  {Sedrakian}(2017)}]{Alford:2017qgh}%
  \BibitemOpen
  \bibfield  {author} {\bibinfo {author} {\bibfnamefont {M.~G.}\ \bibnamefont
  {Alford}}\ and\ \bibinfo {author} {\bibfnamefont {A.}~\bibnamefont
  {Sedrakian}},\ }\href {\doibase 10.1103/PhysRevLett.119.161104} {\bibfield
  {journal} {\bibinfo  {journal} {Phys. Rev. Lett.}\ }\textbf {\bibinfo
  {volume} {119}},\ \bibinfo {pages} {161104} (\bibinfo {year} {2017})},\
  \Eprint {http://arxiv.org/abs/1706.01592} {arXiv:1706.01592 [astro-ph.HE]}
  \BibitemShut {NoStop}%
\bibitem [{\citenamefont {Zacchi}\ \emph {et~al.}(2016)\citenamefont {Zacchi},
  \citenamefont {Hanauske},\ and\ \citenamefont
  {Schaffner-Bielich}}]{Zacchi:2015oma}%
  \BibitemOpen
  \bibfield  {author} {\bibinfo {author} {\bibfnamefont {A.}~\bibnamefont
  {Zacchi}}, \bibinfo {author} {\bibfnamefont {M.}~\bibnamefont {Hanauske}}, \
  and\ \bibinfo {author} {\bibfnamefont {J.}~\bibnamefont
  {Schaffner-Bielich}},\ }\href {\doibase 10.1103/PhysRevD.93.065011}
  {\bibfield  {journal} {\bibinfo  {journal} {Phys. Rev. D}\ }\textbf {\bibinfo
  {volume} {93}},\ \bibinfo {pages} {065011} (\bibinfo {year} {2016})},\
  \Eprint {http://arxiv.org/abs/1510.00180} {arXiv:1510.00180 [nucl-th]}
  \BibitemShut {NoStop}%
\bibitem [{\citenamefont {Alvarez-Castillo}\ \emph {et~al.}(2019)\citenamefont
  {Alvarez-Castillo}, \citenamefont {Blaschke}, \citenamefont {Grunfeld},\ and\
  \citenamefont {Pagura}}]{Alvarez-Castillo:2018pve}%
  \BibitemOpen
  \bibfield  {author} {\bibinfo {author} {\bibfnamefont {D.~E.}\ \bibnamefont
  {Alvarez-Castillo}}, \bibinfo {author} {\bibfnamefont {D.~B.}\ \bibnamefont
  {Blaschke}}, \bibinfo {author} {\bibfnamefont {A.~G.}\ \bibnamefont
  {Grunfeld}}, \ and\ \bibinfo {author} {\bibfnamefont {V.~P.}\ \bibnamefont
  {Pagura}},\ }\href {\doibase 10.1103/PhysRevD.99.063010} {\bibfield
  {journal} {\bibinfo  {journal} {Phys. Rev. D}\ }\textbf {\bibinfo {volume}
  {99}},\ \bibinfo {pages} {063010} (\bibinfo {year} {2019})},\ \Eprint
  {http://arxiv.org/abs/1805.04105} {arXiv:1805.04105 [hep-ph]} \BibitemShut
  {NoStop}%
\bibitem [{\citenamefont {Li}\ \emph {et~al.}(2020)\citenamefont {Li},
  \citenamefont {Sedrakian},\ and\ \citenamefont {Alford}}]{Li:2019fqe}%
  \BibitemOpen
  \bibfield  {author} {\bibinfo {author} {\bibfnamefont {J.~J.}\ \bibnamefont
  {Li}}, \bibinfo {author} {\bibfnamefont {A.}~\bibnamefont {Sedrakian}}, \
  and\ \bibinfo {author} {\bibfnamefont {M.}~\bibnamefont {Alford}},\ }\href
  {\doibase 10.1103/PhysRevD.101.063022} {\bibfield  {journal} {\bibinfo
  {journal} {Phys. Rev. D}\ }\textbf {\bibinfo {volume} {101}},\ \bibinfo
  {pages} {063022} (\bibinfo {year} {2020})},\ \Eprint
  {http://arxiv.org/abs/1911.00276} {arXiv:1911.00276 [astro-ph.HE]}
  \BibitemShut {NoStop}%
\bibitem [{\citenamefont {Wang}\ \emph {et~al.}(2022)\citenamefont {Wang},
  \citenamefont {Shi}, \citenamefont {Yan},\ and\ \citenamefont
  {Zong}}]{Wang:2019npj}%
  \BibitemOpen
  \bibfield  {author} {\bibinfo {author} {\bibfnamefont {Q.-w.}\ \bibnamefont
  {Wang}}, \bibinfo {author} {\bibfnamefont {C.}~\bibnamefont {Shi}}, \bibinfo
  {author} {\bibfnamefont {Y.}~\bibnamefont {Yan}}, \ and\ \bibinfo {author}
  {\bibfnamefont {H.-S.}\ \bibnamefont {Zong}},\ }\href {\doibase
  10.1016/j.nuclphysa.2022.122489} {\bibfield  {journal} {\bibinfo  {journal}
  {Nucl. Phys. A}\ }\textbf {\bibinfo {volume} {1025}},\ \bibinfo {pages}
  {122489} (\bibinfo {year} {2022})},\ \Eprint
  {http://arxiv.org/abs/1912.02312} {arXiv:1912.02312 [hep-ph]} \BibitemShut
  {NoStop}%
\bibitem [{\citenamefont {Bitaghsir~Fadafan}\ \emph {et~al.}(2020)\citenamefont
  {Bitaghsir~Fadafan}, \citenamefont {Cruz~Rojas},\ and\ \citenamefont
  {Evans}}]{Fadafa:2019euu}%
  \BibitemOpen
  \bibfield  {author} {\bibinfo {author} {\bibfnamefont {K.}~\bibnamefont
  {Bitaghsir~Fadafan}}, \bibinfo {author} {\bibfnamefont {J.}~\bibnamefont
  {Cruz~Rojas}}, \ and\ \bibinfo {author} {\bibfnamefont {N.}~\bibnamefont
  {Evans}},\ }\href {\doibase 10.1103/PhysRevD.101.126005} {\bibfield
  {journal} {\bibinfo  {journal} {Phys. Rev. D}\ }\textbf {\bibinfo {volume}
  {101}},\ \bibinfo {pages} {126005} (\bibinfo {year} {2020})},\ \Eprint
  {http://arxiv.org/abs/1911.12705} {arXiv:1911.12705 [hep-ph]} \BibitemShut
  {NoStop}%
\bibitem [{\citenamefont {Xia}\ \emph {et~al.}(2021)\citenamefont {Xia},
  \citenamefont {Zhu}, \citenamefont {Zhou},\ and\ \citenamefont
  {Li}}]{Xia:2019xax}%
  \BibitemOpen
  \bibfield  {author} {\bibinfo {author} {\bibfnamefont {C.}~\bibnamefont
  {Xia}}, \bibinfo {author} {\bibfnamefont {Z.}~\bibnamefont {Zhu}}, \bibinfo
  {author} {\bibfnamefont {X.}~\bibnamefont {Zhou}}, \ and\ \bibinfo {author}
  {\bibfnamefont {A.}~\bibnamefont {Li}},\ }\href {\doibase
  10.1088/1674-1137/abea0d} {\bibfield  {journal} {\bibinfo  {journal} {Chin.
  Phys. C}\ }\textbf {\bibinfo {volume} {45}},\ \bibinfo {pages} {055104}
  (\bibinfo {year} {2021})},\ \Eprint {http://arxiv.org/abs/1906.00826}
  {arXiv:1906.00826 [nucl-th]} \BibitemShut {NoStop}%
\bibitem [{\citenamefont {Yazdizadeh}\ and\ \citenamefont
  {Bordbar}(2019)}]{Yazdizadeh:2019ivy}%
  \BibitemOpen
  \bibfield  {author} {\bibinfo {author} {\bibfnamefont {T.}~\bibnamefont
  {Yazdizadeh}}\ and\ \bibinfo {author} {\bibfnamefont {G.~H.}\ \bibnamefont
  {Bordbar}},\ }\href {\doibase 10.1007/s40995-019-00731-3} {\bibfield
  {journal} {\bibinfo  {journal} {Iran. J. Sci. Technol. A}\ }\textbf {\bibinfo
  {volume} {43}},\ \bibinfo {pages} {2691} (\bibinfo {year} {2019})},\ \Eprint
  {http://arxiv.org/abs/1906.00175} {arXiv:1906.00175 [nucl-th]} \BibitemShut
  {NoStop}%
\bibitem [{\citenamefont {Shahrbaf}\ \emph {et~al.}(2020)\citenamefont
  {Shahrbaf}, \citenamefont {Blaschke}, \citenamefont {Grunfeld},\ and\
  \citenamefont {Moshfegh}}]{Shahrbaf:2019vtf}%
  \BibitemOpen
  \bibfield  {author} {\bibinfo {author} {\bibfnamefont {M.}~\bibnamefont
  {Shahrbaf}}, \bibinfo {author} {\bibfnamefont {D.}~\bibnamefont {Blaschke}},
  \bibinfo {author} {\bibfnamefont {A.~G.}\ \bibnamefont {Grunfeld}}, \ and\
  \bibinfo {author} {\bibfnamefont {H.~R.}\ \bibnamefont {Moshfegh}},\ }\href
  {\doibase 10.1103/PhysRevC.101.025807} {\bibfield  {journal} {\bibinfo
  {journal} {Phys. Rev. C}\ }\textbf {\bibinfo {volume} {101}},\ \bibinfo
  {pages} {025807} (\bibinfo {year} {2020})},\ \Eprint
  {http://arxiv.org/abs/1908.04740} {arXiv:1908.04740 [nucl-th]} \BibitemShut
  {NoStop}%
\bibitem [{\citenamefont {Zacchi}\ and\ \citenamefont
  {Schaffner-Bielich}(2019)}]{Zacchi:2019ayh}%
  \BibitemOpen
  \bibfield  {author} {\bibinfo {author} {\bibfnamefont {A.}~\bibnamefont
  {Zacchi}}\ and\ \bibinfo {author} {\bibfnamefont {J.}~\bibnamefont
  {Schaffner-Bielich}},\ }\href {\doibase 10.1103/PhysRevD.100.123024}
  {\bibfield  {journal} {\bibinfo  {journal} {Phys. Rev. D}\ }\textbf {\bibinfo
  {volume} {100}},\ \bibinfo {pages} {123024} (\bibinfo {year} {2019})},\
  \Eprint {http://arxiv.org/abs/1909.12071} {arXiv:1909.12071 [nucl-th]}
  \BibitemShut {NoStop}%
\bibitem [{\citenamefont {Lopes}\ and\ \citenamefont
  {Menezes}(2021)}]{Lopes:2020rqn}%
  \BibitemOpen
  \bibfield  {author} {\bibinfo {author} {\bibfnamefont {L.~L.}\ \bibnamefont
  {Lopes}}\ and\ \bibinfo {author} {\bibfnamefont {D.~P.}\ \bibnamefont
  {Menezes}},\ }\href {\doibase 10.1016/j.nuclphysa.2021.122171} {\bibfield
  {journal} {\bibinfo  {journal} {Nucl. Phys. A}\ }\textbf {\bibinfo {volume}
  {1009}},\ \bibinfo {pages} {122171} (\bibinfo {year} {2021})},\ \Eprint
  {http://arxiv.org/abs/2004.07909} {arXiv:2004.07909 [astro-ph.HE]}
  \BibitemShut {NoStop}%
\bibitem [{\citenamefont {Blaschke}\ \emph {et~al.}(2020)\citenamefont
  {Blaschke}, \citenamefont {Grigorian},\ and\ \citenamefont
  {R\"opke}}]{Blaschke:2020qrs}%
  \BibitemOpen
  \bibfield  {author} {\bibinfo {author} {\bibfnamefont {D.}~\bibnamefont
  {Blaschke}}, \bibinfo {author} {\bibfnamefont {H.}~\bibnamefont {Grigorian}},
  \ and\ \bibinfo {author} {\bibfnamefont {G.}~\bibnamefont {R\"opke}},\ }\href
  {\doibase 10.3390/particles3020033} {\bibfield  {journal} {\bibinfo
  {journal} {Particles}\ }\textbf {\bibinfo {volume} {3}},\ \bibinfo {pages}
  {477} (\bibinfo {year} {2020})},\ \Eprint {http://arxiv.org/abs/2005.10218}
  {arXiv:2005.10218 [nucl-th]} \BibitemShut {NoStop}%
\bibitem [{\citenamefont {Rho}(2020)}]{Rho:2020eqo}%
  \BibitemOpen
  \bibfield  {author} {\bibinfo {author} {\bibfnamefont {M.}~\bibnamefont
  {Rho}},\ }\href@noop {} {\  (\bibinfo {year} {2020})},\ \Eprint
  {http://arxiv.org/abs/2004.09082} {arXiv:2004.09082 [nucl-th]} \BibitemShut
  {NoStop}%
\bibitem [{\citenamefont {Marczenko}(2020)}]{Marczenko:2020wlc}%
  \BibitemOpen
  \bibfield  {author} {\bibinfo {author} {\bibfnamefont {M.}~\bibnamefont
  {Marczenko}},\ }\href {\doibase 10.1140/epjst/e2020-000093-3} {\bibfield
  {journal} {\bibinfo  {journal} {Eur. Phys. J. ST}\ }\textbf {\bibinfo
  {volume} {229}},\ \bibinfo {pages} {3651} (\bibinfo {year} {2020})},\ \Eprint
  {http://arxiv.org/abs/2005.14535} {arXiv:2005.14535 [nucl-th]} \BibitemShut
  {NoStop}%
\bibitem [{\citenamefont {Minamikawa}\ \emph {et~al.}(2021)\citenamefont
  {Minamikawa}, \citenamefont {Kojo},\ and\ \citenamefont
  {Harada}}]{Minamikawa:2020jfj}%
  \BibitemOpen
  \bibfield  {author} {\bibinfo {author} {\bibfnamefont {T.}~\bibnamefont
  {Minamikawa}}, \bibinfo {author} {\bibfnamefont {T.}~\bibnamefont {Kojo}}, \
  and\ \bibinfo {author} {\bibfnamefont {M.}~\bibnamefont {Harada}},\ }\href
  {\doibase 10.1103/PhysRevC.103.045205} {\bibfield  {journal} {\bibinfo
  {journal} {Phys. Rev. C}\ }\textbf {\bibinfo {volume} {103}},\ \bibinfo
  {pages} {045205} (\bibinfo {year} {2021})},\ \Eprint
  {http://arxiv.org/abs/2011.13684} {arXiv:2011.13684 [nucl-th]} \BibitemShut
  {NoStop}%
\bibitem [{\citenamefont {Hippert}\ \emph {et~al.}(2021)\citenamefont
  {Hippert}, \citenamefont {Fraga},\ and\ \citenamefont
  {Noronha}}]{Hippert:2021gfs}%
  \BibitemOpen
  \bibfield  {author} {\bibinfo {author} {\bibfnamefont {M.}~\bibnamefont
  {Hippert}}, \bibinfo {author} {\bibfnamefont {E.~S.}\ \bibnamefont {Fraga}},
  \ and\ \bibinfo {author} {\bibfnamefont {J.}~\bibnamefont {Noronha}},\ }\href
  {\doibase 10.1103/PhysRevD.104.034011} {\bibfield  {journal} {\bibinfo
  {journal} {Phys. Rev. D}\ }\textbf {\bibinfo {volume} {104}},\ \bibinfo
  {pages} {034011} (\bibinfo {year} {2021})},\ \Eprint
  {http://arxiv.org/abs/2105.04535} {arXiv:2105.04535 [nucl-th]} \BibitemShut
  {NoStop}%
\bibitem [{\citenamefont {Pisarski}(2021)}]{Pisarski:2021aoz}%
  \BibitemOpen
  \bibfield  {author} {\bibinfo {author} {\bibfnamefont {R.~D.}\ \bibnamefont
  {Pisarski}},\ }\href {\doibase 10.1103/PhysRevD.103.L071504} {\bibfield
  {journal} {\bibinfo  {journal} {Phys. Rev. D}\ }\textbf {\bibinfo {volume}
  {103}},\ \bibinfo {pages} {L071504} (\bibinfo {year} {2021})},\ \Eprint
  {http://arxiv.org/abs/2101.05813} {arXiv:2101.05813 [nucl-th]} \BibitemShut
  {NoStop}%
\bibitem [{\citenamefont {Stone}\ \emph {et~al.}(2021)\citenamefont {Stone},
  \citenamefont {Dexheimer}, \citenamefont {Guichon}, \citenamefont {Thomas},\
  and\ \citenamefont {Typel}}]{Stone:2021ngh}%
  \BibitemOpen
  \bibfield  {author} {\bibinfo {author} {\bibfnamefont {J.~R.}\ \bibnamefont
  {Stone}}, \bibinfo {author} {\bibfnamefont {V.}~\bibnamefont {Dexheimer}},
  \bibinfo {author} {\bibfnamefont {P.~A.~M.}\ \bibnamefont {Guichon}},
  \bibinfo {author} {\bibfnamefont {A.~W.}\ \bibnamefont {Thomas}}, \ and\
  \bibinfo {author} {\bibfnamefont {S.}~\bibnamefont {Typel}},\ }\href
  {\doibase 10.1093/mnras/staa4006} {\bibfield  {journal} {\bibinfo  {journal}
  {Mon. Not. Roy. Astron. Soc.}\ }\textbf {\bibinfo {volume} {502}},\ \bibinfo
  {pages} {3476} (\bibinfo {year} {2021})},\ \Eprint
  {http://arxiv.org/abs/1906.11100} {arXiv:1906.11100 [nucl-th]} \BibitemShut
  {NoStop}%
\bibitem [{\citenamefont {Ferreira}\ \emph {et~al.}(2020)\citenamefont
  {Ferreira}, \citenamefont {C\^amara~Pereira},\ and\ \citenamefont
  {Provid\^encia}}]{Ferreira:2020kvu}%
  \BibitemOpen
  \bibfield  {author} {\bibinfo {author} {\bibfnamefont {M.}~\bibnamefont
  {Ferreira}}, \bibinfo {author} {\bibfnamefont {R.}~\bibnamefont
  {C\^amara~Pereira}}, \ and\ \bibinfo {author} {\bibfnamefont
  {C.}~\bibnamefont {Provid\^encia}},\ }\href {\doibase
  10.1103/PhysRevD.102.083030} {\bibfield  {journal} {\bibinfo  {journal}
  {Phys. Rev. D}\ }\textbf {\bibinfo {volume} {102}},\ \bibinfo {pages}
  {083030} (\bibinfo {year} {2020})},\ \Eprint
  {http://arxiv.org/abs/2008.12563} {arXiv:2008.12563 [nucl-th]} \BibitemShut
  {NoStop}%
\bibitem [{\citenamefont {Kapusta}\ and\ \citenamefont
  {Welle}(2021)}]{Kapusta:2021ney}%
  \BibitemOpen
  \bibfield  {author} {\bibinfo {author} {\bibfnamefont {J.~I.}\ \bibnamefont
  {Kapusta}}\ and\ \bibinfo {author} {\bibfnamefont {T.}~\bibnamefont
  {Welle}},\ }\href {\doibase 10.1103/PhysRevC.104.L012801} {\bibfield
  {journal} {\bibinfo  {journal} {Phys. Rev. C}\ }\textbf {\bibinfo {volume}
  {104}},\ \bibinfo {pages} {L012801} (\bibinfo {year} {2021})},\ \Eprint
  {http://arxiv.org/abs/2103.16633} {arXiv:2103.16633 [nucl-th]} \BibitemShut
  {NoStop}%
\bibitem [{\citenamefont {Kojo}(2021)}]{Kojo:2021ugu}%
  \BibitemOpen
  \bibfield  {author} {\bibinfo {author} {\bibfnamefont {T.}~\bibnamefont
  {Kojo}},\ }\href {\doibase 10.1103/PhysRevD.104.074005} {\bibfield  {journal}
  {\bibinfo  {journal} {Phys. Rev. D}\ }\textbf {\bibinfo {volume} {104}},\
  \bibinfo {pages} {074005} (\bibinfo {year} {2021})},\ \Eprint
  {http://arxiv.org/abs/2106.06687} {arXiv:2106.06687 [nucl-th]} \BibitemShut
  {NoStop}%
\bibitem [{\citenamefont {Somasundaram}\ and\ \citenamefont
  {Margueron}(2022)}]{Somasundaram:2021ljr}%
  \BibitemOpen
  \bibfield  {author} {\bibinfo {author} {\bibfnamefont {R.}~\bibnamefont
  {Somasundaram}}\ and\ \bibinfo {author} {\bibfnamefont {J.}~\bibnamefont
  {Margueron}},\ }\href {\doibase 10.1209/0295-5075/ac63de} {\bibfield
  {journal} {\bibinfo  {journal} {EPL}\ }\textbf {\bibinfo {volume} {138}},\
  \bibinfo {pages} {14002} (\bibinfo {year} {2022})},\ \Eprint
  {http://arxiv.org/abs/2104.13612} {arXiv:2104.13612 [astro-ph.HE]}
  \BibitemShut {NoStop}%
\bibitem [{\citenamefont {Zuo}\ \emph {et~al.}(2022)\citenamefont {Zuo},
  \citenamefont {Huang},\ and\ \citenamefont {Feng}}]{Zuo:2022rks}%
  \BibitemOpen
  \bibfield  {author} {\bibinfo {author} {\bibfnamefont {B.-J.}\ \bibnamefont
  {Zuo}}, \bibinfo {author} {\bibfnamefont {Y.-F.}\ \bibnamefont {Huang}}, \
  and\ \bibinfo {author} {\bibfnamefont {H.-T.}\ \bibnamefont {Feng}},\ }\href
  {\doibase 10.1103/PhysRevD.105.074011} {\bibfield  {journal} {\bibinfo
  {journal} {Phys. Rev. D}\ }\textbf {\bibinfo {volume} {105}},\ \bibinfo
  {pages} {074011} (\bibinfo {year} {2022})},\ \Eprint
  {http://arxiv.org/abs/2203.11791} {arXiv:2203.11791 [nucl-th]} \BibitemShut
  {NoStop}%
\bibitem [{\citenamefont {Ivanytskyi}\ and\ \citenamefont
  {Blaschke}(2022)}]{Ivanytskyi:2022oxv}%
  \BibitemOpen
  \bibfield  {author} {\bibinfo {author} {\bibfnamefont {O.}~\bibnamefont
  {Ivanytskyi}}\ and\ \bibinfo {author} {\bibfnamefont {D.}~\bibnamefont
  {Blaschke}},\ }\href {\doibase 10.1103/PhysRevD.105.114042} {\bibfield
  {journal} {\bibinfo  {journal} {Phys. Rev. D}\ }\textbf {\bibinfo {volume}
  {105}},\ \bibinfo {pages} {114042} (\bibinfo {year} {2022})},\ \Eprint
  {http://arxiv.org/abs/2204.03611} {arXiv:2204.03611 [nucl-th]} \BibitemShut
  {NoStop}%
\bibitem [{\citenamefont {Fraga}\ \emph {et~al.}(2022)\citenamefont {Fraga},
  \citenamefont {da~Mata}, \citenamefont {Pitsinigkos},\ and\ \citenamefont
  {Schmitt}}]{Fraga:2022yls}%
  \BibitemOpen
  \bibfield  {author} {\bibinfo {author} {\bibfnamefont {E.~S.}\ \bibnamefont
  {Fraga}}, \bibinfo {author} {\bibfnamefont {R.}~\bibnamefont {da~Mata}},
  \bibinfo {author} {\bibfnamefont {S.}~\bibnamefont {Pitsinigkos}}, \ and\
  \bibinfo {author} {\bibfnamefont {A.}~\bibnamefont {Schmitt}},\ }\href
  {\doibase 10.1103/PhysRevD.106.074018} {\bibfield  {journal} {\bibinfo
  {journal} {Phys. Rev. D}\ }\textbf {\bibinfo {volume} {106}},\ \bibinfo
  {pages} {074018} (\bibinfo {year} {2022})},\ \Eprint
  {http://arxiv.org/abs/2206.09219} {arXiv:2206.09219 [nucl-th]} \BibitemShut
  {NoStop}%
\bibitem [{\citenamefont {Kov\'acs}\ \emph {et~al.}(2022)\citenamefont
  {Kov\'acs}, \citenamefont {Tak\'atsy}, \citenamefont {Schaffner-Bielich},\
  and\ \citenamefont {Wolf}}]{Kovacs:2021ger}%
  \BibitemOpen
  \bibfield  {author} {\bibinfo {author} {\bibfnamefont {P.}~\bibnamefont
  {Kov\'acs}}, \bibinfo {author} {\bibfnamefont {J.}~\bibnamefont {Tak\'atsy}},
  \bibinfo {author} {\bibfnamefont {J.}~\bibnamefont {Schaffner-Bielich}}, \
  and\ \bibinfo {author} {\bibfnamefont {G.}~\bibnamefont {Wolf}},\ }\href
  {\doibase 10.1103/PhysRevD.105.103014} {\bibfield  {journal} {\bibinfo
  {journal} {Phys. Rev. D}\ }\textbf {\bibinfo {volume} {105}},\ \bibinfo
  {pages} {103014} (\bibinfo {year} {2022})},\ \Eprint
  {http://arxiv.org/abs/2111.06127} {arXiv:2111.06127 [nucl-th]} \BibitemShut
  {NoStop}%
\bibitem [{\citenamefont {Rho}(2022)}]{Rho:2022wco}%
  \BibitemOpen
  \bibfield  {author} {\bibinfo {author} {\bibfnamefont {M.}~\bibnamefont
  {Rho}},\ }\href {\doibase 10.3390/sym14102154} {\bibfield  {journal}
  {\bibinfo  {journal} {Symmetry}\ }\textbf {\bibinfo {volume} {14}},\ \bibinfo
  {pages} {2154} (\bibinfo {year} {2022})},\ \Eprint
  {http://arxiv.org/abs/2209.02327} {arXiv:2209.02327 [nucl-th]} \BibitemShut
  {NoStop}%
\bibitem [{\citenamefont {Most}\ \emph {et~al.}(2023)\citenamefont {Most},
  \citenamefont {Motornenko}, \citenamefont {Steinheimer}, \citenamefont
  {Dexheimer}, \citenamefont {Hanauske}, \citenamefont {Rezzolla},\ and\
  \citenamefont {Stoecker}}]{Most:2022wgo}%
  \BibitemOpen
  \bibfield  {author} {\bibinfo {author} {\bibfnamefont {E.~R.}\ \bibnamefont
  {Most}}, \bibinfo {author} {\bibfnamefont {A.}~\bibnamefont {Motornenko}},
  \bibinfo {author} {\bibfnamefont {J.}~\bibnamefont {Steinheimer}}, \bibinfo
  {author} {\bibfnamefont {V.}~\bibnamefont {Dexheimer}}, \bibinfo {author}
  {\bibfnamefont {M.}~\bibnamefont {Hanauske}}, \bibinfo {author}
  {\bibfnamefont {L.}~\bibnamefont {Rezzolla}}, \ and\ \bibinfo {author}
  {\bibfnamefont {H.}~\bibnamefont {Stoecker}},\ }\href {\doibase
  10.1103/PhysRevD.107.043034} {\bibfield  {journal} {\bibinfo  {journal}
  {Phys. Rev. D}\ }\textbf {\bibinfo {volume} {107}},\ \bibinfo {pages}
  {043034} (\bibinfo {year} {2023})},\ \Eprint
  {http://arxiv.org/abs/2201.13150} {arXiv:2201.13150 [nucl-th]} \BibitemShut
  {NoStop}%
\bibitem [{\citenamefont {Shao}\ \emph {et~al.}(2023)\citenamefont {Shao},
  \citenamefont {Yang}, \citenamefont {Xie},\ and\ \citenamefont
  {He}}]{Shao:2023pqu}%
  \BibitemOpen
  \bibfield  {author} {\bibinfo {author} {\bibfnamefont {G.-y.}\ \bibnamefont
  {Shao}}, \bibinfo {author} {\bibfnamefont {X.-r.}\ \bibnamefont {Yang}},
  \bibinfo {author} {\bibfnamefont {C.-l.}\ \bibnamefont {Xie}}, \ and\
  \bibinfo {author} {\bibfnamefont {W.-b.}\ \bibnamefont {He}},\ }\href
  {\doibase 10.1140/epjp/s13360-023-03696-w} {\bibfield  {journal} {\bibinfo
  {journal} {Eur. Phys. J. Plus}\ }\textbf {\bibinfo {volume} {138}},\ \bibinfo
  {pages} {44} (\bibinfo {year} {2023})},\ \Eprint
  {http://arxiv.org/abs/2301.04282} {arXiv:2301.04282 [hep-ph]} \BibitemShut
  {NoStop}%
\bibitem [{\citenamefont {Kumar}\ \emph
  {et~al.}(2023{\natexlab{b}})\citenamefont {Kumar}, \citenamefont {Dey},
  \citenamefont {Haque}, \citenamefont {Mallick},\ and\ \citenamefont
  {Patra}}]{Kumar:2023qyu}%
  \BibitemOpen
  \bibfield  {author} {\bibinfo {author} {\bibfnamefont {A.}~\bibnamefont
  {Kumar}}, \bibinfo {author} {\bibfnamefont {D.}~\bibnamefont {Dey}}, \bibinfo
  {author} {\bibfnamefont {S.}~\bibnamefont {Haque}}, \bibinfo {author}
  {\bibfnamefont {R.}~\bibnamefont {Mallick}}, \ and\ \bibinfo {author}
  {\bibfnamefont {S.~K.}\ \bibnamefont {Patra}},\ }\href@noop {} {\  (\bibinfo
  {year} {2023}{\natexlab{b}})},\ \Eprint {http://arxiv.org/abs/2304.08223}
  {arXiv:2304.08223 [nucl-th]} \BibitemShut {NoStop}%
\bibitem [{\citenamefont {Issifu}\ \emph {et~al.}(2023)\citenamefont {Issifu},
  \citenamefont {da~Silva},\ and\ \citenamefont {Menezes}}]{Issifu:2023ovi}%
  \BibitemOpen
  \bibfield  {author} {\bibinfo {author} {\bibfnamefont {A.}~\bibnamefont
  {Issifu}}, \bibinfo {author} {\bibfnamefont {F.~M.}\ \bibnamefont
  {da~Silva}}, \ and\ \bibinfo {author} {\bibfnamefont {D.~P.}\ \bibnamefont
  {Menezes}},\ }\href {\doibase 10.1093/mnras/stad2509} {\bibfield  {journal}
  {\bibinfo  {journal} {Mon. Not. Roy. Astron. Soc.}\ }\textbf {\bibinfo
  {volume} {525}},\ \bibinfo {pages} {5512} (\bibinfo {year} {2023})},\ \Eprint
  {http://arxiv.org/abs/2307.00386} {arXiv:2307.00386 [nucl-th]} \BibitemShut
  {NoStop}%
\bibitem [{\citenamefont {Yamamoto}\ \emph {et~al.}(2023)\citenamefont
  {Yamamoto}, \citenamefont {Yasutake},\ and\ \citenamefont
  {Rijken}}]{Yamamoto:2023osc}%
  \BibitemOpen
  \bibfield  {author} {\bibinfo {author} {\bibfnamefont {Y.}~\bibnamefont
  {Yamamoto}}, \bibinfo {author} {\bibfnamefont {N.}~\bibnamefont {Yasutake}},
  \ and\ \bibinfo {author} {\bibfnamefont {T.~A.}\ \bibnamefont {Rijken}},\
  }\href {\doibase 10.1103/PhysRevC.108.035811} {\bibfield  {journal} {\bibinfo
   {journal} {Phys. Rev. C}\ }\textbf {\bibinfo {volume} {108}},\ \bibinfo
  {pages} {035811} (\bibinfo {year} {2023})},\ \Eprint
  {http://arxiv.org/abs/2309.10233} {arXiv:2309.10233 [nucl-th]} \BibitemShut
  {NoStop}%
\bibitem [{\citenamefont {Kouno}\ and\ \citenamefont
  {Kashiwa}(2024)}]{Kouno:2023ygw}%
  \BibitemOpen
  \bibfield  {author} {\bibinfo {author} {\bibfnamefont {H.}~\bibnamefont
  {Kouno}}\ and\ \bibinfo {author} {\bibfnamefont {K.}~\bibnamefont
  {Kashiwa}},\ }\href {\doibase 10.1103/PhysRevD.109.054007} {\bibfield
  {journal} {\bibinfo  {journal} {Phys. Rev. D}\ }\textbf {\bibinfo {volume}
  {109}},\ \bibinfo {pages} {054007} (\bibinfo {year} {2024})},\ \Eprint
  {http://arxiv.org/abs/2310.09738} {arXiv:2310.09738 [hep-ph]} \BibitemShut
  {NoStop}%
\bibitem [{\citenamefont {Oliinychenko}\ \emph {et~al.}(2023)\citenamefont
  {Oliinychenko}, \citenamefont {Sorensen}, \citenamefont {Koch},\ and\
  \citenamefont {McLerran}}]{Oliinychenko:2022uvy}%
  \BibitemOpen
  \bibfield  {author} {\bibinfo {author} {\bibfnamefont {D.}~\bibnamefont
  {Oliinychenko}}, \bibinfo {author} {\bibfnamefont {A.}~\bibnamefont
  {Sorensen}}, \bibinfo {author} {\bibfnamefont {V.}~\bibnamefont {Koch}}, \
  and\ \bibinfo {author} {\bibfnamefont {L.}~\bibnamefont {McLerran}},\ }\href
  {\doibase 10.1103/PhysRevC.108.034908} {\bibfield  {journal} {\bibinfo
  {journal} {Phys. Rev. C}\ }\textbf {\bibinfo {volume} {108}},\ \bibinfo
  {pages} {034908} (\bibinfo {year} {2023})},\ \Eprint
  {http://arxiv.org/abs/2208.11996} {arXiv:2208.11996 [nucl-th]} \BibitemShut
  {NoStop}%
\bibitem [{\citenamefont {Weil}\ \emph {et~al.}(2016)\citenamefont {Weil} \emph
  {et~al.}}]{Weil:2016zrk}%
  \BibitemOpen
  \bibfield  {author} {\bibinfo {author} {\bibfnamefont {J.}~\bibnamefont
  {Weil}} \emph {et~al.} (\bibinfo {collaboration} {SMASH}),\ }\href {\doibase
  10.1103/PhysRevC.94.054905} {\bibfield  {journal} {\bibinfo  {journal} {Phys.
  Rev. C}\ }\textbf {\bibinfo {volume} {94}},\ \bibinfo {pages} {054905}
  (\bibinfo {year} {2016})},\ \Eprint {http://arxiv.org/abs/1606.06642}
  {arXiv:1606.06642 [nucl-th]} \BibitemShut {NoStop}%
\bibitem [{\citenamefont {Lovato}\ \emph {et~al.}(2022)\citenamefont {Lovato}
  \emph {et~al.}}]{Lovato:2022vgq}%
  \BibitemOpen
  \bibfield  {author} {\bibinfo {author} {\bibfnamefont {A.}~\bibnamefont
  {Lovato}} \emph {et~al.},\ }\href@noop {} {\  (\bibinfo {year} {2022})},\
  \Eprint {http://arxiv.org/abs/2211.02224} {arXiv:2211.02224 [nucl-th]}
  \BibitemShut {NoStop}%
\bibitem [{\citenamefont {Sorensen}\ \emph {et~al.}(2024)\citenamefont
  {Sorensen} \emph {et~al.}}]{Sorensen:2023zkk}%
  \BibitemOpen
  \bibfield  {author} {\bibinfo {author} {\bibfnamefont {A.}~\bibnamefont
  {Sorensen}} \emph {et~al.},\ }\href {\doibase 10.1016/j.ppnp.2023.104080}
  {\bibfield  {journal} {\bibinfo  {journal} {Prog. Part. Nucl. Phys.}\
  }\textbf {\bibinfo {volume} {134}},\ \bibinfo {pages} {104080} (\bibinfo
  {year} {2024})},\ \Eprint {http://arxiv.org/abs/2301.13253} {arXiv:2301.13253
  [nucl-th]} \BibitemShut {NoStop}%
\bibitem [{\citenamefont {Aryal}\ \emph {et~al.}(2020)\citenamefont {Aryal},
  \citenamefont {Constantinou}, \citenamefont {Farias},\ and\ \citenamefont
  {Dexheimer}}]{Aryal:2020ocm}%
  \BibitemOpen
  \bibfield  {author} {\bibinfo {author} {\bibfnamefont {K.}~\bibnamefont
  {Aryal}}, \bibinfo {author} {\bibfnamefont {C.}~\bibnamefont {Constantinou}},
  \bibinfo {author} {\bibfnamefont {R.~L.~S.}\ \bibnamefont {Farias}}, \ and\
  \bibinfo {author} {\bibfnamefont {V.}~\bibnamefont {Dexheimer}},\ }\href
  {\doibase 10.1103/PhysRevD.102.076016} {\bibfield  {journal} {\bibinfo
  {journal} {Phys. Rev. D}\ }\textbf {\bibinfo {volume} {102}},\ \bibinfo
  {pages} {076016} (\bibinfo {year} {2020})},\ \Eprint
  {http://arxiv.org/abs/2004.03039} {arXiv:2004.03039 [nucl-th]} \BibitemShut
  {NoStop}%
\bibitem [{\citenamefont {Gao}\ \emph {et~al.}(2017)\citenamefont {Gao},
  \citenamefont {Shan}, \citenamefont {Wang},\ and\ \citenamefont
  {Wang}}]{Gao:2017mnl}%
  \BibitemOpen
  \bibfield  {author} {\bibinfo {author} {\bibfnamefont {Z.-F.}\ \bibnamefont
  {Gao}}, \bibinfo {author} {\bibfnamefont {H.}~\bibnamefont {Shan}}, \bibinfo
  {author} {\bibfnamefont {W.}~\bibnamefont {Wang}}, \ and\ \bibinfo {author}
  {\bibfnamefont {N.}~\bibnamefont {Wang}},\ }\href {\doibase
  10.1002/asna.201713437} {\bibfield  {journal} {\bibinfo  {journal} {Astron.
  Nachr.}\ }\textbf {\bibinfo {volume} {338}},\ \bibinfo {pages} {1066}
  (\bibinfo {year} {2017})},\ \Eprint {http://arxiv.org/abs/1709.02186}
  {arXiv:1709.02186 [astro-ph.HE]} \BibitemShut {NoStop}%
\bibitem [{\citenamefont {Bombaci}\ and\ \citenamefont
  {Lombardo}(1991)}]{Bombaci:1991zz}%
  \BibitemOpen
  \bibfield  {author} {\bibinfo {author} {\bibfnamefont {I.}~\bibnamefont
  {Bombaci}}\ and\ \bibinfo {author} {\bibfnamefont {U.}~\bibnamefont
  {Lombardo}},\ }\href {\doibase 10.1103/PhysRevC.44.1892} {\bibfield
  {journal} {\bibinfo  {journal} {Phys. Rev. C}\ }\textbf {\bibinfo {volume}
  {44}},\ \bibinfo {pages} {1892} (\bibinfo {year} {1991})}\BibitemShut
  {NoStop}%
\bibitem [{\citenamefont {Baldo}\ and\ \citenamefont
  {Burgio}(2016)}]{Baldo:2016jhp}%
  \BibitemOpen
  \bibfield  {author} {\bibinfo {author} {\bibfnamefont {M.}~\bibnamefont
  {Baldo}}\ and\ \bibinfo {author} {\bibfnamefont {G.~F.}\ \bibnamefont
  {Burgio}},\ }\href {\doibase 10.1016/j.ppnp.2016.06.006} {\bibfield
  {journal} {\bibinfo  {journal} {Prog. Part. Nucl. Phys.}\ }\textbf {\bibinfo
  {volume} {91}},\ \bibinfo {pages} {203} (\bibinfo {year} {2016})},\ \Eprint
  {http://arxiv.org/abs/1606.08838} {arXiv:1606.08838 [nucl-th]} \BibitemShut
  {NoStop}%
\bibitem [{\citenamefont {Lee}\ and\ \citenamefont {Rho}(2014)}]{Lee:2014nya}%
  \BibitemOpen
  \bibfield  {author} {\bibinfo {author} {\bibfnamefont {H.~K.}\ \bibnamefont
  {Lee}}\ and\ \bibinfo {author} {\bibfnamefont {M.}~\bibnamefont {Rho}},\
  }\href {\doibase 10.1103/PhysRevC.90.045201} {\bibfield  {journal} {\bibinfo
  {journal} {Phys. Rev. C}\ }\textbf {\bibinfo {volume} {90}},\ \bibinfo
  {pages} {045201} (\bibinfo {year} {2014})},\ \Eprint
  {http://arxiv.org/abs/1405.5186} {arXiv:1405.5186 [nucl-th]} \BibitemShut
  {NoStop}%
\bibitem [{\citenamefont {Bedaque}\ and\ \citenamefont
  {Steiner}(2015{\natexlab{b}})}]{Bedaque:2014ada}%
  \BibitemOpen
  \bibfield  {author} {\bibinfo {author} {\bibfnamefont {P.~F.}\ \bibnamefont
  {Bedaque}}\ and\ \bibinfo {author} {\bibfnamefont {A.~W.}\ \bibnamefont
  {Steiner}},\ }\href {\doibase 10.1103/PhysRevC.92.025803} {\bibfield
  {journal} {\bibinfo  {journal} {Phys. Rev. C}\ }\textbf {\bibinfo {volume}
  {92}},\ \bibinfo {pages} {025803} (\bibinfo {year} {2015}{\natexlab{b}})},\
  \Eprint {http://arxiv.org/abs/1412.8686} {arXiv:1412.8686 [nucl-th]}
  \BibitemShut {NoStop}%
\bibitem [{\citenamefont {Zhang}\ and\ \citenamefont
  {Li}(2023)}]{Zhang:2022sep}%
  \BibitemOpen
  \bibfield  {author} {\bibinfo {author} {\bibfnamefont {N.-B.}\ \bibnamefont
  {Zhang}}\ and\ \bibinfo {author} {\bibfnamefont {B.-A.}\ \bibnamefont {Li}},\
  }\href {\doibase 10.1140/epja/s10050-023-01010-x} {\bibfield  {journal}
  {\bibinfo  {journal} {Eur. Phys. J. A}\ }\textbf {\bibinfo {volume} {59}},\
  \bibinfo {pages} {86} (\bibinfo {year} {2023})},\ \Eprint
  {http://arxiv.org/abs/2208.00321} {arXiv:2208.00321 [nucl-th]} \BibitemShut
  {NoStop}%
\bibitem [{sma()}]{smash_version_2.1}%
  \BibitemOpen
  \href@noop {} {\bibinfo  {journal} {SMASH version 2.1 available at
  https://zenodo.org/record/5796168}\ }\BibitemShut {NoStop}%
\bibitem [{\citenamefont {Sorensen}\ and\ \citenamefont
  {Koch}(2021)}]{Sorensen:2020ygf}%
  \BibitemOpen
\bibfield  {journal} {  }\bibfield  {author} {\bibinfo {author} {\bibfnamefont
  {A.}~\bibnamefont {Sorensen}}\ and\ \bibinfo {author} {\bibfnamefont
  {V.}~\bibnamefont {Koch}},\ }\href {\doibase 10.1103/PhysRevC.104.034904}
  {\bibfield  {journal} {\bibinfo  {journal} {Phys. Rev. C}\ }\textbf {\bibinfo
  {volume} {104}},\ \bibinfo {pages} {034904} (\bibinfo {year} {2021})},\
  \Eprint {http://arxiv.org/abs/2011.06635} {arXiv:2011.06635 [nucl-th]}
  \BibitemShut {NoStop}%
\bibitem [{\citenamefont {Adamczewski-Musch}\ \emph {et~al.}(2023)\citenamefont
  {Adamczewski-Musch} \emph {et~al.}}]{HADES:2022osk}%
  \BibitemOpen
  \bibfield  {author} {\bibinfo {author} {\bibfnamefont {J.}~\bibnamefont
  {Adamczewski-Musch}} \emph {et~al.} (\bibinfo {collaboration} {HADES}),\
  }\href {\doibase 10.1140/epja/s10050-023-00936-6} {\bibfield  {journal}
  {\bibinfo  {journal} {Eur. Phys. J. A}\ }\textbf {\bibinfo {volume} {59}},\
  \bibinfo {pages} {80} (\bibinfo {year} {2023})},\ \Eprint
  {http://arxiv.org/abs/2208.02740} {arXiv:2208.02740 [nucl-ex]} \BibitemShut
  {NoStop}%
\bibitem [{\citenamefont {Adam}\ \emph {et~al.}(2021)\citenamefont {Adam} \emph
  {et~al.}}]{STAR:2020dav}%
  \BibitemOpen
  \bibfield  {author} {\bibinfo {author} {\bibfnamefont {J.}~\bibnamefont
  {Adam}} \emph {et~al.} (\bibinfo {collaboration} {STAR}),\ }\href {\doibase
  10.1103/PhysRevC.103.034908} {\bibfield  {journal} {\bibinfo  {journal}
  {Phys. Rev. C}\ }\textbf {\bibinfo {volume} {103}},\ \bibinfo {pages}
  {034908} (\bibinfo {year} {2021})},\ \Eprint
  {http://arxiv.org/abs/2007.14005} {arXiv:2007.14005 [nucl-ex]} \BibitemShut
  {NoStop}%
\bibitem [{\citenamefont {Abdallah}\ \emph {et~al.}(2022)\citenamefont
  {Abdallah} \emph {et~al.}}]{STAR:2021yiu}%
  \BibitemOpen
  \bibfield  {author} {\bibinfo {author} {\bibfnamefont {M.~S.}\ \bibnamefont
  {Abdallah}} \emph {et~al.} (\bibinfo {collaboration} {STAR}),\ }\href
  {\doibase 10.1016/j.physletb.2022.137003} {\bibfield  {journal} {\bibinfo
  {journal} {Phys. Lett. B}\ }\textbf {\bibinfo {volume} {827}},\ \bibinfo
  {pages} {137003} (\bibinfo {year} {2022})},\ \Eprint
  {http://arxiv.org/abs/2108.00908} {arXiv:2108.00908 [nucl-ex]} \BibitemShut
  {NoStop}%
\bibitem [{\citenamefont {Dexheimer}\ and\ \citenamefont
  {Schramm}(2008)}]{Dexheimer:2008ax}%
  \BibitemOpen
  \bibfield  {author} {\bibinfo {author} {\bibfnamefont {V.}~\bibnamefont
  {Dexheimer}}\ and\ \bibinfo {author} {\bibfnamefont {S.}~\bibnamefont
  {Schramm}},\ }\href {\doibase 10.1086/589735} {\bibfield  {journal} {\bibinfo
   {journal} {Astrophys. J.}\ }\textbf {\bibinfo {volume} {683}},\ \bibinfo
  {pages} {943} (\bibinfo {year} {2008})},\ \Eprint
  {http://arxiv.org/abs/0802.1999} {arXiv:0802.1999 [astro-ph]} \BibitemShut
  {NoStop}%
\bibitem [{\citenamefont {Auvinen}\ and\ \citenamefont
  {Petersen}(2013)}]{Auvinen:2013sba}%
  \BibitemOpen
  \bibfield  {author} {\bibinfo {author} {\bibfnamefont {J.}~\bibnamefont
  {Auvinen}}\ and\ \bibinfo {author} {\bibfnamefont {H.}~\bibnamefont
  {Petersen}},\ }\href {\doibase 10.1103/PhysRevC.88.064908} {\bibfield
  {journal} {\bibinfo  {journal} {Phys. Rev. C}\ }\textbf {\bibinfo {volume}
  {88}},\ \bibinfo {pages} {064908} (\bibinfo {year} {2013})},\ \Eprint
  {http://arxiv.org/abs/1310.1764} {arXiv:1310.1764 [nucl-th]} \BibitemShut
  {NoStop}%
\bibitem [{\citenamefont {Zhang}\ \emph
  {et~al.}(2018{\natexlab{a}})\citenamefont {Zhang}, \citenamefont {Li},\ and\
  \citenamefont {Xu}}]{Zhang:2018vrx}%
  \BibitemOpen
  \bibfield  {author} {\bibinfo {author} {\bibfnamefont {N.-B.}\ \bibnamefont
  {Zhang}}, \bibinfo {author} {\bibfnamefont {B.-A.}\ \bibnamefont {Li}}, \
  and\ \bibinfo {author} {\bibfnamefont {J.}~\bibnamefont {Xu}},\ }\href
  {\doibase 10.3847/1538-4357/aac027} {\bibfield  {journal} {\bibinfo
  {journal} {Astrophys. J.}\ }\textbf {\bibinfo {volume} {859}},\ \bibinfo
  {pages} {90} (\bibinfo {year} {2018}{\natexlab{a}})},\ \Eprint
  {http://arxiv.org/abs/1801.06855} {arXiv:1801.06855 [nucl-th]} \BibitemShut
  {NoStop}%
\bibitem [{\citenamefont {Jayasinghe}\ \emph {et~al.}(2021)\citenamefont
  {Jayasinghe} \emph {et~al.}}]{Jayasinghe:2021uqb}%
  \BibitemOpen
  \bibfield  {author} {\bibinfo {author} {\bibfnamefont {T.}~\bibnamefont
  {Jayasinghe}} \emph {et~al.},\ }\href {\doibase 10.1093/mnras/stab907}
  {\bibfield  {journal} {\bibinfo  {journal} {Mon. Not. Roy. Astron. Soc.}\
  }\textbf {\bibinfo {volume} {504}},\ \bibinfo {pages} {2577} (\bibinfo {year}
  {2021})},\ \Eprint {http://arxiv.org/abs/2101.02212} {arXiv:2101.02212
  [astro-ph.SR]} \BibitemShut {NoStop}%
\bibitem [{\citenamefont {Romani}\ \emph {et~al.}(2022)\citenamefont {Romani},
  \citenamefont {Kandel}, \citenamefont {Filippenko}, \citenamefont {Brink},\
  and\ \citenamefont {Zheng}}]{Romani:2022jhd}%
  \BibitemOpen
  \bibfield  {author} {\bibinfo {author} {\bibfnamefont {R.~W.}\ \bibnamefont
  {Romani}}, \bibinfo {author} {\bibfnamefont {D.}~\bibnamefont {Kandel}},
  \bibinfo {author} {\bibfnamefont {A.~V.}\ \bibnamefont {Filippenko}},
  \bibinfo {author} {\bibfnamefont {T.~G.}\ \bibnamefont {Brink}}, \ and\
  \bibinfo {author} {\bibfnamefont {W.}~\bibnamefont {Zheng}},\ }\href
  {\doibase 10.3847/2041-8213/ac8007} {\bibfield  {journal} {\bibinfo
  {journal} {Astrophys. J. Lett.}\ }\textbf {\bibinfo {volume} {934}},\
  \bibinfo {pages} {L17} (\bibinfo {year} {2022})},\ \Eprint
  {http://arxiv.org/abs/2207.05124} {arXiv:2207.05124 [astro-ph.HE]}
  \BibitemShut {NoStop}%
\bibitem [{\citenamefont {Togashi}\ and\ \citenamefont
  {Takano}(2013)}]{Togashi:2013bfg}%
  \BibitemOpen
  \bibfield  {author} {\bibinfo {author} {\bibfnamefont {H.}~\bibnamefont
  {Togashi}}\ and\ \bibinfo {author} {\bibfnamefont {M.}~\bibnamefont
  {Takano}},\ }\href {\doibase 10.1016/j.nuclphysa.2013.02.014} {\bibfield
  {journal} {\bibinfo  {journal} {Nucl. Phys. A}\ }\textbf {\bibinfo {volume}
  {902}},\ \bibinfo {pages} {53} (\bibinfo {year} {2013})},\ \Eprint
  {http://arxiv.org/abs/1302.4261} {arXiv:1302.4261 [nucl-th]} \BibitemShut
  {NoStop}%
\bibitem [{\citenamefont {Togashi}\ \emph {et~al.}(2017)\citenamefont
  {Togashi}, \citenamefont {Nakazato}, \citenamefont {Takehara}, \citenamefont
  {Yamamuro}, \citenamefont {Suzuki},\ and\ \citenamefont
  {Takano}}]{Togashi:2017mjp}%
  \BibitemOpen
  \bibfield  {author} {\bibinfo {author} {\bibfnamefont {H.}~\bibnamefont
  {Togashi}}, \bibinfo {author} {\bibfnamefont {K.}~\bibnamefont {Nakazato}},
  \bibinfo {author} {\bibfnamefont {Y.}~\bibnamefont {Takehara}}, \bibinfo
  {author} {\bibfnamefont {S.}~\bibnamefont {Yamamuro}}, \bibinfo {author}
  {\bibfnamefont {H.}~\bibnamefont {Suzuki}}, \ and\ \bibinfo {author}
  {\bibfnamefont {M.}~\bibnamefont {Takano}},\ }\href {\doibase
  10.1016/j.nuclphysa.2017.02.010} {\bibfield  {journal} {\bibinfo  {journal}
  {Nucl. Phys. A}\ }\textbf {\bibinfo {volume} {961}},\ \bibinfo {pages} {78}
  (\bibinfo {year} {2017})},\ \Eprint {http://arxiv.org/abs/1702.05324}
  {arXiv:1702.05324 [nucl-th]} \BibitemShut {NoStop}%
\bibitem [{\citenamefont {Baym}\ \emph {et~al.}(2019)\citenamefont {Baym},
  \citenamefont {Furusawa}, \citenamefont {Hatsuda}, \citenamefont {Kojo},\
  and\ \citenamefont {Togashi}}]{Baym:2019iky}%
  \BibitemOpen
  \bibfield  {author} {\bibinfo {author} {\bibfnamefont {G.}~\bibnamefont
  {Baym}}, \bibinfo {author} {\bibfnamefont {S.}~\bibnamefont {Furusawa}},
  \bibinfo {author} {\bibfnamefont {T.}~\bibnamefont {Hatsuda}}, \bibinfo
  {author} {\bibfnamefont {T.}~\bibnamefont {Kojo}}, \ and\ \bibinfo {author}
  {\bibfnamefont {H.}~\bibnamefont {Togashi}},\ }\href {\doibase
  10.3847/1538-4357/ab441e} {\bibfield  {journal} {\bibinfo  {journal}
  {Astrophys. J.}\ }\textbf {\bibinfo {volume} {885}},\ \bibinfo {pages} {42}
  (\bibinfo {year} {2019})},\ \Eprint {http://arxiv.org/abs/1903.08963}
  {arXiv:1903.08963 [astro-ph.HE]} \BibitemShut {NoStop}%
\bibitem [{\citenamefont {Komoltsev}\ and\ \citenamefont
  {Kurkela}(2022)}]{Komoltsev:2021jzg}%
  \BibitemOpen
  \bibfield  {author} {\bibinfo {author} {\bibfnamefont {O.}~\bibnamefont
  {Komoltsev}}\ and\ \bibinfo {author} {\bibfnamefont {A.}~\bibnamefont
  {Kurkela}},\ }\href {\doibase 10.1103/PhysRevLett.128.202701} {\bibfield
  {journal} {\bibinfo  {journal} {Phys. Rev. Lett.}\ }\textbf {\bibinfo
  {volume} {128}},\ \bibinfo {pages} {202701} (\bibinfo {year} {2022})},\
  \Eprint {http://arxiv.org/abs/2111.05350} {arXiv:2111.05350 [nucl-th]}
  \BibitemShut {NoStop}%
\bibitem [{\citenamefont {Somasundaram}\ \emph {et~al.}(2023)\citenamefont
  {Somasundaram}, \citenamefont {Tews},\ and\ \citenamefont
  {Margueron}}]{Somasundaram:2022ztm}%
  \BibitemOpen
  \bibfield  {author} {\bibinfo {author} {\bibfnamefont {R.}~\bibnamefont
  {Somasundaram}}, \bibinfo {author} {\bibfnamefont {I.}~\bibnamefont {Tews}},
  \ and\ \bibinfo {author} {\bibfnamefont {J.}~\bibnamefont {Margueron}},\
  }\href {\doibase 10.1103/PhysRevC.107.L052801} {\bibfield  {journal}
  {\bibinfo  {journal} {Phys. Rev. C}\ }\textbf {\bibinfo {volume} {107}},\
  \bibinfo {pages} {L052801} (\bibinfo {year} {2023})},\ \Eprint
  {http://arxiv.org/abs/2204.14039} {arXiv:2204.14039 [nucl-th]} \BibitemShut
  {NoStop}%
\bibitem [{\citenamefont {Mroczek}\ \emph {et~al.}(2023)\citenamefont
  {Mroczek}, \citenamefont {Miller}, \citenamefont {Noronha-Hostler},\ and\
  \citenamefont {Yunes}}]{Mroczek:2023zxo}%
  \BibitemOpen
  \bibfield  {author} {\bibinfo {author} {\bibfnamefont {D.}~\bibnamefont
  {Mroczek}}, \bibinfo {author} {\bibfnamefont {M.~C.}\ \bibnamefont {Miller}},
  \bibinfo {author} {\bibfnamefont {J.}~\bibnamefont {Noronha-Hostler}}, \ and\
  \bibinfo {author} {\bibfnamefont {N.}~\bibnamefont {Yunes}},\ }\href@noop {}
  {\  (\bibinfo {year} {2023})},\ \Eprint {http://arxiv.org/abs/2309.02345}
  {arXiv:2309.02345 [astro-ph.HE]} \BibitemShut {NoStop}%
\bibitem [{\citenamefont {Oppenheimer}\ and\ \citenamefont
  {Volkoff}(1939)}]{Oppenheimer:1939ne}%
  \BibitemOpen
  \bibfield  {author} {\bibinfo {author} {\bibfnamefont {J.~R.}\ \bibnamefont
  {Oppenheimer}}\ and\ \bibinfo {author} {\bibfnamefont {G.~M.}\ \bibnamefont
  {Volkoff}},\ }\href {\doibase 10.1103/PhysRev.55.374} {\bibfield  {journal}
  {\bibinfo  {journal} {Phys. Rev.}\ }\textbf {\bibinfo {volume} {55}},\
  \bibinfo {pages} {374} (\bibinfo {year} {1939})}\BibitemShut {NoStop}%
\bibitem [{\citenamefont {Tolman}(1939)}]{Tolman:1939jz}%
  \BibitemOpen
  \bibfield  {author} {\bibinfo {author} {\bibfnamefont {R.~C.}\ \bibnamefont
  {Tolman}},\ }\href {\doibase 10.1103/PhysRev.55.364} {\bibfield  {journal}
  {\bibinfo  {journal} {Phys. Rev.}\ }\textbf {\bibinfo {volume} {55}},\
  \bibinfo {pages} {364} (\bibinfo {year} {1939})}\BibitemShut {NoStop}%
\bibitem [{\citenamefont {Landry}\ \emph {et~al.}(2020)\citenamefont {Landry},
  \citenamefont {Essick},\ and\ \citenamefont
  {Chatziioannou}}]{Landry:2020vaw}%
  \BibitemOpen
  \bibfield  {author} {\bibinfo {author} {\bibfnamefont {P.}~\bibnamefont
  {Landry}}, \bibinfo {author} {\bibfnamefont {R.}~\bibnamefont {Essick}}, \
  and\ \bibinfo {author} {\bibfnamefont {K.}~\bibnamefont {Chatziioannou}},\
  }\href {\doibase 10.1103/PhysRevD.101.123007} {\bibfield  {journal} {\bibinfo
   {journal} {Phys. Rev. D}\ }\textbf {\bibinfo {volume} {101}},\ \bibinfo
  {pages} {123007} (\bibinfo {year} {2020})},\ \Eprint
  {http://arxiv.org/abs/2003.04880} {arXiv:2003.04880 [astro-ph.HE]}
  \BibitemShut {NoStop}%
\bibitem [{\citenamefont {Nathanail}\ \emph {et~al.}(2021)\citenamefont
  {Nathanail}, \citenamefont {Most},\ and\ \citenamefont
  {Rezzolla}}]{Nathanail:2021tay}%
  \BibitemOpen
  \bibfield  {author} {\bibinfo {author} {\bibfnamefont {A.}~\bibnamefont
  {Nathanail}}, \bibinfo {author} {\bibfnamefont {E.~R.}\ \bibnamefont {Most}},
  \ and\ \bibinfo {author} {\bibfnamefont {L.}~\bibnamefont {Rezzolla}},\
  }\href {\doibase 10.3847/2041-8213/abdfc6} {\bibfield  {journal} {\bibinfo
  {journal} {Astrophys. J. Lett.}\ }\textbf {\bibinfo {volume} {908}},\
  \bibinfo {pages} {L28} (\bibinfo {year} {2021})},\ \Eprint
  {http://arxiv.org/abs/2101.01735} {arXiv:2101.01735 [astro-ph.HE]}
  \BibitemShut {NoStop}%
\bibitem [{\citenamefont {de~Tovar}\ \emph {et~al.}(2021)\citenamefont
  {de~Tovar}, \citenamefont {Ferreira},\ and\ \citenamefont
  {Provid\^encia}}]{deTovar:2021sjo}%
  \BibitemOpen
  \bibfield  {author} {\bibinfo {author} {\bibfnamefont {P.~B.}\ \bibnamefont
  {de~Tovar}}, \bibinfo {author} {\bibfnamefont {M.}~\bibnamefont {Ferreira}},
  \ and\ \bibinfo {author} {\bibfnamefont {C.}~\bibnamefont {Provid\^encia}},\
  }\href {\doibase 10.1103/PhysRevD.104.123036} {\bibfield  {journal} {\bibinfo
   {journal} {Phys. Rev. D}\ }\textbf {\bibinfo {volume} {104}},\ \bibinfo
  {pages} {123036} (\bibinfo {year} {2021})},\ \Eprint
  {http://arxiv.org/abs/2112.05551} {arXiv:2112.05551 [nucl-th]} \BibitemShut
  {NoStop}%
\bibitem [{\citenamefont {Li}\ \emph {et~al.}(2019)\citenamefont {Li},
  \citenamefont {Krastev}, \citenamefont {Wen},\ and\ \citenamefont
  {Zhang}}]{Li:2019xxz}%
  \BibitemOpen
  \bibfield  {author} {\bibinfo {author} {\bibfnamefont {B.-A.}\ \bibnamefont
  {Li}}, \bibinfo {author} {\bibfnamefont {P.~G.}\ \bibnamefont {Krastev}},
  \bibinfo {author} {\bibfnamefont {D.-H.}\ \bibnamefont {Wen}}, \ and\
  \bibinfo {author} {\bibfnamefont {N.-B.}\ \bibnamefont {Zhang}},\ }\href
  {\doibase 10.1140/epja/i2019-12780-8} {\bibfield  {journal} {\bibinfo
  {journal} {Eur. Phys. J. A}\ }\textbf {\bibinfo {volume} {55}},\ \bibinfo
  {pages} {117} (\bibinfo {year} {2019})},\ \Eprint
  {http://arxiv.org/abs/1905.13175} {arXiv:1905.13175 [nucl-th]} \BibitemShut
  {NoStop}%
\bibitem [{\citenamefont {Adhikari}\ \emph {et~al.}(2021)\citenamefont
  {Adhikari} \emph {et~al.}}]{PREX:2021umo}%
  \BibitemOpen
  \bibfield  {author} {\bibinfo {author} {\bibfnamefont {D.}~\bibnamefont
  {Adhikari}} \emph {et~al.} (\bibinfo {collaboration} {PREX}),\ }\href
  {\doibase 10.1103/PhysRevLett.126.172502} {\bibfield  {journal} {\bibinfo
  {journal} {Phys. Rev. Lett.}\ }\textbf {\bibinfo {volume} {126}},\ \bibinfo
  {pages} {172502} (\bibinfo {year} {2021})},\ \Eprint
  {http://arxiv.org/abs/2102.10767} {arXiv:2102.10767 [nucl-ex]} \BibitemShut
  {NoStop}%
\bibitem [{\citenamefont {Reed}\ \emph {et~al.}(2021)\citenamefont {Reed},
  \citenamefont {Fattoyev}, \citenamefont {Horowitz},\ and\ \citenamefont
  {Piekarewicz}}]{Reed:2021nqk}%
  \BibitemOpen
  \bibfield  {author} {\bibinfo {author} {\bibfnamefont {B.~T.}\ \bibnamefont
  {Reed}}, \bibinfo {author} {\bibfnamefont {F.~J.}\ \bibnamefont {Fattoyev}},
  \bibinfo {author} {\bibfnamefont {C.~J.}\ \bibnamefont {Horowitz}}, \ and\
  \bibinfo {author} {\bibfnamefont {J.}~\bibnamefont {Piekarewicz}},\ }\href
  {\doibase 10.1103/PhysRevLett.126.172503} {\bibfield  {journal} {\bibinfo
  {journal} {Phys. Rev. Lett.}\ }\textbf {\bibinfo {volume} {126}},\ \bibinfo
  {pages} {172503} (\bibinfo {year} {2021})},\ \Eprint
  {http://arxiv.org/abs/2101.03193} {arXiv:2101.03193 [nucl-th]} \BibitemShut
  {NoStop}%
\bibitem [{\citenamefont {Xie}\ and\ \citenamefont {Li}(2020)}]{Xie:2020tdo}%
  \BibitemOpen
  \bibfield  {author} {\bibinfo {author} {\bibfnamefont {W.-J.}\ \bibnamefont
  {Xie}}\ and\ \bibinfo {author} {\bibfnamefont {B.-A.}\ \bibnamefont {Li}},\
  }\href {\doibase 10.3847/1538-4357/aba271} {\bibfield  {journal} {\bibinfo
  {journal} {Astrophys. J.}\ }\textbf {\bibinfo {volume} {899}},\ \bibinfo
  {pages} {4} (\bibinfo {year} {2020})},\ \Eprint
  {http://arxiv.org/abs/2005.07216} {arXiv:2005.07216 [astro-ph.HE]}
  \BibitemShut {NoStop}%
\bibitem [{\citenamefont {Tews}\ \emph {et~al.}(2017)\citenamefont {Tews},
  \citenamefont {Lattimer}, \citenamefont {Ohnishi},\ and\ \citenamefont
  {Kolomeitsev}}]{Tews:2016jhi}%
  \BibitemOpen
  \bibfield  {author} {\bibinfo {author} {\bibfnamefont {I.}~\bibnamefont
  {Tews}}, \bibinfo {author} {\bibfnamefont {J.~M.}\ \bibnamefont {Lattimer}},
  \bibinfo {author} {\bibfnamefont {A.}~\bibnamefont {Ohnishi}}, \ and\
  \bibinfo {author} {\bibfnamefont {E.~E.}\ \bibnamefont {Kolomeitsev}},\
  }\href {\doibase 10.3847/1538-4357/aa8db9} {\bibfield  {journal} {\bibinfo
  {journal} {Astrophys. J.}\ }\textbf {\bibinfo {volume} {848}},\ \bibinfo
  {pages} {105} (\bibinfo {year} {2017})},\ \Eprint
  {http://arxiv.org/abs/1611.07133} {arXiv:1611.07133 [nucl-th]} \BibitemShut
  {NoStop}%
\bibitem [{\citenamefont {Typel}\ \emph {et~al.}(2022)\citenamefont {Typel}
  \emph {et~al.}}]{CompOSECoreTeam:2022ddl}%
  \BibitemOpen
  \bibfield  {author} {\bibinfo {author} {\bibfnamefont {S.}~\bibnamefont
  {Typel}} \emph {et~al.} (\bibinfo {collaboration} {CompOSE Core Team}),\
  }\href {\doibase 10.1140/epja/s10050-022-00847-y} {\bibfield  {journal}
  {\bibinfo  {journal} {Eur. Phys. J. A}\ }\textbf {\bibinfo {volume} {58}},\
  \bibinfo {pages} {221} (\bibinfo {year} {2022})},\ \Eprint
  {http://arxiv.org/abs/2203.03209} {arXiv:2203.03209 [astro-ph.HE]}
  \BibitemShut {NoStop}%
\bibitem [{\citenamefont {Sun}\ \emph {et~al.}(2024)\citenamefont {Sun},
  \citenamefont {Bhattiprolu},\ and\ \citenamefont {Lattimer}}]{Sun:2023xkg}%
  \BibitemOpen
  \bibfield  {author} {\bibinfo {author} {\bibfnamefont {B.}~\bibnamefont
  {Sun}}, \bibinfo {author} {\bibfnamefont {S.}~\bibnamefont {Bhattiprolu}}, \
  and\ \bibinfo {author} {\bibfnamefont {J.~M.}\ \bibnamefont {Lattimer}},\
  }\href {\doibase 10.1103/PhysRevC.109.055801} {\bibfield  {journal} {\bibinfo
   {journal} {Phys. Rev. C}\ }\textbf {\bibinfo {volume} {109}},\ \bibinfo
  {pages} {055801} (\bibinfo {year} {2024})},\ \Eprint
  {http://arxiv.org/abs/2311.00843} {arXiv:2311.00843 [nucl-th]} \BibitemShut
  {NoStop}%
\bibitem [{\citenamefont {Imam}\ \emph {et~al.}(2022)\citenamefont {Imam},
  \citenamefont {Patra}, \citenamefont {Mondal}, \citenamefont {Malik},\ and\
  \citenamefont {Agrawal}}]{Imam:2021dbe}%
  \BibitemOpen
  \bibfield  {author} {\bibinfo {author} {\bibfnamefont {S.~M.~A.}\
  \bibnamefont {Imam}}, \bibinfo {author} {\bibfnamefont {N.~K.}\ \bibnamefont
  {Patra}}, \bibinfo {author} {\bibfnamefont {C.}~\bibnamefont {Mondal}},
  \bibinfo {author} {\bibfnamefont {T.}~\bibnamefont {Malik}}, \ and\ \bibinfo
  {author} {\bibfnamefont {B.~K.}\ \bibnamefont {Agrawal}},\ }\href {\doibase
  10.1103/PhysRevC.105.015806} {\bibfield  {journal} {\bibinfo  {journal}
  {Phys. Rev. C}\ }\textbf {\bibinfo {volume} {105}},\ \bibinfo {pages}
  {015806} (\bibinfo {year} {2022})},\ \Eprint
  {http://arxiv.org/abs/2110.15776} {arXiv:2110.15776 [nucl-th]} \BibitemShut
  {NoStop}%
\bibitem [{\citenamefont {Mendes}\ \emph {et~al.}(2021)\citenamefont {Mendes},
  \citenamefont {Cumming}, \citenamefont {Gale},\ and\ \citenamefont
  {Fattoyev}}]{Mendes:2021tos}%
  \BibitemOpen
  \bibfield  {author} {\bibinfo {author} {\bibfnamefont {M.}~\bibnamefont
  {Mendes}}, \bibinfo {author} {\bibfnamefont {A.}~\bibnamefont {Cumming}},
  \bibinfo {author} {\bibfnamefont {C.}~\bibnamefont {Gale}}, \ and\ \bibinfo
  {author} {\bibfnamefont {F.~J.}\ \bibnamefont {Fattoyev}},\ }in\ \href
  {\doibase 10.1142/9789811269776_0309} {\emph {\bibinfo {booktitle} {{16th
  Marcel Grossmann Meeting on~Recent Developments in Theoretical and
  Experimental General Relativity, Astrophysics and Relativistic Field
  Theories}}}}\ (\bibinfo {year} {2021})\ \Eprint
  {http://arxiv.org/abs/2110.11077} {arXiv:2110.11077 [nucl-th]} \BibitemShut
  {NoStop}%
\bibitem [{\citenamefont {Neill}\ \emph {et~al.}(2023)\citenamefont {Neill},
  \citenamefont {Preston}, \citenamefont {Newton},\ and\ \citenamefont
  {Tsang}}]{Neill:2022psd}%
  \BibitemOpen
  \bibfield  {author} {\bibinfo {author} {\bibfnamefont {D.}~\bibnamefont
  {Neill}}, \bibinfo {author} {\bibfnamefont {R.}~\bibnamefont {Preston}},
  \bibinfo {author} {\bibfnamefont {W.~G.}\ \bibnamefont {Newton}}, \ and\
  \bibinfo {author} {\bibfnamefont {D.}~\bibnamefont {Tsang}},\ }\href
  {\doibase 10.1103/PhysRevLett.130.112701} {\bibfield  {journal} {\bibinfo
  {journal} {Phys. Rev. Lett.}\ }\textbf {\bibinfo {volume} {130}},\ \bibinfo
  {pages} {112701} (\bibinfo {year} {2023})},\ \Eprint
  {http://arxiv.org/abs/2208.00994} {arXiv:2208.00994 [astro-ph.HE]}
  \BibitemShut {NoStop}%
\bibitem [{\citenamefont {Lattimer}\ and\ \citenamefont
  {Steiner}(2014)}]{Lattimer:2014sga}%
  \BibitemOpen
  \bibfield  {author} {\bibinfo {author} {\bibfnamefont {J.~M.}\ \bibnamefont
  {Lattimer}}\ and\ \bibinfo {author} {\bibfnamefont {A.~W.}\ \bibnamefont
  {Steiner}},\ }\href {\doibase 10.1140/epja/i2014-14040-y} {\bibfield
  {journal} {\bibinfo  {journal} {Eur. Phys. J. A}\ }\textbf {\bibinfo {volume}
  {50}},\ \bibinfo {pages} {40} (\bibinfo {year} {2014})},\ \Eprint
  {http://arxiv.org/abs/1403.1186} {arXiv:1403.1186 [nucl-th]} \BibitemShut
  {NoStop}%
\bibitem [{\citenamefont {Drischler}\ \emph {et~al.}(2021)\citenamefont
  {Drischler}, \citenamefont {Han}, \citenamefont {Lattimer}, \citenamefont
  {Prakash}, \citenamefont {Reddy},\ and\ \citenamefont
  {Zhao}}]{Drischler:2020fvz}%
  \BibitemOpen
  \bibfield  {author} {\bibinfo {author} {\bibfnamefont {C.}~\bibnamefont
  {Drischler}}, \bibinfo {author} {\bibfnamefont {S.}~\bibnamefont {Han}},
  \bibinfo {author} {\bibfnamefont {J.~M.}\ \bibnamefont {Lattimer}}, \bibinfo
  {author} {\bibfnamefont {M.}~\bibnamefont {Prakash}}, \bibinfo {author}
  {\bibfnamefont {S.}~\bibnamefont {Reddy}}, \ and\ \bibinfo {author}
  {\bibfnamefont {T.}~\bibnamefont {Zhao}},\ }\href {\doibase
  10.1103/PhysRevC.103.045808} {\bibfield  {journal} {\bibinfo  {journal}
  {Phys. Rev. C}\ }\textbf {\bibinfo {volume} {103}},\ \bibinfo {pages}
  {045808} (\bibinfo {year} {2021})},\ \Eprint
  {http://arxiv.org/abs/2009.06441} {arXiv:2009.06441 [nucl-th]} \BibitemShut
  {NoStop}%
\bibitem [{\citenamefont {Drischler}\ \emph {et~al.}(2020)\citenamefont
  {Drischler}, \citenamefont {Melendez}, \citenamefont {Furnstahl},\ and\
  \citenamefont {Phillips}}]{Drischler:2020yad}%
  \BibitemOpen
  \bibfield  {author} {\bibinfo {author} {\bibfnamefont {C.}~\bibnamefont
  {Drischler}}, \bibinfo {author} {\bibfnamefont {J.~A.}\ \bibnamefont
  {Melendez}}, \bibinfo {author} {\bibfnamefont {R.~J.}\ \bibnamefont
  {Furnstahl}}, \ and\ \bibinfo {author} {\bibfnamefont {D.~R.}\ \bibnamefont
  {Phillips}},\ }\href {\doibase 10.1103/PhysRevC.102.054315} {\bibfield
  {journal} {\bibinfo  {journal} {Phys. Rev. C}\ }\textbf {\bibinfo {volume}
  {102}},\ \bibinfo {pages} {054315} (\bibinfo {year} {2020})},\ \Eprint
  {http://arxiv.org/abs/2004.07805} {arXiv:2004.07805 [nucl-th]} \BibitemShut
  {NoStop}%
\bibitem [{\citenamefont {Lynch}\ and\ \citenamefont
  {Tsang}(2022)}]{Lynch:2021xkq}%
  \BibitemOpen
  \bibfield  {author} {\bibinfo {author} {\bibfnamefont {W.~G.}\ \bibnamefont
  {Lynch}}\ and\ \bibinfo {author} {\bibfnamefont {M.~B.}\ \bibnamefont
  {Tsang}},\ }\href {\doibase 10.1016/j.physletb.2022.137098} {\bibfield
  {journal} {\bibinfo  {journal} {Phys. Lett. B}\ }\textbf {\bibinfo {volume}
  {830}},\ \bibinfo {pages} {137098} (\bibinfo {year} {2022})},\ \Eprint
  {http://arxiv.org/abs/2106.10119} {arXiv:2106.10119 [nucl-th]} \BibitemShut
  {NoStop}%
\bibitem [{\citenamefont {Russotto}\ \emph {et~al.}(2016)\citenamefont
  {Russotto} \emph {et~al.}}]{Russotto:2016ucm}%
  \BibitemOpen
  \bibfield  {author} {\bibinfo {author} {\bibfnamefont {P.}~\bibnamefont
  {Russotto}} \emph {et~al.},\ }\href {\doibase 10.1103/PhysRevC.94.034608}
  {\bibfield  {journal} {\bibinfo  {journal} {Phys. Rev. C}\ }\textbf {\bibinfo
  {volume} {94}},\ \bibinfo {pages} {034608} (\bibinfo {year} {2016})},\
  \Eprint {http://arxiv.org/abs/1608.04332} {arXiv:1608.04332 [nucl-ex]}
  \BibitemShut {NoStop}%
\bibitem [{\citenamefont {Russotto}\ \emph {et~al.}(2011)\citenamefont
  {Russotto} \emph {et~al.}}]{Russotto:2011hq}%
  \BibitemOpen
  \bibfield  {author} {\bibinfo {author} {\bibfnamefont {P.}~\bibnamefont
  {Russotto}} \emph {et~al.},\ }\href {\doibase 10.1016/j.physletb.2011.02.033}
  {\bibfield  {journal} {\bibinfo  {journal} {Phys. Lett. B}\ }\textbf
  {\bibinfo {volume} {697}},\ \bibinfo {pages} {471} (\bibinfo {year}
  {2011})},\ \Eprint {http://arxiv.org/abs/1101.2361} {arXiv:1101.2361
  [nucl-ex]} \BibitemShut {NoStop}%
\bibitem [{\citenamefont {Tsang}\ \emph {et~al.}(2009)\citenamefont {Tsang},
  \citenamefont {Zhang}, \citenamefont {Danielewicz}, \citenamefont {Famiano},
  \citenamefont {Li}, \citenamefont {Lynch},\ and\ \citenamefont
  {Steiner}}]{Tsang:2008fd}%
  \BibitemOpen
  \bibfield  {author} {\bibinfo {author} {\bibfnamefont {M.~B.}\ \bibnamefont
  {Tsang}}, \bibinfo {author} {\bibfnamefont {Y.}~\bibnamefont {Zhang}},
  \bibinfo {author} {\bibfnamefont {P.}~\bibnamefont {Danielewicz}}, \bibinfo
  {author} {\bibfnamefont {M.}~\bibnamefont {Famiano}}, \bibinfo {author}
  {\bibfnamefont {Z.}~\bibnamefont {Li}}, \bibinfo {author} {\bibfnamefont
  {W.~G.}\ \bibnamefont {Lynch}}, \ and\ \bibinfo {author} {\bibfnamefont
  {A.~W.}\ \bibnamefont {Steiner}},\ }\href {\doibase
  10.1103/PhysRevLett.102.122701} {\bibfield  {journal} {\bibinfo  {journal}
  {Phys. Rev. Lett.}\ }\textbf {\bibinfo {volume} {102}},\ \bibinfo {pages}
  {122701} (\bibinfo {year} {2009})},\ \Eprint {http://arxiv.org/abs/0811.3107}
  {arXiv:0811.3107 [nucl-ex]} \BibitemShut {NoStop}%
\bibitem [{\citenamefont {Zhang}\ and\ \citenamefont
  {Chen}(2015)}]{Zhang:2015ava}%
  \BibitemOpen
  \bibfield  {author} {\bibinfo {author} {\bibfnamefont {Z.}~\bibnamefont
  {Zhang}}\ and\ \bibinfo {author} {\bibfnamefont {L.-W.}\ \bibnamefont
  {Chen}},\ }\href {\doibase 10.1103/PhysRevC.92.031301} {\bibfield  {journal}
  {\bibinfo  {journal} {Phys. Rev. C}\ }\textbf {\bibinfo {volume} {92}},\
  \bibinfo {pages} {031301} (\bibinfo {year} {2015})},\ \Eprint
  {http://arxiv.org/abs/1504.01077} {arXiv:1504.01077 [nucl-th]} \BibitemShut
  {NoStop}%
\bibitem [{\citenamefont {Kortelainen}\ \emph {et~al.}(2010)\citenamefont
  {Kortelainen}, \citenamefont {Lesinski}, \citenamefont {More}, \citenamefont
  {Nazarewicz}, \citenamefont {Sarich}, \citenamefont {Schunck}, \citenamefont
  {Stoitsov},\ and\ \citenamefont {Wild}}]{Kortelainen:2010hv}%
  \BibitemOpen
  \bibfield  {author} {\bibinfo {author} {\bibfnamefont {M.}~\bibnamefont
  {Kortelainen}}, \bibinfo {author} {\bibfnamefont {T.}~\bibnamefont
  {Lesinski}}, \bibinfo {author} {\bibfnamefont {J.}~\bibnamefont {More}},
  \bibinfo {author} {\bibfnamefont {W.}~\bibnamefont {Nazarewicz}}, \bibinfo
  {author} {\bibfnamefont {J.}~\bibnamefont {Sarich}}, \bibinfo {author}
  {\bibfnamefont {N.}~\bibnamefont {Schunck}}, \bibinfo {author} {\bibfnamefont
  {M.~V.}\ \bibnamefont {Stoitsov}}, \ and\ \bibinfo {author} {\bibfnamefont
  {S.}~\bibnamefont {Wild}},\ }\href {\doibase 10.1103/PhysRevC.82.024313}
  {\bibfield  {journal} {\bibinfo  {journal} {Phys. Rev. C}\ }\textbf {\bibinfo
  {volume} {82}},\ \bibinfo {pages} {024313} (\bibinfo {year} {2010})},\
  \Eprint {http://arxiv.org/abs/1005.5145} {arXiv:1005.5145 [nucl-th]}
  \BibitemShut {NoStop}%
\bibitem [{\citenamefont {Kortelainen}\ \emph {et~al.}(2012)\citenamefont
  {Kortelainen}, \citenamefont {McDonnell}, \citenamefont {Nazarewicz},
  \citenamefont {Reinhard}, \citenamefont {Sarich}, \citenamefont {Schunck},
  \citenamefont {Stoitsov},\ and\ \citenamefont {Wild}}]{Kortelainen:2011ft}%
  \BibitemOpen
  \bibfield  {author} {\bibinfo {author} {\bibfnamefont {M.}~\bibnamefont
  {Kortelainen}}, \bibinfo {author} {\bibfnamefont {J.}~\bibnamefont
  {McDonnell}}, \bibinfo {author} {\bibfnamefont {W.}~\bibnamefont
  {Nazarewicz}}, \bibinfo {author} {\bibfnamefont {P.~G.}\ \bibnamefont
  {Reinhard}}, \bibinfo {author} {\bibfnamefont {J.}~\bibnamefont {Sarich}},
  \bibinfo {author} {\bibfnamefont {N.}~\bibnamefont {Schunck}}, \bibinfo
  {author} {\bibfnamefont {M.~V.}\ \bibnamefont {Stoitsov}}, \ and\ \bibinfo
  {author} {\bibfnamefont {S.~M.}\ \bibnamefont {Wild}},\ }\href {\doibase
  10.1103/PhysRevC.85.024304} {\bibfield  {journal} {\bibinfo  {journal} {Phys.
  Rev. C}\ }\textbf {\bibinfo {volume} {85}},\ \bibinfo {pages} {024304}
  (\bibinfo {year} {2012})},\ \Eprint {http://arxiv.org/abs/1111.4344}
  {arXiv:1111.4344 [nucl-th]} \BibitemShut {NoStop}%
\bibitem [{\citenamefont {Brown}(2013)}]{Brown:2013mga}%
  \BibitemOpen
  \bibfield  {author} {\bibinfo {author} {\bibfnamefont {B.~A.}\ \bibnamefont
  {Brown}},\ }\href {\doibase 10.1103/PhysRevLett.111.232502} {\bibfield
  {journal} {\bibinfo  {journal} {Phys. Rev. Lett.}\ }\textbf {\bibinfo
  {volume} {111}},\ \bibinfo {pages} {232502} (\bibinfo {year} {2013})},\
  \Eprint {http://arxiv.org/abs/1308.3664} {arXiv:1308.3664 [nucl-th]}
  \BibitemShut {NoStop}%
\bibitem [{\citenamefont {Danielewicz}\ and\ \citenamefont
  {Lee}(2014)}]{Danielewicz:2013upa}%
  \BibitemOpen
  \bibfield  {author} {\bibinfo {author} {\bibfnamefont {P.}~\bibnamefont
  {Danielewicz}}\ and\ \bibinfo {author} {\bibfnamefont {J.}~\bibnamefont
  {Lee}},\ }\href {\doibase 10.1016/j.nuclphysa.2013.11.005} {\bibfield
  {journal} {\bibinfo  {journal} {Nucl. Phys. A}\ }\textbf {\bibinfo {volume}
  {922}},\ \bibinfo {pages} {1} (\bibinfo {year} {2014})},\ \Eprint
  {http://arxiv.org/abs/1307.4130} {arXiv:1307.4130 [nucl-th]} \BibitemShut
  {NoStop}%
\bibitem [{\citenamefont {Baym}\ \emph {et~al.}(1971)\citenamefont {Baym},
  \citenamefont {Pethick},\ and\ \citenamefont {Sutherland}}]{Baym:1971pw}%
  \BibitemOpen
  \bibfield  {author} {\bibinfo {author} {\bibfnamefont {G.}~\bibnamefont
  {Baym}}, \bibinfo {author} {\bibfnamefont {C.}~\bibnamefont {Pethick}}, \
  and\ \bibinfo {author} {\bibfnamefont {P.}~\bibnamefont {Sutherland}},\
  }\href {\doibase 10.1086/151216} {\bibfield  {journal} {\bibinfo  {journal}
  {Astrophys. J.}\ }\textbf {\bibinfo {volume} {170}},\ \bibinfo {pages} {299}
  (\bibinfo {year} {1971})}\BibitemShut {NoStop}%
\bibitem [{\citenamefont {{Shapiro}}\ and\ \citenamefont
  {{Teukolsky}}(1986)}]{1986bhwd.book.....S}%
  \BibitemOpen
  \bibfield  {author} {\bibinfo {author} {\bibfnamefont {S.~L.}\ \bibnamefont
  {{Shapiro}}}\ and\ \bibinfo {author} {\bibfnamefont {S.~A.}\ \bibnamefont
  {{Teukolsky}}},\ }\href@noop {} {\emph {\bibinfo {title} {{Black Holes, White
  Dwarfs and Neutron Stars: The Physics of Compact Objects}}}}\ (\bibinfo
  {year} {1986})\BibitemShut {NoStop}%
\bibitem [{\citenamefont {{Glendenning}}(1997)}]{1997csnp.book.....G}%
  \BibitemOpen
  \bibfield  {author} {\bibinfo {author} {\bibfnamefont {N.~K.}\ \bibnamefont
  {{Glendenning}}},\ }\href@noop {} {\emph {\bibinfo {title} {{Compact stars.
  Nuclear physics, particle physics, and general relativity}}}}\ (\bibinfo
  {year} {1997})\BibitemShut {NoStop}%
\bibitem [{\citenamefont {Dorso}\ and\ \citenamefont
  {Aichelin}(1995)}]{Dorso:1994js}%
  \BibitemOpen
  \bibfield  {author} {\bibinfo {author} {\bibfnamefont {C.~O.}\ \bibnamefont
  {Dorso}}\ and\ \bibinfo {author} {\bibfnamefont {J.}~\bibnamefont
  {Aichelin}},\ }\href {\doibase 10.1016/0370-2693(94)01632-M} {\bibfield
  {journal} {\bibinfo  {journal} {Phys. Lett. B}\ }\textbf {\bibinfo {volume}
  {345}},\ \bibinfo {pages} {197} (\bibinfo {year} {1995})}\BibitemShut
  {NoStop}%
\bibitem [{\citenamefont {Puri}\ \emph {et~al.}(1996)\citenamefont {Puri},
  \citenamefont {Hartnack},\ and\ \citenamefont {Aichelin}}]{Puri:1996qv}%
  \BibitemOpen
  \bibfield  {author} {\bibinfo {author} {\bibfnamefont {R.~K.}\ \bibnamefont
  {Puri}}, \bibinfo {author} {\bibfnamefont {C.}~\bibnamefont {Hartnack}}, \
  and\ \bibinfo {author} {\bibfnamefont {J.}~\bibnamefont {Aichelin}},\ }\href
  {\doibase 10.1103/PhysRevC.54.R28} {\bibfield  {journal} {\bibinfo  {journal}
  {Phys. Rev. C}\ }\textbf {\bibinfo {volume} {54}},\ \bibinfo {pages} {R28}
  (\bibinfo {year} {1996})}\BibitemShut {NoStop}%
\bibitem [{\citenamefont {Adam}\ \emph
  {et~al.}(2016{\natexlab{a}})\citenamefont {Adam} \emph
  {et~al.}}]{ALICE:2015wav}%
  \BibitemOpen
  \bibfield  {author} {\bibinfo {author} {\bibfnamefont {J.}~\bibnamefont
  {Adam}} \emph {et~al.} (\bibinfo {collaboration} {ALICE}),\ }\href {\doibase
  10.1103/PhysRevC.93.024917} {\bibfield  {journal} {\bibinfo  {journal} {Phys.
  Rev. C}\ }\textbf {\bibinfo {volume} {93}},\ \bibinfo {pages} {024917}
  (\bibinfo {year} {2016}{\natexlab{a}})},\ \Eprint
  {http://arxiv.org/abs/1506.08951} {arXiv:1506.08951 [nucl-ex]} \BibitemShut
  {NoStop}%
\bibitem [{\citenamefont {Adam}\ \emph {et~al.}(2019)\citenamefont {Adam} \emph
  {et~al.}}]{STAR:2019sjh}%
  \BibitemOpen
  \bibfield  {author} {\bibinfo {author} {\bibfnamefont {J.}~\bibnamefont
  {Adam}} \emph {et~al.} (\bibinfo {collaboration} {STAR}),\ }\href {\doibase
  10.1103/PhysRevC.99.064905} {\bibfield  {journal} {\bibinfo  {journal} {Phys.
  Rev. C}\ }\textbf {\bibinfo {volume} {99}},\ \bibinfo {pages} {064905}
  (\bibinfo {year} {2019})},\ \Eprint {http://arxiv.org/abs/1903.11778}
  {arXiv:1903.11778 [nucl-ex]} \BibitemShut {NoStop}%
\bibitem [{\citenamefont {Adam}\ \emph
  {et~al.}(2016{\natexlab{b}})\citenamefont {Adam} \emph
  {et~al.}}]{ALICE:2015oer}%
  \BibitemOpen
  \bibfield  {author} {\bibinfo {author} {\bibfnamefont {J.}~\bibnamefont
  {Adam}} \emph {et~al.} (\bibinfo {collaboration} {ALICE}),\ }\href {\doibase
  10.1016/j.physletb.2016.01.040} {\bibfield  {journal} {\bibinfo  {journal}
  {Phys. Lett. B}\ }\textbf {\bibinfo {volume} {754}},\ \bibinfo {pages} {360}
  (\bibinfo {year} {2016}{\natexlab{b}})},\ \Eprint
  {http://arxiv.org/abs/1506.08453} {arXiv:1506.08453 [nucl-ex]} \BibitemShut
  {NoStop}%
\bibitem [{\citenamefont {Abelev}\ \emph {et~al.}(2010)\citenamefont {Abelev}
  \emph {et~al.}}]{STAR:2010gyg}%
  \BibitemOpen
  \bibfield  {author} {\bibinfo {author} {\bibfnamefont {B.~I.}\ \bibnamefont
  {Abelev}} \emph {et~al.} (\bibinfo {collaboration} {STAR}),\ }\href {\doibase
  10.1126/science.1183980} {\bibfield  {journal} {\bibinfo  {journal}
  {Science}\ }\textbf {\bibinfo {volume} {328}},\ \bibinfo {pages} {58}
  (\bibinfo {year} {2010})},\ \Eprint {http://arxiv.org/abs/1003.2030}
  {arXiv:1003.2030 [nucl-ex]} \BibitemShut {NoStop}%
\bibitem [{\citenamefont {(STAR)}(2023)}]{STAR:2023fbc2}%
  \BibitemOpen
  \bibfield  {author} {\bibinfo {author} {\bibnamefont {(STAR)}},\ }\href@noop
  {} {\  (\bibinfo {year} {2023})},\ \Eprint {http://arxiv.org/abs/2310.12674}
  {arXiv:2310.12674 [nucl-ex]} \BibitemShut {NoStop}%
\bibitem [{\citenamefont {Noronha-Hostler}\ \emph {et~al.}(2013)\citenamefont
  {Noronha-Hostler}, \citenamefont {Denicol}, \citenamefont {Noronha},
  \citenamefont {Andrade},\ and\ \citenamefont
  {Grassi}}]{Noronha-Hostler:2013gga}%
  \BibitemOpen
  \bibfield  {author} {\bibinfo {author} {\bibfnamefont {J.}~\bibnamefont
  {Noronha-Hostler}}, \bibinfo {author} {\bibfnamefont {G.~S.}\ \bibnamefont
  {Denicol}}, \bibinfo {author} {\bibfnamefont {J.}~\bibnamefont {Noronha}},
  \bibinfo {author} {\bibfnamefont {R.~P.~G.}\ \bibnamefont {Andrade}}, \ and\
  \bibinfo {author} {\bibfnamefont {F.}~\bibnamefont {Grassi}},\ }\href
  {\doibase 10.1103/PhysRevC.88.044916} {\bibfield  {journal} {\bibinfo
  {journal} {Phys. Rev. C}\ }\textbf {\bibinfo {volume} {88}},\ \bibinfo
  {pages} {044916} (\bibinfo {year} {2013})},\ \Eprint
  {http://arxiv.org/abs/1305.1981} {arXiv:1305.1981 [nucl-th]} \BibitemShut
  {NoStop}%
\bibitem [{\citenamefont {Noronha-Hostler}\ \emph {et~al.}(2014)\citenamefont
  {Noronha-Hostler}, \citenamefont {Noronha},\ and\ \citenamefont
  {Grassi}}]{Noronha-Hostler:2014dqa}%
  \BibitemOpen
  \bibfield  {author} {\bibinfo {author} {\bibfnamefont {J.}~\bibnamefont
  {Noronha-Hostler}}, \bibinfo {author} {\bibfnamefont {J.}~\bibnamefont
  {Noronha}}, \ and\ \bibinfo {author} {\bibfnamefont {F.}~\bibnamefont
  {Grassi}},\ }\href {\doibase 10.1103/PhysRevC.90.034907} {\bibfield
  {journal} {\bibinfo  {journal} {Phys. Rev. C}\ }\textbf {\bibinfo {volume}
  {90}},\ \bibinfo {pages} {034907} (\bibinfo {year} {2014})},\ \Eprint
  {http://arxiv.org/abs/1406.3333} {arXiv:1406.3333 [nucl-th]} \BibitemShut
  {NoStop}%
\bibitem [{\citenamefont {Alba}\ \emph {et~al.}(2018)\citenamefont {Alba},
  \citenamefont {Mantovani~Sarti}, \citenamefont {Noronha}, \citenamefont
  {Noronha-Hostler}, \citenamefont {Parotto}, \citenamefont
  {Portillo~Vazquez},\ and\ \citenamefont {Ratti}}]{Alba:2017hhe}%
  \BibitemOpen
  \bibfield  {author} {\bibinfo {author} {\bibfnamefont {P.}~\bibnamefont
  {Alba}}, \bibinfo {author} {\bibfnamefont {V.}~\bibnamefont
  {Mantovani~Sarti}}, \bibinfo {author} {\bibfnamefont {J.}~\bibnamefont
  {Noronha}}, \bibinfo {author} {\bibfnamefont {J.}~\bibnamefont
  {Noronha-Hostler}}, \bibinfo {author} {\bibfnamefont {P.}~\bibnamefont
  {Parotto}}, \bibinfo {author} {\bibfnamefont {I.}~\bibnamefont
  {Portillo~Vazquez}}, \ and\ \bibinfo {author} {\bibfnamefont
  {C.}~\bibnamefont {Ratti}},\ }\href {\doibase 10.1103/PhysRevC.98.034909}
  {\bibfield  {journal} {\bibinfo  {journal} {Phys. Rev. C}\ }\textbf {\bibinfo
  {volume} {98}},\ \bibinfo {pages} {034909} (\bibinfo {year} {2018})},\
  \Eprint {http://arxiv.org/abs/1711.05207} {arXiv:1711.05207 [nucl-th]}
  \BibitemShut {NoStop}%
\bibitem [{\citenamefont {Borsanyi}\ \emph {et~al.}(2010)\citenamefont
  {Borsanyi}, \citenamefont {Fodor}, \citenamefont {Hoelbling}, \citenamefont
  {Katz}, \citenamefont {Krieg}, \citenamefont {Ratti},\ and\ \citenamefont
  {Szabo}}]{Borsanyi:2010bp}%
  \BibitemOpen
  \bibfield  {author} {\bibinfo {author} {\bibfnamefont {S.}~\bibnamefont
  {Borsanyi}}, \bibinfo {author} {\bibfnamefont {Z.}~\bibnamefont {Fodor}},
  \bibinfo {author} {\bibfnamefont {C.}~\bibnamefont {Hoelbling}}, \bibinfo
  {author} {\bibfnamefont {S.~D.}\ \bibnamefont {Katz}}, \bibinfo {author}
  {\bibfnamefont {S.}~\bibnamefont {Krieg}}, \bibinfo {author} {\bibfnamefont
  {C.}~\bibnamefont {Ratti}}, \ and\ \bibinfo {author} {\bibfnamefont {K.~K.}\
  \bibnamefont {Szabo}} (\bibinfo {collaboration} {Wuppertal-Budapest}),\
  }\href {\doibase 10.1007/JHEP09(2010)073} {\bibfield  {journal} {\bibinfo
  {journal} {JHEP}\ }\textbf {\bibinfo {volume} {09}},\ \bibinfo {pages} {073}
  (\bibinfo {year} {2010})},\ \Eprint {http://arxiv.org/abs/1005.3508}
  {arXiv:1005.3508 [hep-lat]} \BibitemShut {NoStop}%
\bibitem [{\citenamefont {Arsene}\ \emph {et~al.}(2005)\citenamefont {Arsene}
  \emph {et~al.}}]{BRAHMS:2004adc}%
  \BibitemOpen
  \bibfield  {author} {\bibinfo {author} {\bibfnamefont {I.}~\bibnamefont
  {Arsene}} \emph {et~al.} (\bibinfo {collaboration} {BRAHMS}),\ }\href
  {\doibase 10.1016/j.nuclphysa.2005.02.130} {\bibfield  {journal} {\bibinfo
  {journal} {Nucl. Phys. A}\ }\textbf {\bibinfo {volume} {757}},\ \bibinfo
  {pages} {1} (\bibinfo {year} {2005})},\ \Eprint
  {http://arxiv.org/abs/nucl-ex/0410020} {arXiv:nucl-ex/0410020} \BibitemShut
  {NoStop}%
\bibitem [{\citenamefont {Back}\ \emph {et~al.}(2005)\citenamefont {Back} \emph
  {et~al.}}]{PHOBOS:2004zne}%
  \BibitemOpen
  \bibfield  {author} {\bibinfo {author} {\bibfnamefont {B.~B.}\ \bibnamefont
  {Back}} \emph {et~al.} (\bibinfo {collaboration} {PHOBOS}),\ }\href {\doibase
  10.1016/j.nuclphysa.2005.03.084} {\bibfield  {journal} {\bibinfo  {journal}
  {Nucl. Phys. A}\ }\textbf {\bibinfo {volume} {757}},\ \bibinfo {pages} {28}
  (\bibinfo {year} {2005})},\ \Eprint {http://arxiv.org/abs/nucl-ex/0410022}
  {arXiv:nucl-ex/0410022} \BibitemShut {NoStop}%
\bibitem [{\citenamefont {Adcox}\ \emph {et~al.}(2005)\citenamefont {Adcox}
  \emph {et~al.}}]{PHENIX:2004vcz}%
  \BibitemOpen
  \bibfield  {author} {\bibinfo {author} {\bibfnamefont {K.}~\bibnamefont
  {Adcox}} \emph {et~al.} (\bibinfo {collaboration} {PHENIX}),\ }\href
  {\doibase 10.1016/j.nuclphysa.2005.03.086} {\bibfield  {journal} {\bibinfo
  {journal} {Nucl. Phys. A}\ }\textbf {\bibinfo {volume} {757}},\ \bibinfo
  {pages} {184} (\bibinfo {year} {2005})},\ \Eprint
  {http://arxiv.org/abs/nucl-ex/0410003} {arXiv:nucl-ex/0410003} \BibitemShut
  {NoStop}%
\bibitem [{\citenamefont {Adams}\ \emph {et~al.}(2005)\citenamefont {Adams}
  \emph {et~al.}}]{STAR:2005gfr}%
  \BibitemOpen
  \bibfield  {author} {\bibinfo {author} {\bibfnamefont {J.}~\bibnamefont
  {Adams}} \emph {et~al.} (\bibinfo {collaboration} {STAR}),\ }\href {\doibase
  10.1016/j.nuclphysa.2005.03.085} {\bibfield  {journal} {\bibinfo  {journal}
  {Nucl. Phys. A}\ }\textbf {\bibinfo {volume} {757}},\ \bibinfo {pages} {102}
  (\bibinfo {year} {2005})},\ \Eprint {http://arxiv.org/abs/nucl-ex/0501009}
  {arXiv:nucl-ex/0501009} \BibitemShut {NoStop}%
\bibitem [{\citenamefont {Noronha-Hostler}\ \emph {et~al.}(2016)\citenamefont
  {Noronha-Hostler}, \citenamefont {Luzum},\ and\ \citenamefont
  {Ollitrault}}]{Noronha-Hostler:2015uye}%
  \BibitemOpen
  \bibfield  {author} {\bibinfo {author} {\bibfnamefont {J.}~\bibnamefont
  {Noronha-Hostler}}, \bibinfo {author} {\bibfnamefont {M.}~\bibnamefont
  {Luzum}}, \ and\ \bibinfo {author} {\bibfnamefont {J.-Y.}\ \bibnamefont
  {Ollitrault}},\ }\href {\doibase 10.1103/PhysRevC.93.034912} {\bibfield
  {journal} {\bibinfo  {journal} {Phys. Rev. C}\ }\textbf {\bibinfo {volume}
  {93}},\ \bibinfo {pages} {034912} (\bibinfo {year} {2016})},\ \Eprint
  {http://arxiv.org/abs/1511.06289} {arXiv:1511.06289 [nucl-th]} \BibitemShut
  {NoStop}%
\bibitem [{\citenamefont {Niemi}\ \emph {et~al.}(2016)\citenamefont {Niemi},
  \citenamefont {Eskola}, \citenamefont {Paatelainen},\ and\ \citenamefont
  {Tuominen}}]{Niemi:2015voa}%
  \BibitemOpen
  \bibfield  {author} {\bibinfo {author} {\bibfnamefont {H.}~\bibnamefont
  {Niemi}}, \bibinfo {author} {\bibfnamefont {K.~J.}\ \bibnamefont {Eskola}},
  \bibinfo {author} {\bibfnamefont {R.}~\bibnamefont {Paatelainen}}, \ and\
  \bibinfo {author} {\bibfnamefont {K.}~\bibnamefont {Tuominen}},\ }\href
  {\doibase 10.1103/PhysRevC.93.014912} {\bibfield  {journal} {\bibinfo
  {journal} {Phys. Rev. C}\ }\textbf {\bibinfo {volume} {93}},\ \bibinfo
  {pages} {014912} (\bibinfo {year} {2016})},\ \Eprint
  {http://arxiv.org/abs/1511.04296} {arXiv:1511.04296 [hep-ph]} \BibitemShut
  {NoStop}%
\bibitem [{\citenamefont {Adam}\ \emph
  {et~al.}(2016{\natexlab{c}})\citenamefont {Adam} \emph
  {et~al.}}]{Adam:2015ptt}%
  \BibitemOpen
  \bibfield  {author} {\bibinfo {author} {\bibfnamefont {J.}~\bibnamefont
  {Adam}} \emph {et~al.} (\bibinfo {collaboration} {ALICE}),\ }\href {\doibase
  10.1103/PhysRevLett.116.222302} {\bibfield  {journal} {\bibinfo  {journal}
  {Phys. Rev. Lett.}\ }\textbf {\bibinfo {volume} {116}},\ \bibinfo {pages}
  {222302} (\bibinfo {year} {2016}{\natexlab{c}})},\ \Eprint
  {http://arxiv.org/abs/1512.06104} {arXiv:1512.06104 [nucl-ex]} \BibitemShut
  {NoStop}%
\bibitem [{\citenamefont {Ryu}\ \emph {et~al.}(2013)\citenamefont {Ryu},
  \citenamefont {Jeon}, \citenamefont {Gale}, \citenamefont {Schenke},\ and\
  \citenamefont {Young}}]{Ryu:2012at}%
  \BibitemOpen
  \bibfield  {author} {\bibinfo {author} {\bibfnamefont {S.}~\bibnamefont
  {Ryu}}, \bibinfo {author} {\bibfnamefont {S.}~\bibnamefont {Jeon}}, \bibinfo
  {author} {\bibfnamefont {C.}~\bibnamefont {Gale}}, \bibinfo {author}
  {\bibfnamefont {B.}~\bibnamefont {Schenke}}, \ and\ \bibinfo {author}
  {\bibfnamefont {C.}~\bibnamefont {Young}},\ }\href {\doibase
  10.1016/j.nuclphysa.2013.02.031} {\bibfield  {journal} {\bibinfo  {journal}
  {Nucl. Phys. A}\ }\textbf {\bibinfo {volume} {904-905}},\ \bibinfo {pages}
  {389c} (\bibinfo {year} {2013})},\ \Eprint {http://arxiv.org/abs/1210.4588}
  {arXiv:1210.4588 [hep-ph]} \BibitemShut {NoStop}%
\bibitem [{\citenamefont {Teaney}\ \emph
  {et~al.}(2001{\natexlab{a}})\citenamefont {Teaney}, \citenamefont {Lauret},\
  and\ \citenamefont {Shuryak}}]{Teaney:2000cw}%
  \BibitemOpen
  \bibfield  {author} {\bibinfo {author} {\bibfnamefont {D.}~\bibnamefont
  {Teaney}}, \bibinfo {author} {\bibfnamefont {J.}~\bibnamefont {Lauret}}, \
  and\ \bibinfo {author} {\bibfnamefont {E.~V.}\ \bibnamefont {Shuryak}},\
  }\href {\doibase 10.1103/PhysRevLett.86.4783} {\bibfield  {journal} {\bibinfo
   {journal} {Phys. Rev. Lett.}\ }\textbf {\bibinfo {volume} {86}},\ \bibinfo
  {pages} {4783} (\bibinfo {year} {2001}{\natexlab{a}})},\ \Eprint
  {http://arxiv.org/abs/nucl-th/0011058} {arXiv:nucl-th/0011058} \BibitemShut
  {NoStop}%
\bibitem [{\citenamefont {Teaney}\ \emph
  {et~al.}(2001{\natexlab{b}})\citenamefont {Teaney}, \citenamefont {Lauret},\
  and\ \citenamefont {Shuryak}}]{Teaney:2001av}%
  \BibitemOpen
  \bibfield  {author} {\bibinfo {author} {\bibfnamefont {D.}~\bibnamefont
  {Teaney}}, \bibinfo {author} {\bibfnamefont {J.}~\bibnamefont {Lauret}}, \
  and\ \bibinfo {author} {\bibfnamefont {E.~V.}\ \bibnamefont {Shuryak}},\
  }\href@noop {} {\  (\bibinfo {year} {2001}{\natexlab{b}})},\ \Eprint
  {http://arxiv.org/abs/nucl-th/0110037} {arXiv:nucl-th/0110037} \BibitemShut
  {NoStop}%
\bibitem [{\citenamefont {Hirano}\ \emph {et~al.}(2006)\citenamefont {Hirano},
  \citenamefont {Heinz}, \citenamefont {Kharzeev}, \citenamefont {Lacey},\ and\
  \citenamefont {Nara}}]{Hirano:2005xf}%
  \BibitemOpen
  \bibfield  {author} {\bibinfo {author} {\bibfnamefont {T.}~\bibnamefont
  {Hirano}}, \bibinfo {author} {\bibfnamefont {U.~W.}\ \bibnamefont {Heinz}},
  \bibinfo {author} {\bibfnamefont {D.}~\bibnamefont {Kharzeev}}, \bibinfo
  {author} {\bibfnamefont {R.}~\bibnamefont {Lacey}}, \ and\ \bibinfo {author}
  {\bibfnamefont {Y.}~\bibnamefont {Nara}},\ }\href {\doibase
  10.1016/j.physletb.2006.03.060} {\bibfield  {journal} {\bibinfo  {journal}
  {Phys. Lett. B}\ }\textbf {\bibinfo {volume} {636}},\ \bibinfo {pages} {299}
  (\bibinfo {year} {2006})},\ \Eprint {http://arxiv.org/abs/nucl-th/0511046}
  {arXiv:nucl-th/0511046} \BibitemShut {NoStop}%
\bibitem [{\citenamefont {Sch\"afer}\ \emph {et~al.}(2022)\citenamefont
  {Sch\"afer}, \citenamefont {Karpenko}, \citenamefont {Wu}, \citenamefont
  {Hammelmann},\ and\ \citenamefont {Elfner}}]{Schafer:2021csj}%
  \BibitemOpen
  \bibfield  {author} {\bibinfo {author} {\bibfnamefont {A.}~\bibnamefont
  {Sch\"afer}}, \bibinfo {author} {\bibfnamefont {I.}~\bibnamefont {Karpenko}},
  \bibinfo {author} {\bibfnamefont {X.-Y.}\ \bibnamefont {Wu}}, \bibinfo
  {author} {\bibfnamefont {J.}~\bibnamefont {Hammelmann}}, \ and\ \bibinfo
  {author} {\bibfnamefont {H.}~\bibnamefont {Elfner}} (\bibinfo {collaboration}
  {SMASH}),\ }\href {\doibase 10.1140/epja/s10050-022-00872-x} {\bibfield
  {journal} {\bibinfo  {journal} {Eur. Phys. J. A}\ }\textbf {\bibinfo {volume}
  {58}},\ \bibinfo {pages} {230} (\bibinfo {year} {2022})},\ \Eprint
  {http://arxiv.org/abs/2112.08724} {arXiv:2112.08724 [hep-ph]} \BibitemShut
  {NoStop}%
\bibitem [{\citenamefont {Shen}\ and\ \citenamefont
  {Schenke}(2022)}]{Shen:2022oyg}%
  \BibitemOpen
  \bibfield  {author} {\bibinfo {author} {\bibfnamefont {C.}~\bibnamefont
  {Shen}}\ and\ \bibinfo {author} {\bibfnamefont {B.}~\bibnamefont {Schenke}},\
  }\href {\doibase 10.1103/PhysRevC.105.064905} {\bibfield  {journal} {\bibinfo
   {journal} {Phys. Rev. C}\ }\textbf {\bibinfo {volume} {105}},\ \bibinfo
  {pages} {064905} (\bibinfo {year} {2022})},\ \Eprint
  {http://arxiv.org/abs/2203.04685} {arXiv:2203.04685 [nucl-th]} \BibitemShut
  {NoStop}%
\bibitem [{\citenamefont {Wolter}\ \emph {et~al.}(2022)\citenamefont {Wolter}
  \emph {et~al.}}]{TMEP:2022xjg}%
  \BibitemOpen
  \bibfield  {author} {\bibinfo {author} {\bibfnamefont {H.}~\bibnamefont
  {Wolter}} \emph {et~al.} (\bibinfo {collaboration} {TMEP}),\ }\href {\doibase
  10.1016/j.ppnp.2022.103962} {\bibfield  {journal} {\bibinfo  {journal} {Prog.
  Part. Nucl. Phys.}\ }\textbf {\bibinfo {volume} {125}},\ \bibinfo {pages}
  {103962} (\bibinfo {year} {2022})},\ \Eprint
  {http://arxiv.org/abs/2202.06672} {arXiv:2202.06672 [nucl-th]} \BibitemShut
  {NoStop}%
\bibitem [{\citenamefont {Inghirami}\ and\ \citenamefont
  {Elfner}(2022)}]{Inghirami:2022afu}%
  \BibitemOpen
  \bibfield  {author} {\bibinfo {author} {\bibfnamefont {G.}~\bibnamefont
  {Inghirami}}\ and\ \bibinfo {author} {\bibfnamefont {H.}~\bibnamefont
  {Elfner}},\ }\href {\doibase 10.1140/epjc/s10052-022-10718-x} {\bibfield
  {journal} {\bibinfo  {journal} {Eur. Phys. J. C}\ }\textbf {\bibinfo {volume}
  {82}},\ \bibinfo {pages} {796} (\bibinfo {year} {2022})},\ \Eprint
  {http://arxiv.org/abs/2201.05934} {arXiv:2201.05934 [hep-ph]} \BibitemShut
  {NoStop}%
\bibitem [{\citenamefont {Bellwied}\ \emph {et~al.}(2015)\citenamefont
  {Bellwied}, \citenamefont {Borsanyi}, \citenamefont {Fodor}, \citenamefont
  {G\"unther}, \citenamefont {Katz}, \citenamefont {Ratti},\ and\ \citenamefont
  {Szabo}}]{Bellwied:2015rza}%
  \BibitemOpen
  \bibfield  {author} {\bibinfo {author} {\bibfnamefont {R.}~\bibnamefont
  {Bellwied}}, \bibinfo {author} {\bibfnamefont {S.}~\bibnamefont {Borsanyi}},
  \bibinfo {author} {\bibfnamefont {Z.}~\bibnamefont {Fodor}}, \bibinfo
  {author} {\bibfnamefont {J.}~\bibnamefont {G\"unther}}, \bibinfo {author}
  {\bibfnamefont {S.~D.}\ \bibnamefont {Katz}}, \bibinfo {author}
  {\bibfnamefont {C.}~\bibnamefont {Ratti}}, \ and\ \bibinfo {author}
  {\bibfnamefont {K.~K.}\ \bibnamefont {Szabo}},\ }\href {\doibase
  10.1016/j.physletb.2015.11.011} {\bibfield  {journal} {\bibinfo  {journal}
  {Phys. Lett. B}\ }\textbf {\bibinfo {volume} {751}},\ \bibinfo {pages} {559}
  (\bibinfo {year} {2015})},\ \Eprint {http://arxiv.org/abs/1507.07510}
  {arXiv:1507.07510 [hep-lat]} \BibitemShut {NoStop}%
\bibitem [{\citenamefont {Borsanyi}\ \emph {et~al.}(2020)\citenamefont
  {Borsanyi}, \citenamefont {Fodor}, \citenamefont {Guenther}, \citenamefont
  {Kara}, \citenamefont {Katz}, \citenamefont {Parotto}, \citenamefont
  {Pasztor}, \citenamefont {Ratti},\ and\ \citenamefont
  {Szabo}}]{Borsanyi:2020fev}%
  \BibitemOpen
  \bibfield  {author} {\bibinfo {author} {\bibfnamefont {S.}~\bibnamefont
  {Borsanyi}}, \bibinfo {author} {\bibfnamefont {Z.}~\bibnamefont {Fodor}},
  \bibinfo {author} {\bibfnamefont {J.~N.}\ \bibnamefont {Guenther}}, \bibinfo
  {author} {\bibfnamefont {R.}~\bibnamefont {Kara}}, \bibinfo {author}
  {\bibfnamefont {S.~D.}\ \bibnamefont {Katz}}, \bibinfo {author}
  {\bibfnamefont {P.}~\bibnamefont {Parotto}}, \bibinfo {author} {\bibfnamefont
  {A.}~\bibnamefont {Pasztor}}, \bibinfo {author} {\bibfnamefont
  {C.}~\bibnamefont {Ratti}}, \ and\ \bibinfo {author} {\bibfnamefont {K.~K.}\
  \bibnamefont {Szabo}},\ }\href {\doibase 10.1103/PhysRevLett.125.052001}
  {\bibfield  {journal} {\bibinfo  {journal} {Phys. Rev. Lett.}\ }\textbf
  {\bibinfo {volume} {125}},\ \bibinfo {pages} {052001} (\bibinfo {year}
  {2020})},\ \Eprint {http://arxiv.org/abs/2002.02821} {arXiv:2002.02821
  [hep-lat]} \BibitemShut {NoStop}%
\bibitem [{\citenamefont {Friese}(2006)}]{Friese:2006dj}%
  \BibitemOpen
  \bibfield  {author} {\bibinfo {author} {\bibfnamefont {V.}~\bibnamefont
  {Friese}},\ }\href {\doibase 10.1016/j.nuclphysa.2006.06.018} {\bibfield
  {journal} {\bibinfo  {journal} {Nucl. Phys. A}\ }\textbf {\bibinfo {volume}
  {774}},\ \bibinfo {pages} {377} (\bibinfo {year} {2006})}\BibitemShut
  {NoStop}%
\bibitem [{\citenamefont {Tahir}\ \emph {et~al.}(2005)\citenamefont {Tahir}
  \emph {et~al.}}]{Tahir:2005zz}%
  \BibitemOpen
  \bibfield  {author} {\bibinfo {author} {\bibfnamefont {N.~A.}\ \bibnamefont
  {Tahir}} \emph {et~al.},\ }\href {\doibase 10.1103/PhysRevLett.95.035001}
  {\bibfield  {journal} {\bibinfo  {journal} {Phys. Rev. Lett.}\ }\textbf
  {\bibinfo {volume} {95}},\ \bibinfo {pages} {035001} (\bibinfo {year}
  {2005})}\BibitemShut {NoStop}%
\bibitem [{\citenamefont {Lutz}\ \emph {et~al.}(2009)\citenamefont {Lutz} \emph
  {et~al.}}]{Lutz:2009ff}%
  \BibitemOpen
  \bibfield  {author} {\bibinfo {author} {\bibfnamefont {M.~F.~M.}\
  \bibnamefont {Lutz}} \emph {et~al.} (\bibinfo {collaboration} {PANDA}),\
  }\href@noop {} {\  (\bibinfo {year} {2009})},\ \Eprint
  {http://arxiv.org/abs/0903.3905} {arXiv:0903.3905 [hep-ex]} \BibitemShut
  {NoStop}%
\bibitem [{\citenamefont {Durante}\ \emph {et~al.}(2019)\citenamefont {Durante}
  \emph {et~al.}}]{Durante:2019hzd}%
  \BibitemOpen
  \bibfield  {author} {\bibinfo {author} {\bibfnamefont {M.}~\bibnamefont
  {Durante}} \emph {et~al.},\ }\href {\doibase 10.1088/1402-4896/aaf93f}
  {\bibfield  {journal} {\bibinfo  {journal} {Phys. Scripta}\ }\textbf
  {\bibinfo {volume} {94}},\ \bibinfo {pages} {033001} (\bibinfo {year}
  {2019})},\ \Eprint {http://arxiv.org/abs/1903.05693} {arXiv:1903.05693
  [nucl-th]} \BibitemShut {NoStop}%
\bibitem [{\citenamefont {Stoecker}\ \emph {et~al.}(1980)\citenamefont
  {Stoecker}, \citenamefont {Maruhn},\ and\ \citenamefont
  {Greiner}}]{Stoecker:1980vf}%
  \BibitemOpen
  \bibfield  {author} {\bibinfo {author} {\bibfnamefont {H.}~\bibnamefont
  {Stoecker}}, \bibinfo {author} {\bibfnamefont {J.~A.}\ \bibnamefont
  {Maruhn}}, \ and\ \bibinfo {author} {\bibfnamefont {W.}~\bibnamefont
  {Greiner}},\ }\href {\doibase 10.1103/PhysRevLett.44.725} {\bibfield
  {journal} {\bibinfo  {journal} {Phys. Rev. Lett.}\ }\textbf {\bibinfo
  {volume} {44}},\ \bibinfo {pages} {725} (\bibinfo {year} {1980})}\BibitemShut
  {NoStop}%
\bibitem [{\citenamefont {Ollitrault}(1992)}]{Ollitrault:1992bk}%
  \BibitemOpen
  \bibfield  {author} {\bibinfo {author} {\bibfnamefont {J.-Y.}\ \bibnamefont
  {Ollitrault}},\ }\href {\doibase 10.1103/PhysRevD.46.229} {\bibfield
  {journal} {\bibinfo  {journal} {Phys. Rev. D}\ }\textbf {\bibinfo {volume}
  {46}},\ \bibinfo {pages} {229} (\bibinfo {year} {1992})}\BibitemShut
  {NoStop}%
\bibitem [{\citenamefont {Rischke}\ \emph {et~al.}(1995)\citenamefont
  {Rischke}, \citenamefont {P\"urs\"un}, \citenamefont {Maruhn}, \citenamefont
  {Stoecker},\ and\ \citenamefont {Greiner}}]{Rischke:1995pe}%
  \BibitemOpen
  \bibfield  {author} {\bibinfo {author} {\bibfnamefont {D.~H.}\ \bibnamefont
  {Rischke}}, \bibinfo {author} {\bibfnamefont {Y.}~\bibnamefont {P\"urs\"un}},
  \bibinfo {author} {\bibfnamefont {J.~A.}\ \bibnamefont {Maruhn}}, \bibinfo
  {author} {\bibfnamefont {H.}~\bibnamefont {Stoecker}}, \ and\ \bibinfo
  {author} {\bibfnamefont {W.}~\bibnamefont {Greiner}},\ }\href {\doibase
  10.1007/BF03053749} {\bibfield  {journal} {\bibinfo  {journal} {Acta Phys.
  Hung. A}\ }\textbf {\bibinfo {volume} {1}},\ \bibinfo {pages} {309} (\bibinfo
  {year} {1995})},\ \Eprint {http://arxiv.org/abs/nucl-th/9505014}
  {arXiv:nucl-th/9505014} \BibitemShut {NoStop}%
\bibitem [{\citenamefont {Stoecker}(2005)}]{Stoecker:2004qu}%
  \BibitemOpen
  \bibfield  {author} {\bibinfo {author} {\bibfnamefont {H.}~\bibnamefont
  {Stoecker}},\ }\href {\doibase 10.1016/j.nuclphysa.2004.12.074} {\bibfield
  {journal} {\bibinfo  {journal} {Nucl. Phys. A}\ }\textbf {\bibinfo {volume}
  {750}},\ \bibinfo {pages} {121} (\bibinfo {year} {2005})},\ \Eprint
  {http://arxiv.org/abs/nucl-th/0406018} {arXiv:nucl-th/0406018} \BibitemShut
  {NoStop}%
\bibitem [{\citenamefont {Brachmann}\ \emph {et~al.}(2000)\citenamefont
  {Brachmann}, \citenamefont {Soff}, \citenamefont {Dumitru}, \citenamefont
  {Stoecker}, \citenamefont {Maruhn}, \citenamefont {Greiner}, \citenamefont
  {Bravina},\ and\ \citenamefont {Rischke}}]{Brachmann:1999xt}%
  \BibitemOpen
  \bibfield  {author} {\bibinfo {author} {\bibfnamefont {J.}~\bibnamefont
  {Brachmann}}, \bibinfo {author} {\bibfnamefont {S.}~\bibnamefont {Soff}},
  \bibinfo {author} {\bibfnamefont {A.}~\bibnamefont {Dumitru}}, \bibinfo
  {author} {\bibfnamefont {H.}~\bibnamefont {Stoecker}}, \bibinfo {author}
  {\bibfnamefont {J.~A.}\ \bibnamefont {Maruhn}}, \bibinfo {author}
  {\bibfnamefont {W.}~\bibnamefont {Greiner}}, \bibinfo {author} {\bibfnamefont
  {L.~V.}\ \bibnamefont {Bravina}}, \ and\ \bibinfo {author} {\bibfnamefont
  {D.~H.}\ \bibnamefont {Rischke}},\ }\href {\doibase
  10.1103/PhysRevC.61.024909} {\bibfield  {journal} {\bibinfo  {journal} {Phys.
  Rev. C}\ }\textbf {\bibinfo {volume} {61}},\ \bibinfo {pages} {024909}
  (\bibinfo {year} {2000})},\ \Eprint {http://arxiv.org/abs/nucl-th/9908010}
  {arXiv:nucl-th/9908010} \BibitemShut {NoStop}%
\bibitem [{\citenamefont {Csernai}\ and\ \citenamefont
  {Rohrich}(1999)}]{Csernai:1999nf}%
  \BibitemOpen
  \bibfield  {author} {\bibinfo {author} {\bibfnamefont {L.~P.}\ \bibnamefont
  {Csernai}}\ and\ \bibinfo {author} {\bibfnamefont {D.}~\bibnamefont
  {Rohrich}},\ }\href {\doibase 10.1016/S0370-2693(99)00615-2} {\bibfield
  {journal} {\bibinfo  {journal} {Phys. Lett. B}\ }\textbf {\bibinfo {volume}
  {458}},\ \bibinfo {pages} {454} (\bibinfo {year} {1999})},\ \Eprint
  {http://arxiv.org/abs/nucl-th/9908034} {arXiv:nucl-th/9908034} \BibitemShut
  {NoStop}%
\bibitem [{\citenamefont {Ivanov}\ and\ \citenamefont
  {Soldatov}(2015)}]{Ivanov:2014ioa}%
  \BibitemOpen
  \bibfield  {author} {\bibinfo {author} {\bibfnamefont {Y.~B.}\ \bibnamefont
  {Ivanov}}\ and\ \bibinfo {author} {\bibfnamefont {A.~A.}\ \bibnamefont
  {Soldatov}},\ }\href {\doibase 10.1103/PhysRevC.91.024915} {\bibfield
  {journal} {\bibinfo  {journal} {Phys. Rev. C}\ }\textbf {\bibinfo {volume}
  {91}},\ \bibinfo {pages} {024915} (\bibinfo {year} {2015})},\ \Eprint
  {http://arxiv.org/abs/1412.1669} {arXiv:1412.1669 [nucl-th]} \BibitemShut
  {NoStop}%
\bibitem [{\citenamefont {Hartnack}\ \emph {et~al.}(1994)\citenamefont
  {Hartnack}, \citenamefont {Aichelin}, \citenamefont {Stoecker},\ and\
  \citenamefont {Greiner}}]{Hartnack:1994ce}%
  \BibitemOpen
  \bibfield  {author} {\bibinfo {author} {\bibfnamefont {C.}~\bibnamefont
  {Hartnack}}, \bibinfo {author} {\bibfnamefont {J.}~\bibnamefont {Aichelin}},
  \bibinfo {author} {\bibfnamefont {H.}~\bibnamefont {Stoecker}}, \ and\
  \bibinfo {author} {\bibfnamefont {W.}~\bibnamefont {Greiner}},\ }\href
  {\doibase 10.1016/0370-2693(94)90237-2} {\bibfield  {journal} {\bibinfo
  {journal} {Phys. Lett. B}\ }\textbf {\bibinfo {volume} {336}},\ \bibinfo
  {pages} {131} (\bibinfo {year} {1994})}\BibitemShut {NoStop}%
\bibitem [{\citenamefont {Li}\ and\ \citenamefont {Ko}(1998)}]{Li:1998ze}%
  \BibitemOpen
  \bibfield  {author} {\bibinfo {author} {\bibfnamefont {B.-A.}\ \bibnamefont
  {Li}}\ and\ \bibinfo {author} {\bibfnamefont {C.~M.}\ \bibnamefont {Ko}},\
  }\href {\doibase 10.1103/PhysRevC.58.R1382} {\bibfield  {journal} {\bibinfo
  {journal} {Phys. Rev. C}\ }\textbf {\bibinfo {volume} {58}},\ \bibinfo
  {pages} {R1382} (\bibinfo {year} {1998})},\ \Eprint
  {http://arxiv.org/abs/nucl-th/9807088} {arXiv:nucl-th/9807088} \BibitemShut
  {NoStop}%
\bibitem [{\citenamefont {Le~F\`evre}\ \emph {et~al.}(2016)\citenamefont
  {Le~F\`evre}, \citenamefont {Leifels}, \citenamefont {Reisdorf},
  \citenamefont {Aichelin},\ and\ \citenamefont {Hartnack}}]{LeFevre:2015paj}%
  \BibitemOpen
  \bibfield  {author} {\bibinfo {author} {\bibfnamefont {A.}~\bibnamefont
  {Le~F\`evre}}, \bibinfo {author} {\bibfnamefont {Y.}~\bibnamefont {Leifels}},
  \bibinfo {author} {\bibfnamefont {W.}~\bibnamefont {Reisdorf}}, \bibinfo
  {author} {\bibfnamefont {J.}~\bibnamefont {Aichelin}}, \ and\ \bibinfo
  {author} {\bibfnamefont {C.}~\bibnamefont {Hartnack}},\ }\href {\doibase
  10.1016/j.nuclphysa.2015.09.015} {\bibfield  {journal} {\bibinfo  {journal}
  {Nucl. Phys. A}\ }\textbf {\bibinfo {volume} {945}},\ \bibinfo {pages} {112}
  (\bibinfo {year} {2016})},\ \Eprint {http://arxiv.org/abs/1501.05246}
  {arXiv:1501.05246 [nucl-ex]} \BibitemShut {NoStop}%
\bibitem [{\citenamefont {Wang}\ \emph {et~al.}(2018)\citenamefont {Wang},
  \citenamefont {Guo}, \citenamefont {Li}, \citenamefont {Le~F\`evre},
  \citenamefont {Leifels},\ and\ \citenamefont {Trautmann}}]{Wang:2018hsw}%
  \BibitemOpen
  \bibfield  {author} {\bibinfo {author} {\bibfnamefont {Y.}~\bibnamefont
  {Wang}}, \bibinfo {author} {\bibfnamefont {C.}~\bibnamefont {Guo}}, \bibinfo
  {author} {\bibfnamefont {Q.}~\bibnamefont {Li}}, \bibinfo {author}
  {\bibfnamefont {A.}~\bibnamefont {Le~F\`evre}}, \bibinfo {author}
  {\bibfnamefont {Y.}~\bibnamefont {Leifels}}, \ and\ \bibinfo {author}
  {\bibfnamefont {W.}~\bibnamefont {Trautmann}},\ }\href {\doibase
  10.1016/j.physletb.2018.01.035} {\bibfield  {journal} {\bibinfo  {journal}
  {Phys. Lett. B}\ }\textbf {\bibinfo {volume} {778}},\ \bibinfo {pages} {207}
  (\bibinfo {year} {2018})},\ \Eprint {http://arxiv.org/abs/1804.04293}
  {arXiv:1804.04293 [nucl-th]} \BibitemShut {NoStop}%
\bibitem [{\citenamefont {Nara}\ and\ \citenamefont
  {Ohnishi}(2022)}]{Nara:2021fuu}%
  \BibitemOpen
  \bibfield  {author} {\bibinfo {author} {\bibfnamefont {Y.}~\bibnamefont
  {Nara}}\ and\ \bibinfo {author} {\bibfnamefont {A.}~\bibnamefont {Ohnishi}},\
  }\href {\doibase 10.1103/PhysRevC.105.014911} {\bibfield  {journal} {\bibinfo
   {journal} {Phys. Rev. C}\ }\textbf {\bibinfo {volume} {105}},\ \bibinfo
  {pages} {014911} (\bibinfo {year} {2022})},\ \Eprint
  {http://arxiv.org/abs/2109.07594} {arXiv:2109.07594 [nucl-th]} \BibitemShut
  {NoStop}%
\bibitem [{\citenamefont {Steinheimer}\ \emph {et~al.}(2022)\citenamefont
  {Steinheimer}, \citenamefont {Motornenko}, \citenamefont {Sorensen},
  \citenamefont {Nara}, \citenamefont {Koch},\ and\ \citenamefont
  {Bleicher}}]{Steinheimer:2022gqb}%
  \BibitemOpen
  \bibfield  {author} {\bibinfo {author} {\bibfnamefont {J.}~\bibnamefont
  {Steinheimer}}, \bibinfo {author} {\bibfnamefont {A.}~\bibnamefont
  {Motornenko}}, \bibinfo {author} {\bibfnamefont {A.}~\bibnamefont
  {Sorensen}}, \bibinfo {author} {\bibfnamefont {Y.}~\bibnamefont {Nara}},
  \bibinfo {author} {\bibfnamefont {V.}~\bibnamefont {Koch}}, \ and\ \bibinfo
  {author} {\bibfnamefont {M.}~\bibnamefont {Bleicher}},\ }\href {\doibase
  10.1140/epjc/s10052-022-10894-w} {\bibfield  {journal} {\bibinfo  {journal}
  {Eur. Phys. J. C}\ }\textbf {\bibinfo {volume} {82}},\ \bibinfo {pages} {911}
  (\bibinfo {year} {2022})},\ \Eprint {http://arxiv.org/abs/2208.12091}
  {arXiv:2208.12091 [nucl-th]} \BibitemShut {NoStop}%
\bibitem [{\citenamefont {Zhang}\ \emph
  {et~al.}(2018{\natexlab{b}})\citenamefont {Zhang}, \citenamefont {Chen},
  \citenamefont {Luo}, \citenamefont {Liu},\ and\ \citenamefont
  {Nara}}]{Zhang:2018wlk}%
  \BibitemOpen
  \bibfield  {author} {\bibinfo {author} {\bibfnamefont {C.}~\bibnamefont
  {Zhang}}, \bibinfo {author} {\bibfnamefont {J.}~\bibnamefont {Chen}},
  \bibinfo {author} {\bibfnamefont {X.}~\bibnamefont {Luo}}, \bibinfo {author}
  {\bibfnamefont {F.}~\bibnamefont {Liu}}, \ and\ \bibinfo {author}
  {\bibfnamefont {Y.}~\bibnamefont {Nara}},\ }\href {\doibase
  10.1103/PhysRevC.97.064913} {\bibfield  {journal} {\bibinfo  {journal} {Phys.
  Rev. C}\ }\textbf {\bibinfo {volume} {97}},\ \bibinfo {pages} {064913}
  (\bibinfo {year} {2018}{\natexlab{b}})},\ \Eprint
  {http://arxiv.org/abs/1803.02053} {arXiv:1803.02053 [nucl-ex]} \BibitemShut
  {NoStop}%
\bibitem [{\citenamefont {Dutra}\ \emph {et~al.}(2012)\citenamefont {Dutra},
  \citenamefont {Lourenco}, \citenamefont {Sa~Martins}, \citenamefont
  {Delfino}, \citenamefont {Stone},\ and\ \citenamefont
  {Stevenson}}]{Dutra:2012mb}%
  \BibitemOpen
  \bibfield  {author} {\bibinfo {author} {\bibfnamefont {M.}~\bibnamefont
  {Dutra}}, \bibinfo {author} {\bibfnamefont {O.}~\bibnamefont {Lourenco}},
  \bibinfo {author} {\bibfnamefont {J.~S.}\ \bibnamefont {Sa~Martins}},
  \bibinfo {author} {\bibfnamefont {A.}~\bibnamefont {Delfino}}, \bibinfo
  {author} {\bibfnamefont {J.~R.}\ \bibnamefont {Stone}}, \ and\ \bibinfo
  {author} {\bibfnamefont {P.~D.}\ \bibnamefont {Stevenson}},\ }\href {\doibase
  10.1103/PhysRevC.85.035201} {\bibfield  {journal} {\bibinfo  {journal} {Phys.
  Rev. C}\ }\textbf {\bibinfo {volume} {85}},\ \bibinfo {pages} {035201}
  (\bibinfo {year} {2012})},\ \Eprint {http://arxiv.org/abs/1202.3902}
  {arXiv:1202.3902 [nucl-th]} \BibitemShut {NoStop}%
\bibitem [{\citenamefont {Omana~Kuttan}\ \emph {et~al.}(2023)\citenamefont
  {Omana~Kuttan}, \citenamefont {Steinheimer}, \citenamefont {Zhou},\ and\
  \citenamefont {Stoecker}}]{OmanaKuttan:2022aml}%
  \BibitemOpen
  \bibfield  {author} {\bibinfo {author} {\bibfnamefont {M.}~\bibnamefont
  {Omana~Kuttan}}, \bibinfo {author} {\bibfnamefont {J.}~\bibnamefont
  {Steinheimer}}, \bibinfo {author} {\bibfnamefont {K.}~\bibnamefont {Zhou}}, \
  and\ \bibinfo {author} {\bibfnamefont {H.}~\bibnamefont {Stoecker}},\ }\href
  {\doibase 10.1103/PhysRevLett.131.202303} {\bibfield  {journal} {\bibinfo
  {journal} {Phys. Rev. Lett.}\ }\textbf {\bibinfo {volume} {131}},\ \bibinfo
  {pages} {202303} (\bibinfo {year} {2023})},\ \Eprint
  {http://arxiv.org/abs/2211.11670} {arXiv:2211.11670 [hep-ph]} \BibitemShut
  {NoStop}%
\bibitem [{\citenamefont {Botermans}\ and\ \citenamefont
  {Malfliet}(1990)}]{Botermans:1990qi}%
  \BibitemOpen
  \bibfield  {author} {\bibinfo {author} {\bibfnamefont {W.}~\bibnamefont
  {Botermans}}\ and\ \bibinfo {author} {\bibfnamefont {R.}~\bibnamefont
  {Malfliet}},\ }\href {\doibase 10.1016/0370-1573(90)90174-Z} {\bibfield
  {journal} {\bibinfo  {journal} {Phys. Rept.}\ }\textbf {\bibinfo {volume}
  {198}},\ \bibinfo {pages} {115} (\bibinfo {year} {1990})}\BibitemShut
  {NoStop}%
\bibitem [{\citenamefont {Sammarruca}\ and\ \citenamefont
  {Millerson}(2021)}]{Sammarruca:2021bpn}%
  \BibitemOpen
  \bibfield  {author} {\bibinfo {author} {\bibfnamefont {F.}~\bibnamefont
  {Sammarruca}}\ and\ \bibinfo {author} {\bibfnamefont {R.}~\bibnamefont
  {Millerson}},\ }\href {\doibase 10.1103/PhysRevC.104.064312} {\bibfield
  {journal} {\bibinfo  {journal} {Phys. Rev. C}\ }\textbf {\bibinfo {volume}
  {104}},\ \bibinfo {pages} {064312} (\bibinfo {year} {2021})},\ \Eprint
  {http://arxiv.org/abs/2109.01985} {arXiv:2109.01985 [nucl-th]} \BibitemShut
  {NoStop}%
\bibitem [{\citenamefont {{FRIB Science Community}}(2023)}]{FRIB400}%
  \BibitemOpen
  \bibfield  {author} {\bibinfo {author} {\bibnamefont {{FRIB Science
  Community}}},\ }\href@noop {} {\enquote {\bibinfo {title} {{FRIB400: The
  Scientific Case for the 400 MeV/u Energy Upgrade of FRIB}},}\ }\bibinfo
  {howpublished} {\url{https://frib.msu.edu/_files/pdfs/frib400_final.pdf}}
  (\bibinfo {year} {2019, updated February 2023}),\ \bibinfo {note} {accessed:
  2023-07-16}\BibitemShut {NoStop}%
\bibitem [{\citenamefont {{B.~A.~Brown \textit{et
  al.}}}()}]{FRIB-TA_white_paper_motivations}%
  \BibitemOpen
  \bibfield  {author} {\bibinfo {author} {\bibnamefont {{B.~A.~Brown \textit{et
  al.}}}},\ }\href@noop {} {\enquote {\bibinfo {title} {{Motivations for Early
  High-Profile FRIB Experiments}},}\ }\bibinfo {note} {In
  preparation}\BibitemShut {NoStop}%
\bibitem [{\citenamefont {Das}\ \emph {et~al.}(2003)\citenamefont {Das},
  \citenamefont {Gupta}, \citenamefont {Gale},\ and\ \citenamefont
  {Li}}]{Das:2002fr}%
  \BibitemOpen
  \bibfield  {author} {\bibinfo {author} {\bibfnamefont {C.~B.}\ \bibnamefont
  {Das}}, \bibinfo {author} {\bibfnamefont {S.~D.}\ \bibnamefont {Gupta}},
  \bibinfo {author} {\bibfnamefont {C.}~\bibnamefont {Gale}}, \ and\ \bibinfo
  {author} {\bibfnamefont {B.-A.}\ \bibnamefont {Li}},\ }\href {\doibase
  10.1103/PhysRevC.67.034611} {\bibfield  {journal} {\bibinfo  {journal} {Phys.
  Rev. C}\ }\textbf {\bibinfo {volume} {67}},\ \bibinfo {pages} {034611}
  (\bibinfo {year} {2003})},\ \Eprint {http://arxiv.org/abs/nucl-th/0212090}
  {arXiv:nucl-th/0212090} \BibitemShut {NoStop}%
\bibitem [{\citenamefont {Li}\ \emph {et~al.}(2004)\citenamefont {Li},
  \citenamefont {Das}, \citenamefont {Das~Gupta},\ and\ \citenamefont
  {Gale}}]{Li:2003zg}%
  \BibitemOpen
  \bibfield  {author} {\bibinfo {author} {\bibfnamefont {B.-A.}\ \bibnamefont
  {Li}}, \bibinfo {author} {\bibfnamefont {C.~B.}\ \bibnamefont {Das}},
  \bibinfo {author} {\bibfnamefont {S.}~\bibnamefont {Das~Gupta}}, \ and\
  \bibinfo {author} {\bibfnamefont {C.}~\bibnamefont {Gale}},\ }\href {\doibase
  10.1103/PhysRevC.69.011603} {\bibfield  {journal} {\bibinfo  {journal} {Phys.
  Rev. C}\ }\textbf {\bibinfo {volume} {69}},\ \bibinfo {pages} {011603}
  (\bibinfo {year} {2004})},\ \Eprint {http://arxiv.org/abs/nucl-th/0312032}
  {arXiv:nucl-th/0312032} \BibitemShut {NoStop}%
\bibitem [{\citenamefont {Li}\ \emph {et~al.}(2018)\citenamefont {Li},
  \citenamefont {Cai}, \citenamefont {Chen},\ and\ \citenamefont
  {Xu}}]{Li:2018lpy}%
  \BibitemOpen
  \bibfield  {author} {\bibinfo {author} {\bibfnamefont {B.-A.}\ \bibnamefont
  {Li}}, \bibinfo {author} {\bibfnamefont {B.-J.}\ \bibnamefont {Cai}},
  \bibinfo {author} {\bibfnamefont {L.-W.}\ \bibnamefont {Chen}}, \ and\
  \bibinfo {author} {\bibfnamefont {J.}~\bibnamefont {Xu}},\ }\href {\doibase
  10.1016/j.ppnp.2018.01.001} {\bibfield  {journal} {\bibinfo  {journal} {Prog.
  Part. Nucl. Phys.}\ }\textbf {\bibinfo {volume} {99}},\ \bibinfo {pages} {29}
  (\bibinfo {year} {2018})},\ \Eprint {http://arxiv.org/abs/1801.01213}
  {arXiv:1801.01213 [nucl-th]} \BibitemShut {NoStop}%
\bibitem [{\citenamefont {STARcollaboration}(2014)}]{STARnote}%
  \BibitemOpen
  \bibfield  {author} {\bibinfo {author} {\bibnamefont {STARcollaboration}},\
  }\href@noop {} {\enquote {\bibinfo {title} {{Studying the Phase Diagram of
  QCD Matter at RHIC}},}\ } (\bibinfo {year} {2014})\BibitemShut {NoStop}%
\bibitem [{\citenamefont {Cebra}\ \emph {et~al.}(2014)\citenamefont {Cebra},
  \citenamefont {Brovko}, \citenamefont {Flores}, \citenamefont {Haag},\ and\
  \citenamefont {Klay}}]{Cebra:2014sxa}%
  \BibitemOpen
  \bibfield  {author} {\bibinfo {author} {\bibfnamefont {D.}~\bibnamefont
  {Cebra}}, \bibinfo {author} {\bibfnamefont {S.~G.}\ \bibnamefont {Brovko}},
  \bibinfo {author} {\bibfnamefont {C.~E.}\ \bibnamefont {Flores}}, \bibinfo
  {author} {\bibfnamefont {B.~A.}\ \bibnamefont {Haag}}, \ and\ \bibinfo
  {author} {\bibfnamefont {J.~L.}\ \bibnamefont {Klay}},\ }\href@noop {} {\
  (\bibinfo {year} {2014})},\ \Eprint {http://arxiv.org/abs/1408.1369}
  {arXiv:1408.1369 [nucl-ex]} \BibitemShut {NoStop}%
\bibitem [{\citenamefont {Agakishiev}\ \emph {et~al.}(2009)\citenamefont
  {Agakishiev} \emph {et~al.}}]{HADES:2009aat}%
  \BibitemOpen
  \bibfield  {author} {\bibinfo {author} {\bibfnamefont {G.}~\bibnamefont
  {Agakishiev}} \emph {et~al.} (\bibinfo {collaboration} {HADES}),\ }\href
  {\doibase 10.1140/epja/i2009-10807-5} {\bibfield  {journal} {\bibinfo
  {journal} {Eur. Phys. J. A}\ }\textbf {\bibinfo {volume} {41}},\ \bibinfo
  {pages} {243} (\bibinfo {year} {2009})},\ \Eprint
  {http://arxiv.org/abs/0902.3478} {arXiv:0902.3478 [nucl-ex]} \BibitemShut
  {NoStop}%
\bibitem [{\citenamefont {Sorensen}(2021)}]{Sorensen:2021zxd}%
  \BibitemOpen
  \bibfield  {author} {\bibinfo {author} {\bibfnamefont {A.~M.}\ \bibnamefont
  {Sorensen}},\ }\emph {\bibinfo {title} {{Density Functional Equation of State
  and Its Application to the Phenomenology of Heavy-Ion Collisions}}},\
  \href@noop {} {Ph.D. thesis},\ \bibinfo  {school} {UCLA, Los Angeles (main),
  UCLA} (\bibinfo {year} {2021}),\ \Eprint {http://arxiv.org/abs/2109.08105}
  {arXiv:2109.08105 [nucl-th]} \BibitemShut {NoStop}%
\bibitem [{\citenamefont {Landau}\ and\ \citenamefont
  {Lifshitz}(1980)}]{Landau_Stat}%
  \BibitemOpen
  \bibfield  {author} {\bibinfo {author} {\bibfnamefont {L.~D.}\ \bibnamefont
  {Landau}}\ and\ \bibinfo {author} {\bibfnamefont {E.~M.}\ \bibnamefont
  {Lifshitz}},\ }\href@noop {} {\emph {\bibinfo {title} {Statistical Physics,
  Course of Theoretical Physics, Vol.V}}}\ (\bibinfo  {publisher} {Pergamon
  Press, New York},\ \bibinfo {year} {1980})\BibitemShut {NoStop}%
\bibitem [{\citenamefont {Baym}\ and\ \citenamefont
  {Chin}(1976)}]{Baym:1975va}%
  \BibitemOpen
  \bibfield  {author} {\bibinfo {author} {\bibfnamefont {G.}~\bibnamefont
  {Baym}}\ and\ \bibinfo {author} {\bibfnamefont {S.~A.}\ \bibnamefont
  {Chin}},\ }\href {\doibase 10.1016/0375-9474(76)90513-3} {\bibfield
  {journal} {\bibinfo  {journal} {Nucl. Phys. A}\ }\textbf {\bibinfo {volume}
  {262}},\ \bibinfo {pages} {527} (\bibinfo {year} {1976})}\BibitemShut
  {NoStop}%
\end{thebibliography}%

\appendix
\section{Definition of the asymmetry parameter~$\delta$}
\label{app:delta_definition}

For a hadronic system consisting of only protons and neutrons, the usual definition of the asymmetry parameter $\delta$ in terms of proton and neutron densities, $n_p$ and $n_n$, can be rewritten in terms of the isospin fraction $Y_I$ as
\begin{equation}
    \delta=\frac{n_n-n_p}{n_B}=-2 \left(\frac{\sum_i n_i I_i}{\sum_i n_i}\right)=-2Y_I\ ,
    \label{delta}
\end{equation}
where $n_B = n_p + n_n$, and $n_i$ and $I_i$ are number density and isospin of the $i$-th species, respectively. The parameter $\delta$ can also be easily described in terms of the charge fraction $Y_{Q,\rm{QCD}}$, given that for nonstrange matter (i.e., matter where $Y_S ={\sum_i n_i S_i}/{\big(\sum_i n_i\big)}=0$, where $S_i$ is the strangeness of particle species $i$) one has  $Y_I=Y_{Q,\rm{QCD}}-0.5$~\cite{Aryal:2020ocm}, where $Y_{Q,\rm{QCD}} ={\sum_i n_i Q_i}/{\sum_i n_i}$; as a result, one can write
\begin{equation}
    \delta=-2(Y_{Q,\rm{QCD}}-0.5)=1-2Y_{Q,QCD} \hspace{.5cm} (Y_S=0)\ .
\end{equation}

\section{Derivation of the proton fraction~$Y_p$}
\label{app:YQ}

We now derive the proton fraction $Y_p$ from Ref.~\cite{Mendes:2021tos}, following Ref.~\cite{Lattimer:2014sga}, but staying consistent with the notation used in this paper. At $T=0$, in $\beta$ equilibrium, and with charge neutrality we have
\begin{equation}
\frac{\partial\left(E_{\rm{QCD}}+E_{e}\right)}{\partial Y_p}=\mu_{p}-\mu_{n}+\mu_{e}-\left(m_{n}-m_{p}\right)=0\ ,
\label{eq:partial_Yp}
\end{equation}
where $E_{\rm{QCD}}$ and $E_{e}$ are the (kinetic) energy of nucleons and electrons, respectively, and $\mu_p$ and $\mu_n$ are proton and neutron chemical potentials measured with respect to their rest masses, $m_p$ and $m_n$. Neglecting the difference between the proton and neutron mass in Eq.~\eqref{eq:partial_Yp} leads to
\begin{align}
\mu_{e} =  - (\mu_{p}- \mu_{n} ) ~.
\label{eq:Yp_deriv0}
\end{align}
Since we have charge neutrality, it must be that $n_e = Y_p n_B$, and furthermore, in the ultra-relativistic approximation, we have 
\begin{align}
\mu_e \approx k_{F,e} = (3 \pi^2 n_B Y_p)^{1/3}   ~.
\label{eq:Yp_deriv1}
\end{align}
Taking the derivative of Eq.~\eqref{eqn:symExpan} (up to quadratic terms), combined with Eq.~\eqref{eqn:delta}, gives
\begin{equation}
\frac{\partial E_{\rm{QCD}}}{\partial Y_p}= -4E_{\rm{sym}}\big(1-2 Y_p\big) ~ ,
\end{equation}
which, again neglecting the difference between the proton and neutron mass, so that $\inparr{E_{\rm{QCD}}}{Y_p} \approx \mu_p - \mu_n$, leads to
\begin{eqnarray}
\mu_p - \mu_n \approx -4E_{\rm{sym}}\big(1-2 Y_p\big) ~.
\label{eq:Yp_deriv2}
\end{eqnarray}

Put together, Eqs.\ \eqref{eq:Yp_deriv0}, \eqref{eq:Yp_deriv1}, and \eqref{eq:Yp_deriv2} yield
\begin{align}
 (3 \pi^2 n_B Y_p)^{1/3}  =  4E_{\rm{sym}}\big(1-2 Y_p\big)  ~.
\end{align}
We can then solve for $Y_{p}$ exactly and take the only real solution,
\begin{equation}
    Y_{p}=\frac{1}{16}\left[8-\frac{\pi^{4/3}n_{B}}{2^{1/3}X}+\left(\frac{\pi}{2}\right)^{2/3}\frac{X}{E_{\rm{sym}}^3}\right]~,
\end{equation} 
where $X$ is given by
\begin{equation}
    X=
(-24 E_{\rm{sym}}^6 n_B+\sqrt{2} \sqrt{288 E_{\rm{sym}}^{12} n_B^2+\pi^2 E_{\rm{sym}}^9 n_B^3})^{1 / 3}~.
\end{equation}
As a reference, we also show the two complex solutions here:
\begin{equation}
    Y_{p,\pm}=\frac{1}{2}+\frac{1\pm i \sqrt{3}\pi^{4/3}n_{B}}{32\times 2^{1/3}X}-\frac{(1\pm i\sqrt{3})\pi^{2/3}X
    }{32\times 2^{2/3}E_{\rm{sym}}^{3}}~.
\end{equation}

\begin{figure}
\centering
\includegraphics[width=\linewidth]{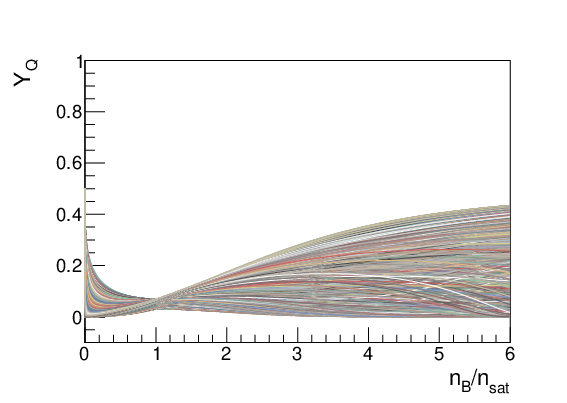} 
\caption{Different values of $Y_{Q}$ as a function of $n_B$ obtained using different symmetry energy coefficients.
}
\label{fig:YQ}
\end{figure}
In Fig.~\ref{fig:YQ}, we show the band of possible values of $Y_{Q,\rm{QCD}}$ as a function of density $n_B$ as obtained for all values of the symmetry energy expansion coefficients considered in this work, see Tab.~\ref{tab:symmetry_energy_expansion_coefficients_ranges}. Generally, we anticipate an increase in $Y_{Q,\rm{QCD}}$ at $n_B\approx n_{\txt{sat}}$, followed by a decrease at large densities if a phase transition to deconfined quark matter takes place.

\section{Symmetry expansion of eos1 and eos3}\label{sec:SymEn_13}              

\begin{figure}
\centering
\includegraphics[width=\linewidth]{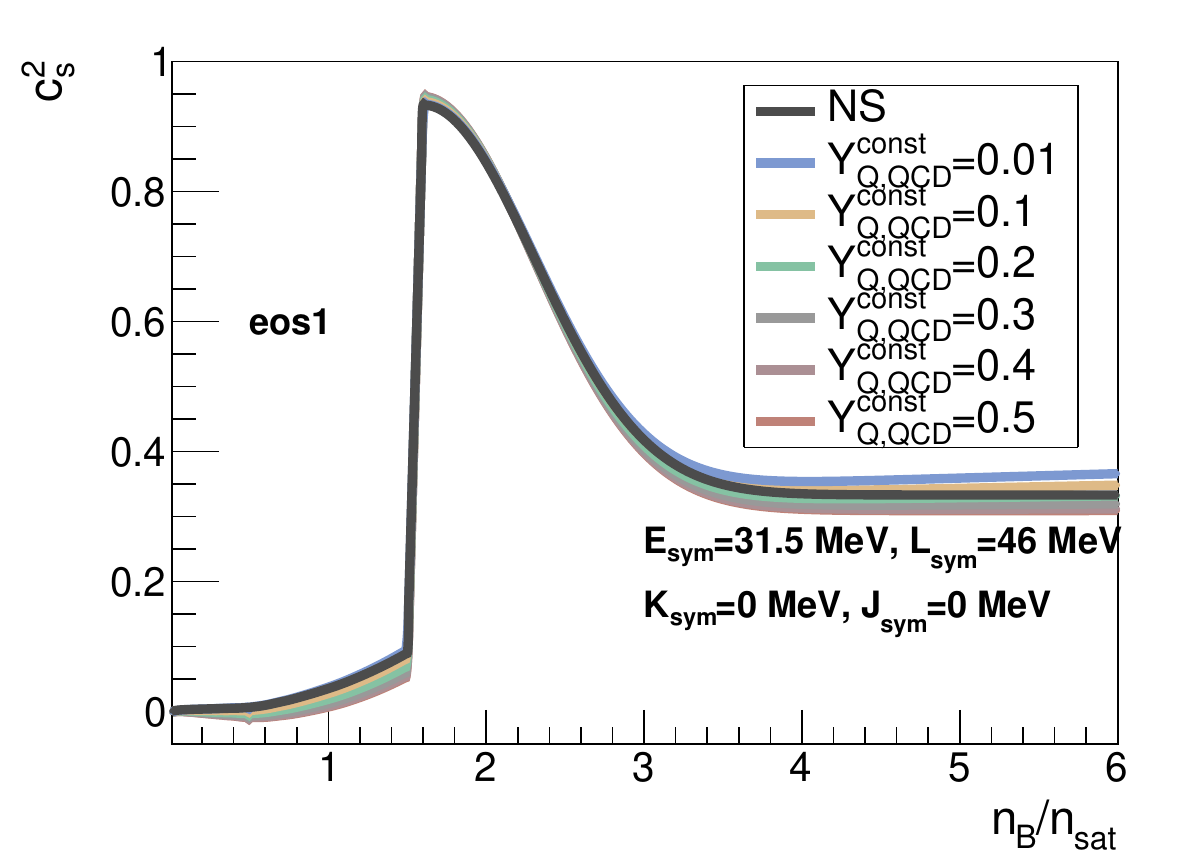}  \\  \includegraphics[width=\linewidth]{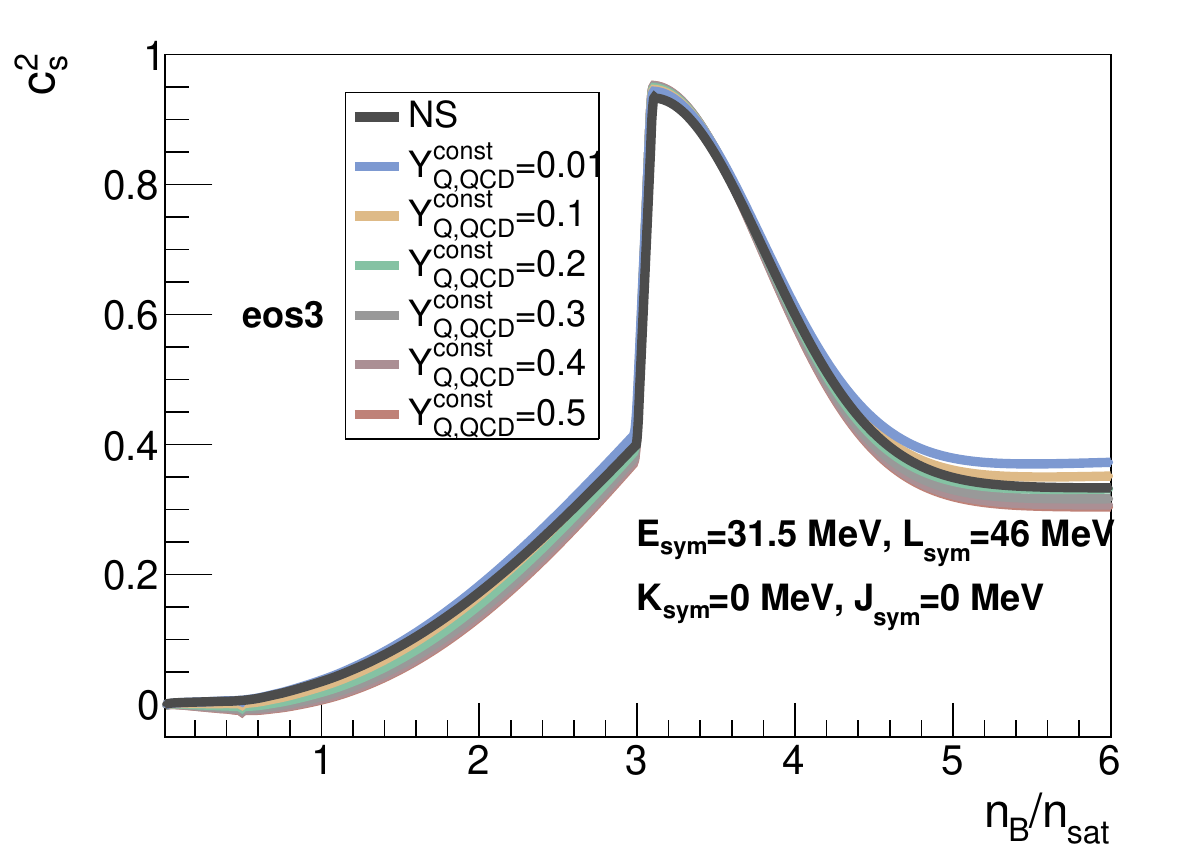}
\caption{\textit{Top}: $c^{2}_{s}$ dependence on $Y_{Q,\rm{QCD}}^{\txt{const}}$ for eos1 (peak in $c_s^2$ at $n_B = 1.5n_{\rm{sat}}$). \textit{Bottom}: $c^{2}_{s}$ dependence on $Y_{Q,\rm{QCD}}^{\txt{const}}$ for eos3 (peak in $c_s^2$ at $n_B = 3n_{\rm{sat}}$). For EOSs in both panels, the central values of $E_{\txt{sym,sat}}$ and $L_{\txt{sym,sat}}$ from~\cite{Li:2019xxz} were used.}
\label{fig:YQexpan_Appen}
\end{figure}

In the main text, we described in detail the steps to apply the symmetry energy expansion to eos2. Here, we apply the same method to eos1 and eos3. In Fig.~\ref{fig:YQexpan_Appen}, we compare different $Y_{Q,\rm{QCD}}^{\txt{const}}$ slices for the same fixed set of the symmetry energy expansion coefficients, similarly as previously shown in 
Fig.~\ref{fig:YQexpan} for eos2. Interestingly, we find that the location of the peak in $c_s^2$ for eos1 and eos3 does not change at all when the symmetry energy expansion is applied, while the height of both peaks increases as matter becomes more symmetric. Qualitatively, we find very similar behavior as in Fig.~\ref{fig:YQexpan}, i.e., that the low $n_B$ region has a smaller effect from the change in $Y_{Q,\rm{QCD}}^{\txt{const}}$. For eos3, we see more of a dependence on $Y_{Q,\rm{QCD}}^{\txt{const}}$ before the peak, which is due to the fact that the peak occurs at a large $n_B$.

\begin{figure*} 
\centering
\begin{tabular}{cc}
\includegraphics[width=0.5\linewidth]{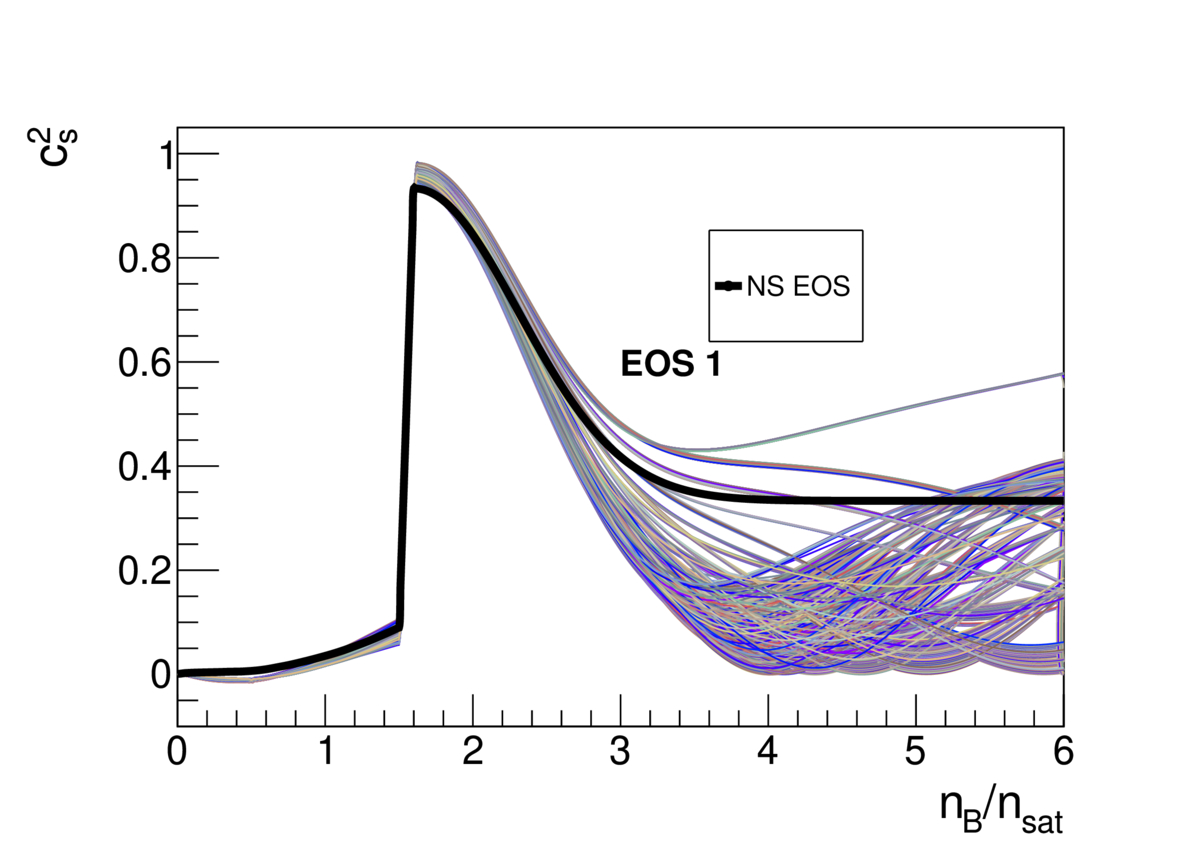}  &  \includegraphics[width=0.5\linewidth]{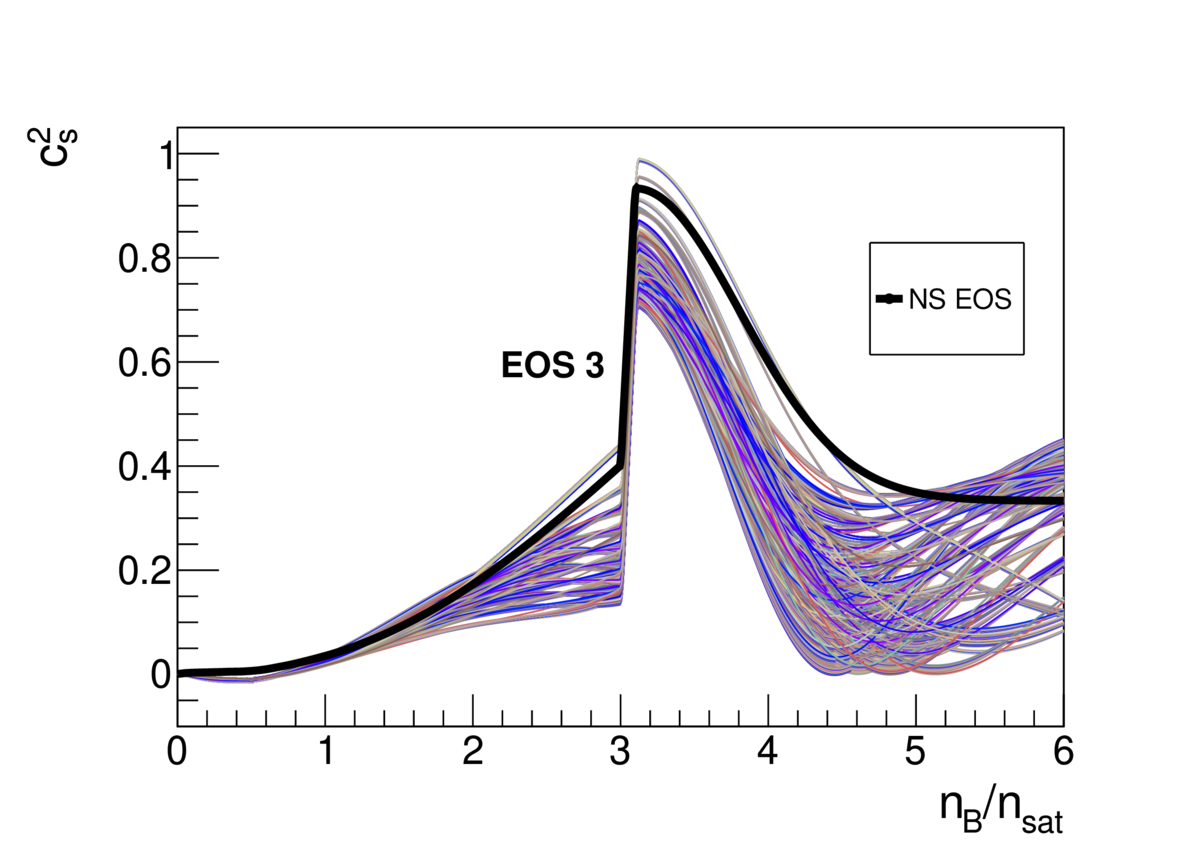}\\
\includegraphics[width=0.5\linewidth]{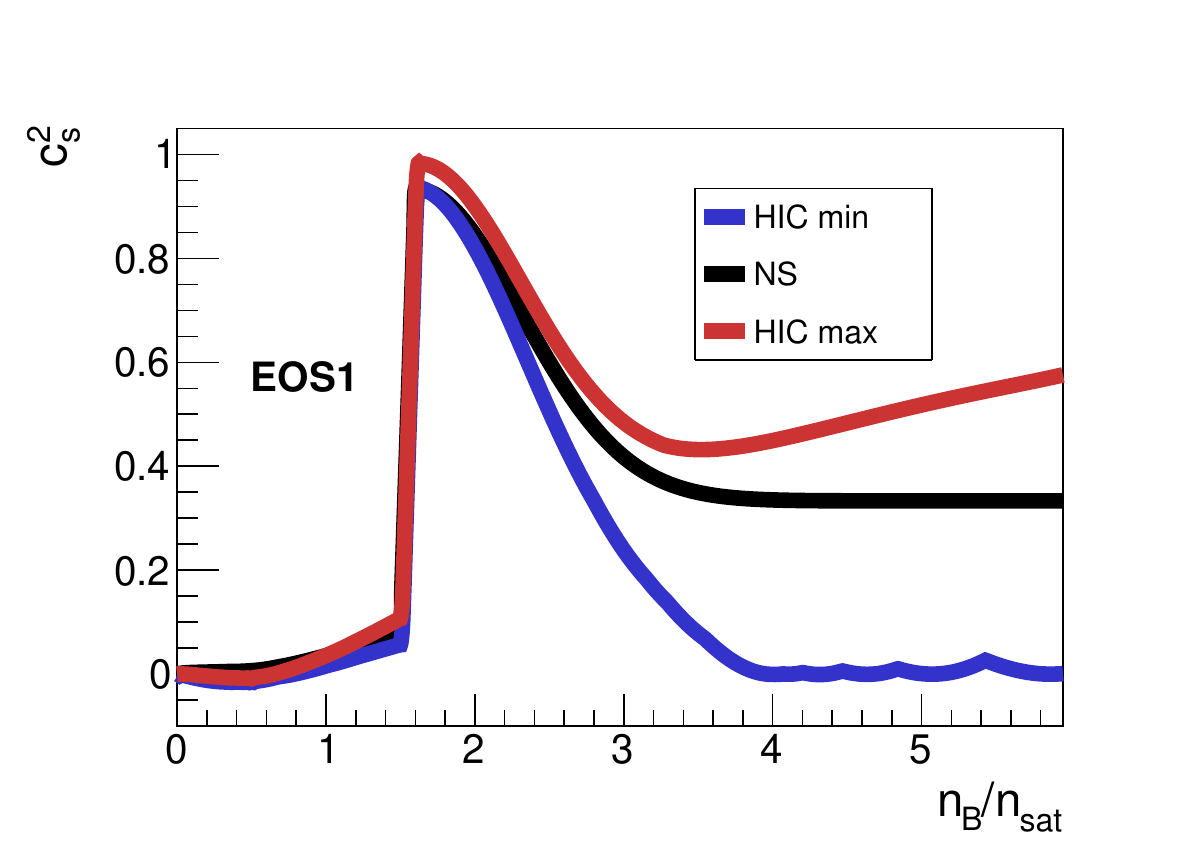}  &  \includegraphics[width=0.5\linewidth]{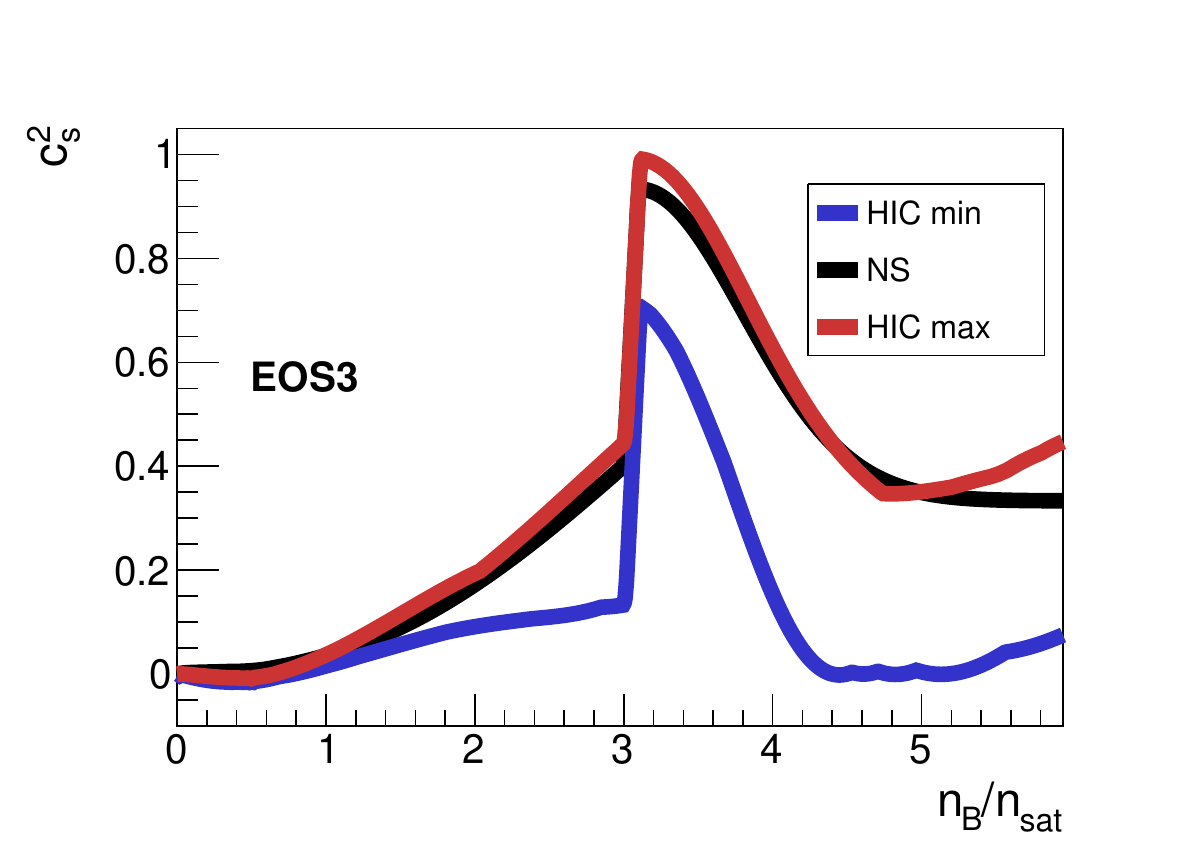}
\end{tabular}
\caption{$c_{s}^{2}$ as a function of $n_B$ for HIC EOSs obtained using the symmetry energy expansion with varying coefficients. The left column shows the results for the low-density peak (eos1) and the right column shows the results for the high-density peak (eos3). The top row shows all EOSs obtained for specific combinations of the symmetry energy expansion coefficients (after applying the causality and stability constraints), while the bottom row shows the extracted minimum and maximum envelopes.}
\label{fig:dens} 
\end{figure*}

Next, in Fig.~\ref{fig:dens} we show $c^{2}_{s}$ calculated from the converted HIC EOSs ($Y_{Q,\rm{QCD}}^{\rm{const}}=0.5$) for eos1 (left panels) and eos3 (right panels). For both eos1 and eos2, the location of the peak in $c_s^2$ stays the same after the expansion to SNM and the magnitude is only slightly dependent on the coefficients. For eos3, since the peak is at a larger~$n_B$, the wide (i.e., poorly constrained) ranges of the allowed higher order symmetry energy expansion coefficients lead to the result that the magnitude of the $c_s^2$ peak in the HIC EOS can either become significantly larger or smaller than in the corresponding NS EOS. The same wide ranges of $K_{\rm{sym,sat}}$ and $J_{\rm{sym,sat}}$ are also responsible for the fact that for all considered NS EOSs, there are wide variations in the converted HIC EOSs for densities larger than the one at which the peak occurs. The main takeaway here is that peaks in $c_s^2$ that appear at large densities may be more strongly affected by the high-density behavior of the symmetry energy expansion, including the possibility that they may be somewhat washed out.

\begin{figure*}
    \centering
    \includegraphics[width=0.8\linewidth]{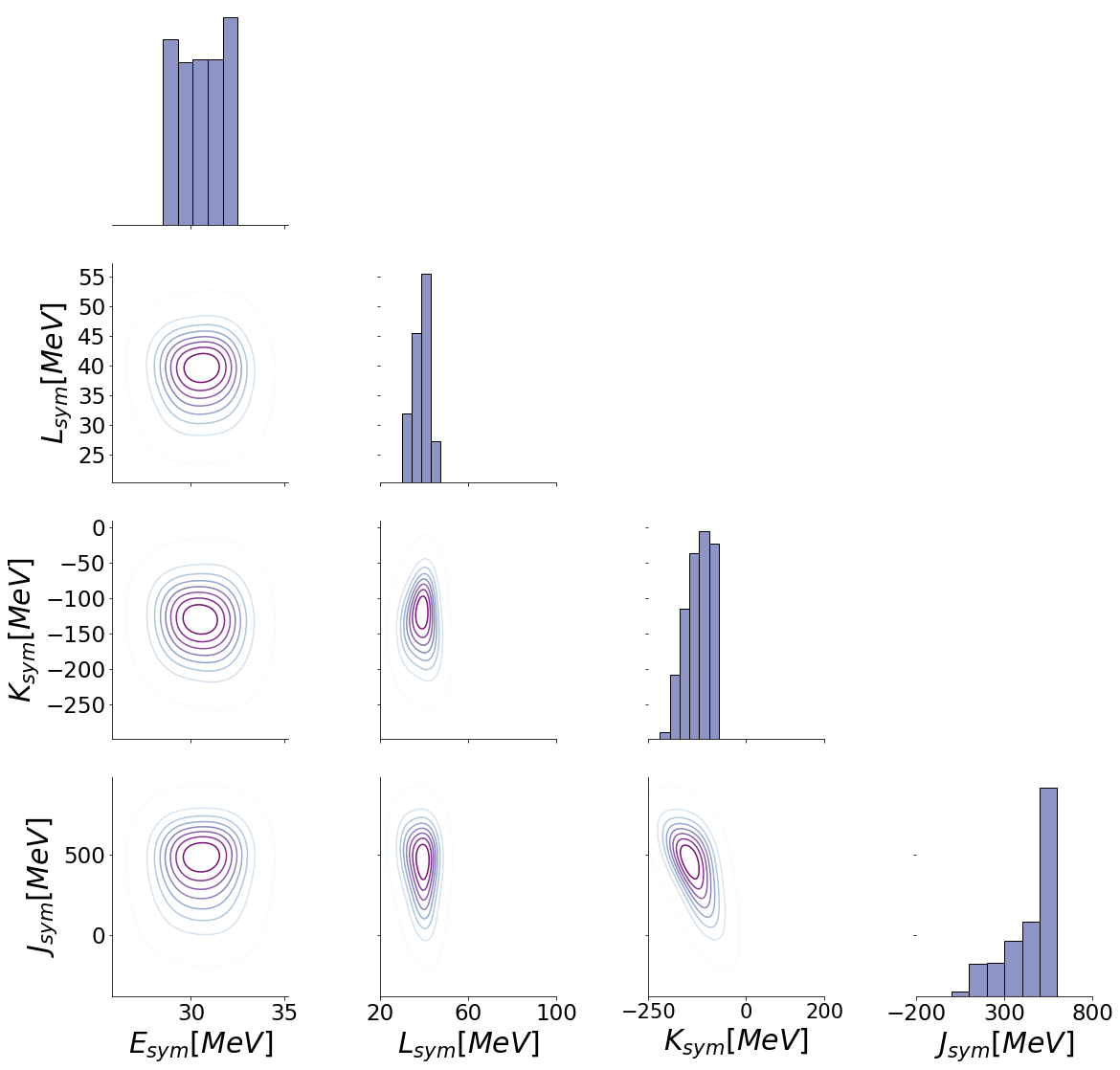}
    \caption{Same as Fig.~\ref{fig:corner_med}, but for eos1.}
    \label{fig:corner_low}
\end{figure*}

\begin{figure*}
    \centering
    \includegraphics[width=0.8\linewidth]{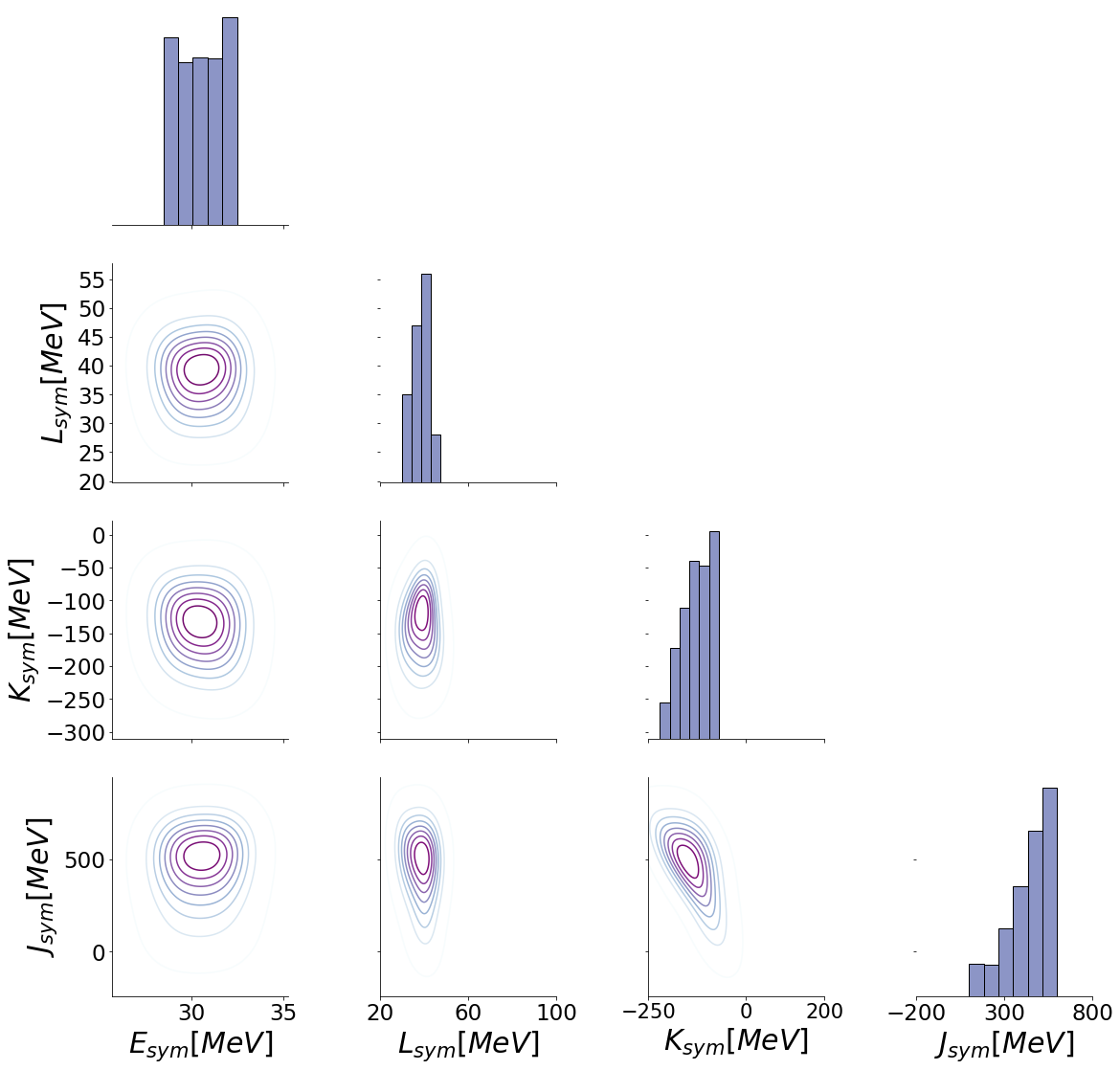}
    \caption{Same as Fig.~\ref{fig:corner_med}, but for eos3.}
    \label{fig:corner_high}
\end{figure*}

Next, in Figs.~\ref{fig:corner_low} and \ref{fig:corner_high} we show corner plots for eos1 and eos3, respectively. We find similar results for the symmetry energy expansion coefficients as we did for eos2, see Fig.~\ref{fig:corner_med}. For instance, $E_{\txt{sym},\rm{sat}}$ has a fairly flat posterior with a possible slight preference for small values, and there is a preference for a small $L_{\txt{sym},\rm{sat}}$, regardless of the NS EOS considered. Similarly as before, $K_{\txt{sym,sat}}$ appears to prefer values $\approx - 100~\rm{MeV}$ and $J_{\txt{sym,sat}}$ demonstrates a preference for large positive values.

\begin{figure}
    \centering
    \includegraphics[width=0.99\linewidth]{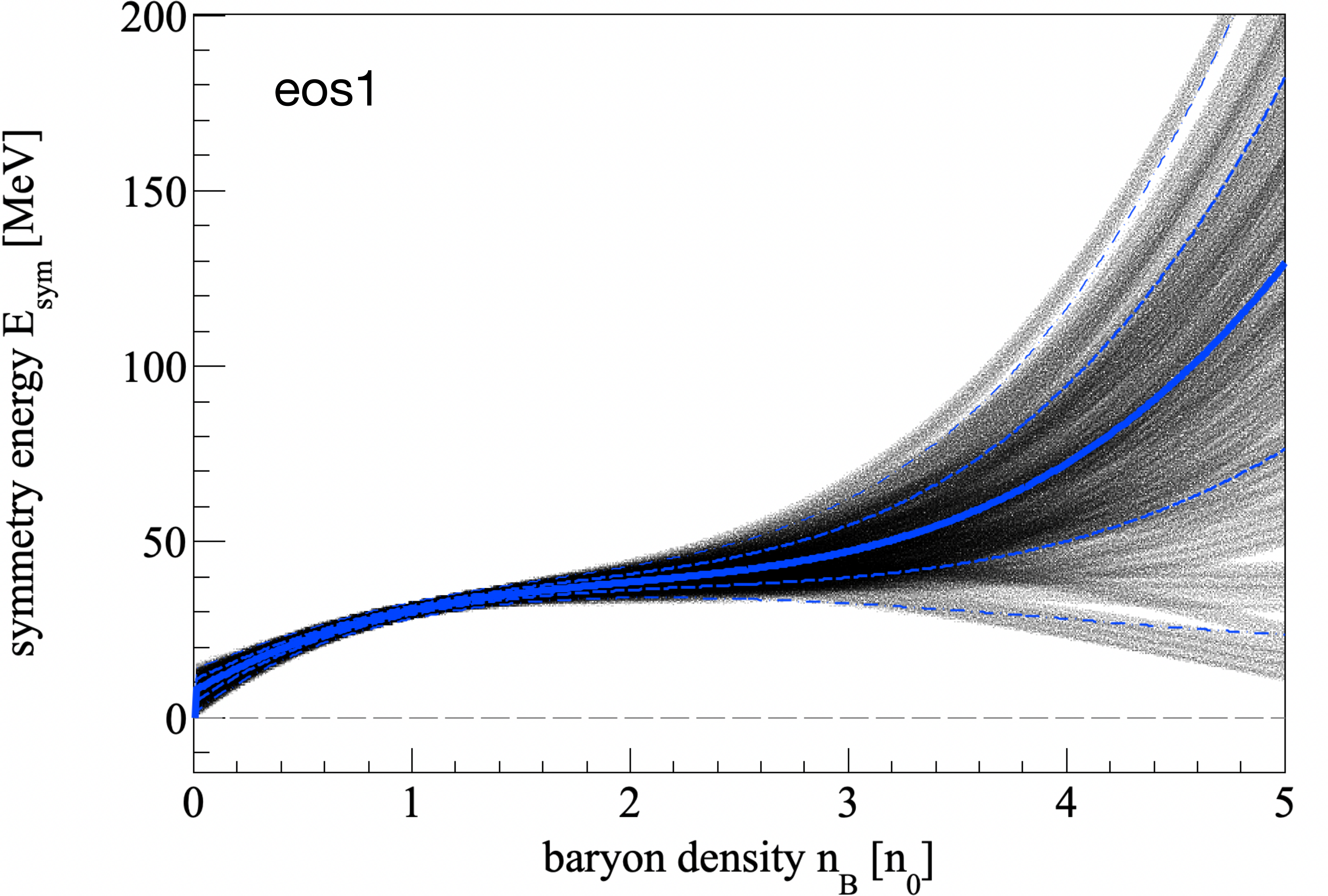} 
    \includegraphics[width=0.99\linewidth]{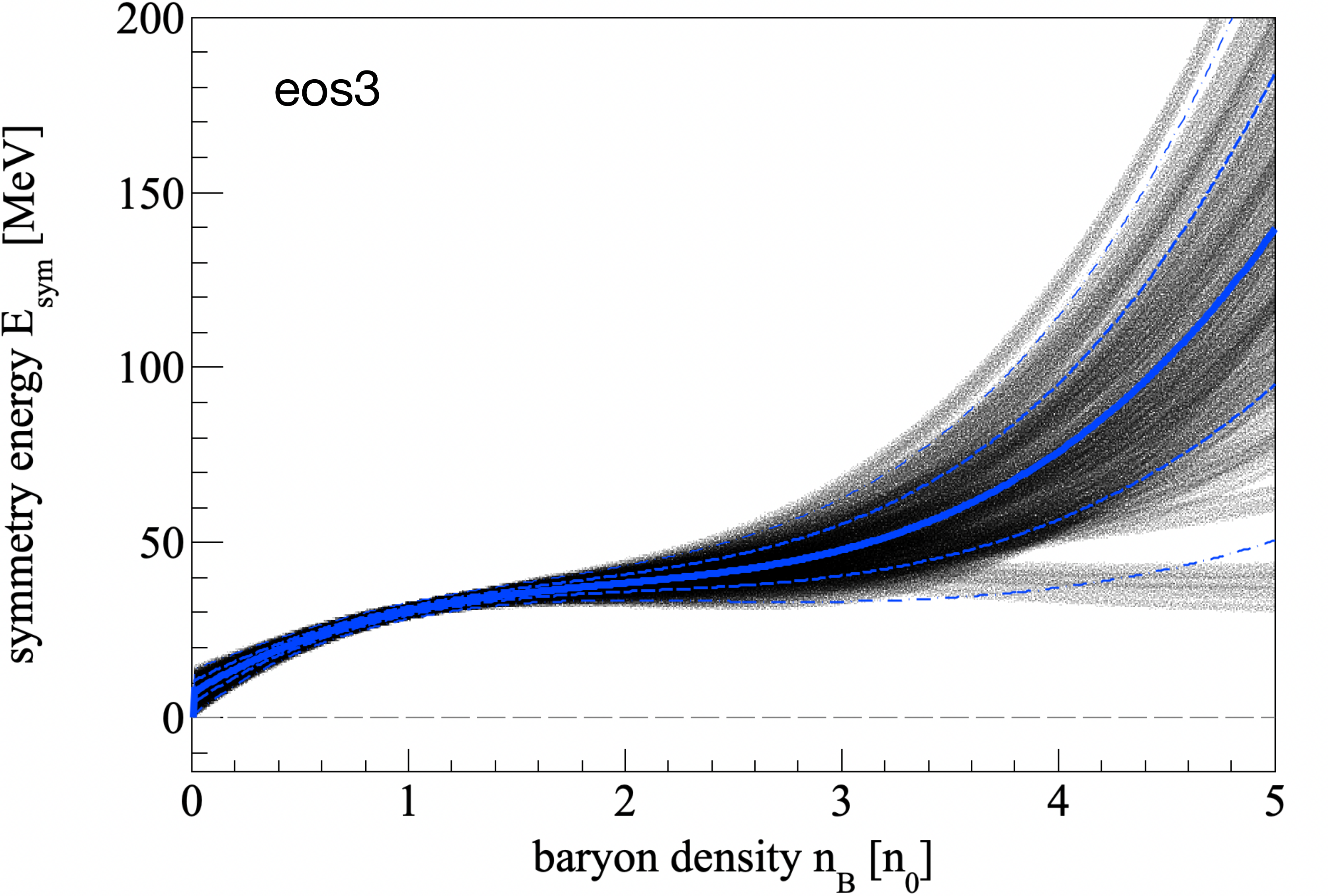} 
    \caption{Same as Fig.~\ref{fig:symmetry_energy_scatter_eos2}, but for eos1 (\textit{top}) and eos3 (\textit{bottom}).
    }    \label{fig:symmetry_energy_scatter_eos1_and_3}
\end{figure}

Finally, we also show scatter plots of the symmetry energy as a function of density, based on all accepted symmetry energy expansion coefficients, for eos1 and eos3 in Fig.~\ref{fig:symmetry_energy_scatter_eos1_and_3}. Similarly as in the case of eos2, see Fig.~\ref{fig:symmetry_energy_scatter_eos2}, we see a tight constraint around $n_{\rm{sat}}$, reflecting the tight constraint on the lowest two coefficients of the symmetry energy, followed by a soft rise of the symmetry energy at moderate densities which is a direct consequence of the simultaneous preference for negative values of $K_{\rm{sym, sat}}$ and positive values of $J_{\rm{sym, sat}}$ or, equivalently, a consequence of the causality and stability constraints.

\section{Temperature dependence of the EOS in transport simulations}
\label{sec:T-dependence}    

While within our formalism the EOS is determined for nucleons at $T = 0$, it still displays nontrivial behavior as a function of $T$ and its implementation in hadronic transport is appropriate for describing complex systems of many baryonic species that inevitably arise in considerations at finite temperature as well as in heavy-ion collisions.

To see that this is the case, let us first consider a simple model of nuclear matter in which nucleons are relativistic fermions interacting through a simple vector-density--dependent potential (for example of the Skyrme type). At $T=0$, the parameters of the potential are fixed so that the model describes nuclear matter with its usual relatively well-established properties: saturation density, binding energy, and incompressibility. The fact that nuclear matter is in equilibrium at a finite density $n_{\rm{sat}}$ means that at $n_B = n_{\rm{sat}}$, $P=0$. Moreover, $dP/dn_B > 0$ for $n_B = n_{\rm{sat}}$ (because incompressibility is positive), and we also naturally have $P (T = 0, n_B = 0) = 0$. This means that at $T=0$, there is a region between $n_B = 0$ and $n_B = n_{\rm{sat}}$ where $P < 0$, and, consequently, there is also a region where $dP/dn_B <0$, that is a region where the matter is mechanically unstable. In other words, the considered model describes matter with a first-order phase transition.

The behavior of this simple model at $T=0$ is almost entirely driven by the interaction terms, which do not change as a function of temperature since the density does not depend on temperature. In this situation, the only term that changes with temperature is the kinetic contribution to the pressure. However, this is enough to produce nontrivial features: As the temperature is increased, the momenta of particles increase and so does the kinetic contribution to the pressure. At some point, for high enough temperatures, this kinetic contribution overwhelms the contribution from the interaction terms in such a way that not only there isn't any region where $P<0$, but also there isn't any region where $dP/dn_B < 0$. The temperature at which this happens is the critical temperature, while critical density is the density at which the region with $dP/dn_B < 0$ becomes, with increasing temperature, a point where $dP/dn_B = 0$ (for a more extensive discussion of features of first-order phase transitions the Reader may further consult Appendix~B of Ref.~\cite{Sorensen:2021zxd} or a standard resource such as Ref.~\cite{Landau_Stat}). 

All of the above features are preserved in transport simulations. Our model for the EOS has density-dependent terms, which in particular lead to density-dependent terms in single-particle energies of particles carrying baryon number. In the simulations, baryon density can be calculated simply by summing contributions from all baryons in the vicinity of a chosen point, which includes baryons other than nucleons produced in the collisions through scatterings and decays. The equations of motion for baryons are based on Hamilton's equations of motion, that is include gradients of the single-particle energy. The single-particle energy, which can be directly derived from the EOS of the system~\cite{Baym:1975va,Sorensen:2020ygf}, is comprised out of two terms: the kinetic term and the interaction term. Gradients of the interaction term carry information about the influence of density-dependent interactions. Gradients of the kinetic term are proportional to the particle velocity, which reflects the amount of energy that the particles have. In equilibrium, this energy can be explicitly connected to the temperature of the system. \textit{Consequently, since the evolution of the particles involves both their velocities and gradients of mean-field interactions (where the former, in equilibrium, can be directly connected to temperature), transport simulations can exactly reproduce the nontrivial thermodynamic behavior of the underlying EOS with density-dependent interactions even though there are no explicit $T$-dependent terms in the single-particle energy.} All that is needed to describe systems at vastly different temperatures (if equilibrium is available so that temperature can be well-defined) is differences in the kinetic momenta of particles. These differences can stem, for example, from different energies of the collisions or simply from considering different stages of collisions at a given energy. Since the interactions enter microscopic transport on the level of single-particle energies (as opposed to on the level of the EOS as in, e.g., hydrodynamic simulations), the effects of the interactions both extend to and are well-defined in situations away from equilibrium.

Overall, the used simulation framework takes into account all nontrivial changes with temperature that can be described within a model employing interactions dependent on density. Extensions of this approach applicable to transport simulations include using interactions of scalar type in which the momentum distribution (in equilibrium directly related to the temperature) affects the effective masses of particles and thus introduces nontrivial changes also in the interaction terms. Such changes, in particular, could lead to a more complex behavior of the EOS with temperature. For example, the speed of sound that does not have a minimum for a given $n_B$ at $T=0$ could possibly develop such a minimum at the same $n_B$ for $T>0$. However, here of note is the \textit{significantly} higher, almost prohibitive, computational cost of simulations using interactions beyond vector (density) type. Studies such as the one pursued in this work will become much harder to perform once such developments are included. Nevertheless, efforts in this direction, aimed in particular at enabling comprehensive Bayesian analyses, are ongoing~\cite{Sorensen:2021zxd}.

\end{document}